\documentclass[12pt,onecolumn,draftcls]{IEEEtran}
\usepackage[utf8]{inputenc}

\usepackage{cite}
\usepackage{amsfonts, amsmath, amssymb}
\usepackage{indentfirst}
\usepackage{verbatim}
\usepackage[pdftex]{graphicx}
\usepackage{caption}
\usepackage{subcaption}
\usepackage{epstopdf}
\usepackage{algorithm}
\usepackage{algorithmic}
\usepackage{soul}

\usepackage[normalem]{ulem}
\usepackage[usenames,dvipsnames]{color}
\usepackage{amsfonts}
\newcommand{\rd}{\textcolor{Red}}

\usepackage[nomarkers,figuresonly,nofiglist]{endfloat}
\usepackage{morefloats}
\usepackage{float}
\usepackage[pdfauthor={Qiwei Sheng, Kun Wang, Jun Xia, Liren Zhu, Lihong V. Wang and Mark A. Anastasio},pdftitle={A constrained variable projection reconstruction method for photoacoustic computed tomography without accurate knowledge of transducer responses},pdftex,unicode=true,colorlinks,linkcolor=blue]{hyperref}

\DeclareMathOperator*{\argmin}{argmin}

\DeclareMathOperator{\sinc}{sinc}
\title{A constrained variable projection reconstruction method for photoacoustic computed tomography {without accurate knowledge of} transducer responses}
\author{Qiwei Sheng, Kun~Wang,~\IEEEmembership{Member,~IEEE,} Thomas P. Matthews, Jun Xia,\\ Liren Zhu, Lihong~V.~Wang,~\IEEEmembership{Fellow,~IEEE,}
 \\ and~Mark~A.~Anastasio,~\IEEEmembership{Senior Member,~IEEE}
\thanks{
Department of Biomedical Engineering, Washington University in St. Louis,
St. Louis, MO 63130
}}

\begin{document}

\maketitle

\begin{abstract}

Photoacoustic computed tomography (PACT) is an  emerging computed imaging modality that exploits optical contrast and ultrasonic detection principles to form images of the absorbed optical energy density within tissue. 
 When the imaging system employs conventional piezoelectric ultrasonic transducers, the
ideal photoacoustic (PA) signals are degraded by the transducers'
 acousto-electric impulse responses (EIRs) during the measurement process.
 If unaccounted for, 
this can degrade the accuracy of the reconstructed image.
  In principle, the effect of the EIRs on the measured PA signals can be {ameliorated} 
via deconvolution; images can be reconstructed subsequently by application of a reconstruction method that assumes an idealized EIR.  Alternatively, the effect of the  EIR can be  incorporated into {an imaging} model and implicitly compensated for during reconstruction. In either case, 
{the efficacy of the correction can be limited by errors in the assumed EIRs}. In this work, 
 a joint optimization approach to PACT image reconstruction is proposed for mitigating errors in reconstructed images that are caused by use of an inaccurate EIR.  The method exploits the bi-linear nature of the imaging model and seeks to refine the measured  EIR during the process of reconstructing the sought-after absorbed optical energy density. 
Computer-simulation and experimental studies are conducted to investigate the numerical properties of the method and demonstrate its value for mitigating image distortions and enhancing the visibility of fine structures.


\end{abstract}

\section{Introduction}
Photoacoustic computed tomography (PACT), also known as thermoacoustic or optoacoustic tomography, is an emerging imaging modality that holds great promise for biomedical imaging \cite{OK2003, Wang2008,KRK1999,CAKB2006,MarkBookChapter}. It is a hybrid modality that exploits the high optical contrast of soft tissue and the high spatial resolution of ultrasonic methods. In PACT, short laser pulses (typically nanosecond-duration) are employed to illuminate tissue. Absorption of the optical energy results in local heating followed by thermal expansion, which generates internal broadband photoacoustic (PA) wavefields via the photoacoustic effect \cite{Wang2008,OK2003}. From measurements of the PA wavefields acquired outside the object,  an image reconstruction method can be employed to estimate the spatially variant absorbed optical energy density within the tissue.

When the imaging system employs piezoelectric transducers,
 the PA signals at the transducer locations
 are convolved with the  transducers' acousto-electric impulse responses (EIRs) during the measurement process.  If unaccounted for, this degradation of the measurement data will result in a modulation of the spatial frequency components of the estimated absorbed optical energy density distribution \cite{FourierShell:07}.  In principle, the effect of the EIRs on the {PA} signals can be removed via deconvolution
 {if the EIR is accurately known}; 
{subsequently,} images can be reconstructed by application of a reconstruction {method
that neglects the EIR.}  Alternatively, the effect of the  EIR can be  incorporated into {an imaging} model and compensated for {implicitly} during image reconstruction \cite{WESB2011,RNVR2011,KRAM2012}.

Unlike the spatial impulse response (SIR) of a transducer, which can be described accurately by use of a relatively simple physics-based model \cite{S1971, H1981}, a transducer's EIR poses challenges to an analytical description \cite{RNVR2011,KKZP2004}. Various theoretical models \cite{LKM1971,RNVR2011,KKZP2004} have been proposed for describing a transducer's EIR. The parameters employed in such models, however, are either difficult or even impossible to  measure accurately in practice. Consequently,
 in applications of PACT,  it is common for the EIR to be measured experimentally.

Although a conceptually simple task, measurement of the EIR is subject to noise and other errors \cite{RNR2011}, which can limit image quality in PACT. Several techniques for measuring the EIR have been developed \cite{conjusteau2009measurement,GAMH2008,CRBR2013,CESS2009}. It was suggested \cite{GAMH2008} that the impulse response could be measured by illuminating the transducer with an ultra-short laser pulse. However, the impulse response measured in this way represents the convolution of the photoacoustic pressure produced by parasitic sources on the surface of the transducer and the sought-after EIR \cite{CRBR2013}. Alternatively, the derivative of the EIR can be estimated by measuring the signal produced by optically illuminating an
absorber that is small {relative to the acoustic wavelength}. In practice, signals produced
by  small absorbers can be weak \cite{RNR2011} and errors in their low frequencies can be amplified if the signals are integrated to estimate the EIR. Recently, an alternate method to estimate the EIR was proposed to circumvent this \cite{RNR2011,CRBR2013}. All of these methods require precise alignment of the acoustic source with respect to the transducer axis.  When focused transducers are employed, the acoustic source must be aligned at the focal point. Misalignment of the acoustic source can result in errors in the measured EIR.  In effect, the measured EIR can be contaminated by the SIR. For characterizing the spectral directivity of flat transducers, an optoacoustic source that produces quasi-plane waves was produced \cite{conjusteau2009measurement}; however, it cannot be readily utilized to characterize the EIR of focused transducers. Finally, when transducer arrays are purchased, although they may differ, the EIRs of individual elements are not typically provided, and it can be an arduous task to characterize each EIR.

In this work, a joint optimization approach to PACT image reconstruction is developed for mitigating errors in reconstructed images that are caused by use of an inaccurate EIR. To accomplish this, a variable projection method \cite{CN2010,AL2012,GP1973} is employed to refine the measured EIR during the process of reconstructing the sought-after absorbed optical energy density distribution. This method exploits the separable nature of the PACT imaging model.
{When an array of transducers is employed that is characterized by a collection
of EIRs, the reconstruction method will determine a single effective EIR. 
Similarly, if other modeling errors are present, the response function produced by the method
can be interpreted as an effective system response that minimizes the inconsistency
between the measured data and the imaging model.}
 Computer-simulation and experimental studies are conducted to investigate the numerical properties of the method and demonstrate its value for mitigating image distortions and enhancing the visibility of fine structures.

The remainder of the article is organized as follows. In section \ref{Sect:background},  the relevant physics and PACT imaging model are reviewed. The proposed image reconstruction method is described in Section \ref{Sect:ImageRecon}. The numerical studies and results are presented in Sections \ref{Sect:ComputerStudies}--\ref{Sect:Exp}. Finally, a summary of the work is provided in Section \ref{Sect:Summary}.

\section{Background}\label{Sect:background}

Below we review the basic imaging physics and discrete PACT imaging model. The reader is referred to \cite{OK2003, Wang2008, XW2006, Wang2009,MarkBookChapter} for comprehensive reviews of PACT.

\subsection{Canonical imaging model in continuous form}
In PACT, a short laser pulse is employed to irradiate an object at time $t=0$
and an internal pressure wavefield $p(\mathbf{r},t)$
is established according to the photoacoustic (PA) effect. Here, $\mathbf{r}\in \mathbb{R}^3$ and $t\in [0,\infty)$.
In this work, the to-be-imaged object and surrounding medium are
 assumed to have homogeneous and lossless acoustic properties.
 Additionally, the width of the laser pulse is assumed to be negligible.
Under these assumptions, the PA wavefield at a location  $\mathbf{r}_0\in \Omega_0$, where $\Omega_0 \subset \mathbb{R}^{3}$ is
the measurement aperture, satisfies
\begin{equation}\label{wave_eq_sol}
p(\mathbf{r}_0,t)=\frac{\beta}{4\pi C_p}\int_V A(\mathbf{r})\frac{d}{dt}\frac{\delta(t-\frac{|\mathbf{r}_0-\mathbf{r}|}{c_0})}{|\mathbf{r}_0-\mathbf{r}|} \,d\mathbf{r}.
\end{equation}
Here,  $A(\mathbf{r})$ is a  compactly supported and bounded function, referred to as the object function,  which represents the absorbed optical energy density. 
The quantity $c_0$ denotes the (constant) speed-of-sound (SOS) in the object and the background medium; $\beta$ and $C_p$ denote the thermal coefficient of volume expansion and the specific heat capacity of the medium at constant pressure, respectively; and $V$ denotes the object's support volume.

Equation \eqref{wave_eq_sol}, which neglects the response of the imaging system as well as other physical factors \cite{MarkBookChapter}, represents an idealized imaging model for PACT in its continuous form. The associated image reconstruction problem is to determine an estimate of $A(\mathbf{r})$ from knowledge of $p(\mathbf{r}_0,t)$.

\subsection{Discrete imaging models that include transducer responses}

When piezoelectric ultrasonic transducers are employed, the photoacoustic signal $p(\mathbf r_0,t)$ is converted to a voltage signal that is subsequently sampled. Consider the case in which the transducers collect data at $Q$ locations, specified by the index $q=0,1,\cdots,Q-1$, and at each location $S$ temporal samples are acquired, specified by the index $s=0,1,\cdots,S-1$. The data are acquired at each location with a sampling interval $\Delta T$. The vector $\mathbf{u}\in \mathbb{R}^M$ represents a lexicographically ordered representation of the sampled voltage data, where $M=QS$. The notation $ [\mathbf{u}]_n$ will be employed to denote the $n$-th element of $\mathbf{u}$.

Under the same assumptions regarding the imaging physics that are required to establish Eqn.\ (\ref{wave_eq_sol}), the measured data vector $\mathbf u$ is related to $A(\mathbf r)$ as
\begin{equation}\label{eq_CD}
 [\mathbf{u}]_{qS+s}=u_q(t)|_{t=s\Delta T}=h^e(t)*_t \frac{1}{\Omega_q(\mathbf{r}_q)}\int_{\Omega_q(\mathbf{r}_q)} p(\mathbf{r}_0, t) \,d\mathbf{r}_0 \bigg|_{t=s\Delta T},
\end{equation}
where $u_q(t)$ is the pre-sampled electric voltage signal corresponding to the $q$-th transducer whose active area $\Omega_q(\mathbf{r}_q)$ is centered at $\mathbf{r}_{q}$, and $h^e(t)$ is the EIR. $p(\mathbf{r}_0,t)$ is given by Eqn.~\eqref{wave_eq_sol}. The notation $*_t$ denotes a 1-dimensional (1D)  temporal convolution. 
Equation (\ref{eq_CD}) represents a continuous-to-discrete (C-D) imaging model for PACT.
 {Note that Eqn.~\eqref{eq_CD} assumes that no acoustic lenses are attached to the piezoelectric surfaces of the transducers.} 

{When point-like transducers are employed}, Eqn.~\eqref{eq_CD} {degenerates} to
\begin{equation}\label{eq1}
 [\mathbf{u}]_{qS+s}=h^e(t)*_t p(\mathbf{r}_q,t)\Big|_{t=s\Delta T},
\end{equation}
where $p(\mathbf{r}_q,t)$ is specified by Eqn.~\eqref{wave_eq_sol}.

In order to formulate the image reconstruction task as a numerical optimization problem, the C-D imaging model in Eqn.\ (\ref{eq_CD})
 is typically approximated in practice by a discrete-to-discrete (D-D) imaging model \cite{wang2014discrete}. To establish a D-D imaging model, the object function $A(\mathbf r)$ can be approximated as
\begin{equation}\label{rep_A}
 A_a(\mathbf{r})= \sum^{N-1}_{n=0} [\boldsymbol{\theta}]_n\phi_n(\mathbf{r}),
\end{equation}
where the subscript $a$ indicates that $A_a(\mathbf{r})$ is an approximation of $A(\mathbf{r})$, $[\boldsymbol{\theta}]_n$ is the $n$-th component of the coefficient  vector $\boldsymbol{\theta}$, and $\{\phi_n(\mathbf{r})\}^{N-1}_{n=0}$ are expansion functions. In this work, interpolation-based expansion functions \cite{KRAM2012,wang2013accelerating} are employed.

On substitution from Eqn.\ (\ref{rep_A}) into Eqn.\ (\ref{eq_CD}), a D-D imaging model can be established as \cite{KRAM2012,MarkBookChapter}
\begin{equation}\label{DDD}
 \mathbf{u}=\mathbf{H}\boldsymbol{\theta},
\end{equation}
where the  $M\times N$ matrix is commonly known  as the system matrix.
The system matrix $\mathbf{H}$
depends on the   EIR, SIR, and the choice of expansion functions.
Specifically, the elements of $\mathbf{H}$ are a function  of the sampled EIR values,
which will be represented by the vector $\mathbf{h}\in \mathbb{R}^I$. Namely, $[\mathbf{h}]_i\equiv h^e(i\Delta T)$ for $ i=0,1,\cdots,I-1$,
where $I$ denotes the number of samples required to represent the EIR.
 In practice, $I\ll S$.
The explicit forms of  $\mathbf{H}$ that is employed in this study are provided in Appendix \ref{Sect:AppendixA}. 

 To emphasize the dependence of $\mathbf{H}$ on $\mathbf{h}$, the D-D imaging model will be expressed as
\begin{equation}\label{DD}
 \mathbf{u}=\mathbf{H}(\mathbf{h})\boldsymbol{\theta}.
\end{equation}
The accuracy of the system matrix will be degraded when the measured EIR contains errors.
When an inaccurate system matrix is employed in an iterative image reconstruction method, 
the resulting images can contain distortions and artifacts  \cite{QH2005}.
Below, we propose a method to circumvent this.

\section{PACT image reconstruction without accurate {knowledge of} transducer responses}\label{Sect:ImageRecon}
\subsection{Formulation of the image reconstruction problem}

We formulate image reconstruction as a numerical optimization problem
\begin{equation}\label{min_prob}
 (\boldsymbol{\hat{\theta}},\mathbf{\hat{h}})=
 \argmin_{\boldsymbol{\theta}\geq 0,\mathbf{h}}\varphi(\boldsymbol{\theta},\mathbf{h}),
\end{equation}
where the cost function $\varphi(\boldsymbol\theta, \mathbf h)$ is defined as
\begin{equation}\label{min_cost}
 \varphi(\boldsymbol{\theta},\mathbf{h})\equiv\|\mathbf{u}-\mathbf{H}(\mathbf{h})\boldsymbol{\theta}\|^2+\lambda R_1(\boldsymbol{\theta}) + \alpha R_2(\mathbf{h}).
\end{equation}
Here, $R_1(\boldsymbol{\theta})$ and $R_2(\mathbf{h})$ represent penalty terms, whose impacts are controlled by the regularization parameters $\lambda$ and $\alpha$, respectively.
The constraint $\boldsymbol{\theta}\geq 0$ in Eqn.\ (\ref{min_prob}) reflects that  $A(\mathbf{r})\ge 0$ and $\phi_n(\mathbf{r})\ge 0$ for
the interpolation-based expansion functions employed in this work.
 If the expansion functions are not non-negative, this constraint should not be enforced.

Equation \eqref{min_prob} is fundamentally different from the conventional formulation of PACT image reconstruction
\cite{KRAM2012,RNR2013} in that the EIR is treated as an unknown to be estimated along with the
approximation of $A(\mathbf{r})$. This provides the opportunity for the experimentally-measured EIR to be refined during image reconstruction. Since Eqn.~\eqref{min_cost} is non-convex, determining the solution to Eqn.~\eqref{min_prob} can present challenges.
As demonstrated below, the use of experimentally measured EIRs can provide relatively good initial estimates of  $\mathbf{h}$ 
 that will help the optimization algorithm {avoid} local minima. It is also important to properly {design the} penalties---$R_1(\boldsymbol\theta)$ and $R_2(\mathbf h)$---to regularize the solution.

\subsection{Variable projection method}
A variable projection (VP) method \cite{GV2003} 
is employed to reformulate the minimization problem given in Eqn.~\eqref{min_prob}.
This approach is motivated by previous studies in which the VP method was employed successfully to estimate
unknown parameters of a system matrix in separable inverse problems \cite{B1999,SG2013,CN2010}.


The VP method is based on the observation that
\begin{equation}\label{eqn:OptCond1}
  \mathbf {\hat h} = \arg\min_{\mathbf h} \varphi(\boldsymbol{\hat\theta}, \mathbf h),
\end{equation}
where $(\boldsymbol{\hat\theta},\mathbf {\hat h})$ is defined in Eqn.~\eqref{min_prob}.
 Inspired by this observation, $\mathbf h$ can be parameterized  as
\begin{equation}\label{eqn:ParaModel}
  \mathbf h^*(\boldsymbol\theta) = \arg\min_{\mathbf h}
     \varphi(\boldsymbol\theta, \mathbf h).
\end{equation}
By use of this parameterization, it can be verified \cite{BBHH2008, AL2012} that  $\hat{\boldsymbol\theta}$ 
can be computed as
\begin{equation}\label{eqn:SubProb2}
  \hat{\boldsymbol\theta} = \arg\min_{\boldsymbol\theta\geq 0} \varphi(\boldsymbol\theta,
     \mathbf h^*(\boldsymbol\theta)),
\end{equation}
and, subsequently, 
$\hat{\mathbf h}$ can be computed via Eqn.~\eqref{eqn:OptCond1}. In this way, the original optimization problem
in Eqn.~\eqref{min_prob} can be solved by consideration of the two subproblems in Eqns.~\eqref{eqn:OptCond1} and~\eqref{eqn:SubProb2}.

It will serve useful to note that the gradient of $\varphi(\boldsymbol\theta, \mathbf h^*(\boldsymbol\theta))$ with respect to $\boldsymbol\theta$ can be computed as
\begin{equation}\label{eqn:Grad}
  \nabla_{\boldsymbol\theta} \varphi(\boldsymbol{\theta}, \mathbf h^*(\boldsymbol\theta))
 = \nabla_{\boldsymbol\theta}\varphi(\boldsymbol\theta, \mathbf h)\Big|_{\mathbf h = \mathbf h^*(\boldsymbol\theta)},
\end{equation}
where $\nabla_{\boldsymbol\theta}$ denotes the discrete gradient operator with respect to $\boldsymbol\theta$. The derivation of Eqn.~\eqref{eqn:Grad} makes use of the optimality condition for Eqn.~\eqref{eqn:ParaModel}; namely, $\nabla_{\mathbf{h}}\varphi(\boldsymbol\theta, \mathbf h)\equiv 0$ at the point $\mathbf h =\mathbf h^*$.
Equation \eqref{eqn:ParaModel} simplifies the gradient calculation; the gradient computation prescribed by Eqn.~\eqref{eqn:Grad} is
 identical to that employed  by standard gradient descent methods for  penalized least squares reconstruction problems.

\begin{algorithm}
\caption{Variable projection (VP) algorithm for joint estimation of $\boldsymbol{\theta}$ and $\mathbf{h}$}
\label{VP_alg}
\begin{algorithmic}[1]
\STATE {$\mathbf{h}^0\gets$ experimentally-measured EIR.}
\STATE {$\boldsymbol{\theta}^0 \gets \arg\min_{\boldsymbol\theta\geq0}\varphi(\boldsymbol\theta, \mathbf h^0)$}
\STATE{$k\gets0$} \COMMENT{$k$ is the number of algorithm iteration}
\WHILE {stopping criterion is not satisfied}
\STATE $k\gets k+1$
\STATE{ $ \mathbf{h}^{k}\gets\arg\min_{\mathbf{h}} \varphi(\boldsymbol\theta^{k-1},\mathbf h)$}
\STATE{$
 \boldsymbol{\theta}^k\gets
        \mathcal P_{\rm c}\big\{
           \boldsymbol{\theta}^{k-1}-\gamma^k \nabla_{\boldsymbol\theta}
        \varphi(\boldsymbol\theta^{k-1}, \mathbf h^k )\big\}$}
\ENDWHILE
\STATE{$\hat{\boldsymbol\theta}\gets\boldsymbol\theta^k$ and $\hat{\mathbf h}\gets\mathbf h^k$. }
\end{algorithmic}
\end{algorithm}

\subsection{VP algorithm}
A VP algorithm for solving Eqns.~\eqref{eqn:OptCond1} and~\eqref{eqn:SubProb2} is provided in Algorithm~\ref{VP_alg}.
An experimentally-measured EIR, denoted by $\mathbf h^0$,  is utilized to initialize $\mathbf h$.
 Initialization of $\boldsymbol\theta$ is achieved by solving a constrained optimization problem in Line-2. A variety of established iterative image reconstruction algorithms can be employed to accomplish this \cite{KRAM2012,RNR2013}.
In Line-6, the estimate of $\mathbf{h}$ at the $k$-th iteration, {denoted by} $\mathbf{h}^k$, is obtained according to Eqn.~\eqref{eqn:ParaModel} by use of the previously-estimated $\boldsymbol\theta^{k-1}$.
 The updating scheme for $\boldsymbol\theta^k$ in Line-7 represents a gradient descent step for solving Eqn.~\eqref{eqn:SubProb2}, where $\mathcal P_{\rm c}$ denotes the operator that projects all negative values to $0$. The gradient is computed
as specified in  Eqn.~\eqref{eqn:Grad}. Note that Line-7 does not fully solve Eqn.~\eqref{eqn:SubProb2}. Instead, Line-7 moves $\boldsymbol\theta^k$ along the negative gradient of $\varphi$ by a small step size, denoted by $\gamma^k$. This distinguishes the VP algorithm from a
block-coordinate descent algorithm \cite{GV2003}, in which $\boldsymbol\theta^k\gets\arg\min_{\boldsymbol\theta\geq0}\varphi(\boldsymbol\theta, \mathbf h^k)$. Note that Line-7 
can be computed much more efficiently than the problem $\arg\min_{\boldsymbol\theta\geq0}\varphi(\boldsymbol\theta, \mathbf h^k)$.
 In addition,  VP algorithms have been reported to possess  faster convergence rates and may be less likely to be become trapped
by local minima as compared to block-coordinate descent algorithms \cite{DB2013}.

\subsection{Implementation of the VP algorithm}
Numerical details regarding the solution of the  sub-problems defined by  Lines-2 and -6 in Algorithm \ref{VP_alg} are provided below.

A method for solving  the constrained minimization problem in Line-2 has been described in \cite{KRAM2012,wang2013accelerating}. In this study, we assume that $R_1(\boldsymbol\theta)$ is differentiable and therefore, the constrained optimization problem can be solved by use of a projected gradient descent algorithm. In particular, we employ the updating scheme
\begin{equation}\label{eqn:updatingSub}
 \boldsymbol{\theta}^{0,j}\gets
        \mathcal P_{\rm c}\big\{
           \boldsymbol{\theta}^{0,j-1}-\gamma^j \nabla_{\boldsymbol\theta}
        \varphi(\boldsymbol\theta^{0,j-1}, \mathbf h^0 )\big\},
\end{equation}
where $\boldsymbol\theta^{0,j}$ denotes the estimate of $\boldsymbol\theta^0$ after the $j$-th iteration and $\gamma^j$ denotes an updating step size. The step size is determined by use of a line search method \cite{GNS2009}, and the gradient is calculated as
\begin{equation}\label{eqn:GradSubProb1}
 \nabla_{\boldsymbol\theta}\varphi(\boldsymbol\theta, \mathbf h) =
     \big (\mathbf{H}(\mathbf{h})\big)^{\rm T}
     \big (\mathbf{u}-\mathbf{H}(\mathbf{h})\boldsymbol{\theta}\big)
     + \nabla_{\boldsymbol{\theta}}R_1(\boldsymbol{\theta}),
\end{equation}
where $(\cdot)^{\rm T}$ denotes the matrix transpose operator. For $R_1(\boldsymbol\theta)$ with a typical quadratic form, the computation of $\nabla_{\boldsymbol\theta}R_1(\boldsymbol{\theta})$ is straightforward.
 Note that  Eqn.~\eqref{eqn:updatingSub} is of the same form as the updating scheme in Line-7 of Algorithm \ref{VP_alg},
 suggesting the same numerical procedure can be employed to implement both lines.

The second sub-problem in Line-6 of Algorithm \ref{VP_alg} can be efficiently implemented due to relatively low dimension of $\mathbf h$---less than $100$, typically. To be specific, we assume that $R_2(\mathbf h)=\|\mathbf{D}\mathbf{h}\|^2$, where
 $\mathbf{D}\in \mathbb{R}^{I\times I}$ is given by
\begin{equation}
 \mathbf{D}=\begin{pmatrix}
             1 & 0 & 0 & \cdots & 0 & 0 \\
             -1 & 1 & 0 & \cdots & 0 & 0 \\
             0 & -1 & 1 & \cdots & 0 & 0 \\
             \vdots & \vdots & \vdots & \ddots & \vdots & \vdots \\
             0 & 0 & 0 & \cdots & -1 & 1
            \end{pmatrix}.
\end{equation}
In this case, Line-6 can be implemented as
\begin{equation}
  \mathbf h^k = (\mathbf{P}^{\rm T}\mathbf{P}+\alpha\mathbf{D}^{\rm T}\mathbf{D})^{-1}
     \mathbf{P}^{\rm T} \mathbf{u},
\end{equation}
where the matrix $\mathbf P$ satisfies
\begin{equation}
  \mathbf P(\boldsymbol\theta^{k-1}) \mathbf h = \mathbf H(\mathbf h)\boldsymbol\theta^{k-1}.
\end{equation}
The matrix $\mathbf P$ is described in Appendix-\ref{appd:B}. Because of its small size, $\mathbf{P}^t\mathbf{P}+\alpha\mathbf{D}^t\mathbf{D}$ can be stored in random access memory and efficiently inverted by use of established algorithms. 
In the studies below, this was accomplished by use of the LU decomposition method\cite{GL2012}.

\section{Description of computer-simulation studies}\label{Sect:ComputerStudies}
Computer-simulation studies were conducted to investigate the numerical properties of the VP algorithm. 

\subsection{Simulation of noise-free data}

\begin{figure}[!htb]
 \centering
 \begin{subfigure}[!htb]{0.4\textwidth}
 \includegraphics[width=2.5in]{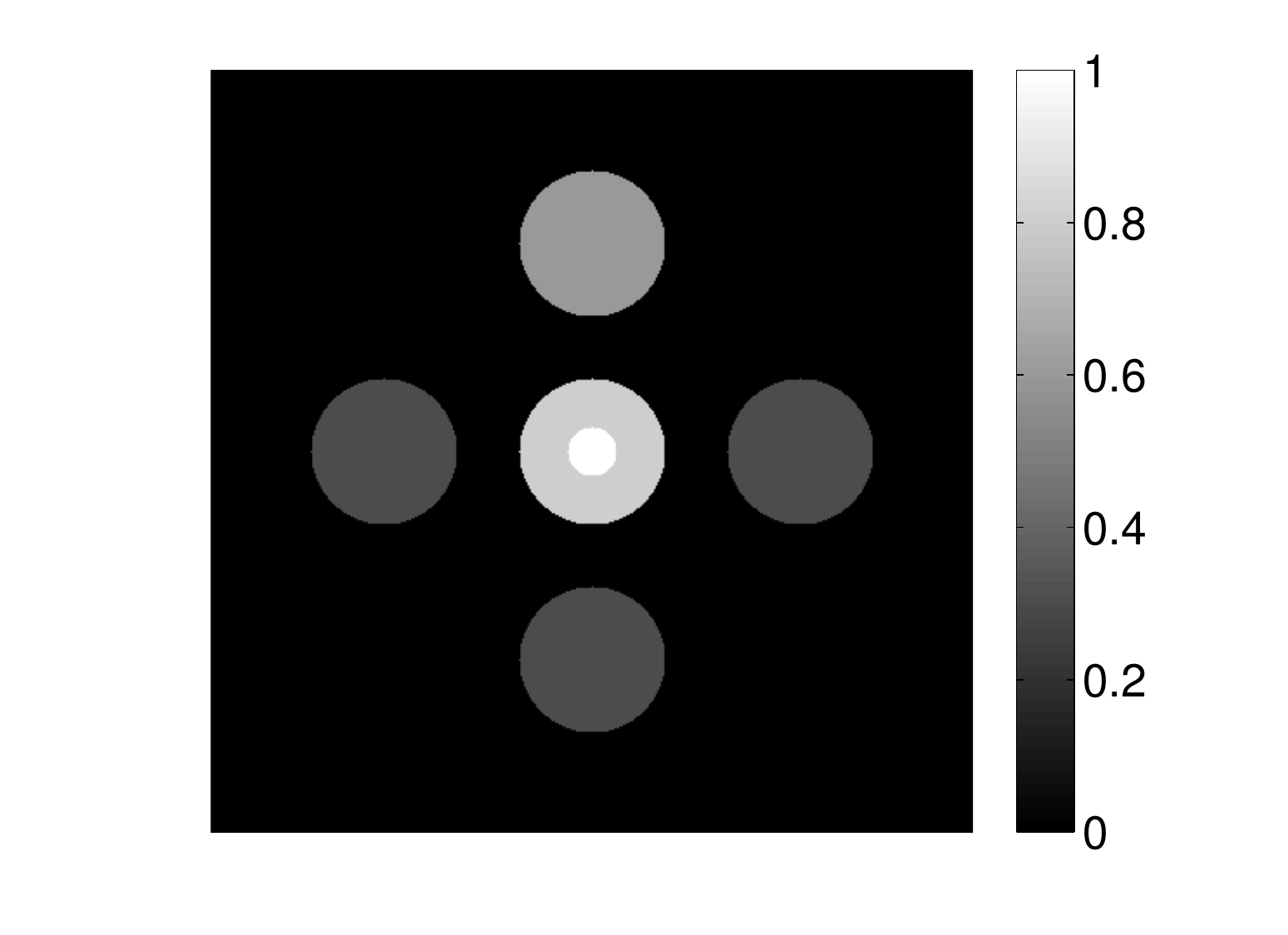}
 \caption{Phantom}\label{Phantom2D}
 \end{subfigure}
 \begin{subfigure}[!htb]{0.4\textwidth}
 \includegraphics[width=2.5in]{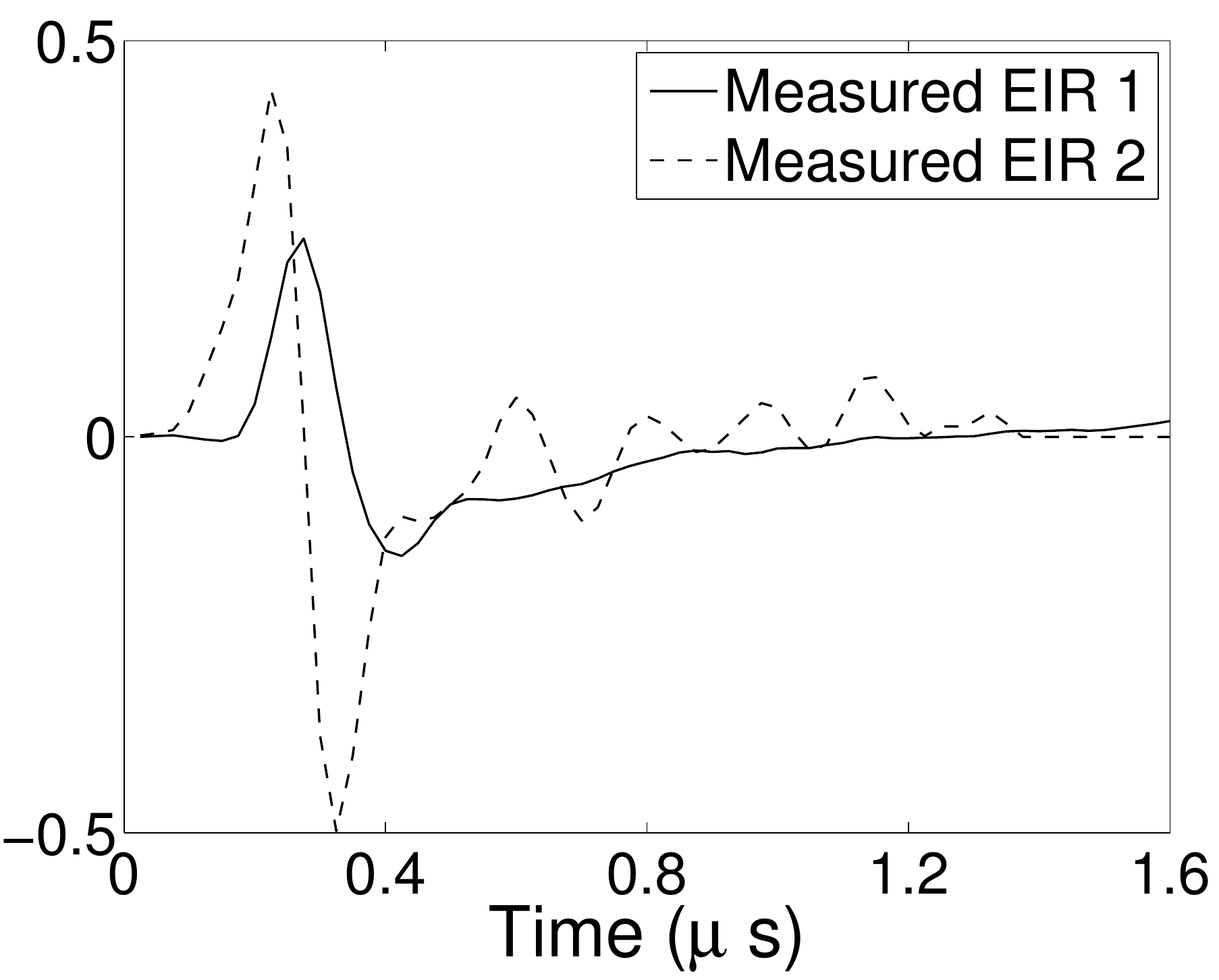}
 \caption{EIRs}\label{EIR_plots}
 \end{subfigure}
 \caption{(a) The absorbed energy density map employed in the computer-simulation studies. (b) EIRs employed for 2D experiments.}
\end{figure}

The numerical phantom shown in Figure~\ref{Phantom2D} was employed. The phantom had a support area of $22.0\times 22.0$ mm$^2$ and contained six uniform disks that were assigned different values of absorbed optical energy density.

A 2D circular measurement geometry was employed. $Q=128$ transducers were evenly distributed on a ring of radius $25$ mm that enclosed the phantom. The SOS was assumed to be constant and set at $c_0=1.5$ mm/$\mu$s.
{Since the simulated  data were formed by use of the C-D imaging model in Eqn.~\eqref{eq_CD}, no inverse crime was committed.} The components of this vector corresponded to  $T=600$ equally spaced temporal samples over the interval $[10,25)$ $\mu$s.
 Subsequently, the noiseless voltage vector $\mathbf{u}_q$ was obtained by convolving the pressure data with  EIR-1
 in Figure~\ref{EIR_plots}.

\subsection{Simulation of noisy data}
Noisy measurement data were computed as
\begin{equation}
 \tilde{\mathbf{u}}= \mathbf{u}_q + \tilde{\mathbf{n}},
\end{equation}
where $\tilde{\mathbf{n}}$ is a random vector whose components were independent and identically distributed Gaussian random variables. The standard deviation of each element of  $\tilde{\mathbf{n}}$ was $3\%$ of $[\mathbf{u}_q]_{\max}$, where $[\mathbf{u}_q]_{\max}$ denotes the maximum value contained in $\mathbf{u}_q$.

\subsection{Implementation of image reconstruction algorithms}
From the simulated noiseless and noisy data, images were reconstructed by solving the minimization problem in Eqn.~\eqref{min_prob} by use of Algorithm \ref{VP_alg}.
Conventional quadratic smoothness penalties were employed:
\begin{align}
 R_1(\boldsymbol{\theta})&=\sum^{N-1}_{n=0}\sum_{k\in\mathcal{N}_n}\left([\boldsymbol{\theta}]_n-[\boldsymbol{\theta}]_k\right)^2,\label{penalty1}\\
 R_2(\mathbf{h})&=\sum^{I-2}_{i=0}([\mathbf{h}]_{i+1}-[\mathbf{h}]_{i})^2\label{penalty2},
\end{align}
where $\mathcal{N}_n$ is the index set of four neighboring pixels of the $n$-th pixel.

The reconstruction region ($22.0\times 22.0$ mm$^2$) was represented by $440\times 440$ pixels with pixel size $0.05$ mm in each dimension.
The initial guess of the EIR employed in the VP algorithm was different than the EIR that was assumed when generating the
 simulated data. This served to simulate a situation in which an experimentally measured EIR contained errors.

Each element in a real-world transducer array possesses its own EIR.
In practice, the differences between the EIRs are sometimes neglected and an EIR corresponding to a single
element may be used to represent all elements in the array.
In some of the studies below, the EIR employed to initialize the VP algorithm (EIR-2 in Figure~\ref{EIR_plots})
and the EIR employed to produce the simulated measurements  (EIR-1 in Figure~\ref{EIR_plots})
were experimentally measured from two different transducer
elements in a circular transducer array (see Sec. VI-B).
EIR-1 was measured by temporally integrating the PA signal produced by a point source positioned
at the focus of the transducer.
EIR-2 was measured by use of the method reported in \cite{RNR2011}.
In order to investigate the sensitivity of the VP algorithm 
to the initialization of the EIR, we employed different EIRs
 obtained by degrading EIR-1 as described later. When solving the sub-problem in Line-2 of Algorithm \ref{VP_alg}, $\boldsymbol{\theta}$
was initialized as the zero vector. 
Algorithm \ref{VP_alg} was terminated after 500 iterations, since it was observed that the changes in the reconstructed images with more iterations were negligible.
 {When implemented by use of a single core of an Intel Xeon E5-2640 CPU, each
iteration required approximately 7s to complete.} 

For comparison, we also reconstructed images by use of a conventional gradient-based iterative image reconstruction algorithm that considered the EIR to be fixed.
This algorithm  was the same as the one employed to compute the initial guess of $\boldsymbol{\theta}$ in Line-2 of Algorithm \ref{VP_alg}, which was described in Section III-D.
 {As with Algorithm 1, each iteration required approximately 
7s to complete.  The reconstruction algorithm was run for 150 iterations, since the changes in the reconstructed images with more iterations were negligible. Note that the computational cost of Line-6 in Algorithm \ref{VP_alg} was about 5\% of the Line-7 in Algorithm \ref{VP_alg}, which is why each iteration took almost the same time in both the conventional iterative method and the VP algorithm.}


\subsection{Image accuracy assessment}
The accuracy of the reconstructed images was assessed in terms of the root-mean-squared-error (RMSE) between the reconstructed image and the true phantom as
\begin{equation}
 RMSE=\sqrt{\frac{1}{N}\sum_{n=1}^N ([\boldsymbol{\theta}]_n-[\boldsymbol{\theta}_0]_n)^2},
\end{equation}
where $N$ is the number of pixels, and $[\boldsymbol{\theta}]_n$ {and $[\boldsymbol{\theta}_0]_n$ are the $n$-th pixel
 values of the reconstructed image and  phantom}, respectively.

\section{Computer-simulation results}\label{Sect:ComputerResults}

\subsection{Images reconstructed from noise-free data}

\subsubsection{Mitigation of artifacts and distortions caused by errors in the assumed EIR}
\begin{figure}[!htb]
 \centering
 \begin{subfigure}[!htb]{0.3\textwidth}
 \includegraphics[width=\textwidth]{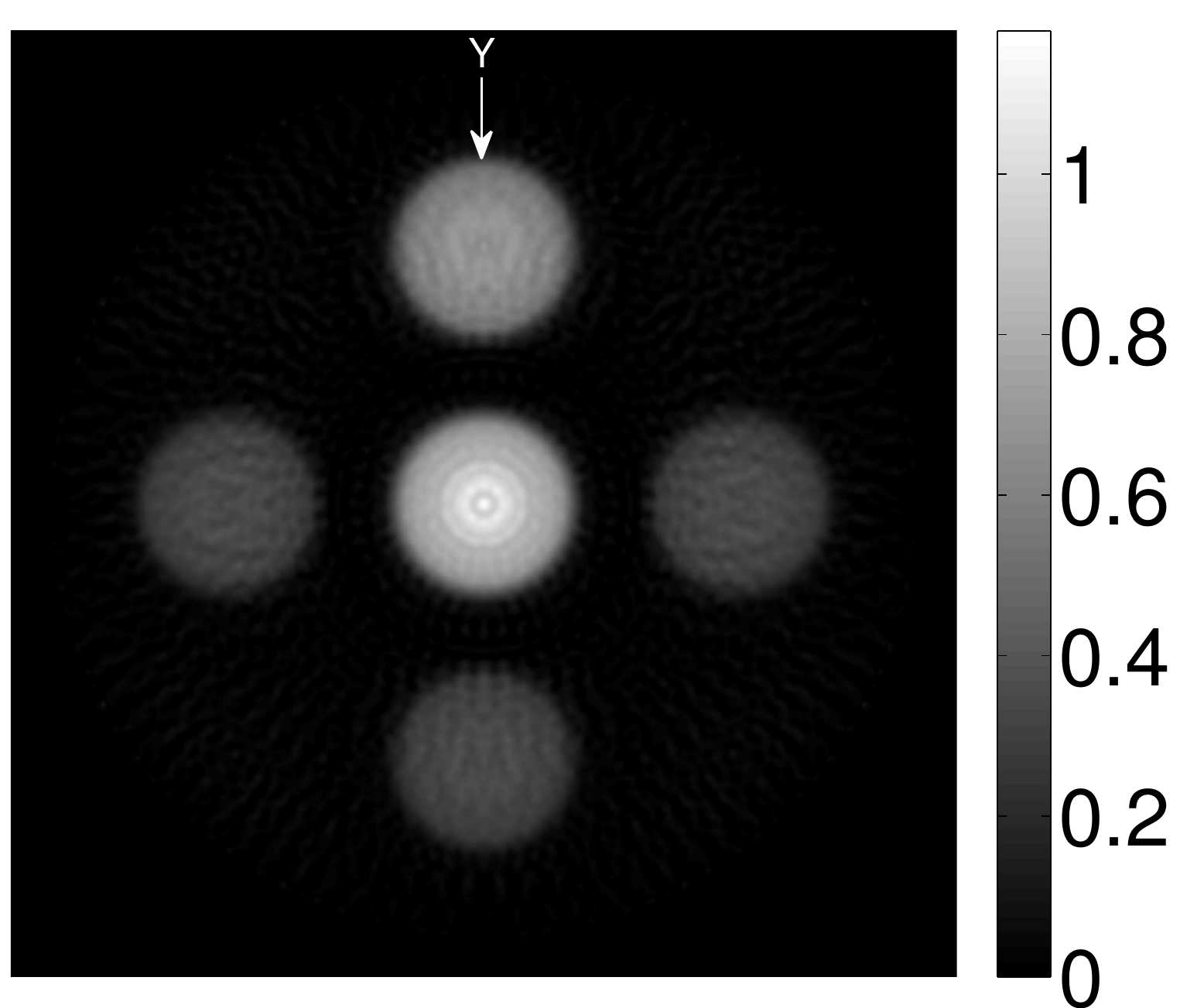}
  \caption{RMSE=$0.0445$}\label{2d_0ns_B_NoVPM}
 \end{subfigure}
  \begin{subfigure}[!htb]{0.3\textwidth}
 \includegraphics[width=\textwidth]{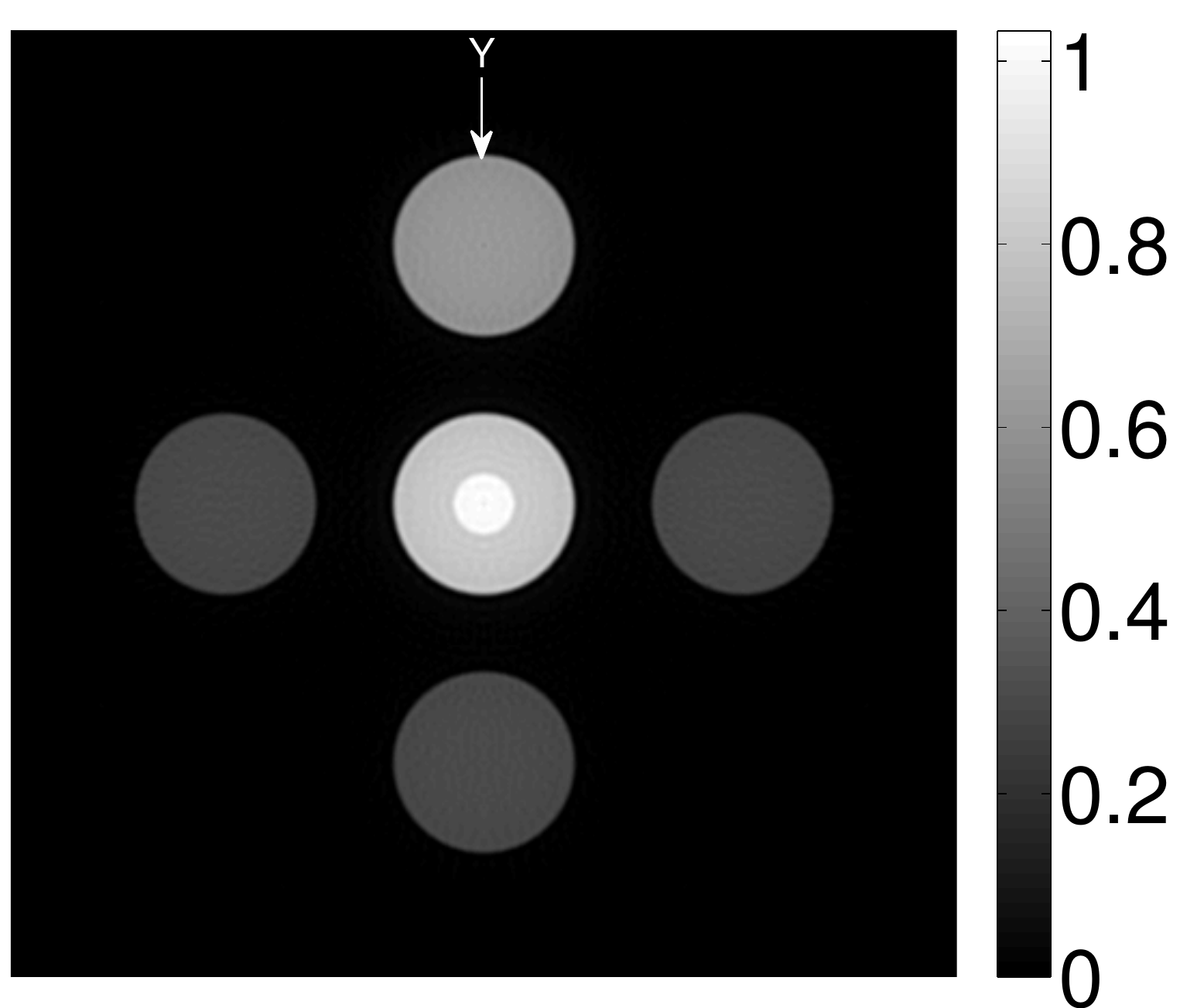}
  \caption{RMSE=$0.0105$}\label{2d_0ns_B_VPM}
 \end{subfigure}
 \begin{subfigure}[!htb]{0.3\textwidth}
 \includegraphics[width=\textwidth]{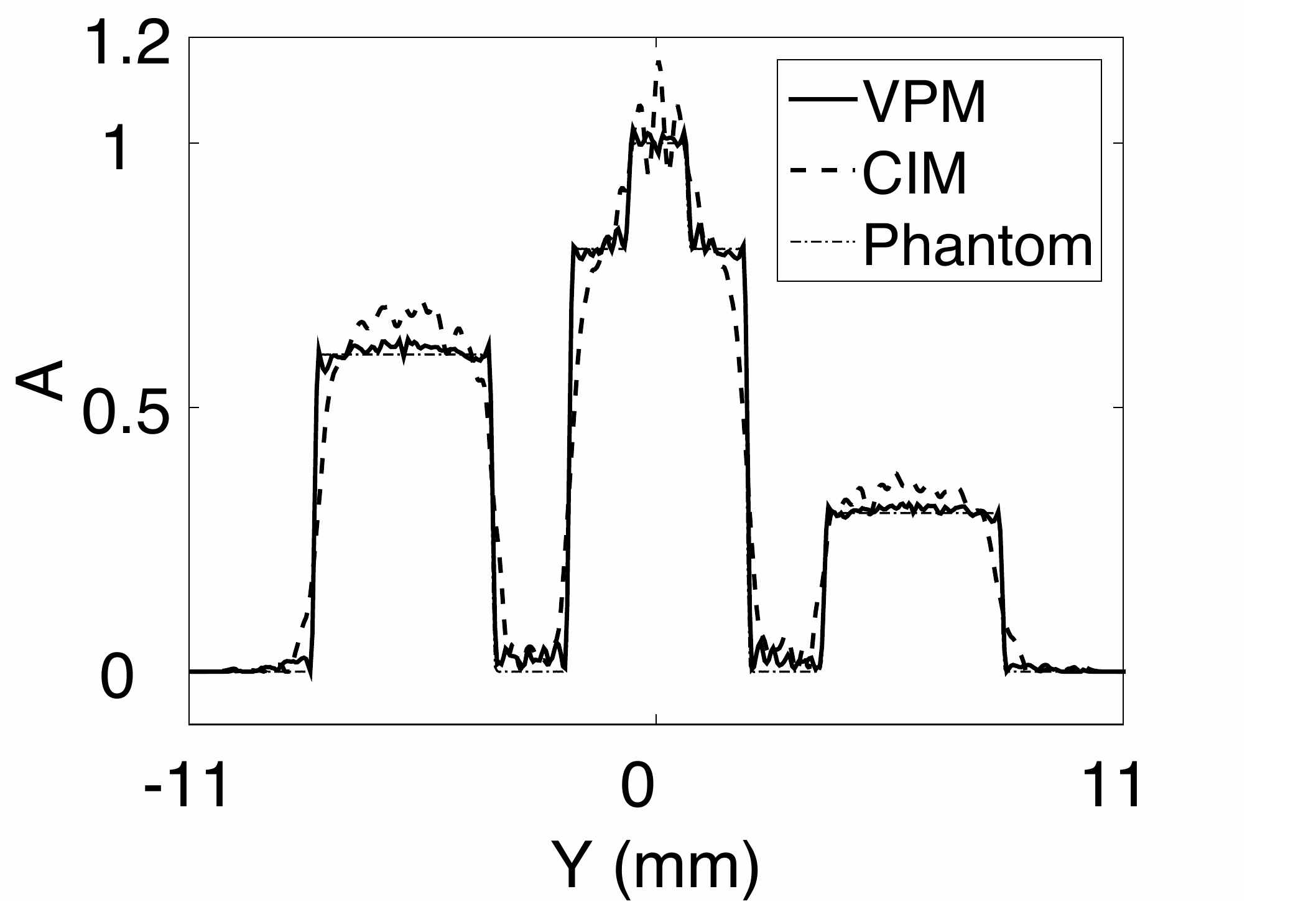}
 \caption{}\label{0ns_B_VPM_plots}
 \end{subfigure}
 \caption{(a) and (b) Images reconstructed from noiseless data using the conventional iterative method with $\lambda=1.0\times10^{-3}$ and the VP algorithm with $\lambda=1.0\times10^{-4}$ and $\alpha=2000$, respectively. (c) Image profiles through noiseless images seen in Fig.~\ref{0ns_B_improve} corresponding to the reconstructions with the VP algorithm (solid line), with the conventional iterative method (dashed line), and phantom (dash-dot line). The locations of the profiles are indicated by the ``Y'' arrows in the Fig.~\ref{2d_0ns_B_NoVPM}, and \ref{2d_0ns_B_VPM}, respectively.}\label{0ns_B_improve}
\end{figure}

Figure~\ref{2d_0ns_B_NoVPM} shows the image reconstructed by use of the conventional iterative method that
utilized a system matrix based on EIR-2.
 Different values of the regularization parameter $\lambda$ from the interval $[0,10^{-1}]$ were considered. The reconstructed image with the value of $\lambda$ that minimized the RMSE was chosen to represent {the best performance of} the conventional iterative method. Figure~\ref{2d_0ns_B_NoVPM} and the profile in Figure~\ref{0ns_B_VPM_plots} demonstrate that {the} use of an inaccurate EIR can result in strong artifacts and distortions in  images reconstructed by use of the conventional methods.

When the VP algorithm was applied,  different values of the regularization parameter $\lambda$ from the interval $[10^{-8},10^{-1}]$ and $\alpha$ from the interval $[200, 20000]$ were considered. The image that minimized the RMSE was chosen and displayed in Figure~\ref{2d_0ns_B_VPM}.
 As revealed by this image and the profiles in \ref{0ns_B_VPM_plots}, the VP algorithm yielded an image with fewer artifacts and distortions, and image fidelity was improved as reflected by the reduced RMSE.

\subsubsection{Effect of frequency contents of the objects and EIR} Since the voltage signal is generated through convolution of the pressure data and EIR, the {EIR} 
serves as a bandpass filter. Thus, the information contained in the high frequency components is lost in the resulting voltage signal.
We conducted a series of computer-simulations to show that the accuracy of the reconstructed $\boldsymbol{\theta}$ and $\mathbf{h}$ will be {affected} by this loss of information.

\begin{figure}[!htb]
 \centering
 \includegraphics[width=0.5\textwidth]{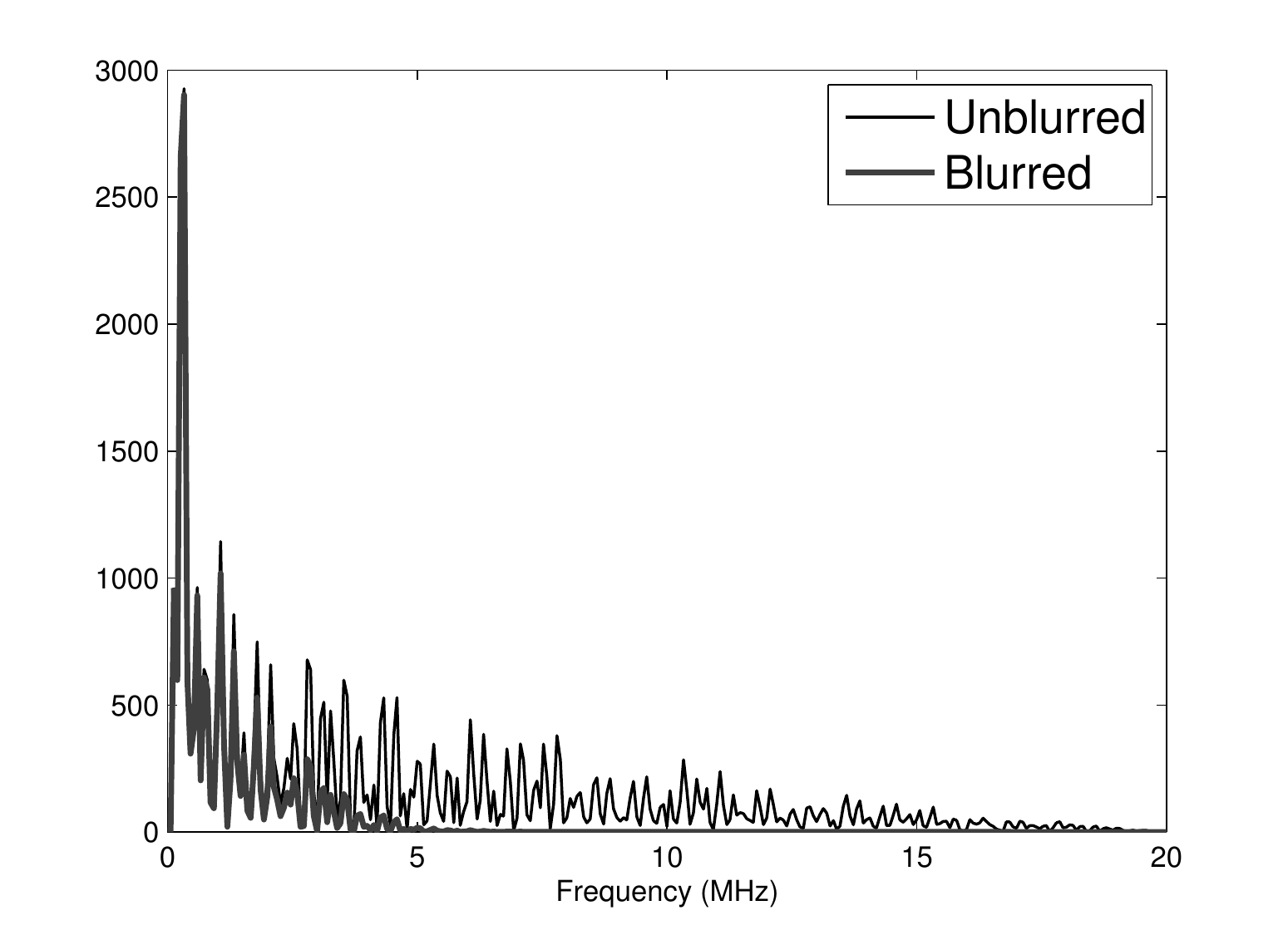}
 \caption{Spectrums of the pressure data generated by the sharp and smoothed objects.}\label{pres_bandwidth}
\end{figure}

The original sharp phantom shown in Figure~\ref{Phantom2D} was convolved with a Gaussian blurring kernel to generate a smoothed phantom that possessed smaller relative spatial bandwidths. 
We employed the sharp and the smoothed phantoms to generate pressure data;  the pressure data generated by the sharp phantom had {a larger bandwidth} than that generated by the smoothed one, as shown in Figure~\ref{pres_bandwidth}.

\begin{figure}[!htb]
 \centering
 \begin{subfigure}[!htb]{0.23\textwidth}
 \includegraphics[width=\textwidth]{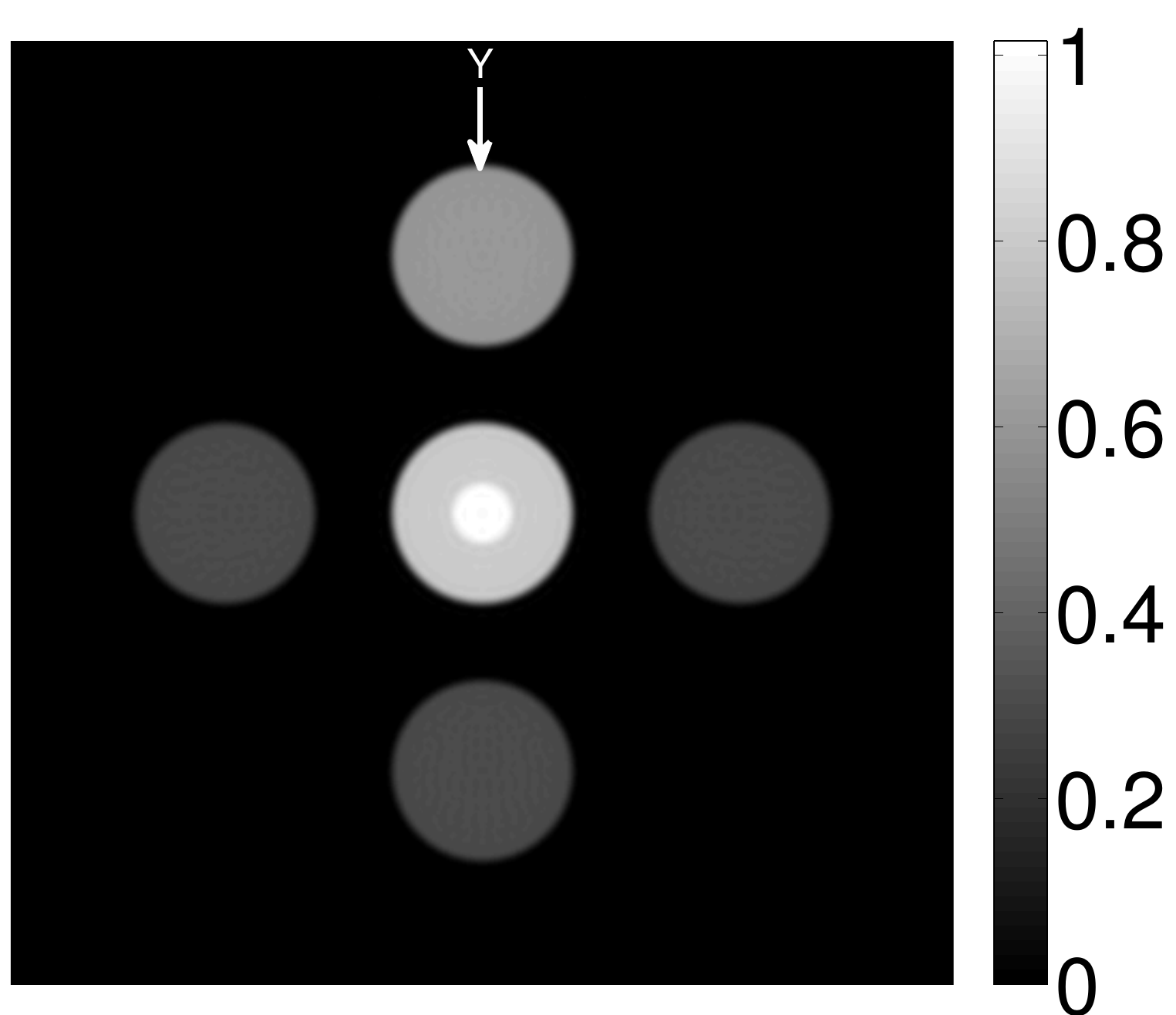}
  \caption{}\label{0ns_bandwidth_a1}
 \end{subfigure}
 \begin{subfigure}[!htb]{0.24\textwidth}
 \includegraphics[width=\textwidth]{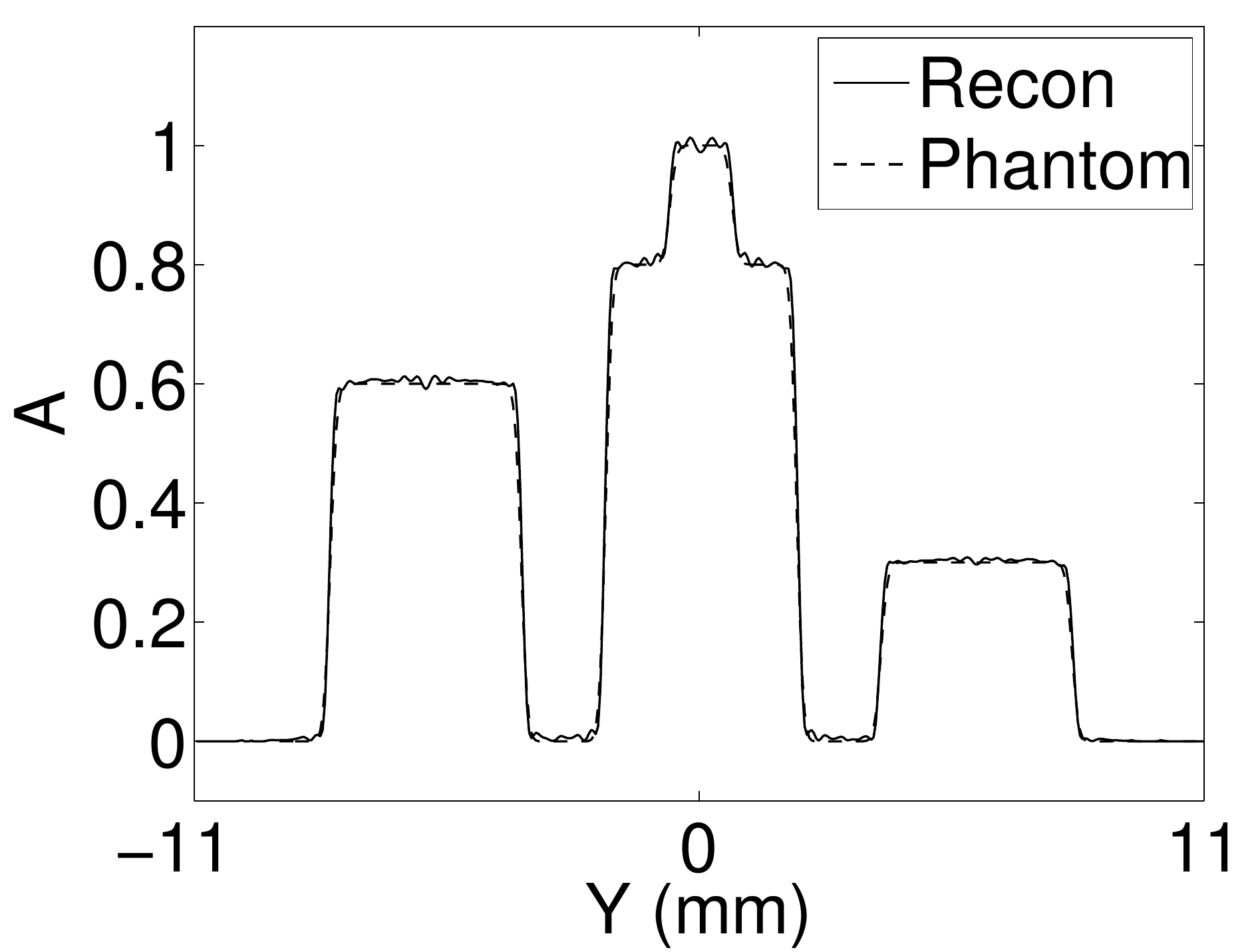}
  \caption{}\label{0ns_bandwidth_a2}
 \end{subfigure}
 \begin{subfigure}[!htb]{0.24\textwidth}
 \includegraphics[width=\textwidth]{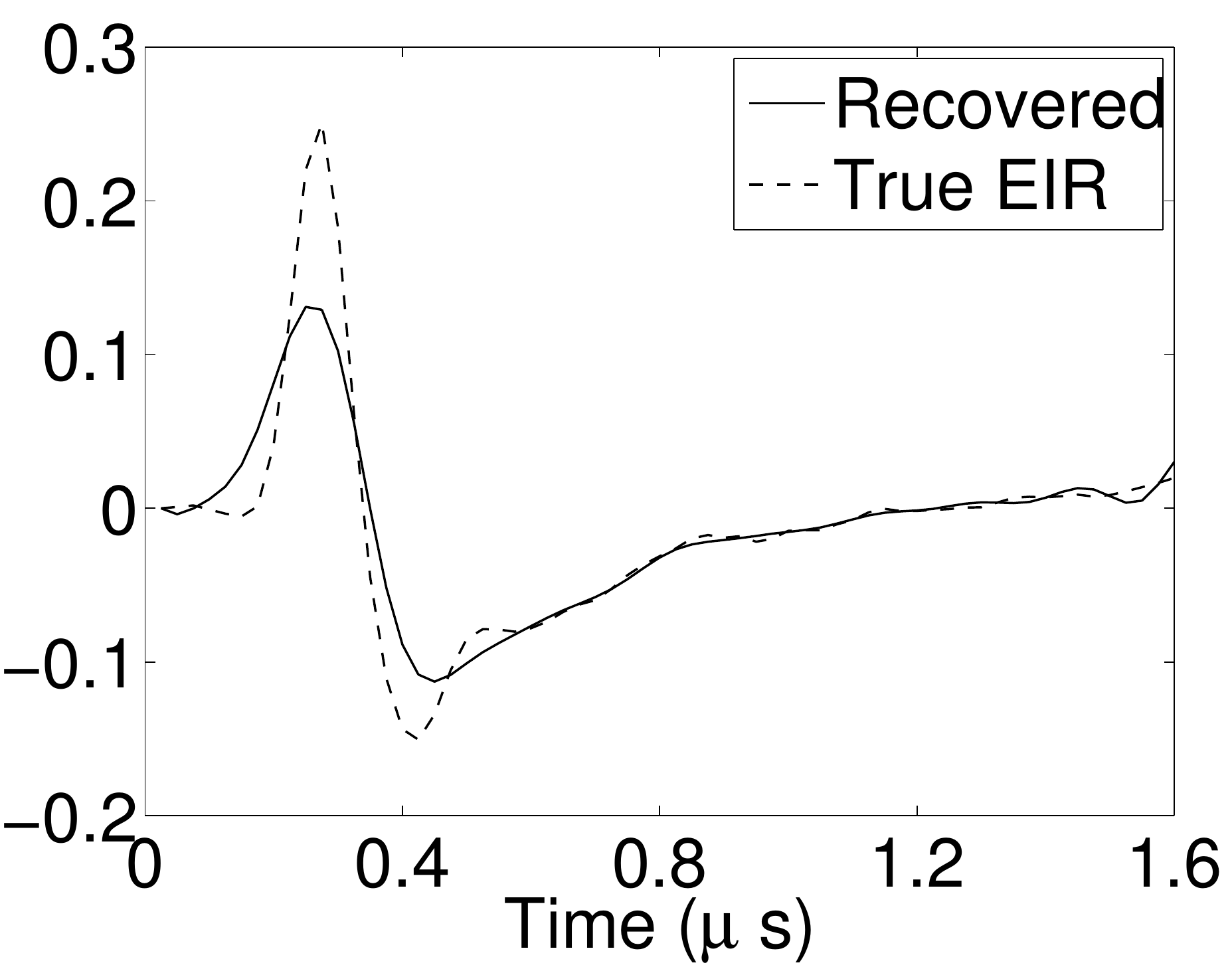}
  \caption{}
 \end{subfigure}
 \begin{subfigure}[!htb]{0.24\textwidth}
 \includegraphics[width=\textwidth]{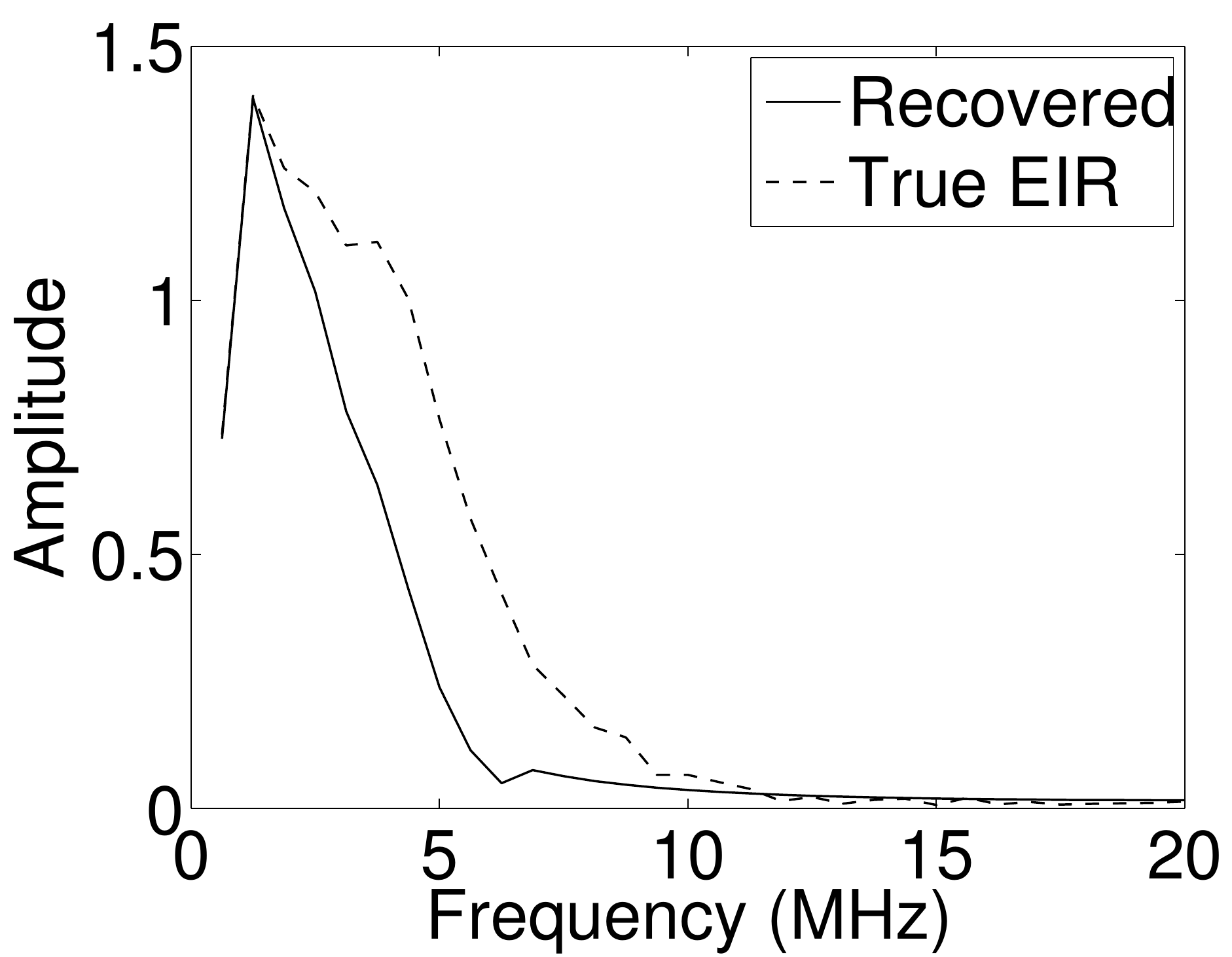}
  \caption{}
 \end{subfigure}
 \begin{subfigure}[]{0.23\textwidth}
 \includegraphics[width=\textwidth]{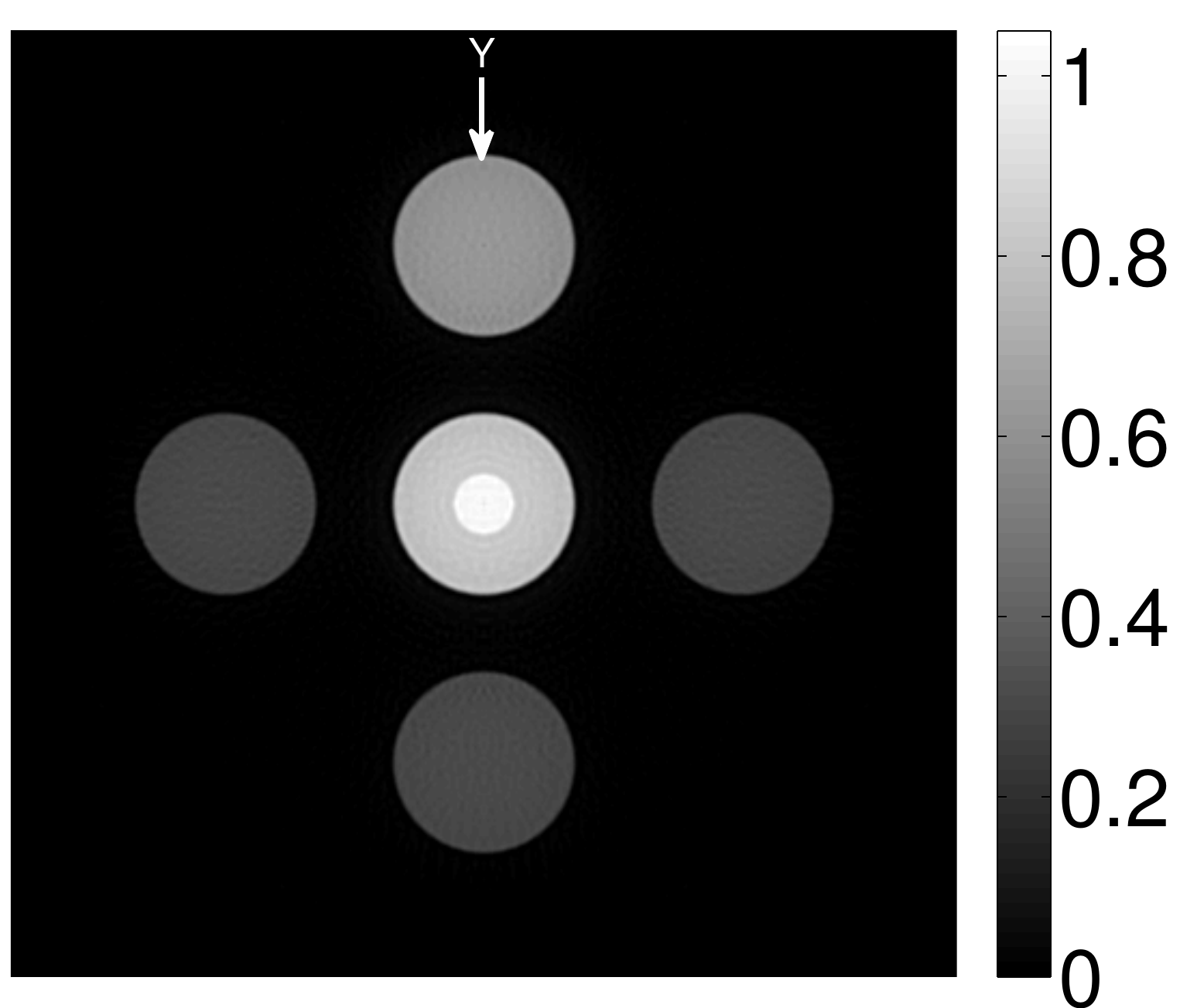}
  \caption{}
 \end{subfigure}
  \begin{subfigure}[!htb]{0.24\textwidth}
 \includegraphics[width=\textwidth]{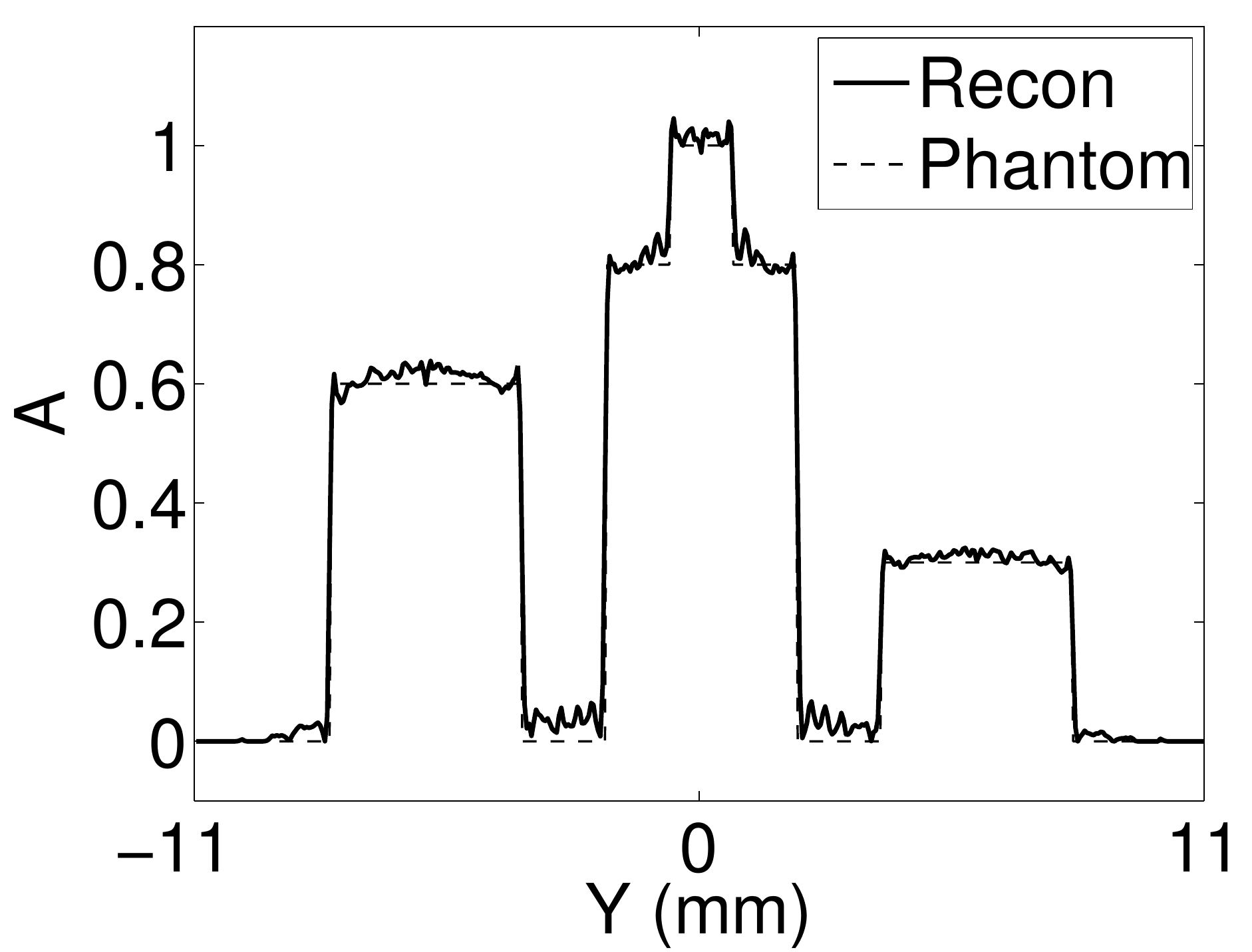}
  \caption{}
 \end{subfigure}
 \begin{subfigure}[!htb]{0.24\textwidth}
 \includegraphics[width=\textwidth]{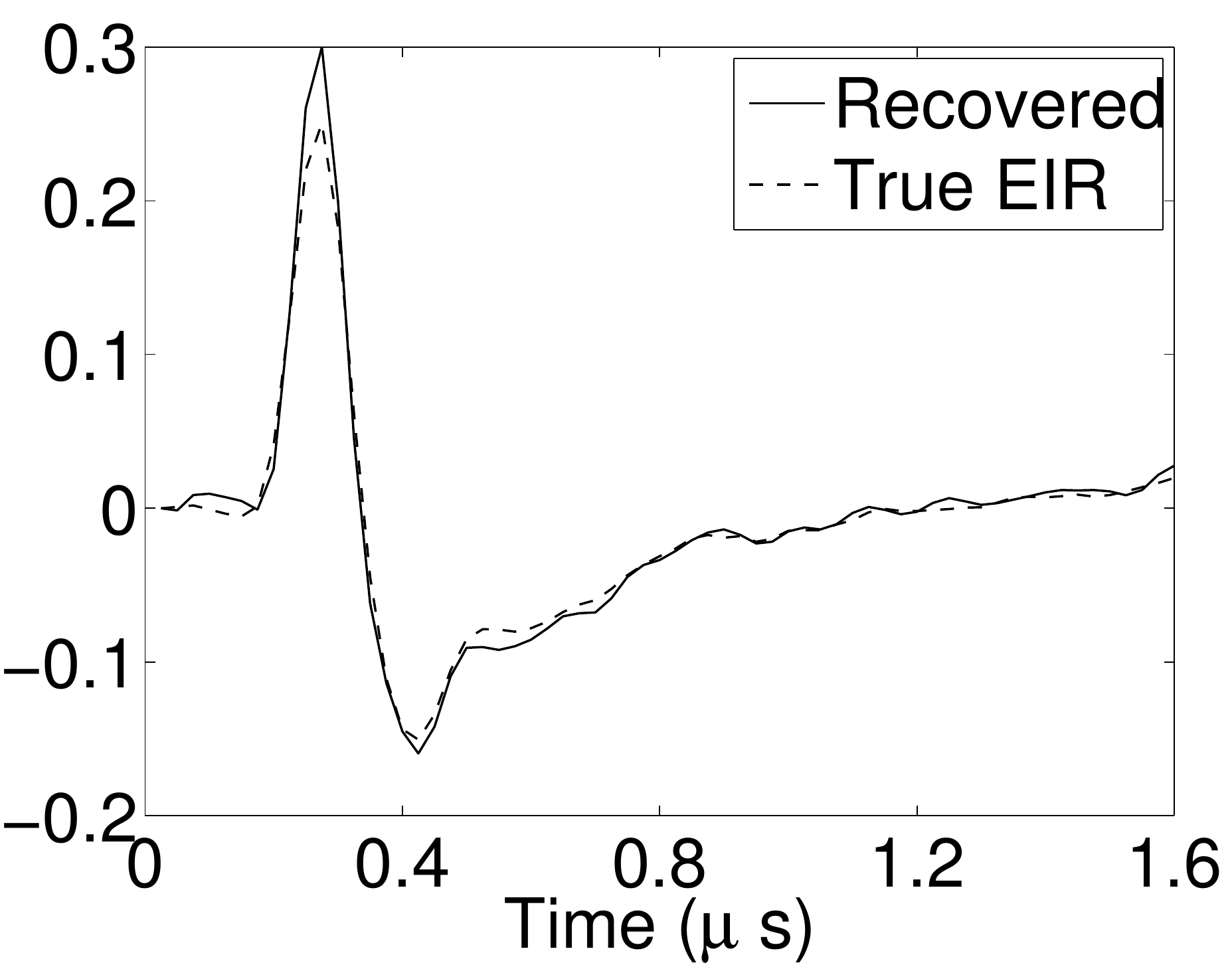}
  \caption{}\label{0ns_bandwidth_eir}
 \end{subfigure}
 \begin{subfigure}[!htb]{0.24\textwidth}
 \includegraphics[width=\textwidth]{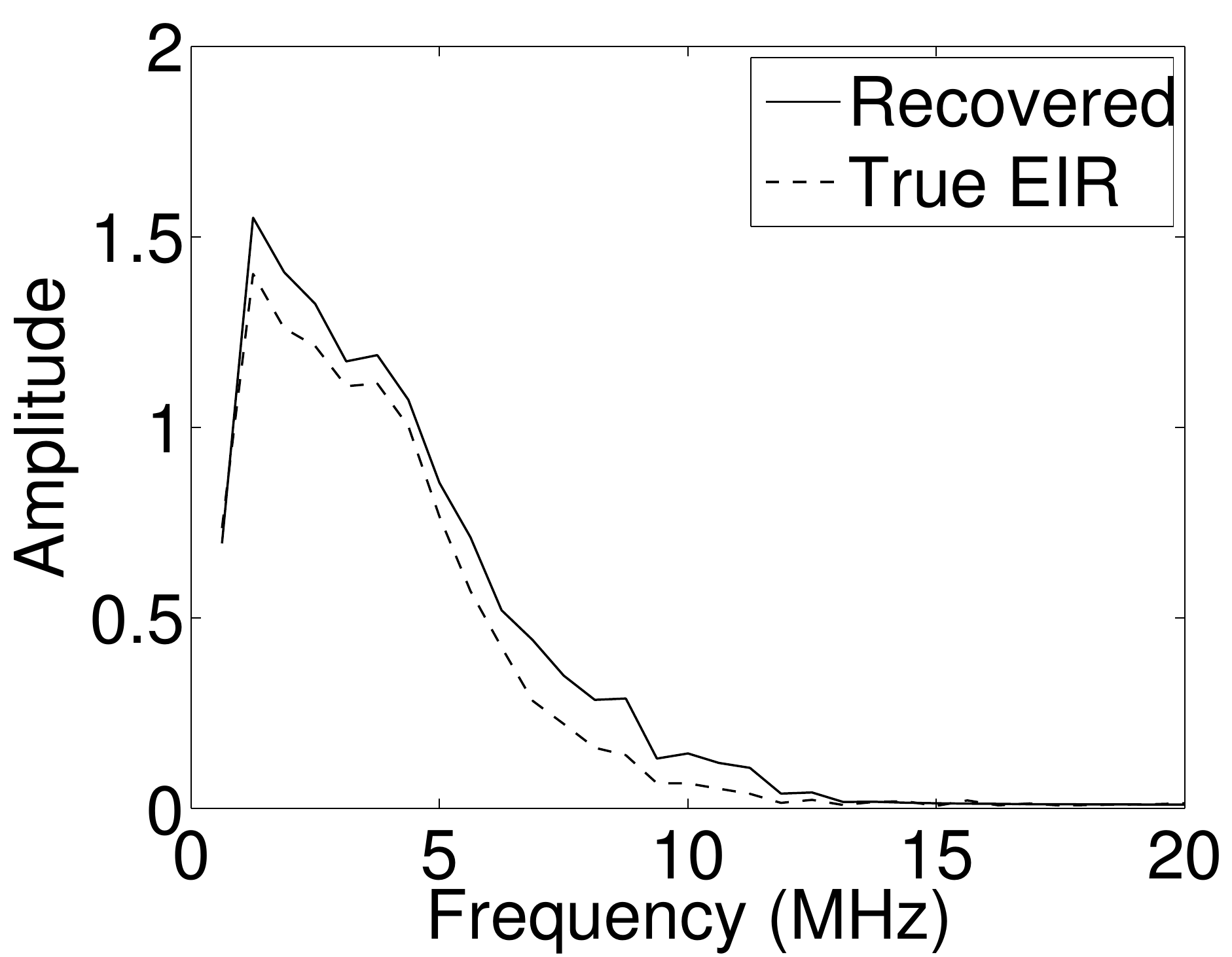}
  \caption{}\label{0ns_bandwidth_eir2}
 \end{subfigure}
  \caption{The first row shows the reconstructed image and the EIR from the smoothed object, where the spectrums of the generated pressure data are smaller than that of the EIR. The second row shows the ones from the sharp object, where the spectrums of the generated pressure data are larger than that of the EIR. Fig.~(a, e) are the reconstructed images, Fig.~(b, f) are the profile plots, Fig.~(c, g) are the recovered EIRs, and Fig.~(d, h) are the recovered EIRs in the frequency domain.}\label{0ns_bandwidth}
\end{figure}

The results shown in Figure~\ref{0ns_bandwidth} suggest that the reconstructed estimates of the EIR
become more accurate when 
   the bandwidth of the $A(\mathbf r)$ is {increased}  (Figure~\ref{0ns_bandwidth_eir} and \ref{0ns_bandwidth_eir2}).
On the other hand, the reconstructed
estimates of $A(\mathbf r)$ become more accurate when the bandwidth of the EIR is increased  (Figure~\ref{0ns_bandwidth_a1} and \ref{0ns_bandwidth_a2}).
 For a given EIR, the reconstructed estimates of $A(\mathbf r)$ that contain sharp features contain
 more oscillations than the estimates corresponding to the smoothed versions of $A(\mathbf r)$. 
This is because more high frequency information is lost during the convolution.

\subsubsection{Effect of data incompleteness} 
\begin{figure}[!htb]
 \centering
 \begin{subfigure}[!htb]{0.32\textwidth}
 \includegraphics[width=\textwidth]{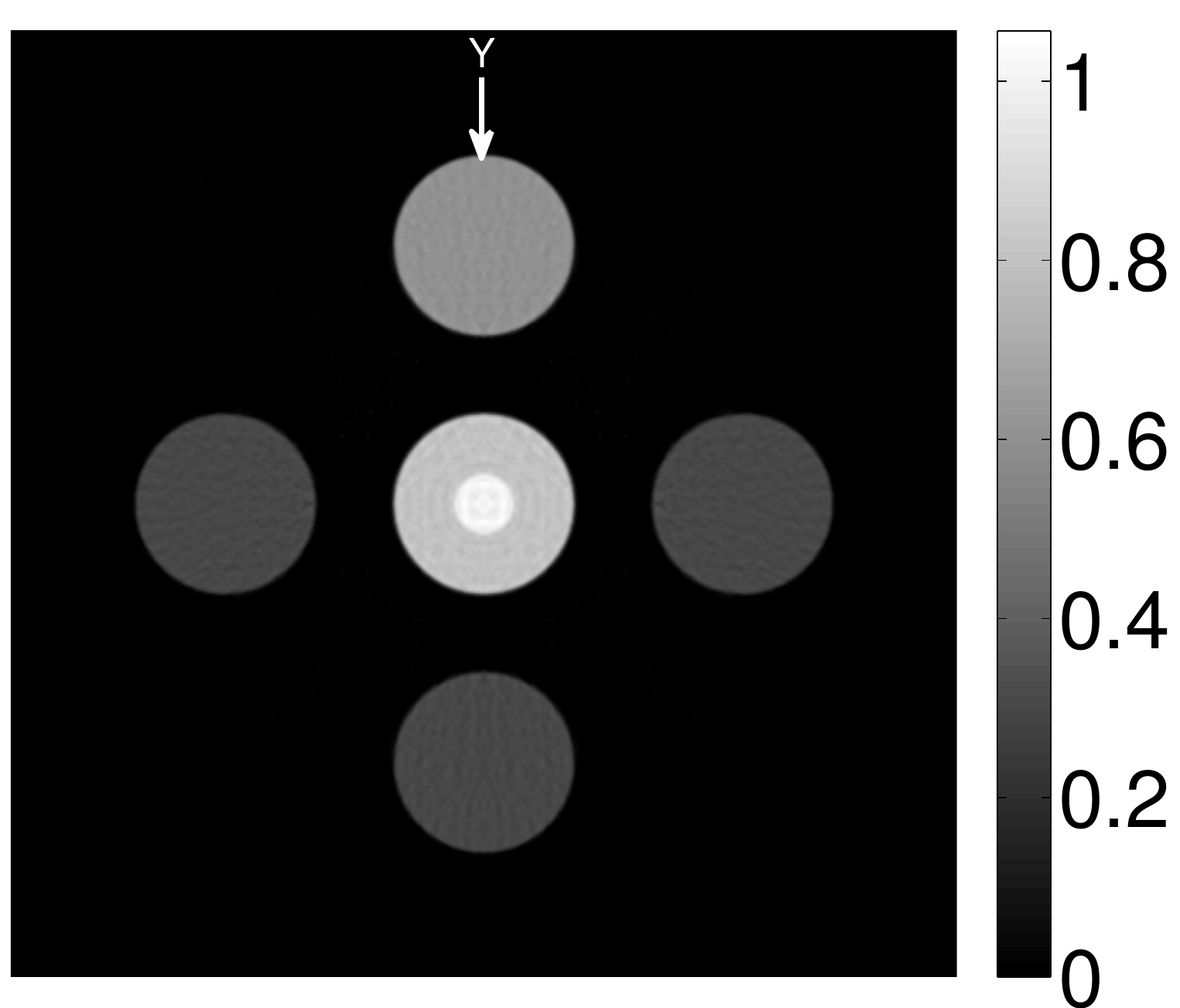}
  \caption{$\lambda=0$}\label{0ns_incomplete_true_full}
 \end{subfigure}
  \begin{subfigure}[!htb]{0.32\textwidth}
 \includegraphics[width=\textwidth]{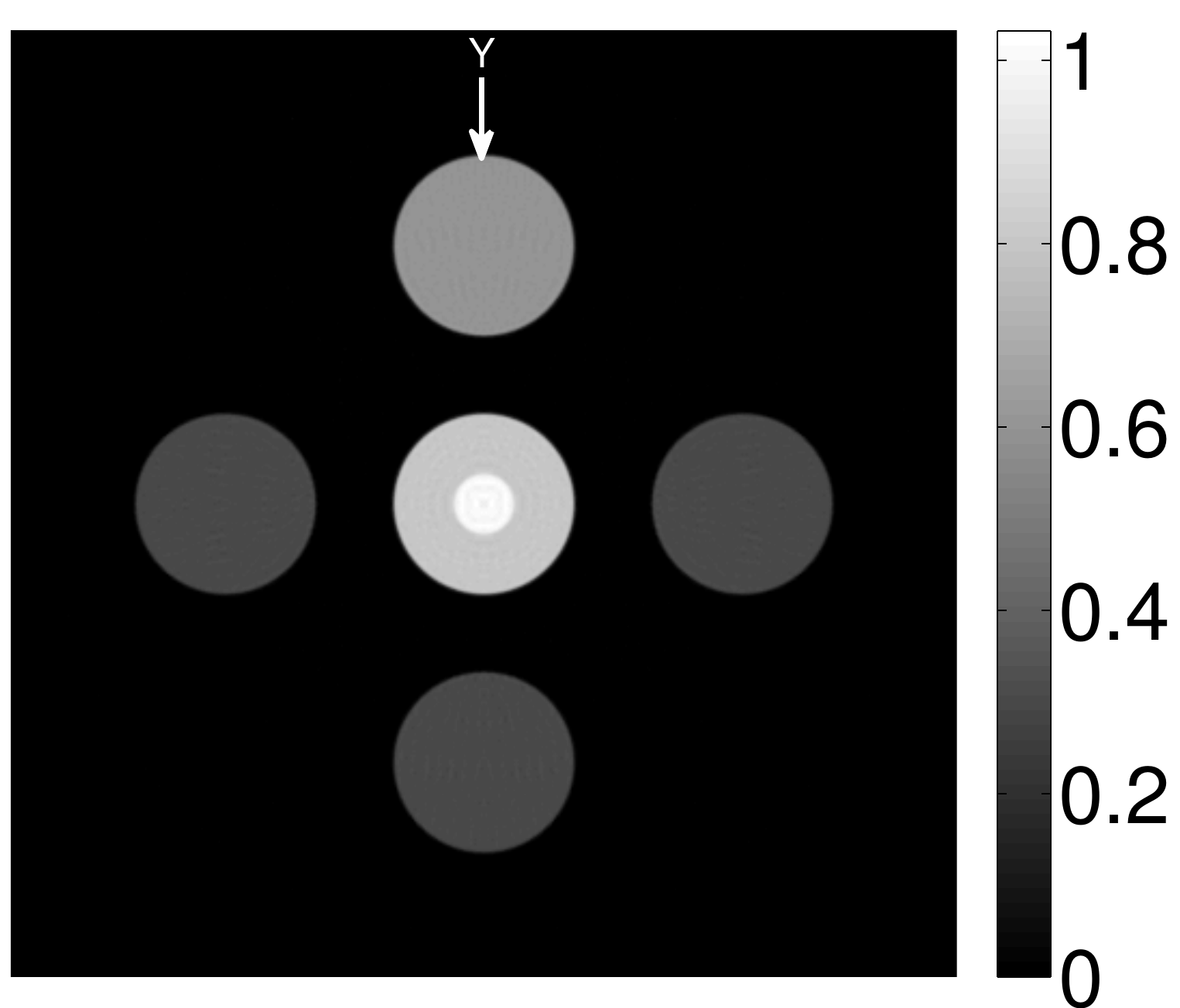}
  \caption{$\lambda=0$}\label{0ns_incomplete_true_half}
 \end{subfigure}
   \begin{subfigure}[!htb]{0.33\textwidth}
 \includegraphics[width=\textwidth]{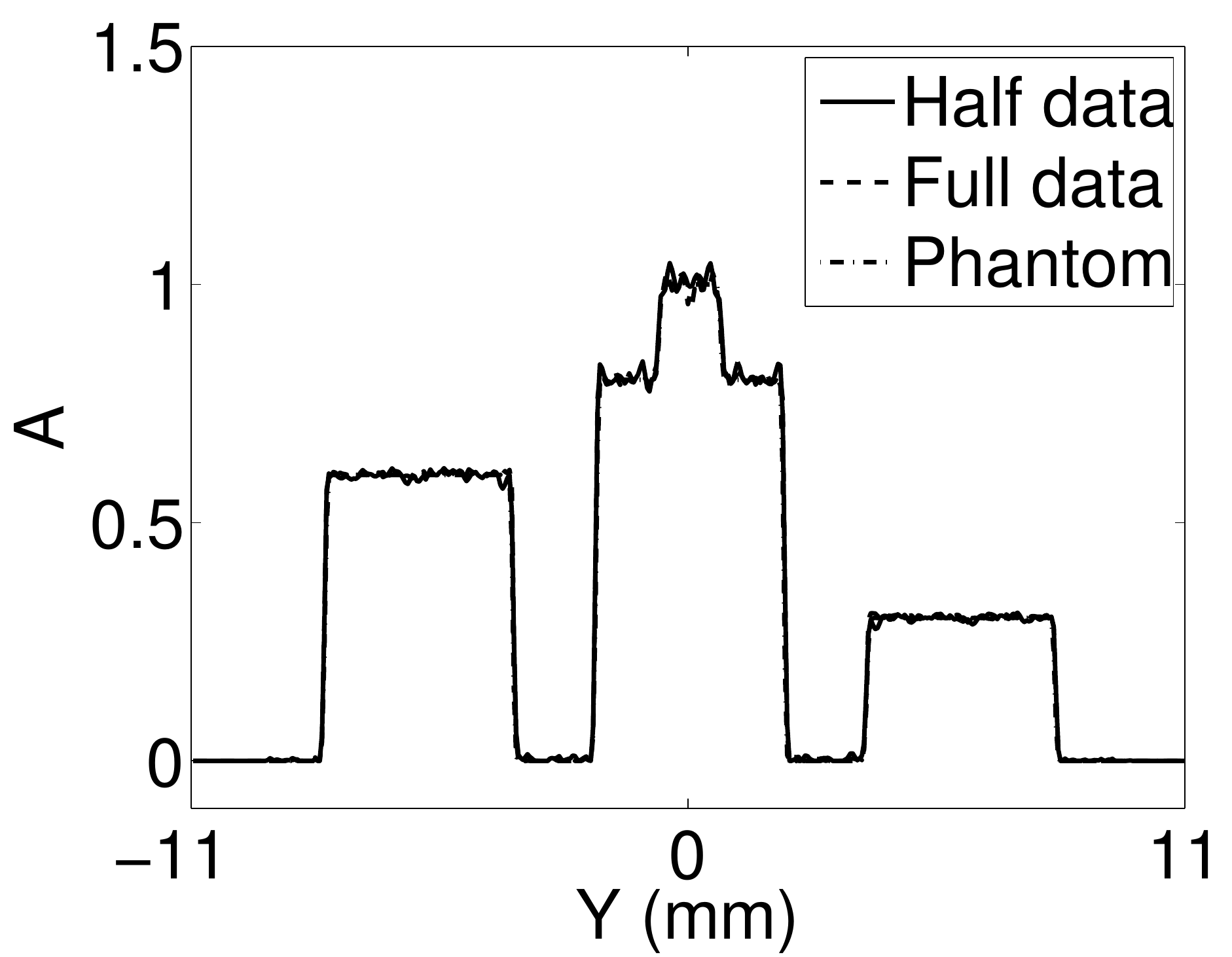}
  \caption{}\label{0ns_incomplete_true_plot}
 \end{subfigure}
 \begin{subfigure}[!htb]{0.32\textwidth}
 \includegraphics[width=\textwidth]{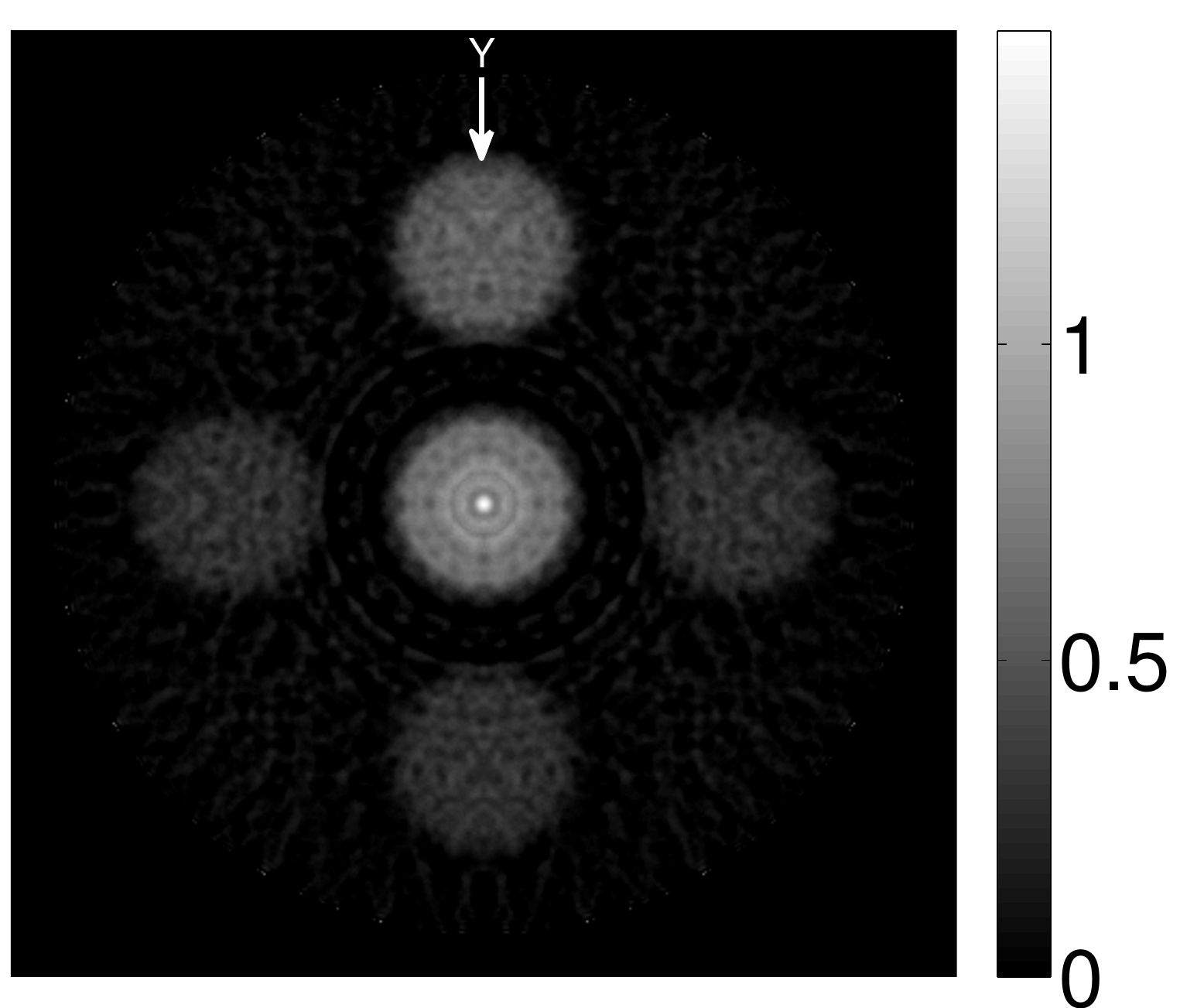}
  \caption{$\lambda=0$}\label{0ns_incomplete_wrong_full}
 \end{subfigure}
 \begin{subfigure}[!htb]{0.32\textwidth}
 \includegraphics[width=\textwidth]{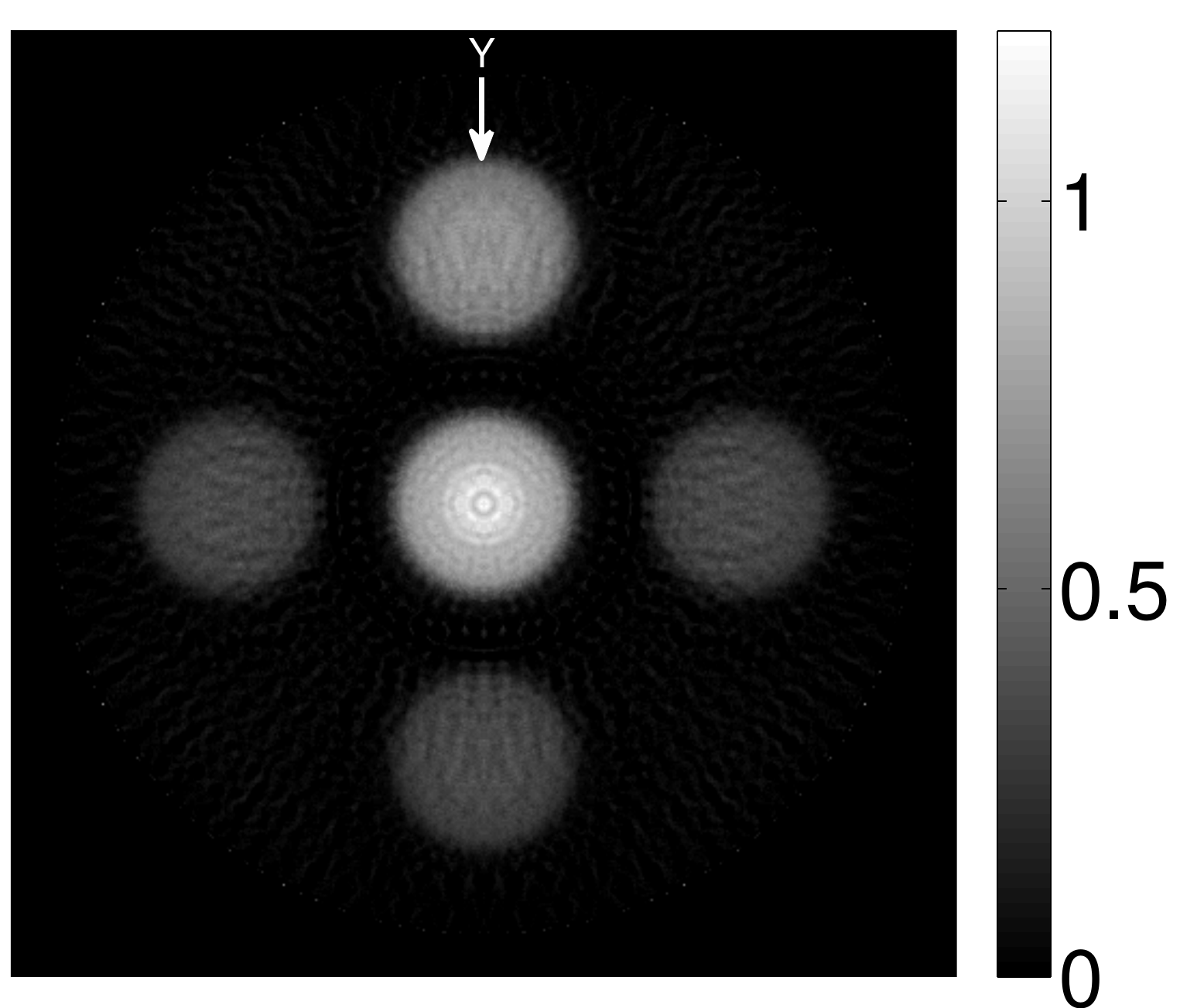}
  \caption{$\lambda=0$}\label{0ns_incomplete_wrong_half}
 \end{subfigure}
 \begin{subfigure}[!htb]{0.33\textwidth}
 \includegraphics[width=\textwidth]{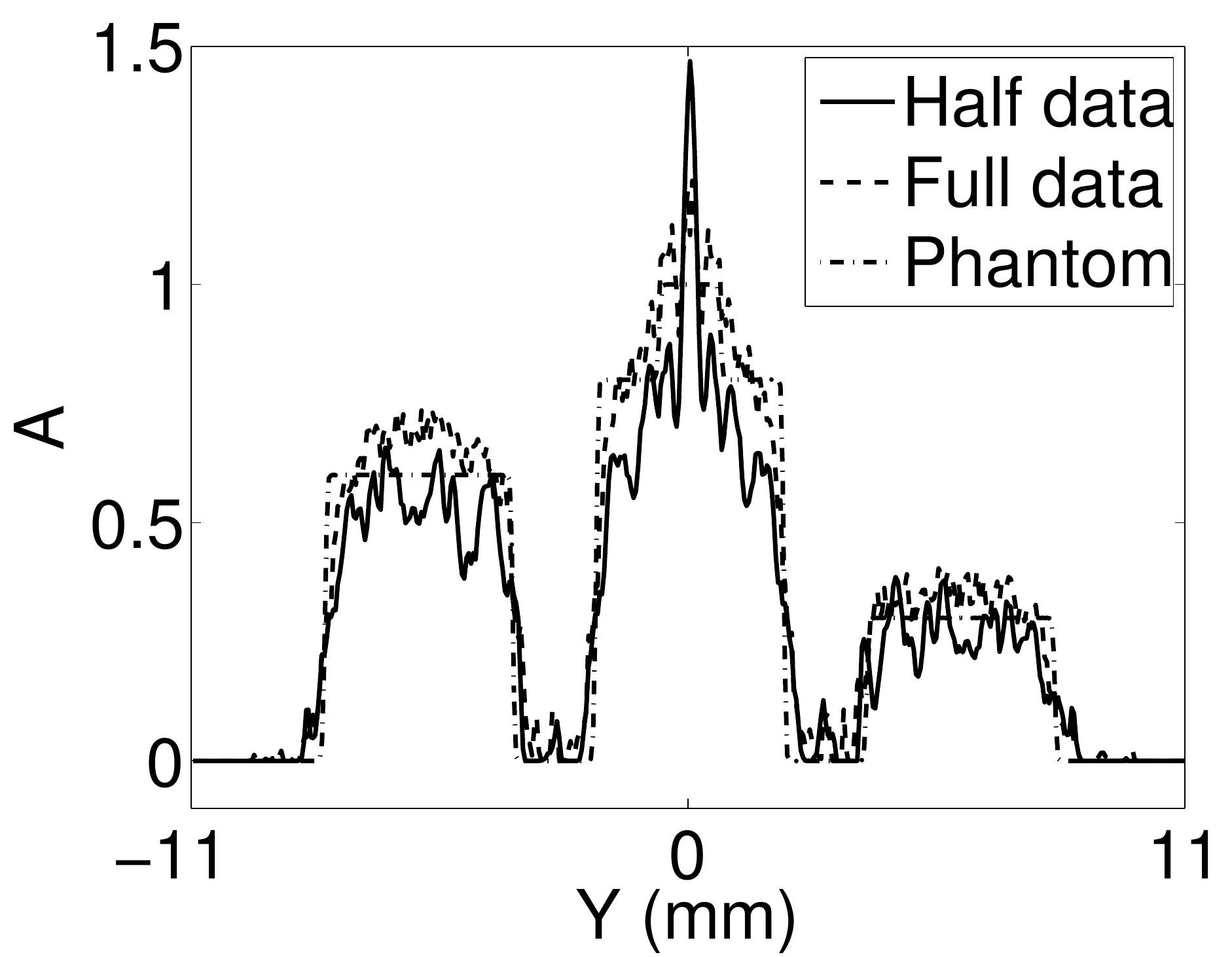}
  \caption{}\label{0ns_incomplete_wrong_plot}
 \end{subfigure}
 \begin{subfigure}[!htb]{0.32\textwidth}
 \includegraphics[width=\textwidth]{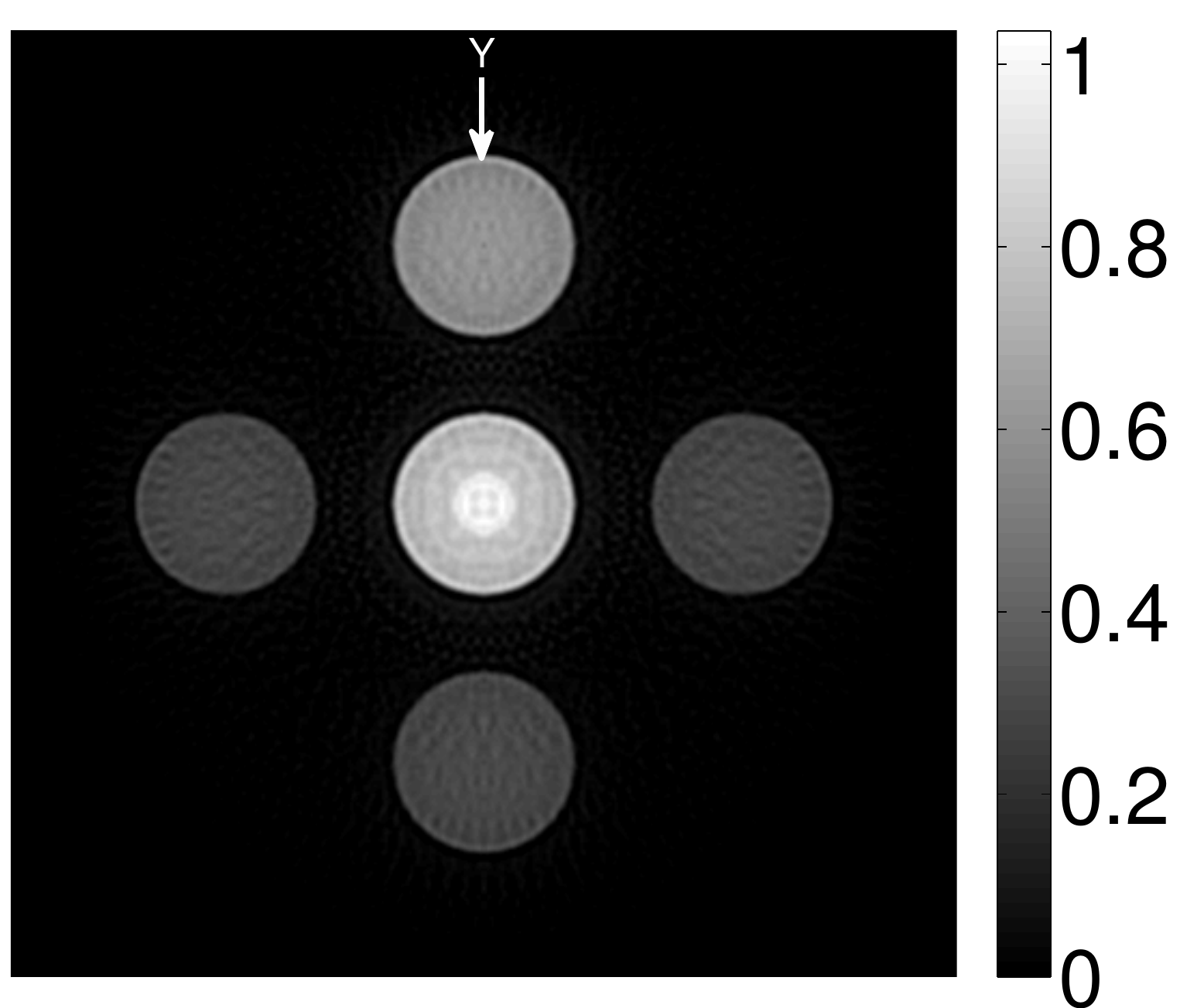}
  \caption{$\lambda=1.0\times 10^{-4}, \alpha=200$}\label{0ns_incomplete_VPM_full}
 \end{subfigure}
 \begin{subfigure}[!htb]{0.32\textwidth}
 \includegraphics[width=\textwidth]{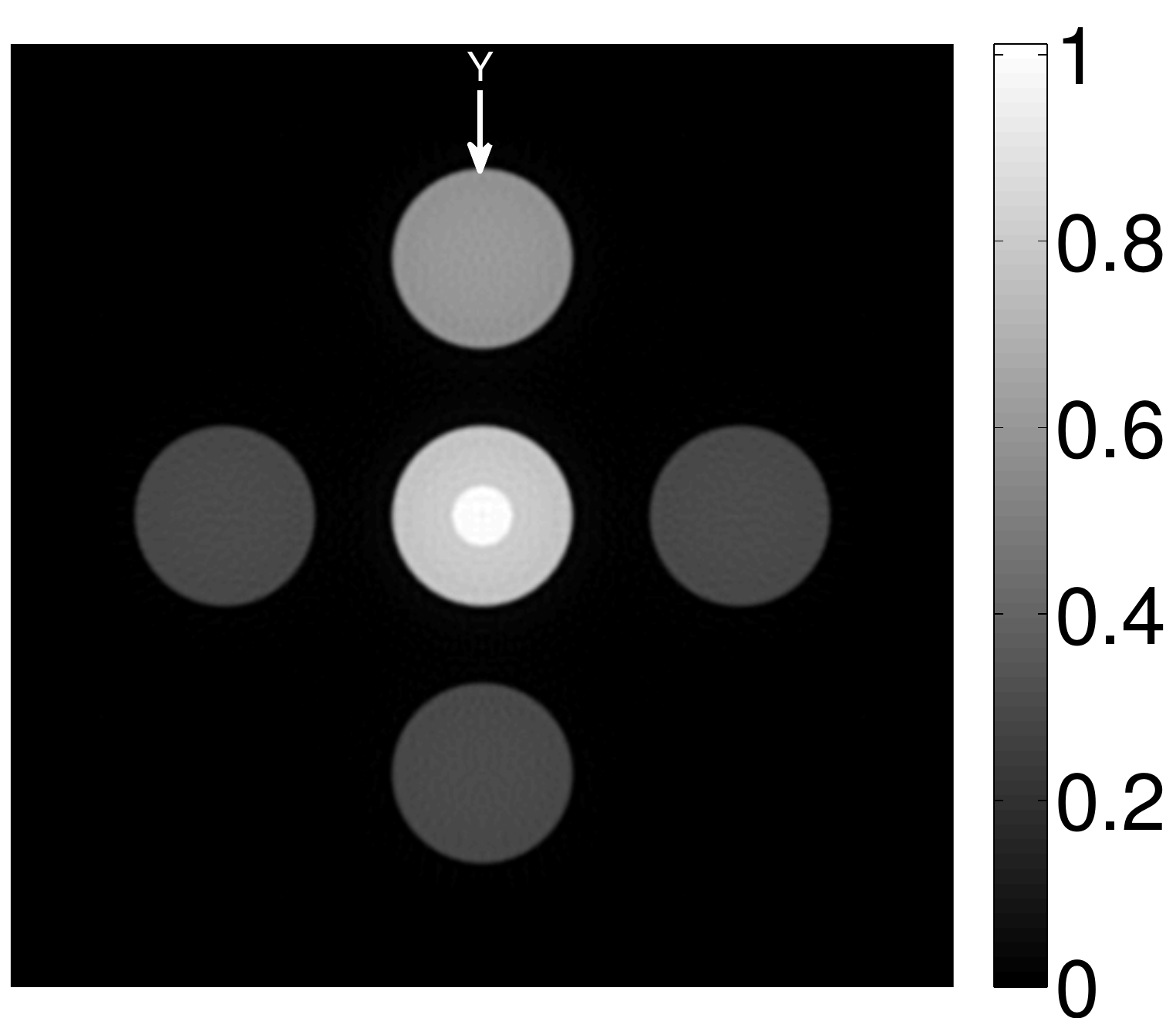}
  \caption{$\lambda=1.0\times 10^{-3}, \alpha=20000$}\label{0ns_incomplete_VPM_half}
 \end{subfigure}
 \begin{subfigure}[!htb]{0.34\textwidth}
 \includegraphics[width=\textwidth]{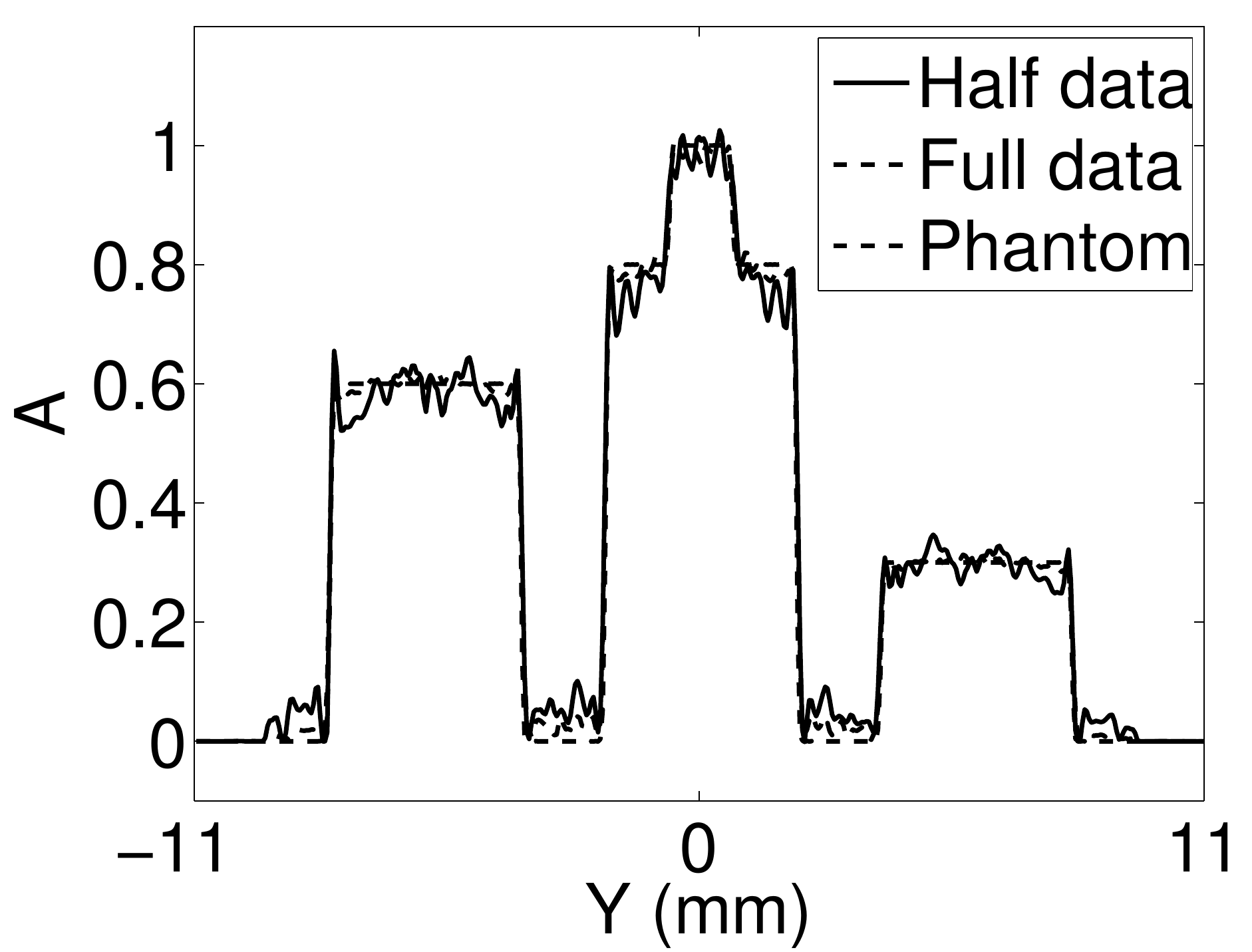}
  \caption{}\label{0ns_incomplete_VPM_plot}
 \end{subfigure}
  \caption{Images reconstructed from incomplete data sets. The first row shows the images reconstructed using the true EIR and the conventional iterative method. The second row shows the images reconstructed using the wrong EIR and the conventional iterative method. The third row shows the images reconstructed using the VP algorithm. Images from left to right in each row are reconstructed using incomplete (half) data, full data, and the profile plots, respectively. The locations of the profiles are indicated by the ``Y'' arrows in the images.}\label{0ns_incomplete}
\end{figure}

Incomplete, or sparsely sampled, data sets are sometimes
 acquired in practice. 
 To study the effect of data incompleteness on the VP algorithm,
  we reconstructed images from data corresponding to half of the equally spaced transducers ($Q=64$).
Because the data were noiseless, no explicit regularization was employed ($\lambda=0$) in the
 conventional reconstruction algorithm.
 However, the explicit regularization was still employed in the VP algorithm because of the ill-posed nature of the joint reconstruction problem. The results are shown in Figure~\ref{0ns_incomplete}.
As expected, use of the incomplete data set resulted in less accurate reconstructed images for both the conventional iterative reconstruction method and the VP algorithm. However, this effect was more pronounced for the VP algorithm. Note that for the VP algorithm, larger values of the regularization parameters were applied when the incomplete data set was employed than when the complete data set was employed (Figure~\ref{0ns_incomplete_VPM_half} and \ref{0ns_incomplete_VPM_full}).

 \subsubsection{Effect of initial estimate of EIR}
The robustness of the VP algorithm with respect to perturbations in the EIR was investigated.
 Perturbed EIRs were generated by adding different levels of random noise to the low frequency components (first 10\% of the total bandwidth, except for the DC component) of the true EIR. The similarity of a perturbed EIR to the true EIR
 was quantified by the correlation coefficient, which is defined by
 \begin{equation}
  \rho
  =\frac{(\mathbf{h}_1-\mu_{\mathbf{h}_1})^T(\mathbf{h}_2-\mu_{\mathbf{h}_2})}{\sigma_{\mathbf{h}_1}\sigma_{\mathbf{h}_2}I},
 \end{equation}
where $\sigma_{\mathbf{h}_k}$ is the standard deviation of $\mathbf{h}_k$, $\mu_{\mathbf{h}_k}$ is the mean of $\mathbf{h}_k$, $k=1,2$, and $I$ is the length of $\mathbf{h}_1$ and $\mathbf{h}_2$. The value of $\rho$ ranges from $-1$ to $1$. The maximum value of $\rho$ is achieved when one EIR is  linear with respect to the other EIR with a positive slope (i.e., $\mathbf{h}_1=a\mathbf{h}_2+b$, for some constant $a>0$ and $b$), which indicates that the two EIRs are `identical' to each other in terms of similarity. On the other hand, $\rho$ equals $-1$ when $\mathbf{h}_1=a\mathbf{h}_2+b$, for some constant $a<0$ and $b$. 

\begin{figure}[!htb]
 \centering
 \begin{subfigure}[]{0.33\textwidth}
  \includegraphics[width=\textwidth]{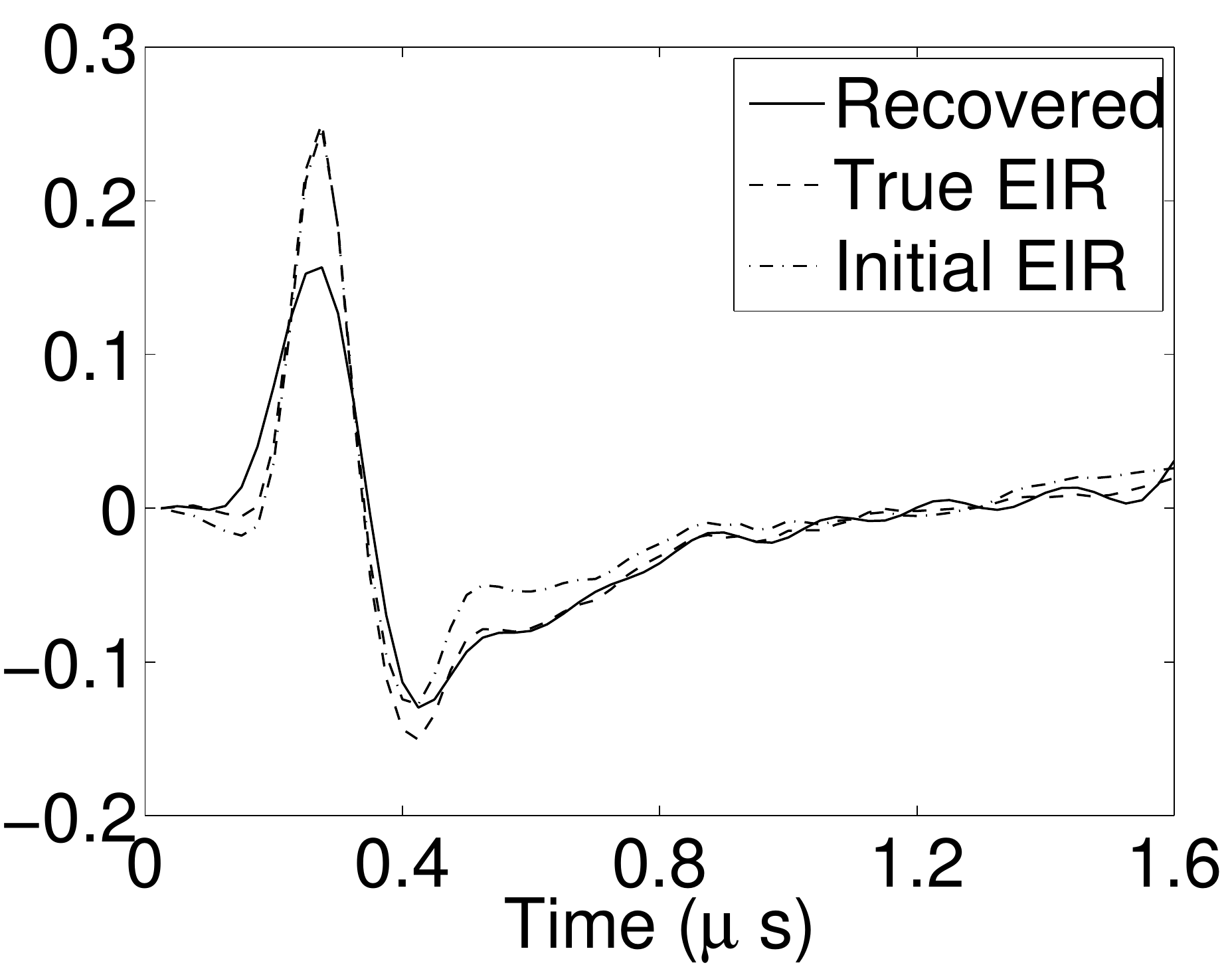}
  \caption{$\rho=0.9914$}\label{0ns_initGuess_180_EIR}
 \end{subfigure}
 \begin{subfigure}[]{0.30\textwidth}
  \includegraphics[width=\textwidth]{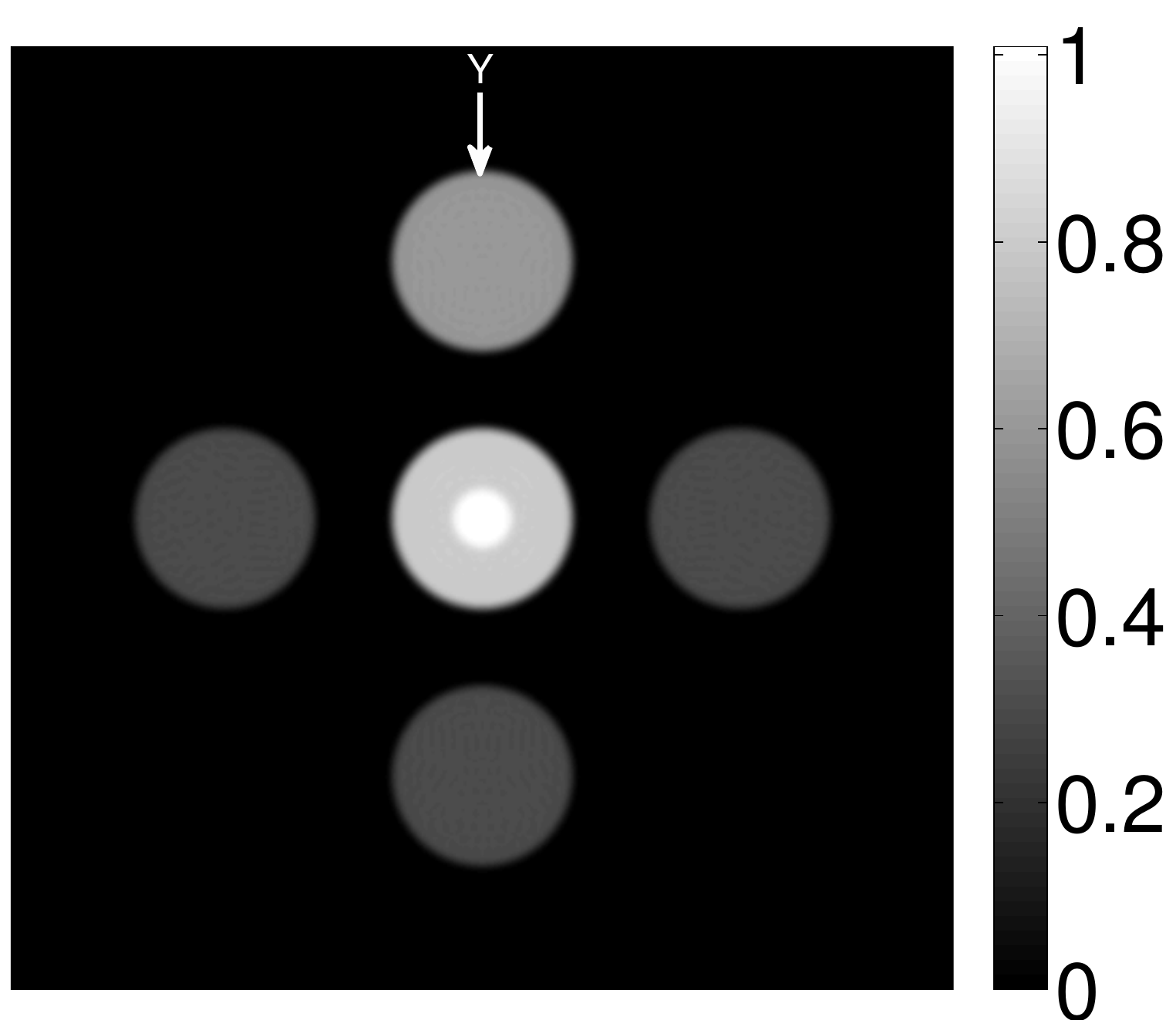}
  \caption{$\lambda=10^{-5},\alpha=2000$}\label{0ns_initGuess_180}
 \end{subfigure}

 


  \begin{subfigure}[]{0.33\textwidth}
  \includegraphics[width=\textwidth]{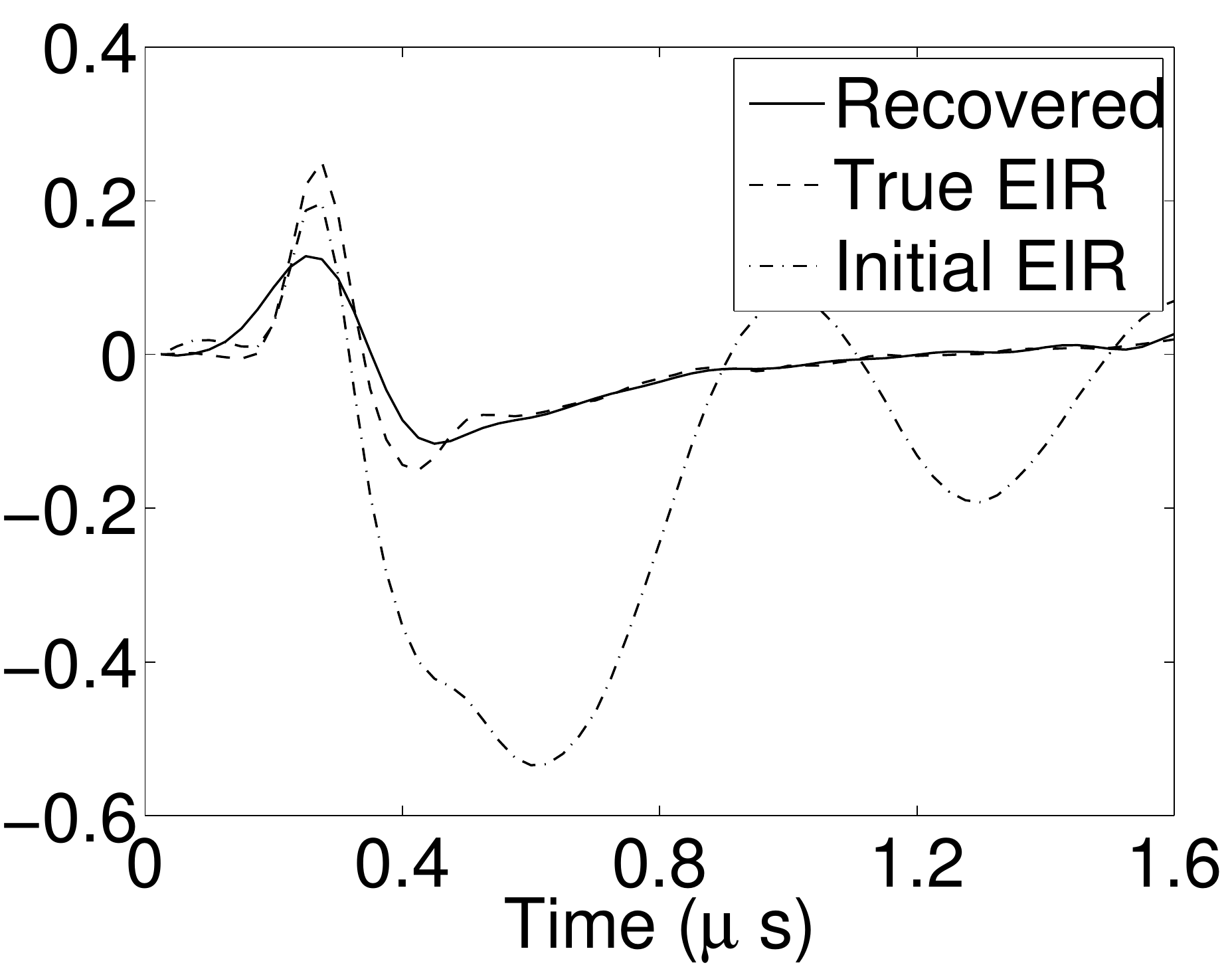}
  \caption{$\rho=0.7271$}\label{0ns_initGuess_3000_EIR}
 \end{subfigure}
 \begin{subfigure}[]{0.30\textwidth}
  \includegraphics[width=\textwidth]{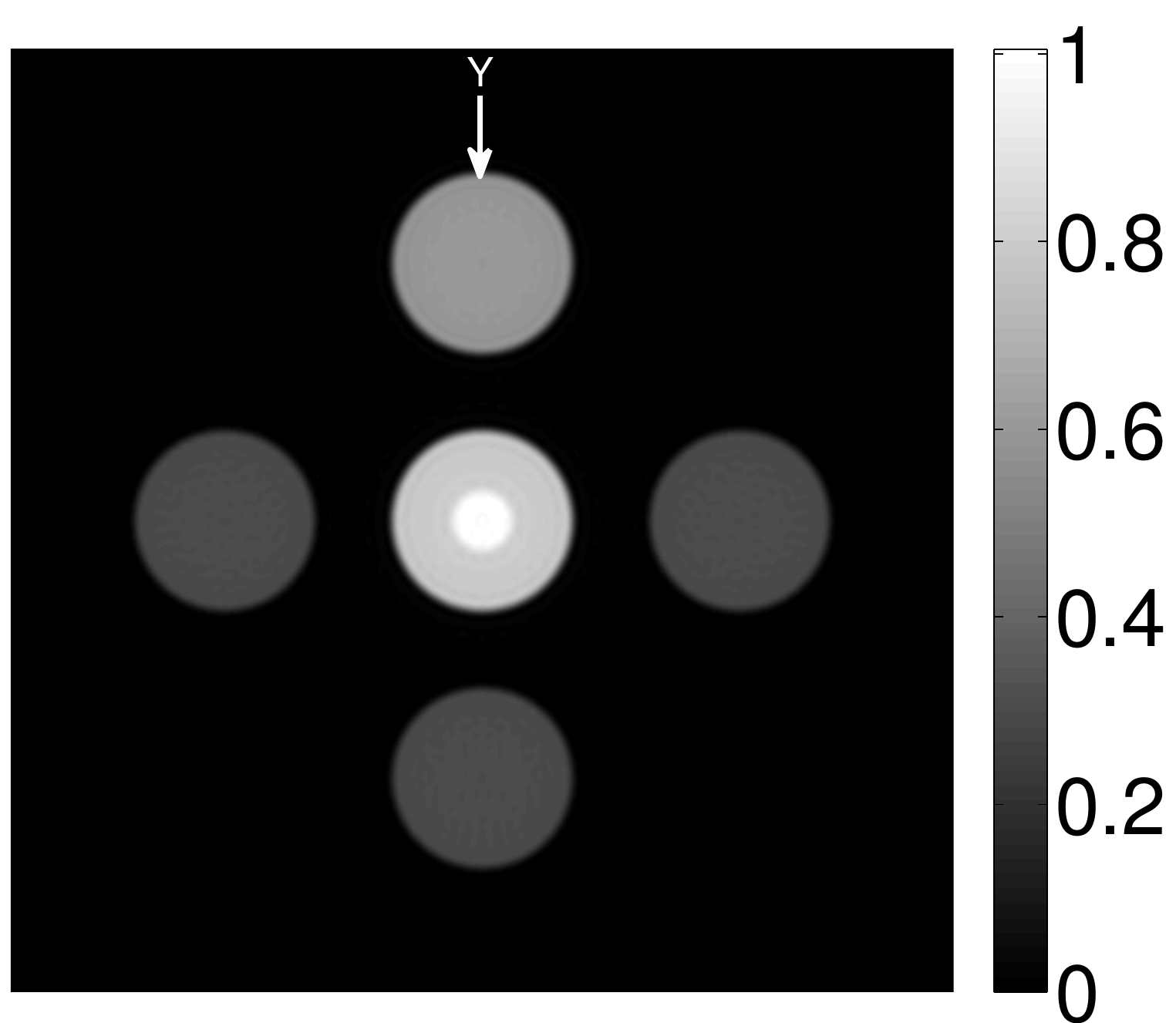}
  \caption{$\lambda=2.0\times 10^{-3},\alpha=2000$}\label{0ns_initGuess_3000}
 \end{subfigure}

   \begin{subfigure}[]{0.33\textwidth}
  \includegraphics[width=\textwidth]{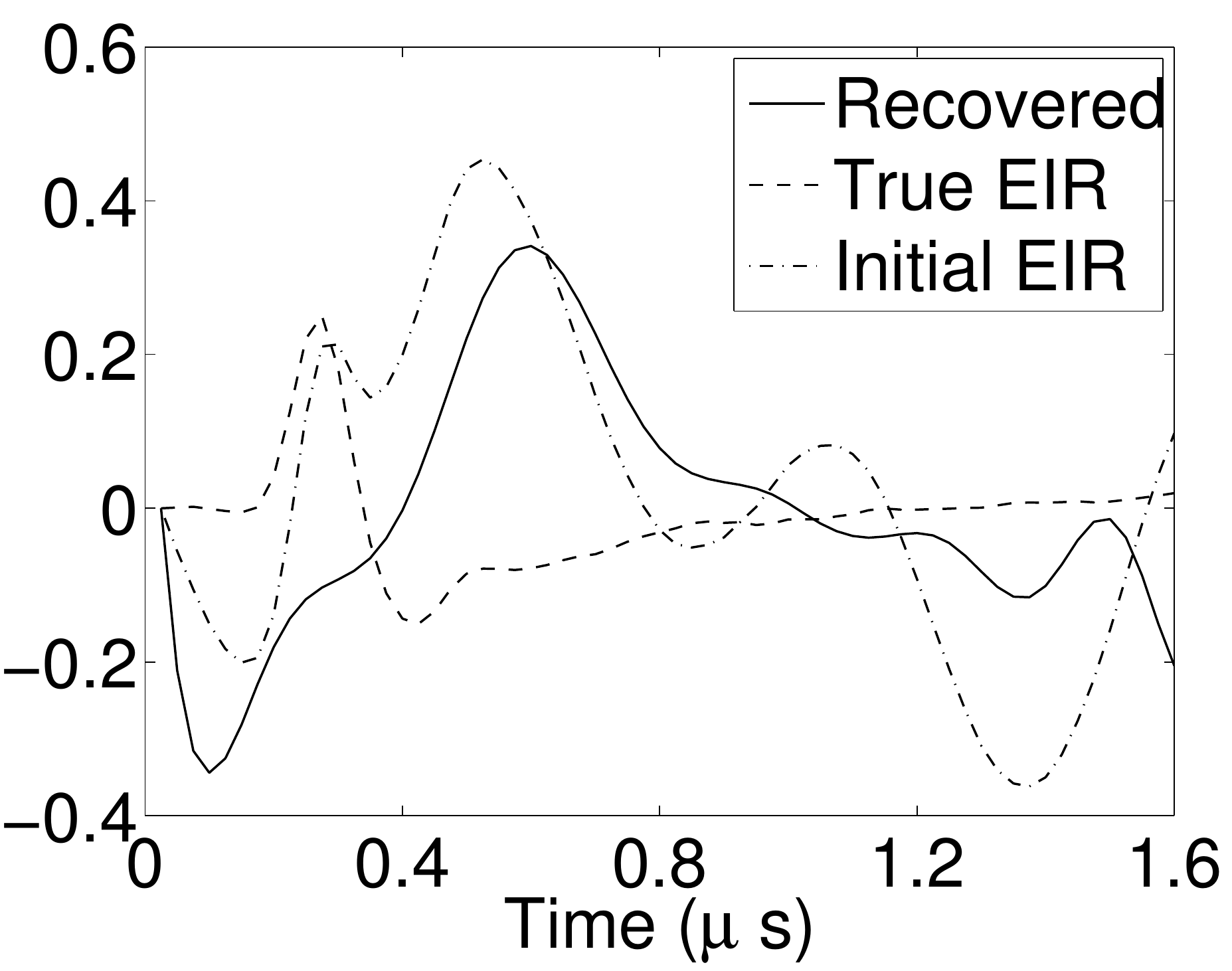}\vfill
  \caption{$\rho=-0.3177$}\label{0ns_initGuess_4000_EIR}
 \end{subfigure}
 \begin{subfigure}[]{0.30\textwidth}
  \includegraphics[width=\textwidth]{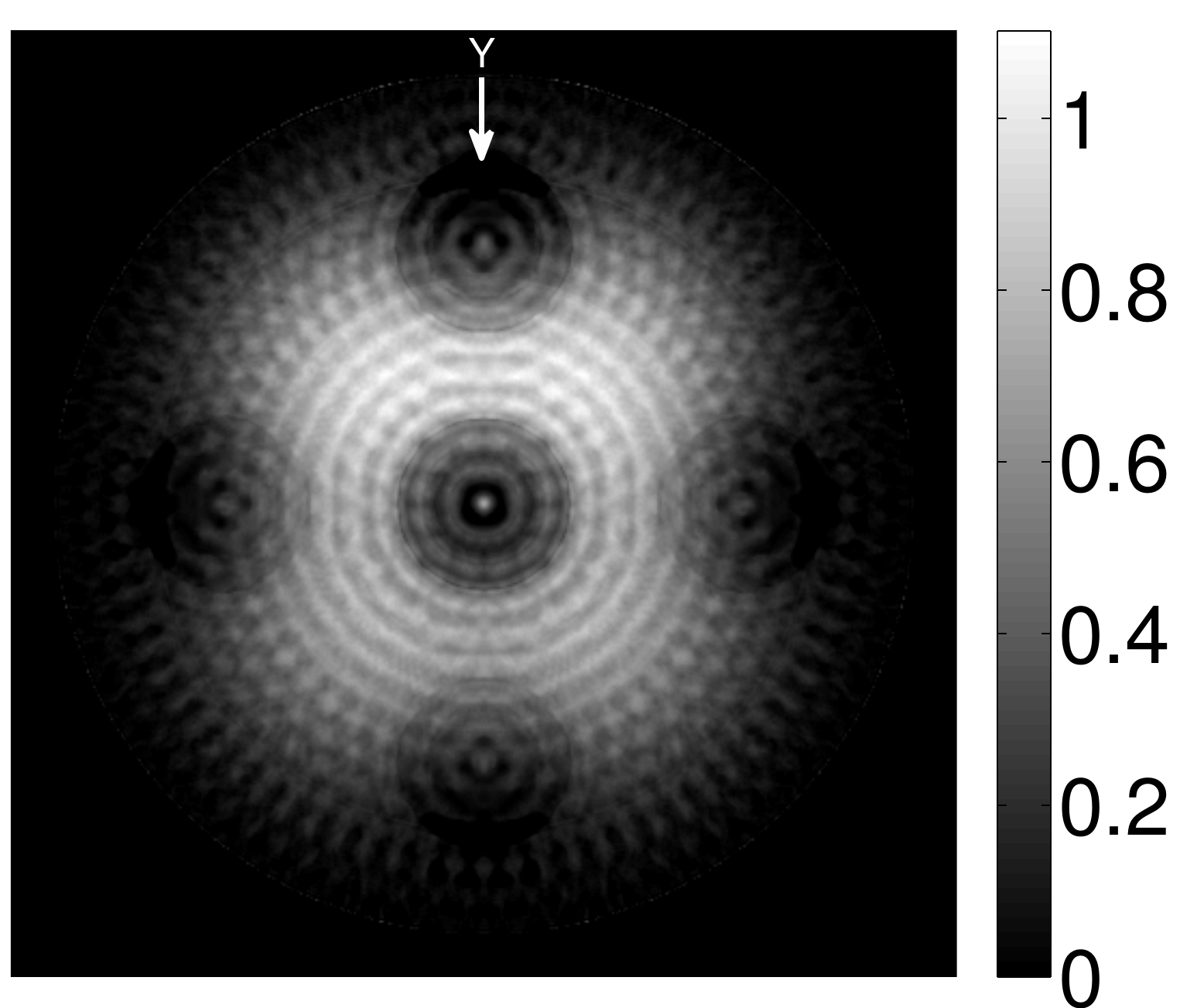}
  \caption{$\lambda=10^{-4},\alpha=2000$}\label{0ns_initGuess_4000}
 \end{subfigure}
 \caption{Images reconstructed by use of different initial guesses of EIR. Each row shows the results using the same initial EIR. The first column shows the true , initial , and recovered EIR. The second column shows the images reconstructed using the VP algorithm. 
 }\label{0ns_initGuess}
\end{figure}
As shown in  Figure~\ref{0ns_initGuess}, when the error in the EIR was small (e.g., as with the  EIR in Figure~\ref{0ns_initGuess_180_EIR}), images were reconstructed with high accuracy using the VP algorithm.  When the perturbations in the EIR were stronger
(e.g, as in Figure~\ref{0ns_initGuess_3000_EIR}), artifacts and distortions in the reconstructed images were still significantly reduced by use of the VP algorithm; however, larger values of the  regularization parameters had to be applied. When $\rho<0$ as in the initial EIR in Figure~\ref{0ns_initGuess_4000_EIR}, no improvement was observed in the image reconstructed by use of the VP algorithm. 

 \subsection{Images reconstructed from noisy data}


\subsubsection{Mitigation of artifacts and distortions caused by an inaccurate EIR}
\begin{figure}[!htb]
 \centering
 \begin{subfigure}[]{0.32\textwidth}
 \includegraphics[width=\textwidth]{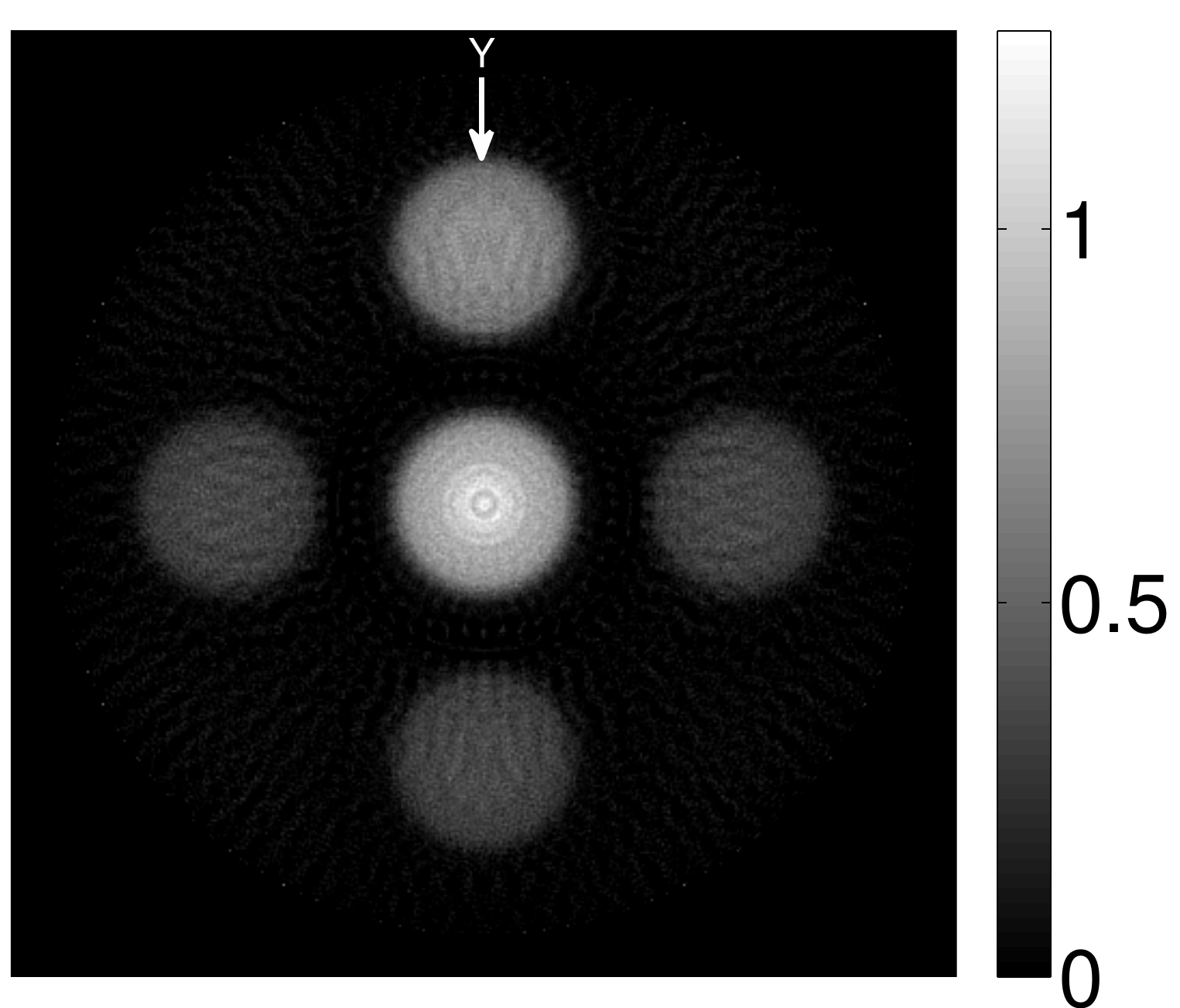}
  \caption{}\label{2d_3ns_B2_NoVPM}
 \end{subfigure}
  \begin{subfigure}[]{0.32\textwidth}
 \includegraphics[width=\textwidth]{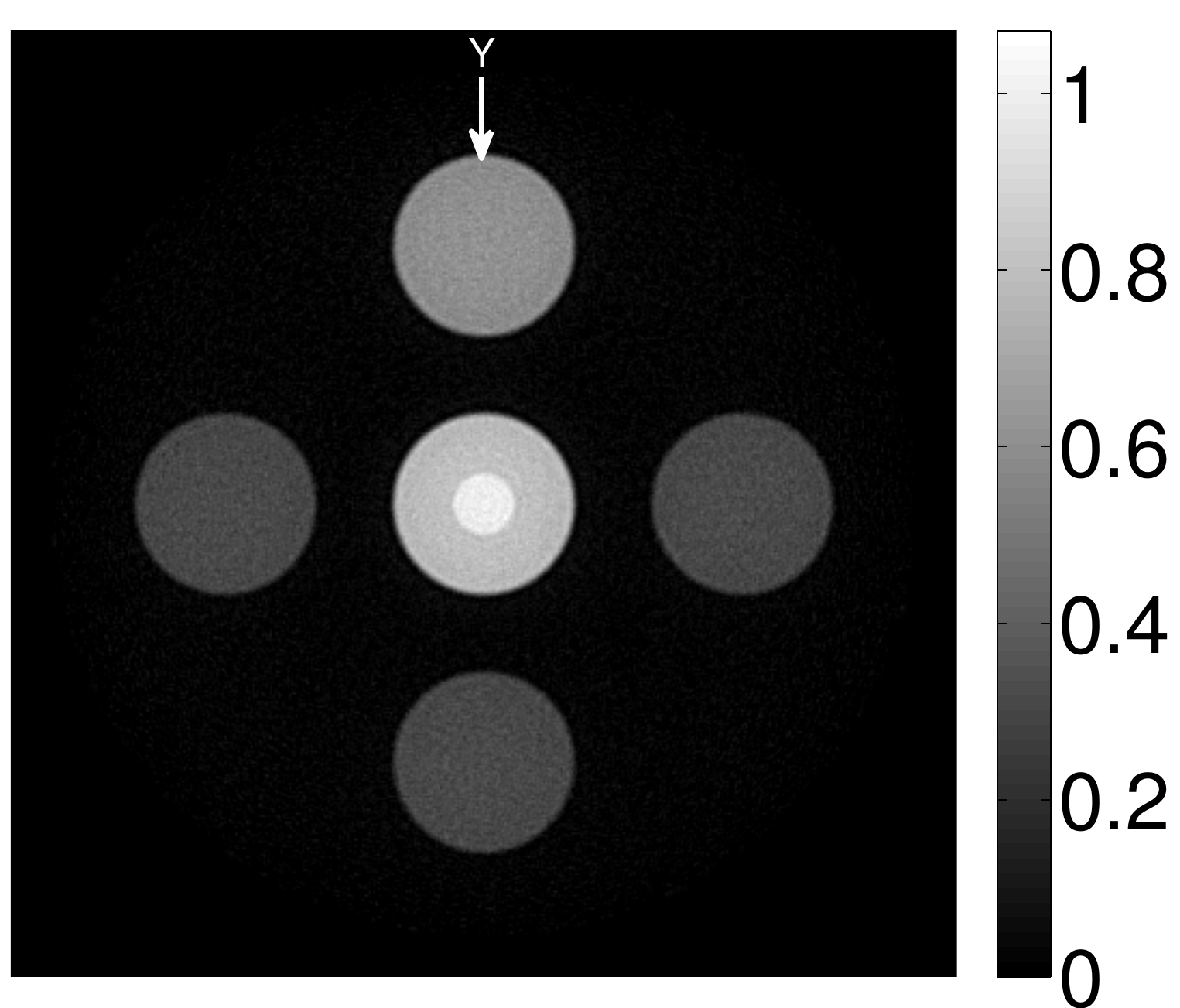}
  \caption{}\label{2d_3ns_B2_VPM}
 \end{subfigure}
 \begin{subfigure}[]{0.32\textwidth}
 \includegraphics[width=\textwidth]{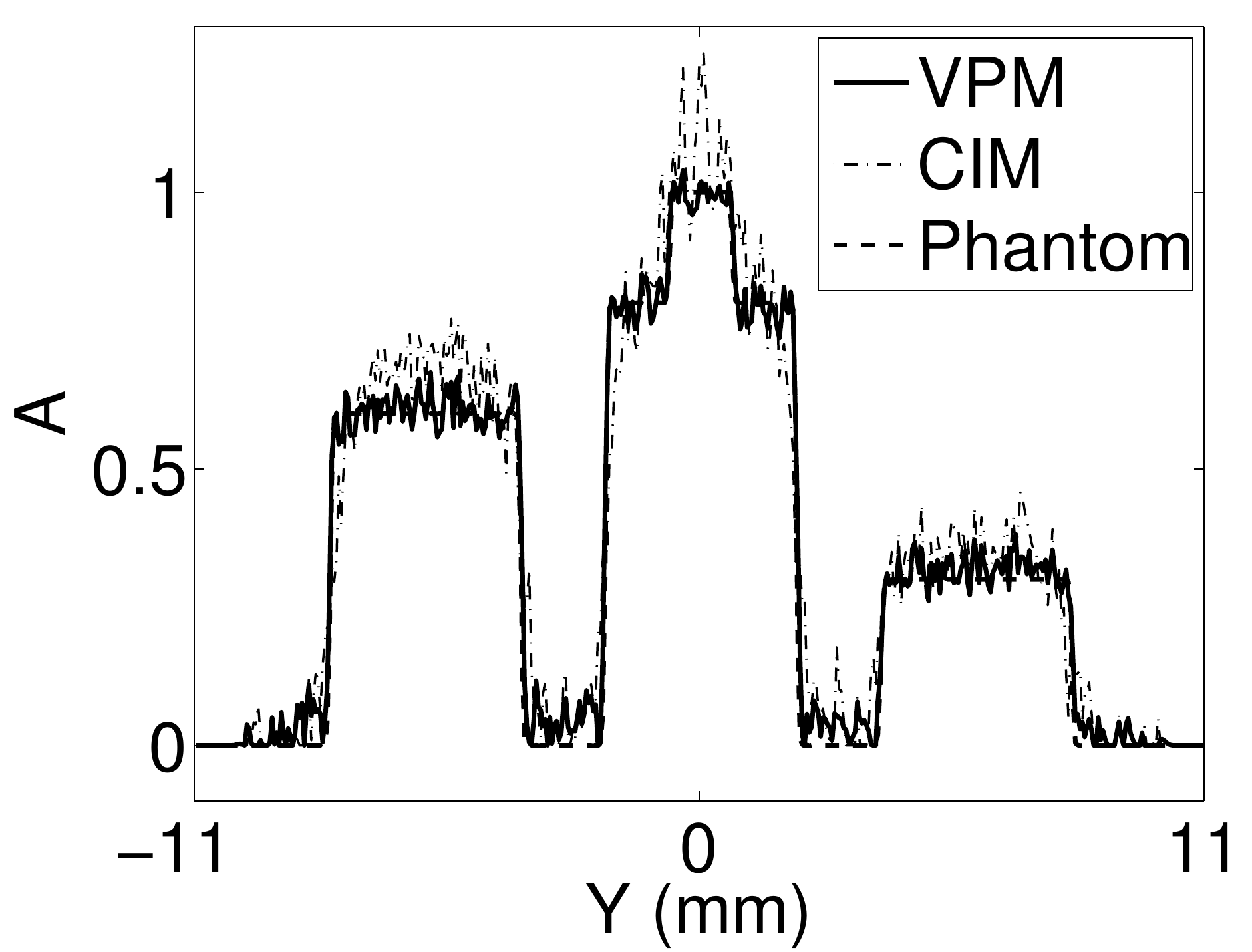}
  \caption{}\label{3ns_B2_VPM_plots}
 \end{subfigure}

 \begin{subfigure}[!htb]{0.45\textwidth}
 \includegraphics[width=\textwidth]{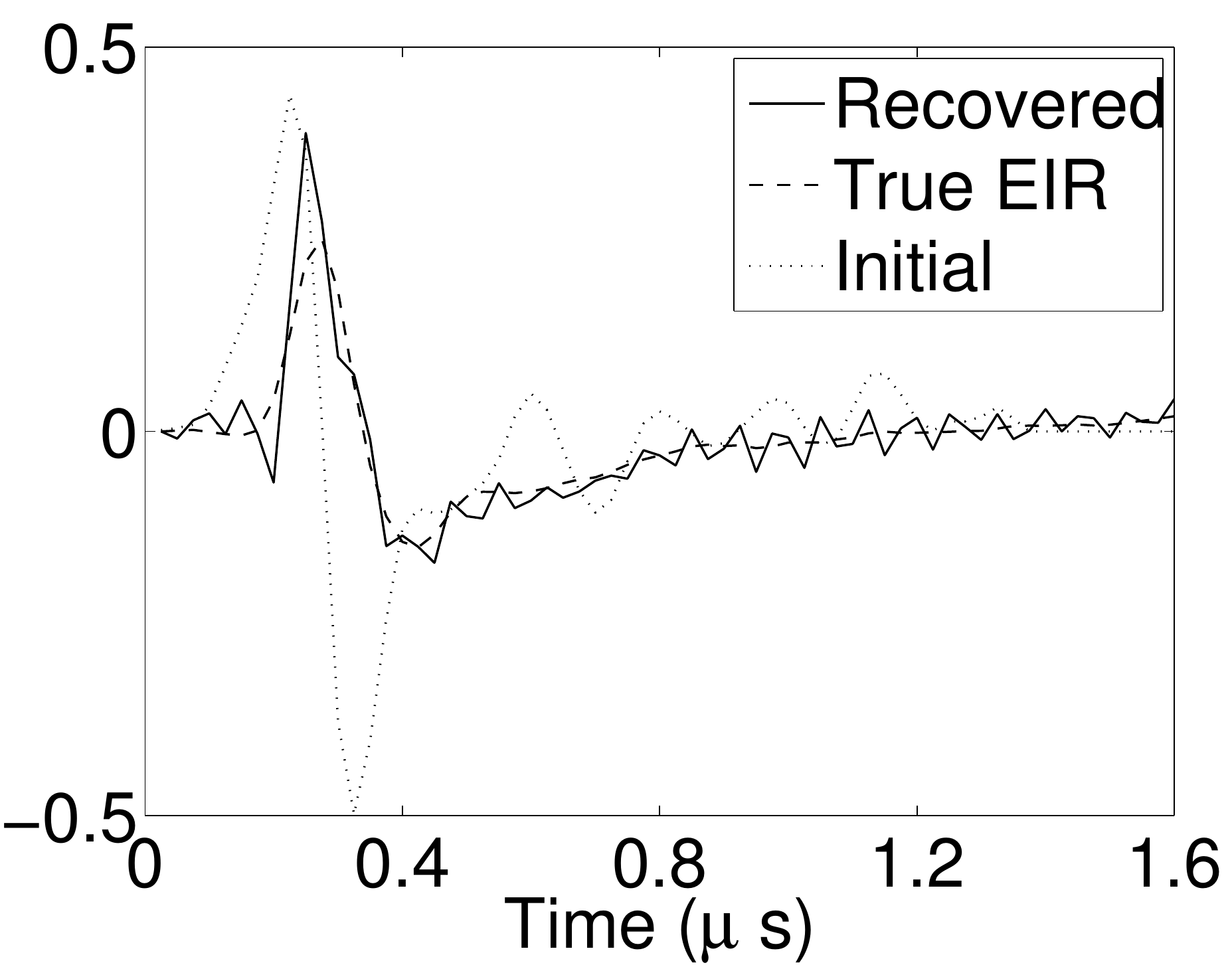}
 \caption{}\label{2d_3ns_B2_eir}
 \end{subfigure}
  \begin{subfigure}[!htb]{0.45\textwidth}
 \includegraphics[width=\textwidth]{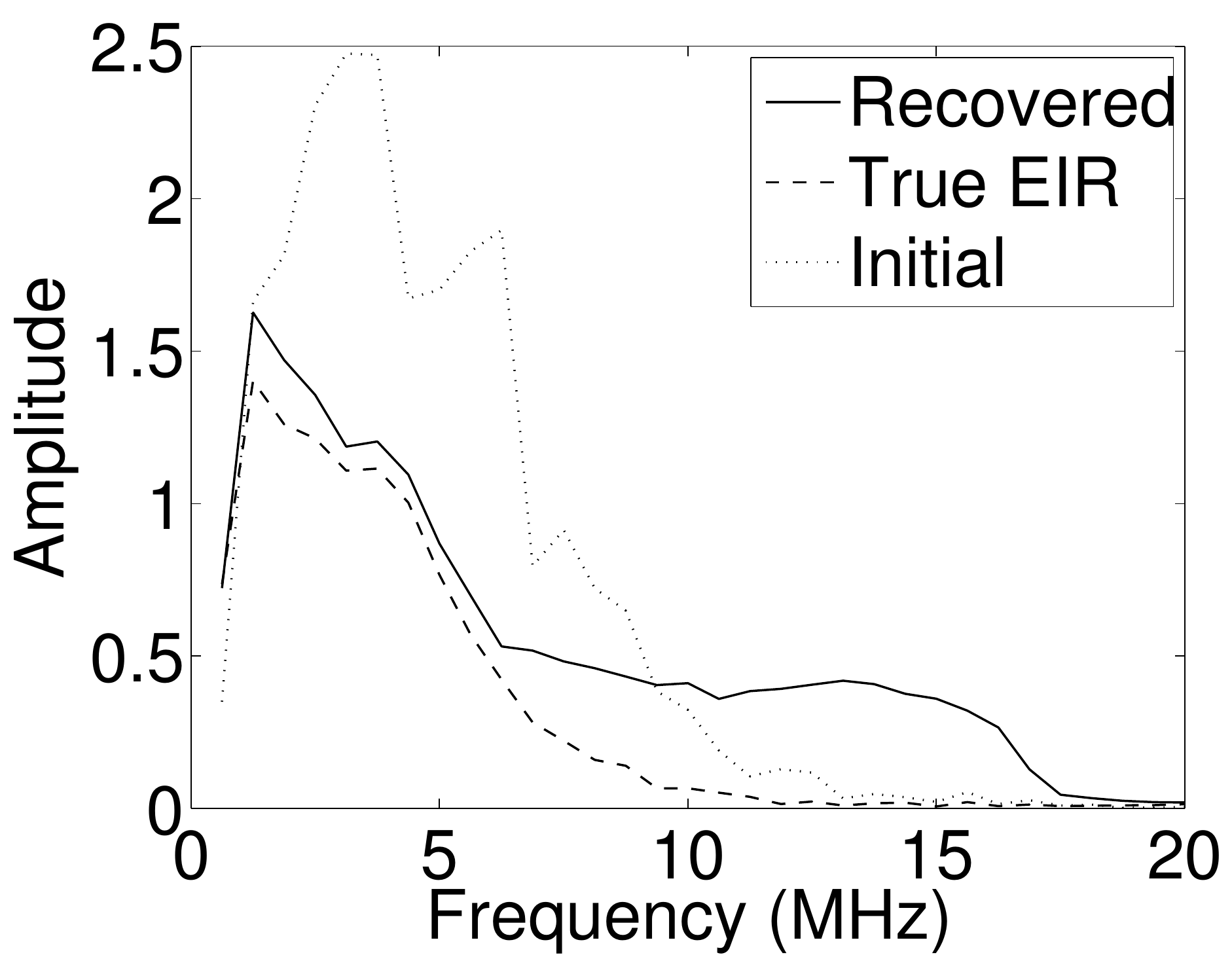}
 \caption{}\label{2d_3ns_B2_eir_freq}
 \end{subfigure}
 \caption{(a) and (b) Images reconstructed from noisy data using the conventional iterative method and using the VP algorithm, respectively. (c) Image profiles through the noisy images in \rd{Fig.~\ref{2d_3ns_B2_NoVPM} and \ref{2d_3ns_B2_VPM}}, corresponding to the reconstructions using the VP algorithm (solid line), the conventional iterative method (dash-dot line), and the phantom (dashed line). The locations of the profiles are indicated by the ``Y'' arrows in the Fig.~\ref{2d_3ns_B2_NoVPM}, and \ref{2d_3ns_B2_VPM}, respectively. (d) The recovered EIR corresponding to the reconstruction using noisy data with the VP algorithm (solid line), the true EIR (dashed line), and the initial EIR (dash-dot line). (e) The spectrums of the corresponding EIRs.}\label{3ns_B2_VPM}
\end{figure}

 Figure~\ref{2d_3ns_B2_NoVPM}
 reveals that use of the inaccurate EIR in the conventional iterative method created strong artifacts and distortions.
 Figures~\ref{2d_3ns_B2_VPM}  confirms that the artifacts
 and distortions were significantly mitigated when the VP method was employed.
Image profiles for both cases are shown in  Figures~\ref{3ns_B2_VPM_plots}.
 The overall accuracy of the recovered EIR, shown in Figure~\ref{2d_3ns_B2_eir} and \ref{2d_3ns_B2_eir_freq},
 was improved, but it contained spurious oscillations. 

\subsubsection{Continuous dependency on regularization parameters} 
\begin{figure}[!htb]
 \centering
 \begin{subfigure}[!htb]{0.325\textwidth}
 \includegraphics[width=\textwidth]{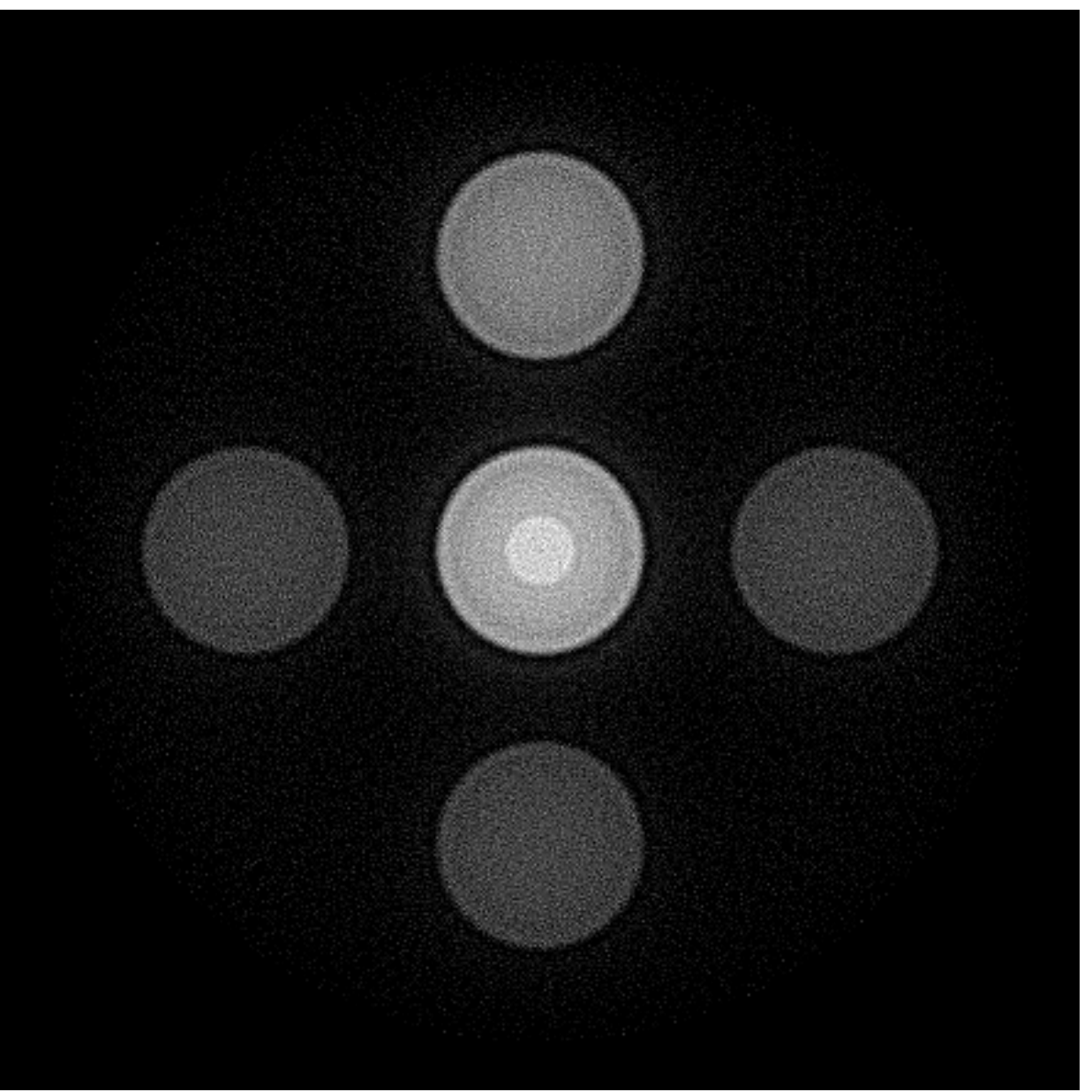}
  \caption{$\lambda=10^{-5}$, $\alpha=2000$, $RMSE=0.0623$}\label{3ns_B2_lam1_5_alp2e3}
 \end{subfigure}
 \begin{subfigure}[!htb]{0.325\textwidth}
 \includegraphics[width=\textwidth]{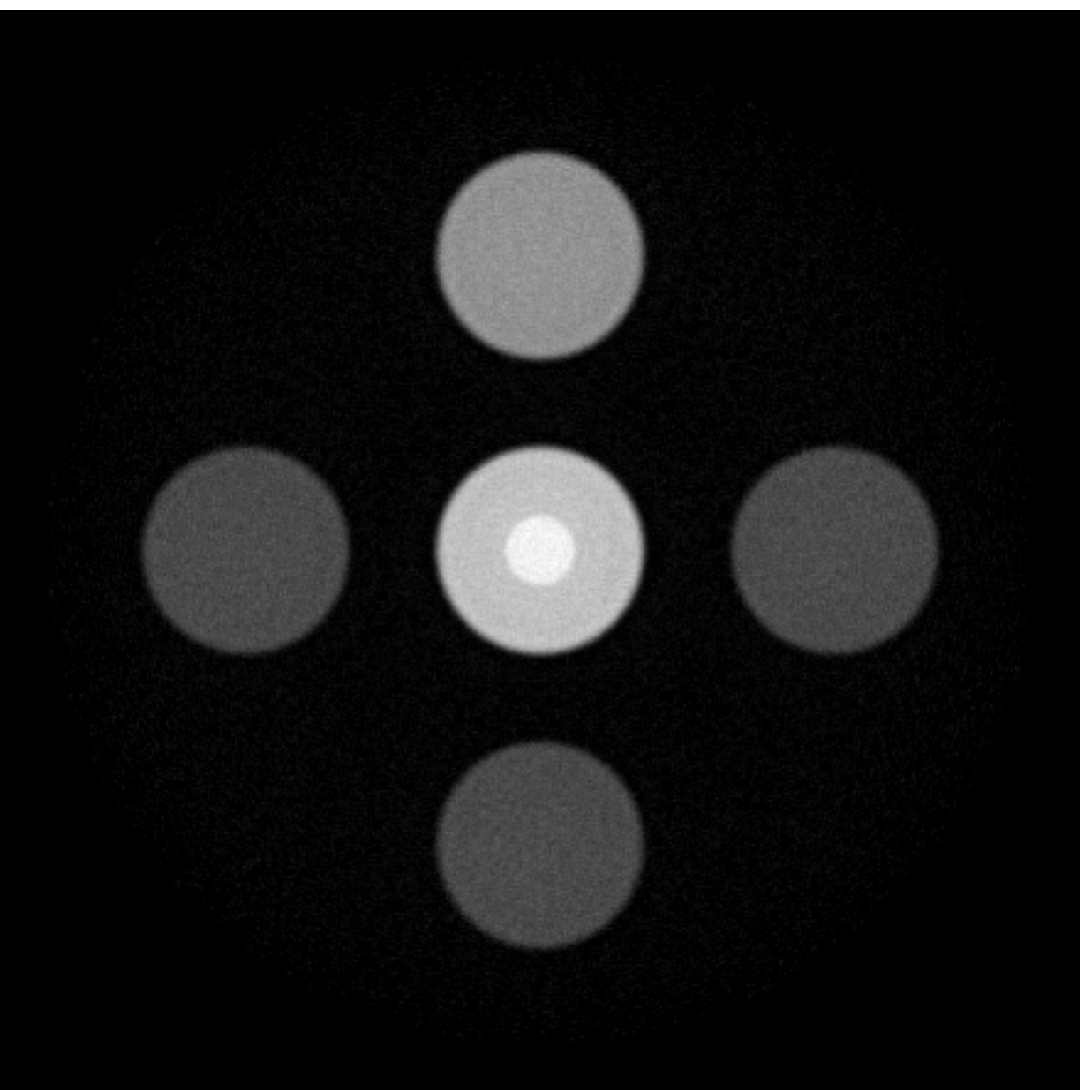}
  \caption{$\lambda=10^{-3}$, $\alpha=2000$, $RMSE=0.0270$}\label{3ns_B2_lam1_3_alp2e3}
 \end{subfigure}
 \begin{subfigure}[!htb]{0.325\textwidth}
 \includegraphics[width=\textwidth]{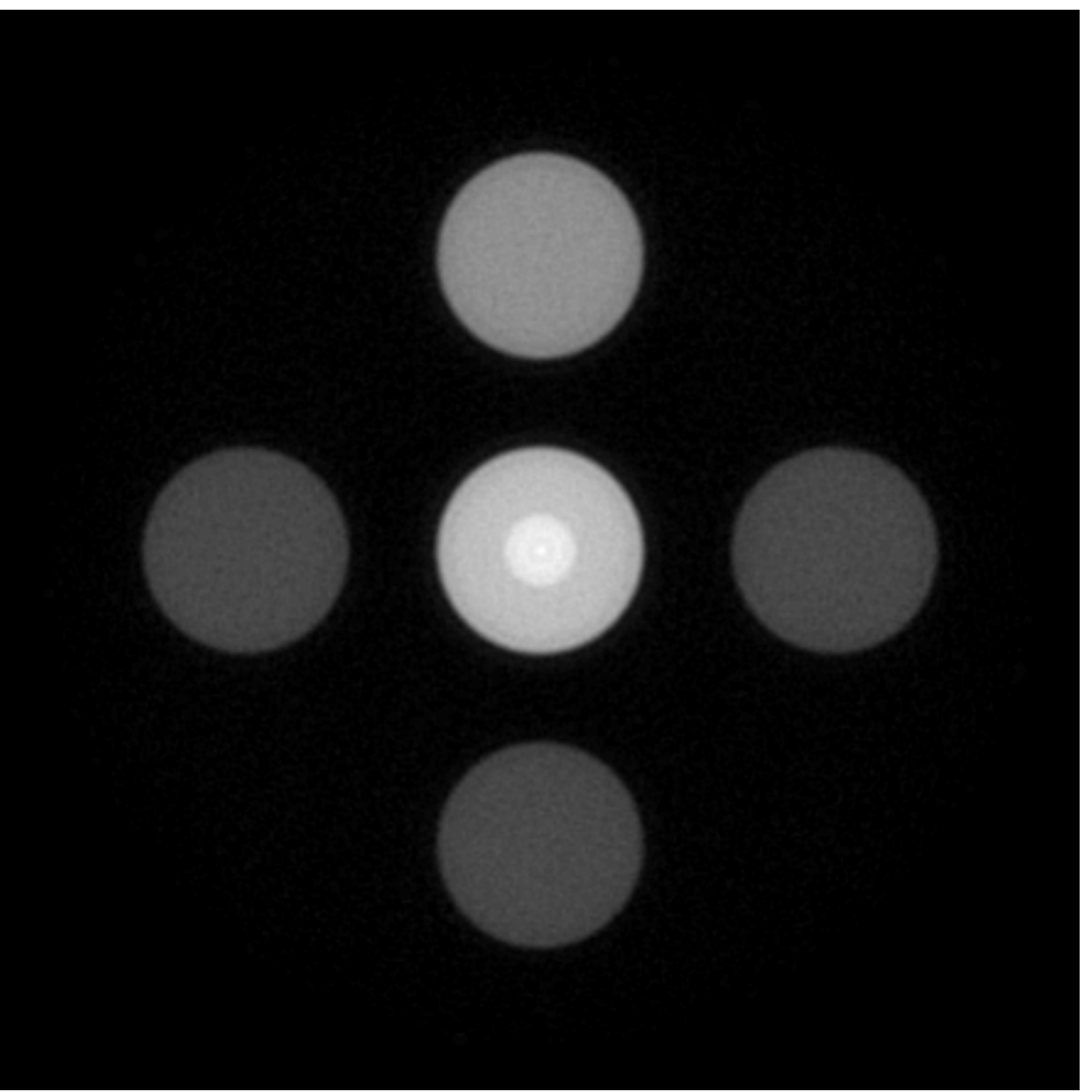}
  \caption{$\lambda=10^{-1}$, $\alpha=2000$, $RMSE=0.0238$}\label{3ns_B2_lam1e_1_alp2e3}
 \end{subfigure}
 \begin{subfigure}[!htb]{0.325\textwidth}
 \includegraphics[width=\textwidth]{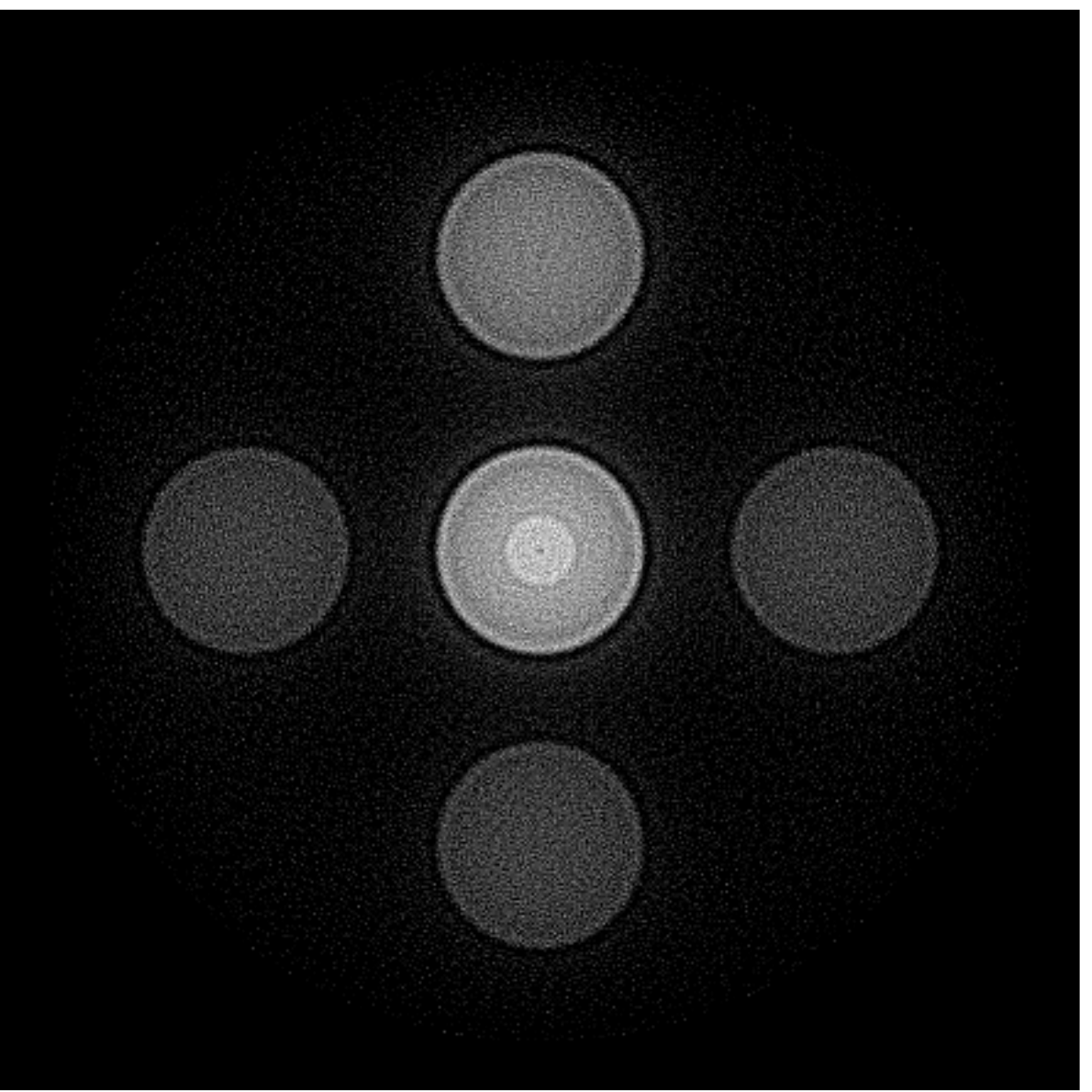}
  \caption{$\lambda=10^{-5}$, $\alpha=20000$, $RMSE=0.0792$}\label{3ns_B2_lam1_5_alp2e4}
 \end{subfigure}
 \begin{subfigure}[!htb]{0.325\textwidth}
 \includegraphics[width=\textwidth]{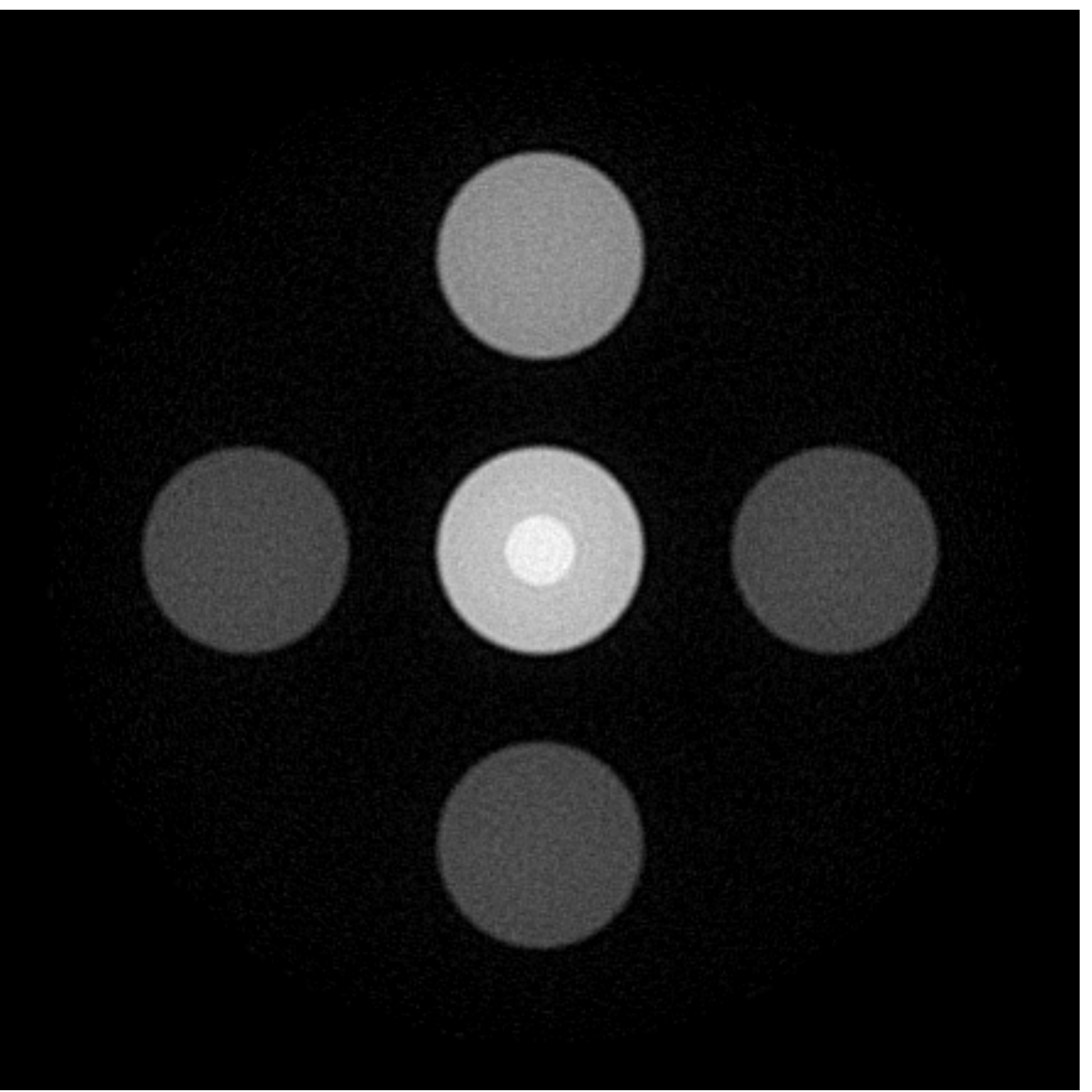}
  \caption{$\lambda=10^{-3}$, $\alpha=20000$, $RMSE=0.0275$}\label{3ns_B2_lam1_3_alp2e4}
 \end{subfigure}
 \begin{subfigure}[!htb]{0.325\textwidth}
 \includegraphics[width=\textwidth]{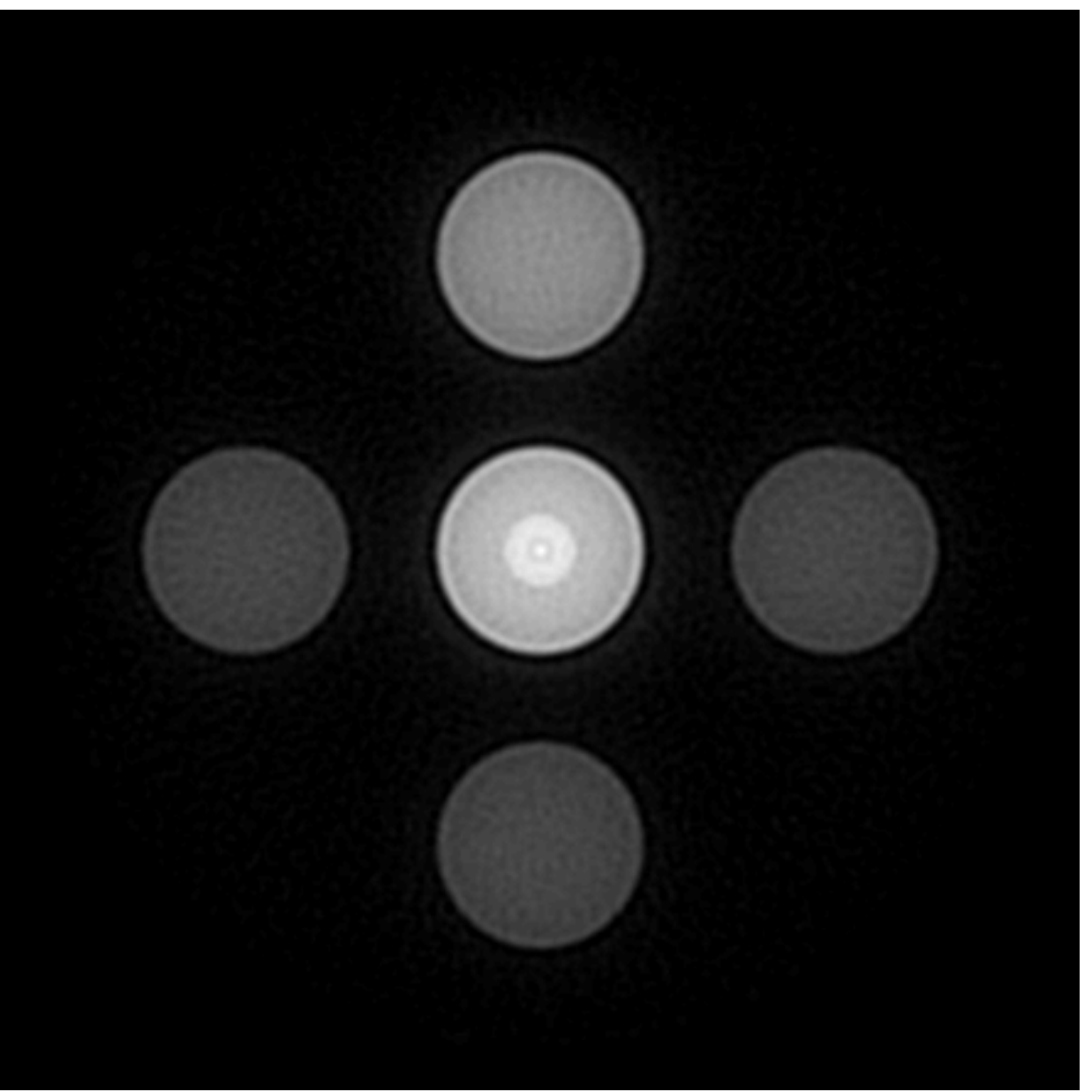}
  \caption{$\lambda=10^{-1}$, $\alpha=20000$, $RMSE=0.0286$}\label{3ns_B2_lam1_1_alp2e4}
 \end{subfigure}
 \begin{subfigure}[!htb]{0.325\textwidth}
 \includegraphics[width=\textwidth]{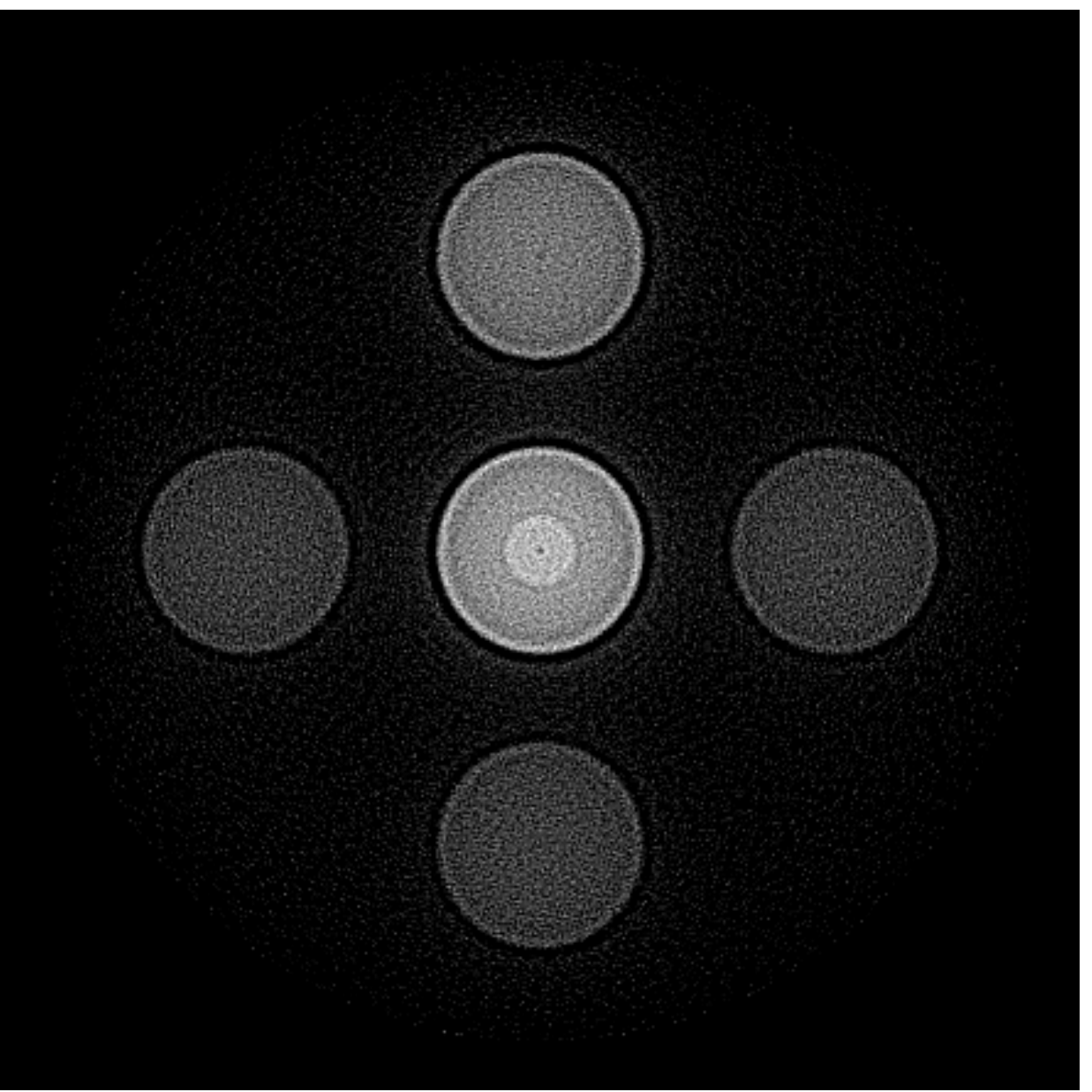}
  \caption{$\lambda=10^{-5}$, $\alpha=2.0\times 10^5$, $RMSE=0.0807$}\label{3ns_B2_lam1_5_alp2e5}
 \end{subfigure}
 \begin{subfigure}[!htb]{0.325\textwidth}
 \includegraphics[width=\textwidth]{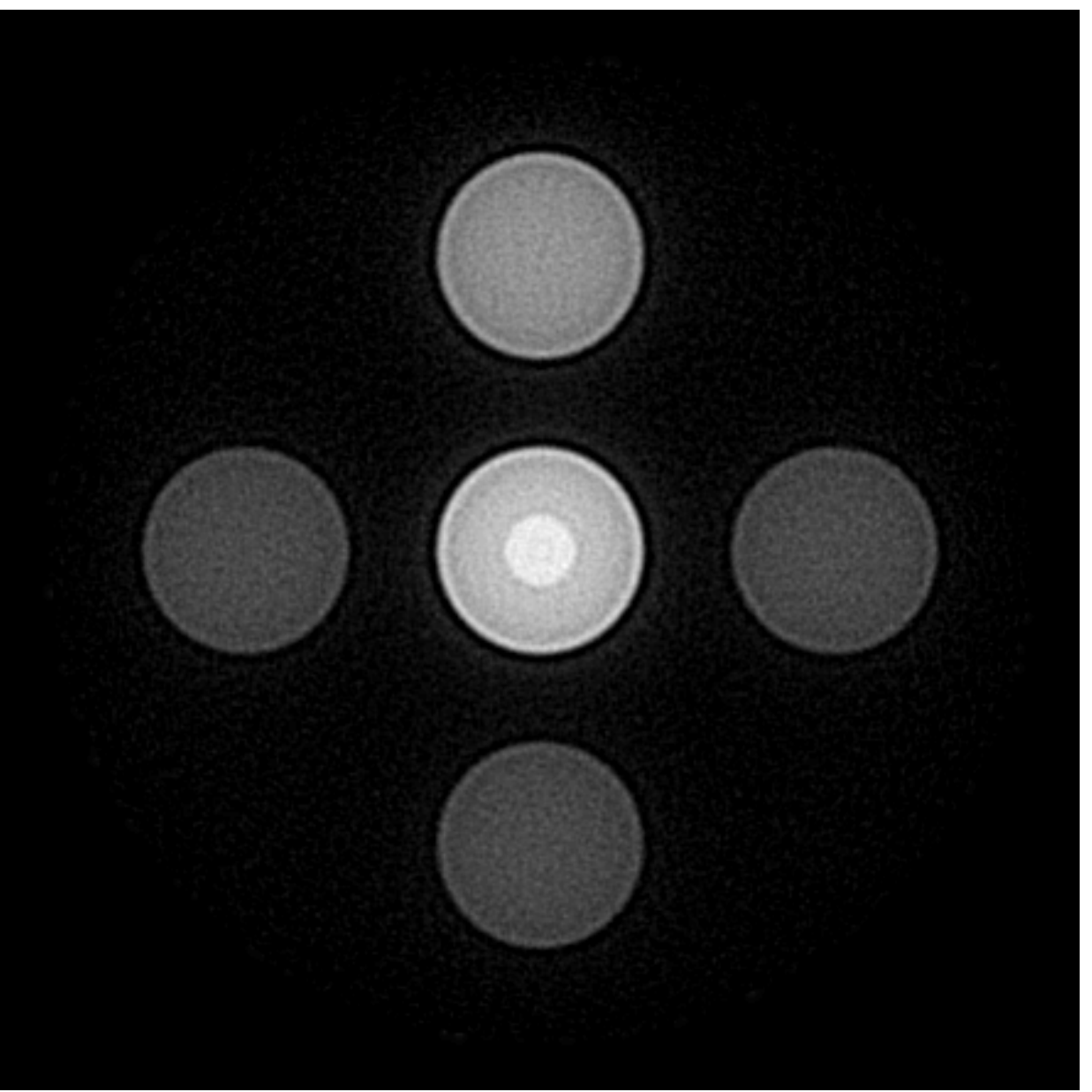}
  \caption{$\lambda=10^{-3}$, $\alpha=2.0\times 10^5$, $RMSE=0.0353$}\label{3ns_B2_lam1_3_alp2e5}
 \end{subfigure}
 \begin{subfigure}[!htb]{0.325\textwidth}
 \includegraphics[width=\textwidth]{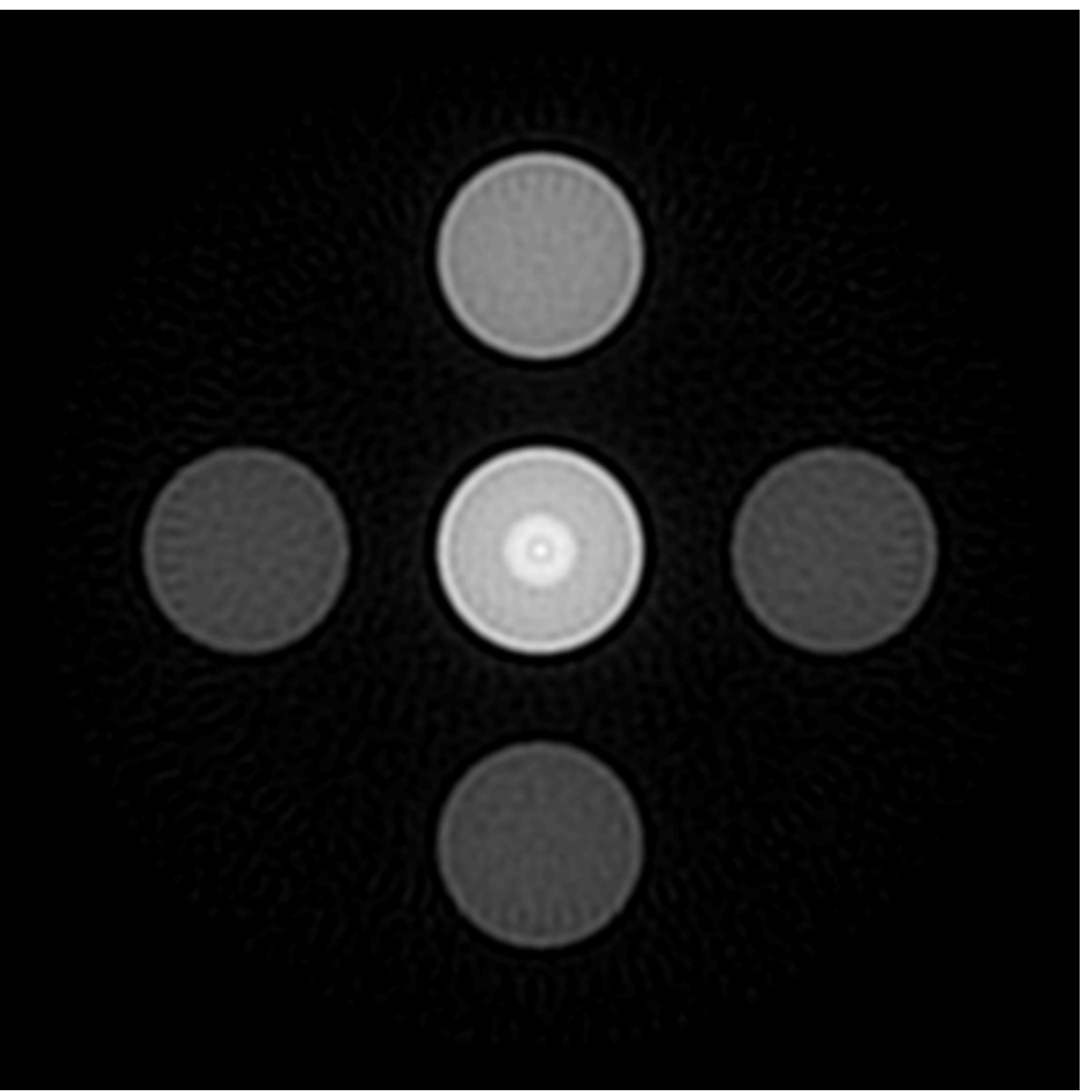}
  \caption{$\lambda=10^{-1}$, $\alpha=2.0\times 10^5$, $RMSE=0.0270$}\label{3ns_B2_lam1_1_alp2e5}
 \end{subfigure}
\caption{Images reconstructed from noisy data by use of the VP algorithm corresponding to regularization parameters $\lambda=\{10^{-5},10^{-3},10^{-1}\}$ and $\alpha=\{2.0\times 10^3, 2.0\times 10^4, 2.0\times 10^5\}$. RMSE values are also shown. 
Images are displayed in their full dynamic ranges respectively.}\label{3ns_B2_reg_para}
\end{figure}

\begin{figure}[!htb]
 \centering
 \begin{subfigure}[!htb]{0.32\textwidth}
 \includegraphics[width=\textwidth]{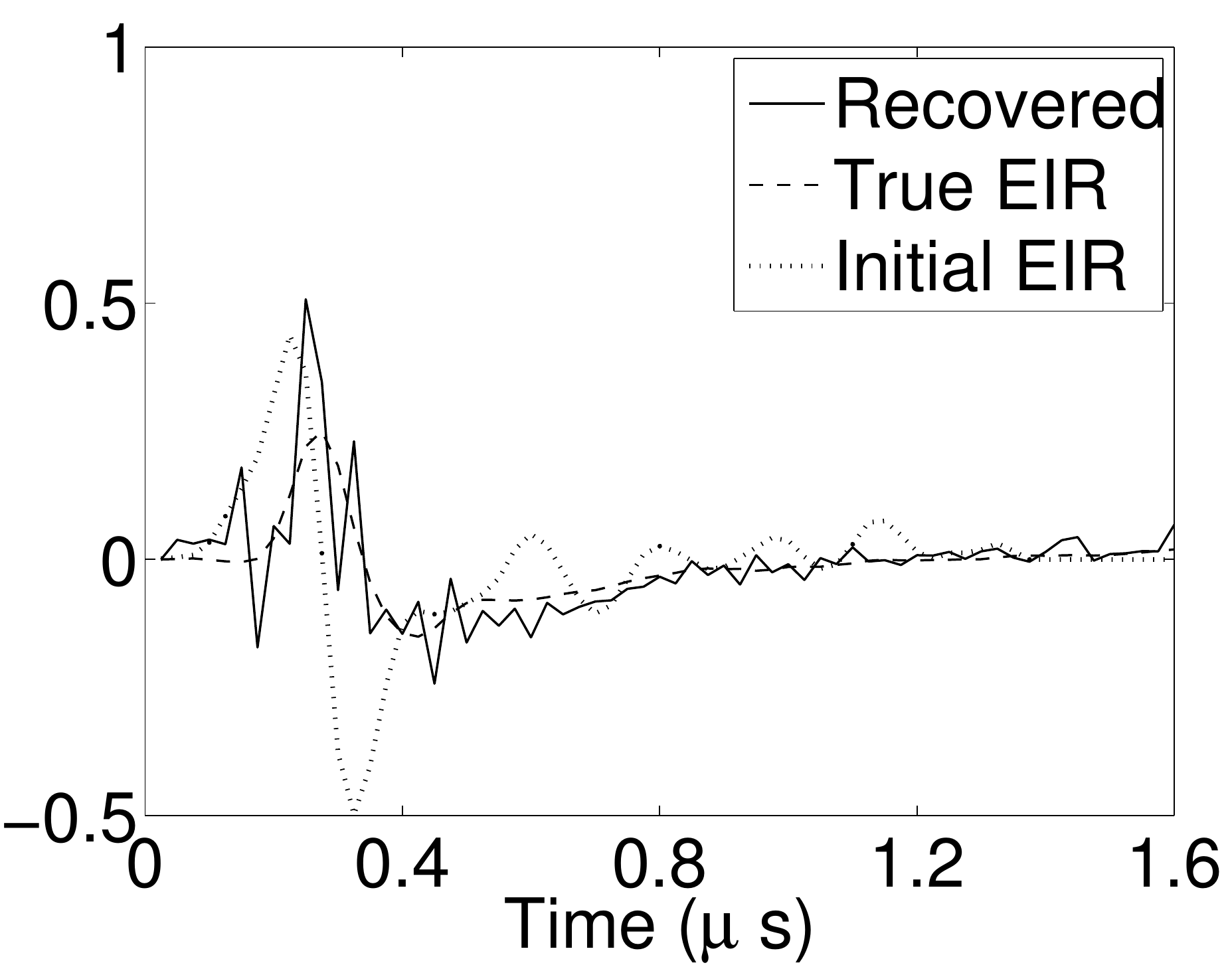}
  \caption{$\lambda=10^{-5}$, $\alpha=2000$}\label{3ns_B2_EIR_lam1_5_alp2e3}
 \end{subfigure}
 \begin{subfigure}[!htb]{0.32\textwidth}
 \includegraphics[width=\textwidth]{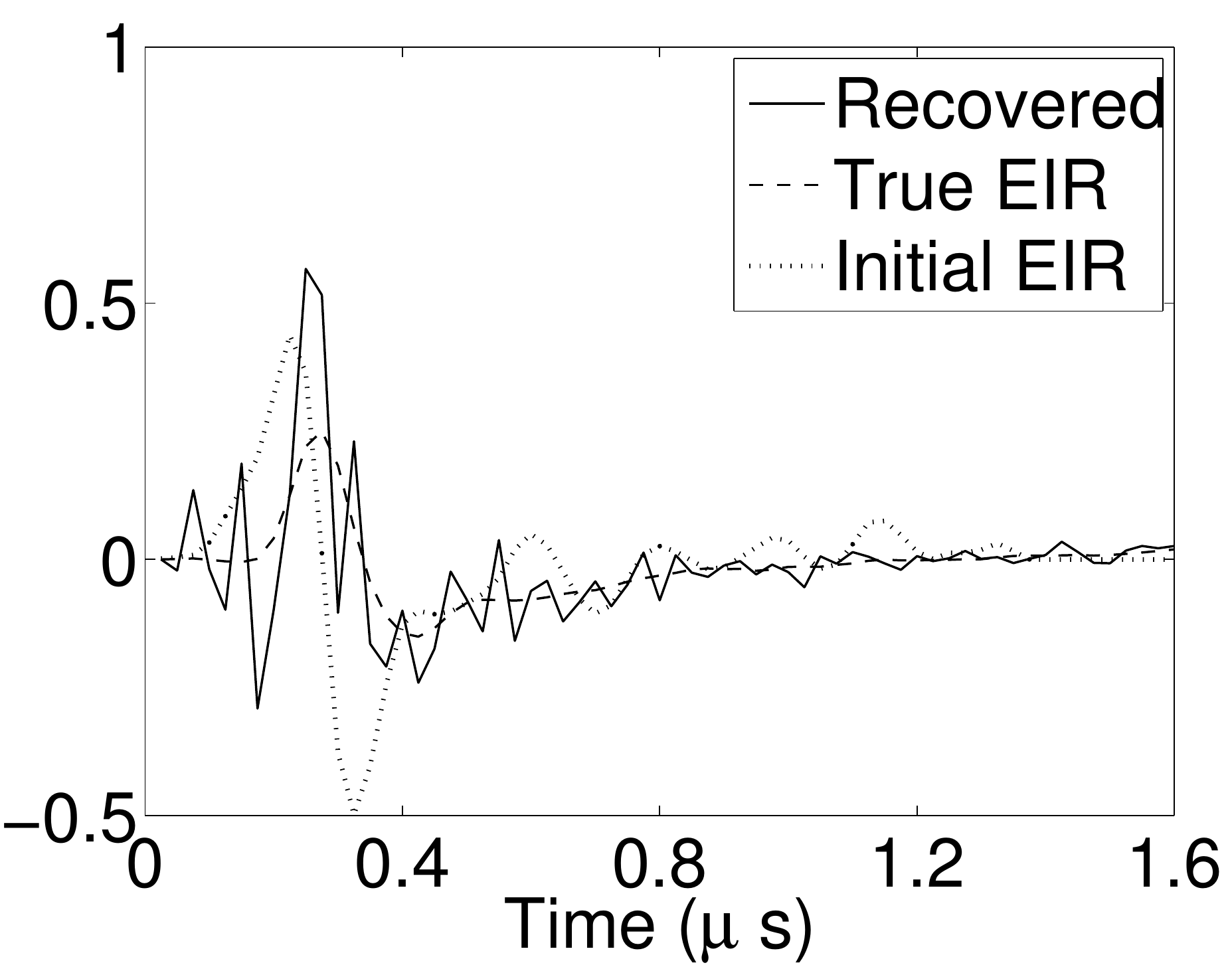}
  \caption{$\lambda=10^{-3}$, $\alpha=2000$}\label{3ns_B2_EIR_lam1_3_alp2e3}
 \end{subfigure}
 \begin{subfigure}[!htb]{0.32\textwidth}
 \includegraphics[width=\textwidth]{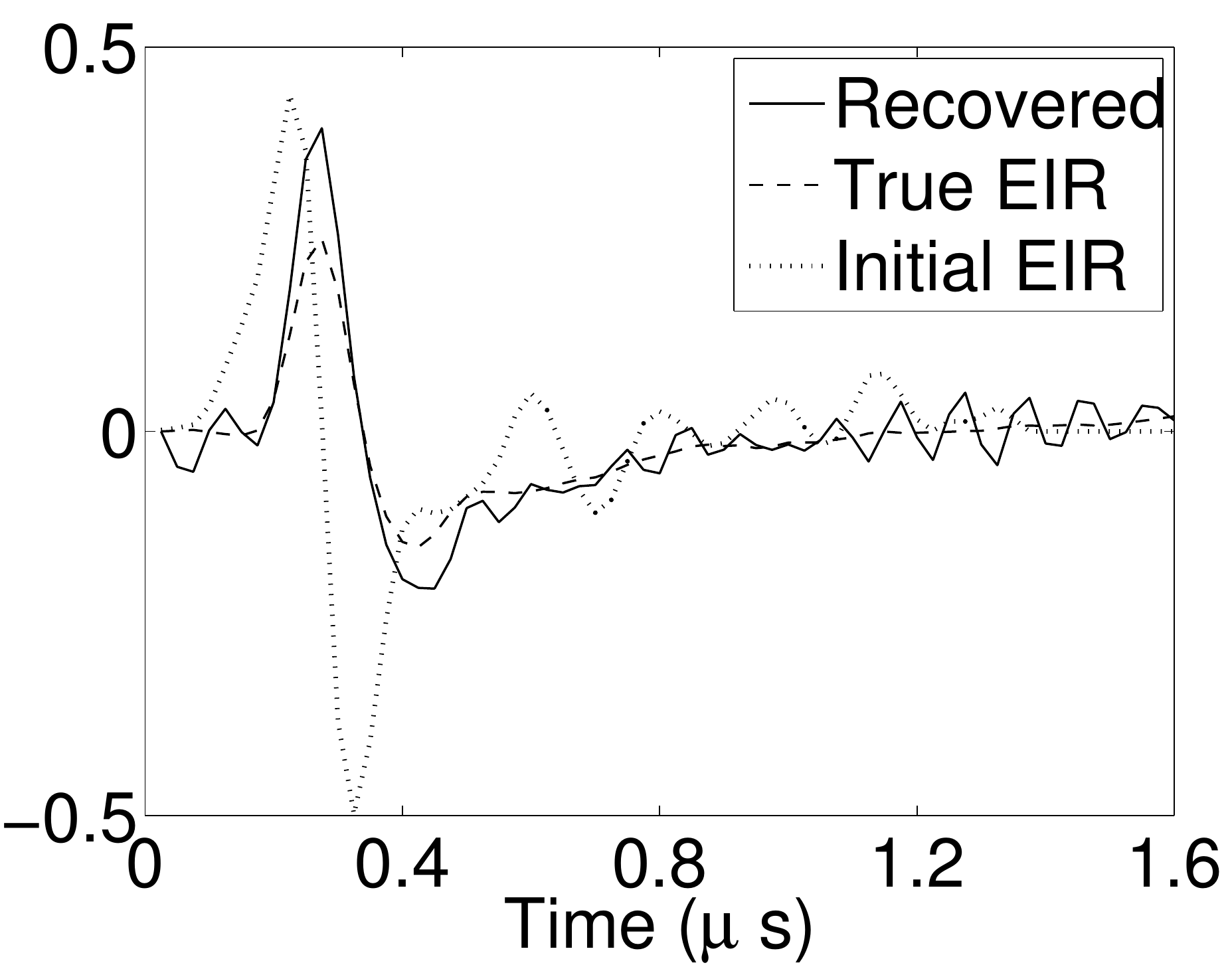}
  \caption{$\lambda=10^{-1}$, $\alpha=2000$}\label{3ns_B2_EIR_lam1e_1_alp2e3}
 \end{subfigure}
 \begin{subfigure}[!htb]{0.32\textwidth}
 \includegraphics[width=\textwidth]{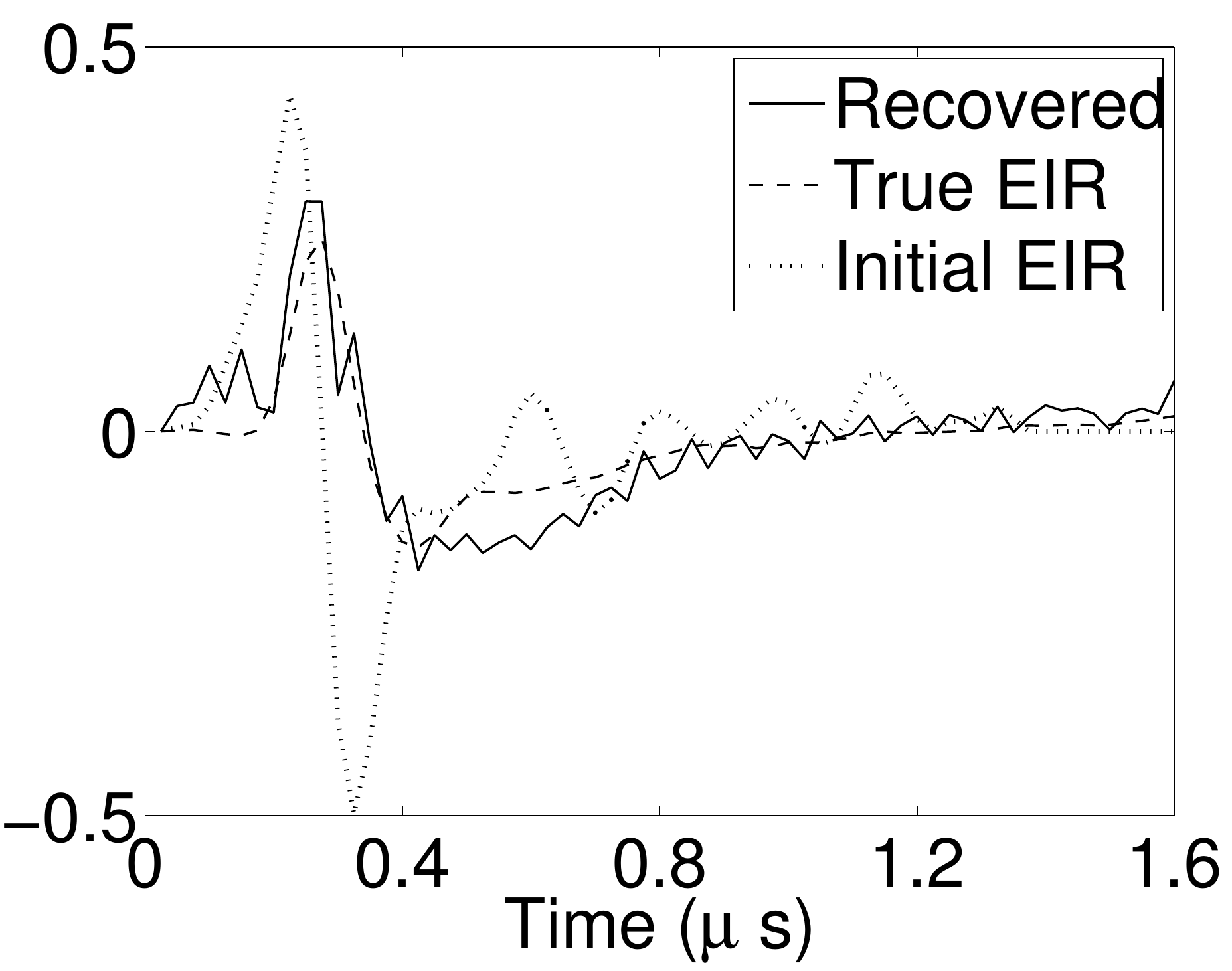}
  \caption{$\lambda=10^{-5}$, $\alpha=20000$}\label{3ns_B2_EIR_lam1_5_alp2e4}
 \end{subfigure}
 \begin{subfigure}[!htb]{0.32\textwidth}
 \includegraphics[width=\textwidth]{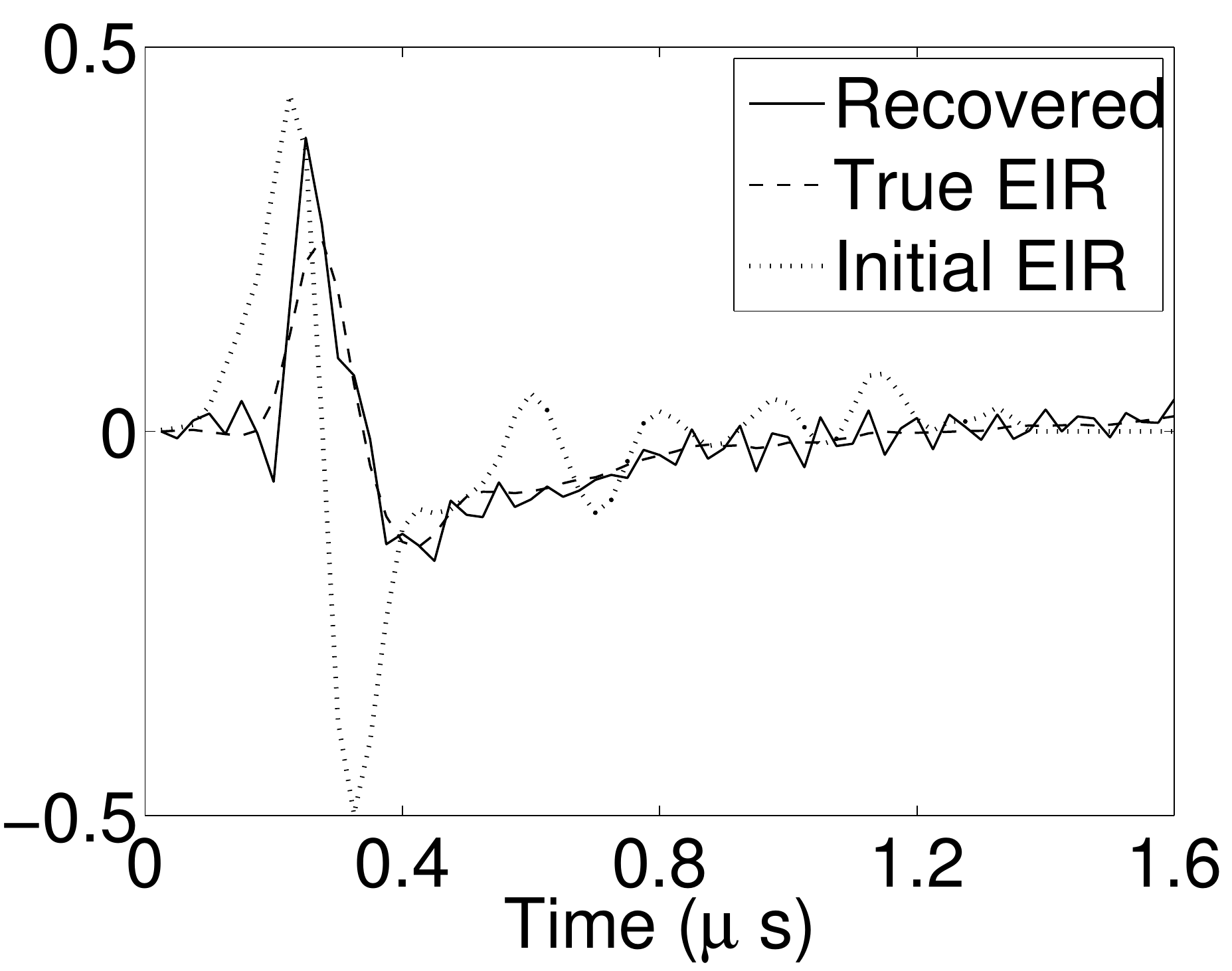}
  \caption{$\lambda=10^{-3}$, $\alpha=20000$}\label{3ns_B2_EIR_lam1_3_alp2e4}
 \end{subfigure}
 \begin{subfigure}[!htb]{0.32\textwidth}
 \includegraphics[width=\textwidth]{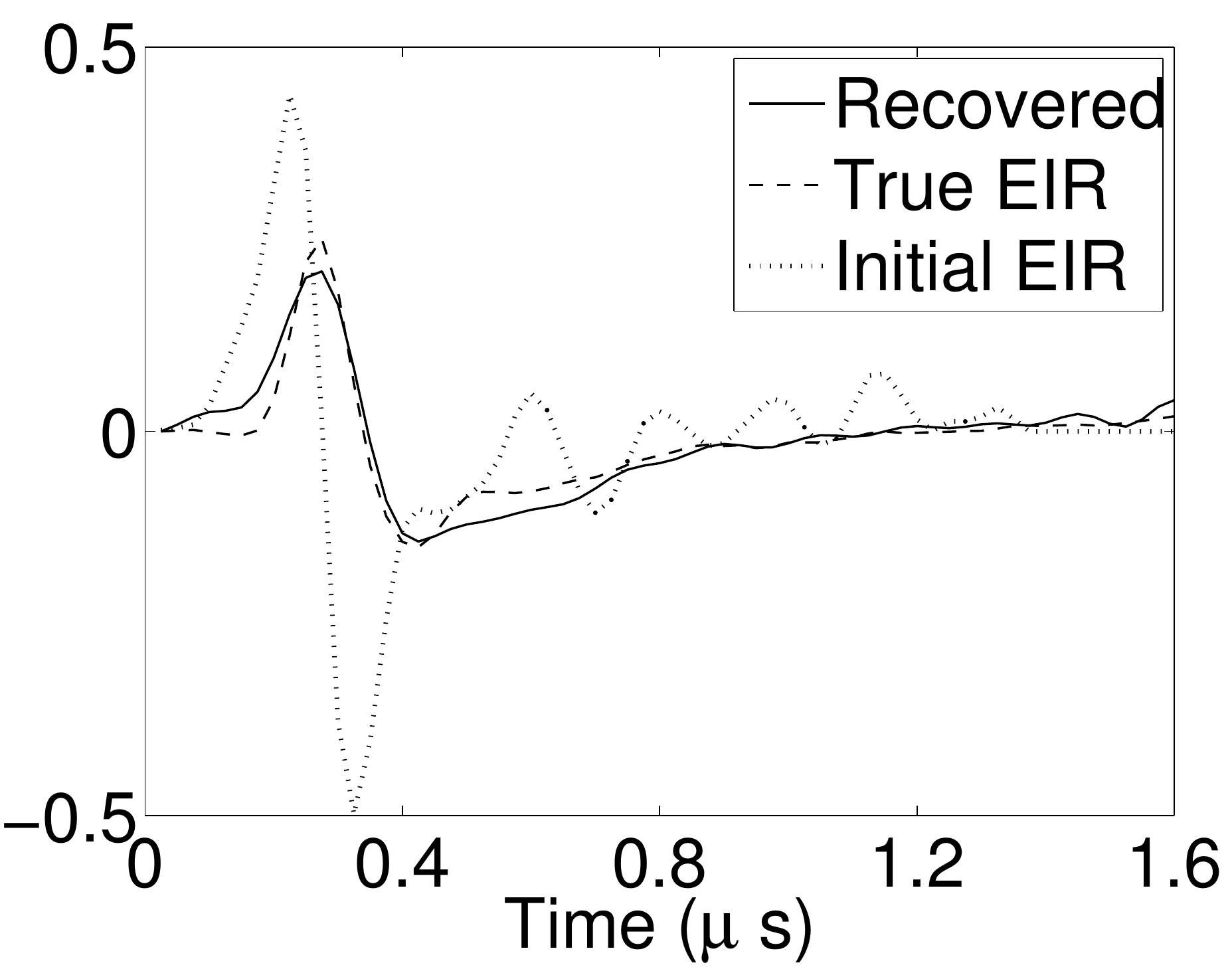}
  \caption{$\lambda=10^{-1}$, $\alpha=20000$}\label{3ns_B2_EIR_lam1_1_alp2e4}
 \end{subfigure}
 \begin{subfigure}[!htb]{0.32\textwidth}
 \includegraphics[width=\textwidth]{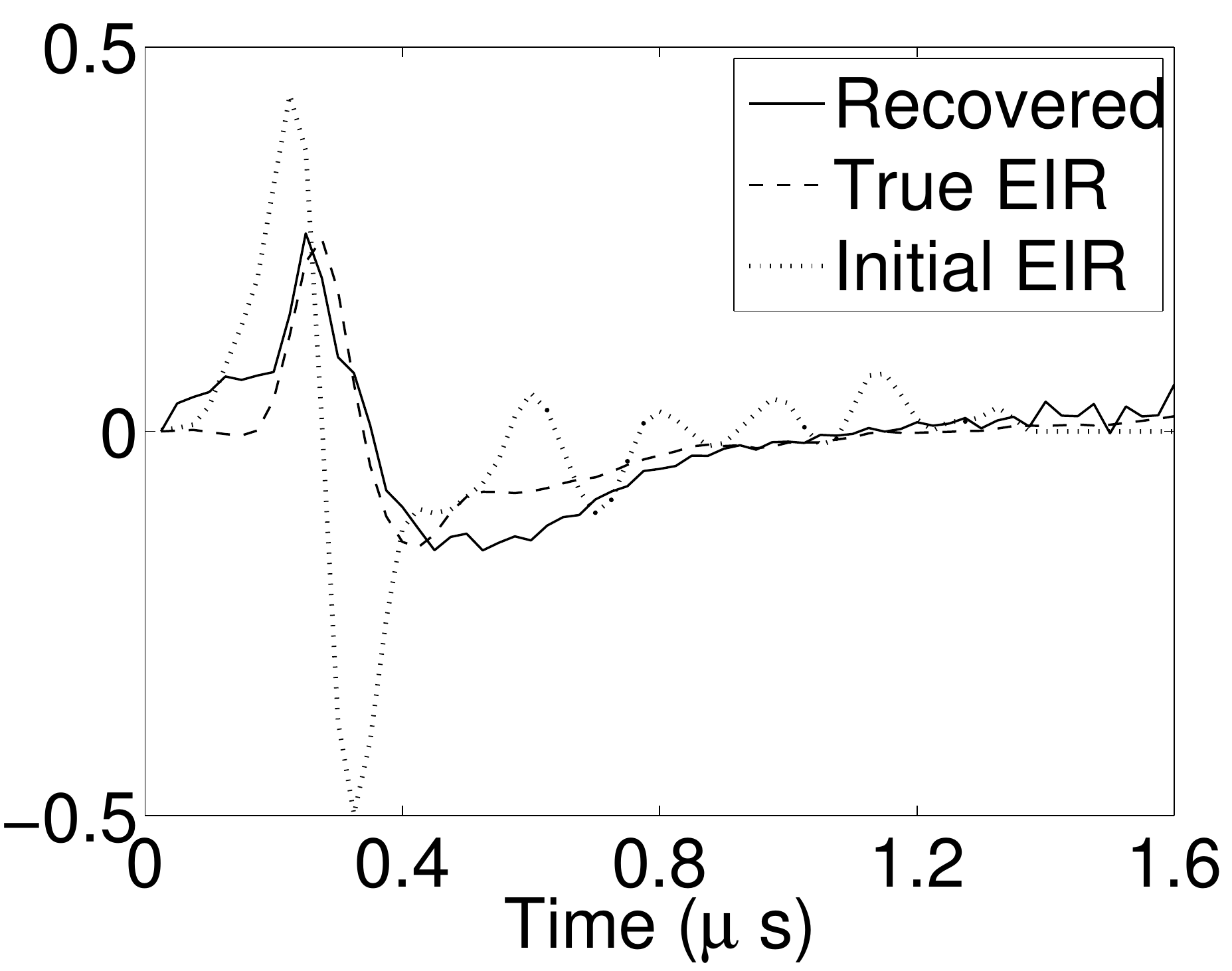}
  \caption{$\lambda=10^{-5}$, $\alpha=2.0\times 10^5$}\label{3ns_B2_EIR_lam1_5_alp2e5}
 \end{subfigure}
 \begin{subfigure}[!htb]{0.32\textwidth}
 \includegraphics[width=\textwidth]{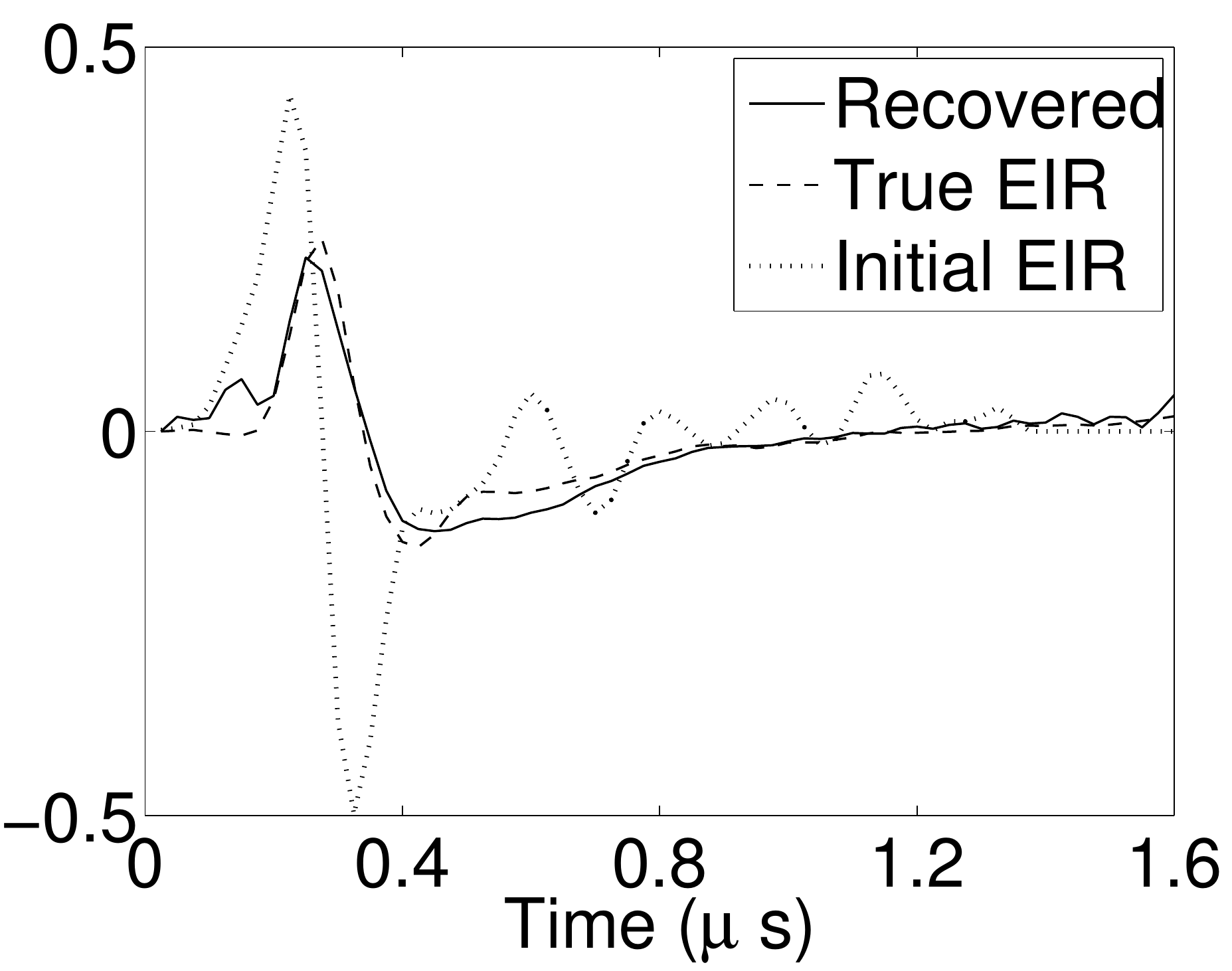}
  \caption{$\lambda=10^{-3}$, $\alpha=2.0\times 10^5$}\label{3ns_B2_EIR_lam1_3_alp2e5}
 \end{subfigure}
 \begin{subfigure}[!htb]{0.32\textwidth}
 \includegraphics[width=\textwidth]{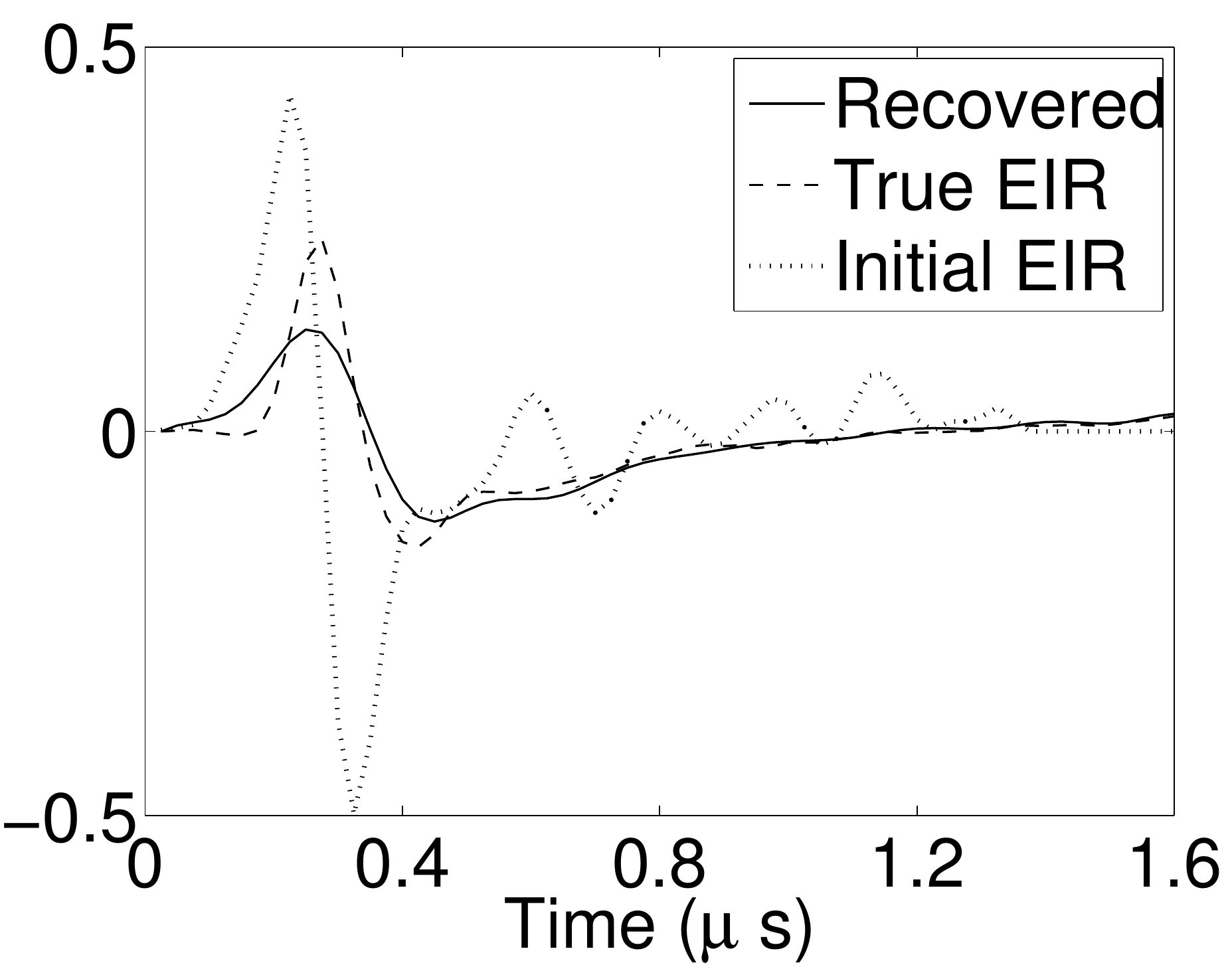}
  \caption{$\lambda=10^{-1}$, $\alpha=2.0\times 10^5$}\label{3ns_B2_EIR_lam1_1_alp2e5}
 \end{subfigure}
\caption{The recovered EIRs corresponding to the reconstructed images in Fig.~\ref{3ns_B2_reg_para}. $\lambda=\{10^{-5}$, $10^{-3}$, $10^{-1}\}$ and $\alpha=\{2.0\times 10^3, 2.0\times 10^4, 2.0\times 10^5\}$.}\label{3ns_B2_EIR_reg_para}
\end{figure}

\begin{figure}[!htb]
 \centering
 \begin{subfigure}[!htb]{0.32\textwidth}
 \includegraphics[width=\textwidth]{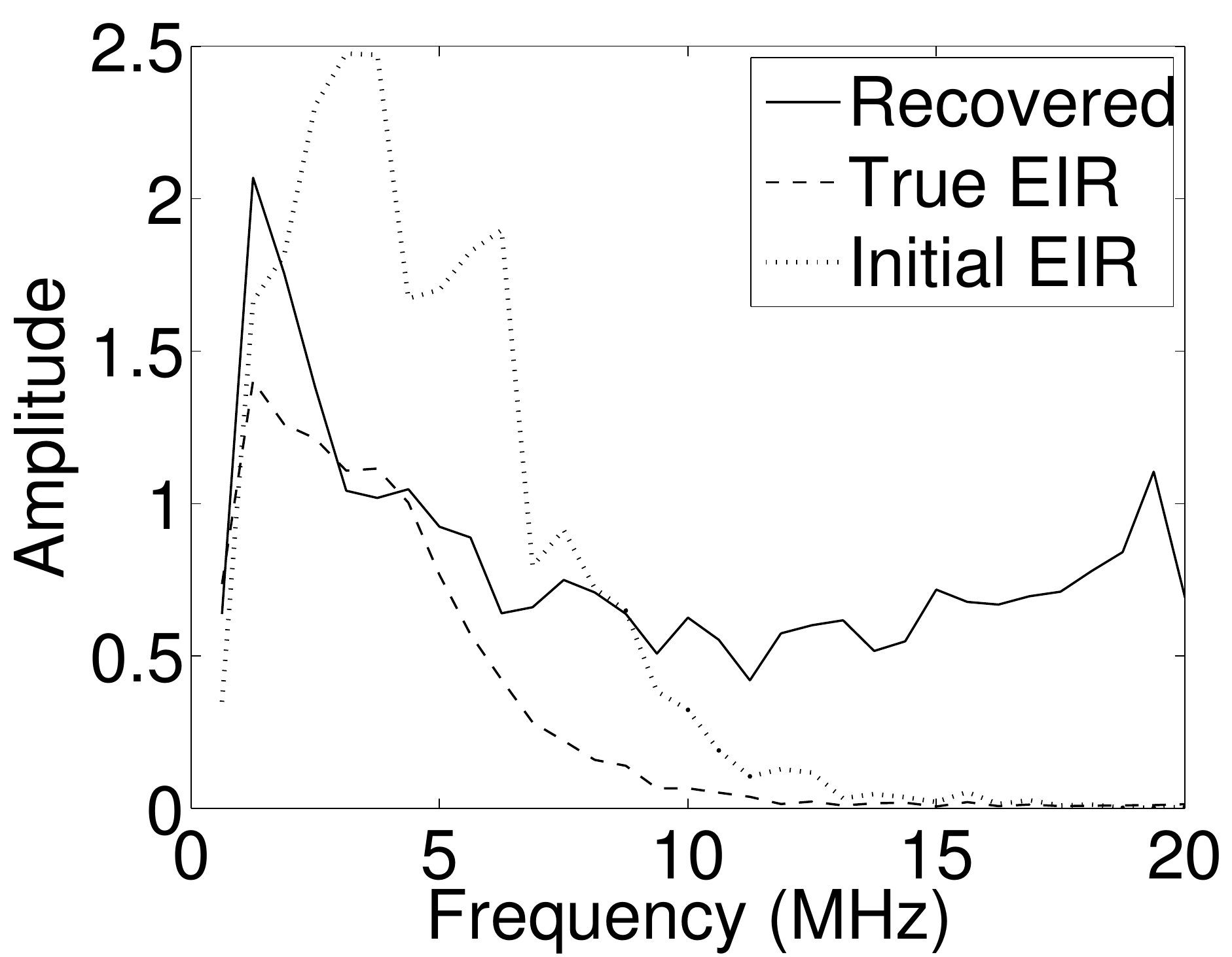}
  \caption{$\lambda=10^{-5}$, $\alpha=2000$}\label{3ns_B2_EIR_freq_lam1_5_alp2e3}
 \end{subfigure}
 \begin{subfigure}[!htb]{0.32\textwidth}
 \includegraphics[width=\textwidth]{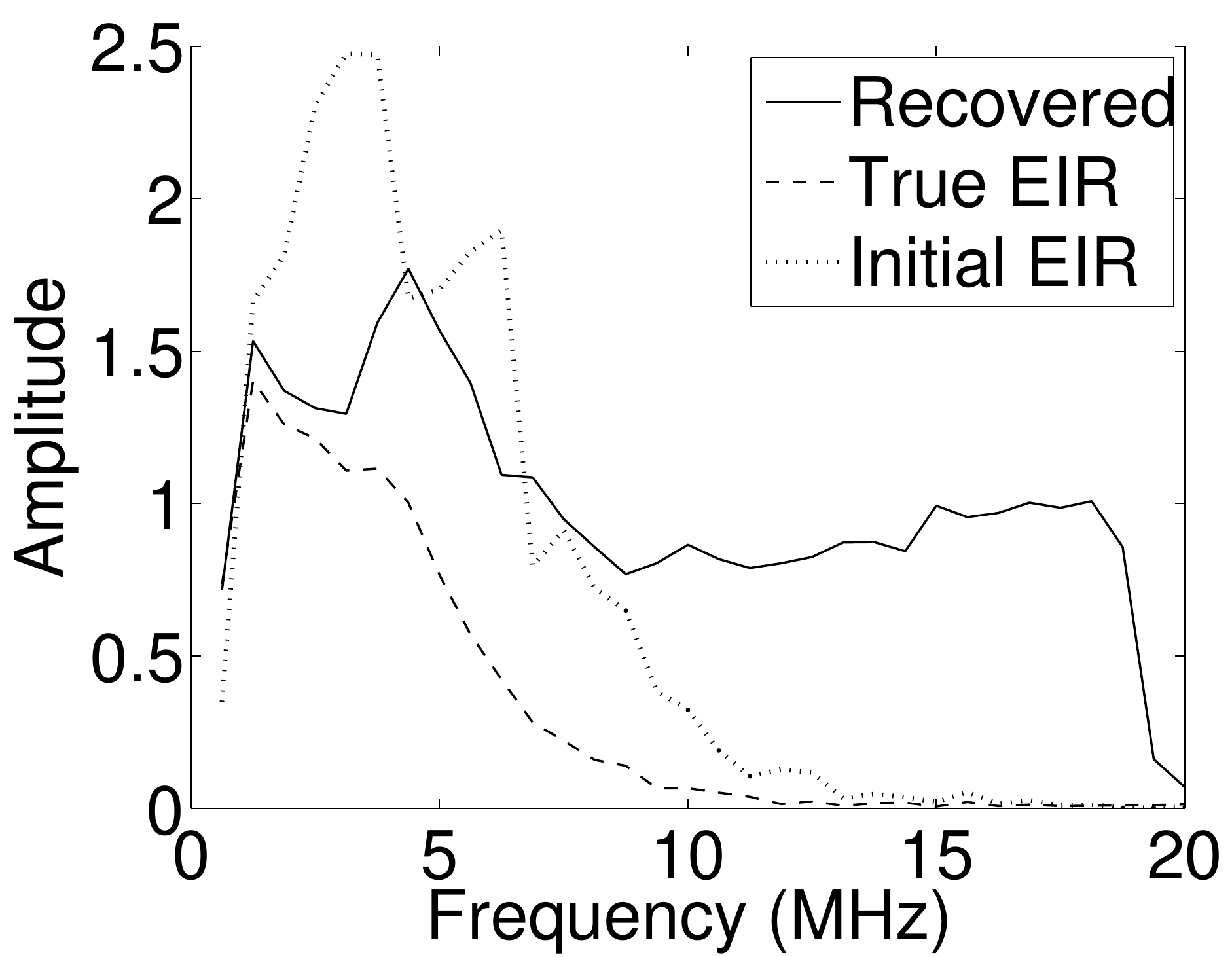}
  \caption{$\lambda=10^{-3}$, $\alpha=2000$}\label{3ns_B2_EIR_freq_lam1_3_alp2e3}
 \end{subfigure}
 \begin{subfigure}[!htb]{0.32\textwidth}
 \includegraphics[width=\textwidth]{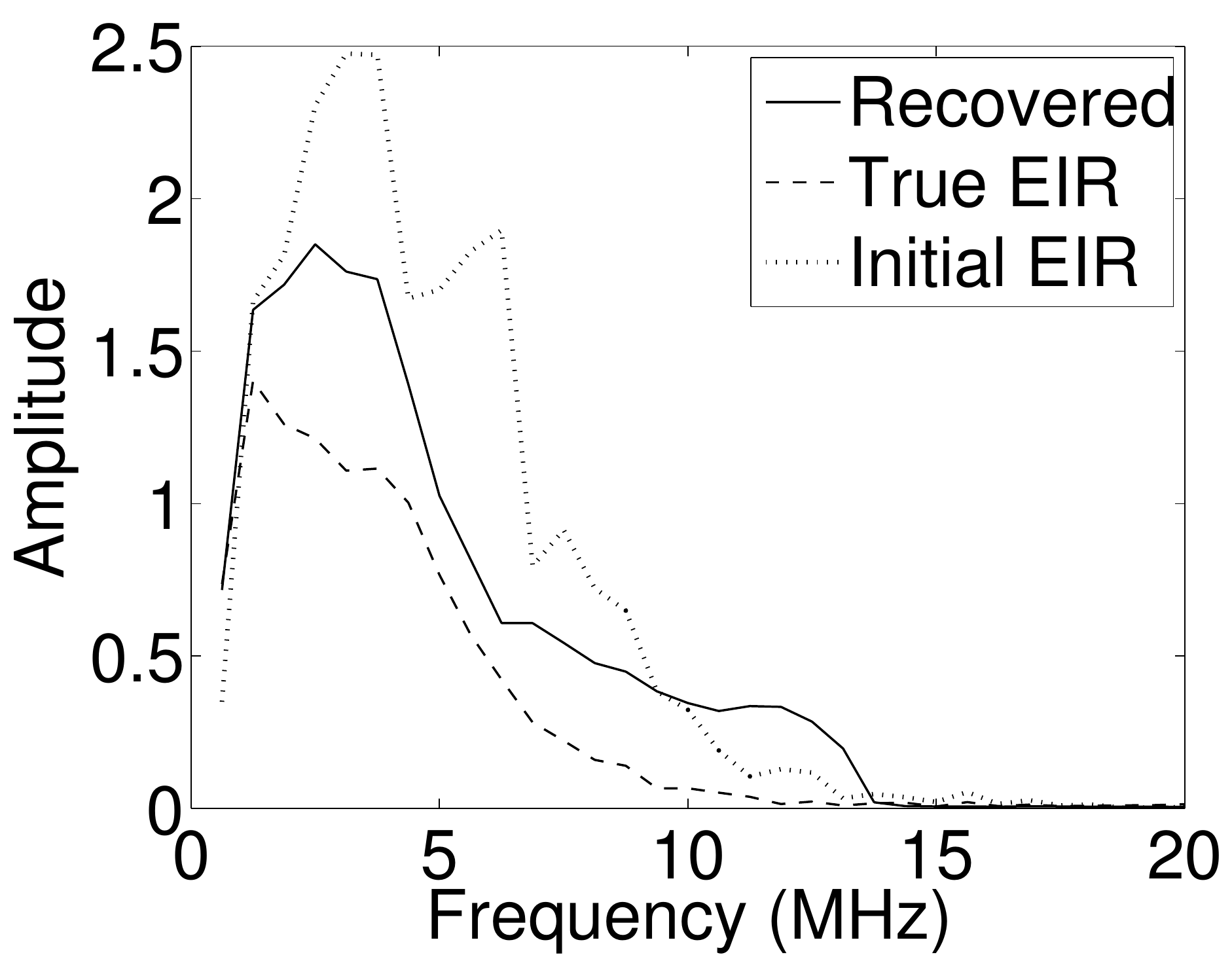}
  \caption{$\lambda=10^{-1}$, $\alpha=2000$}\label{3ns_B2_EIR_freq_lam1e_1_alp2e3}
 \end{subfigure}
 \begin{subfigure}[!htb]{0.32\textwidth}
 \includegraphics[width=\textwidth]{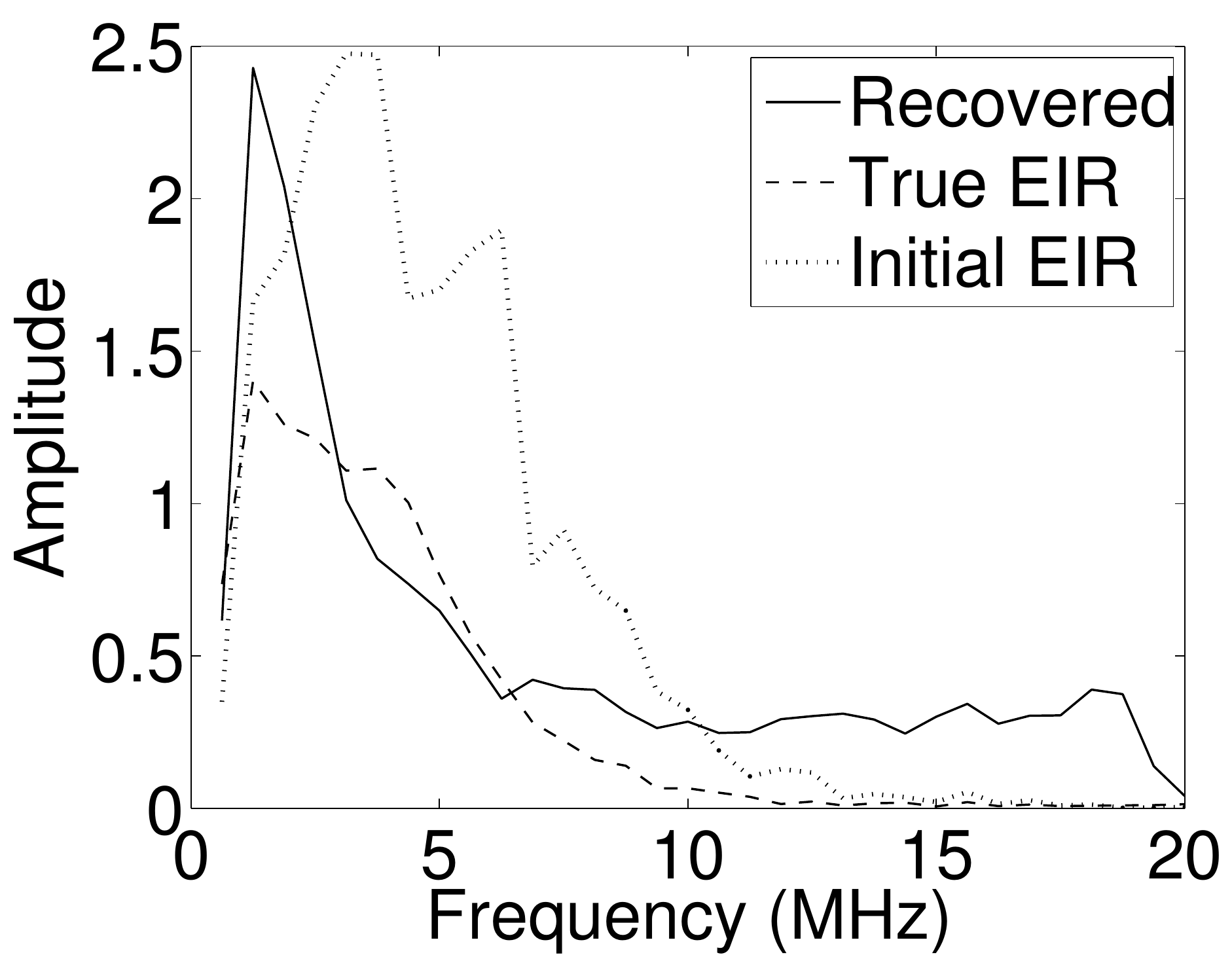}
  \caption{$\lambda=10^{-5}$, $\alpha=20000$}\label{3ns_B2_EIR_freq_lam1_5_alp2e4}
 \end{subfigure}
 \begin{subfigure}[!htb]{0.32\textwidth}
 \includegraphics[width=\textwidth]{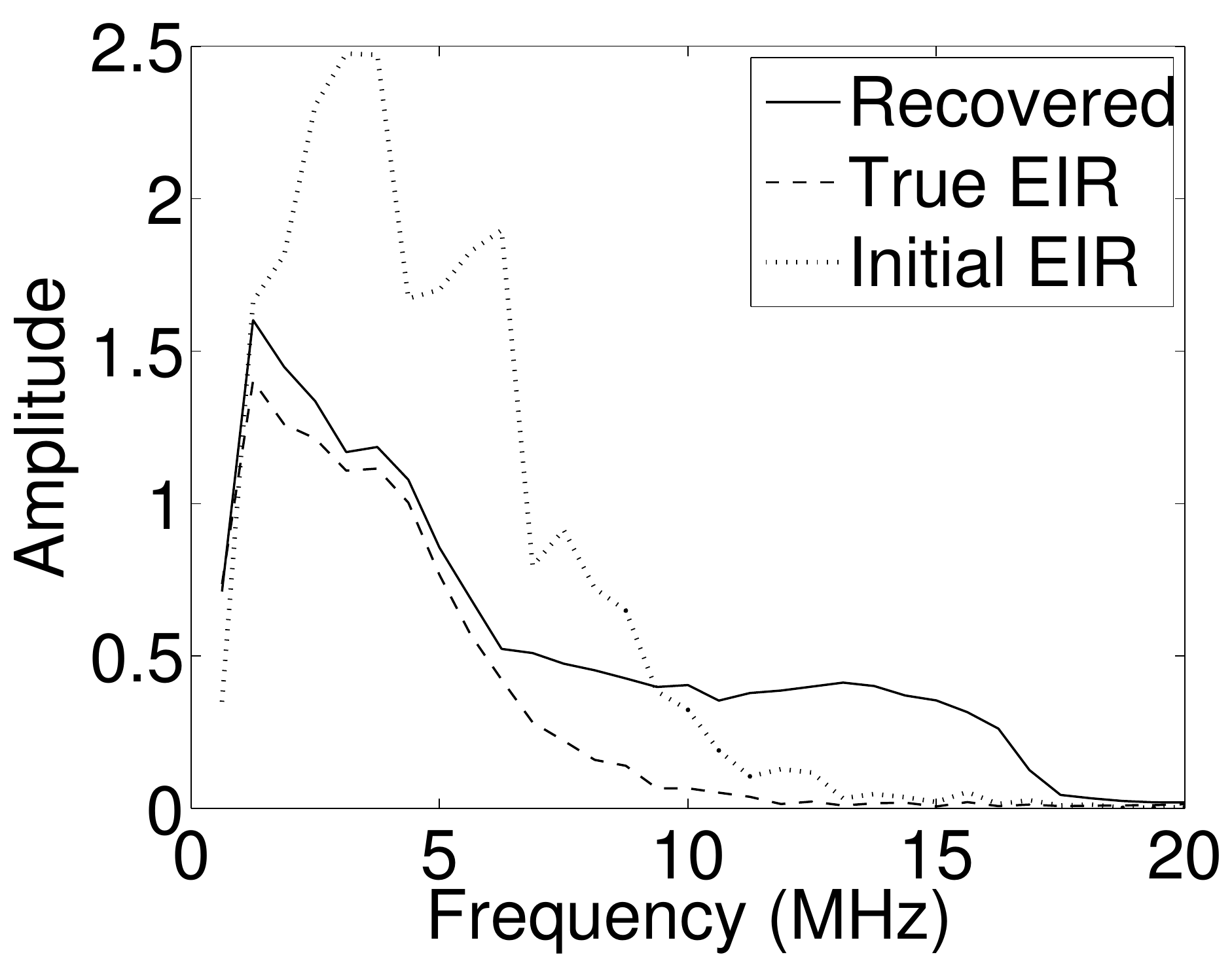}
  \caption{$\lambda=10^{-3}$, $\alpha=20000$}\label{3ns_B2_EIR_freq_lam1_3_alp2e4}
 \end{subfigure}
 \begin{subfigure}[!htb]{0.32\textwidth}
 \includegraphics[width=\textwidth]{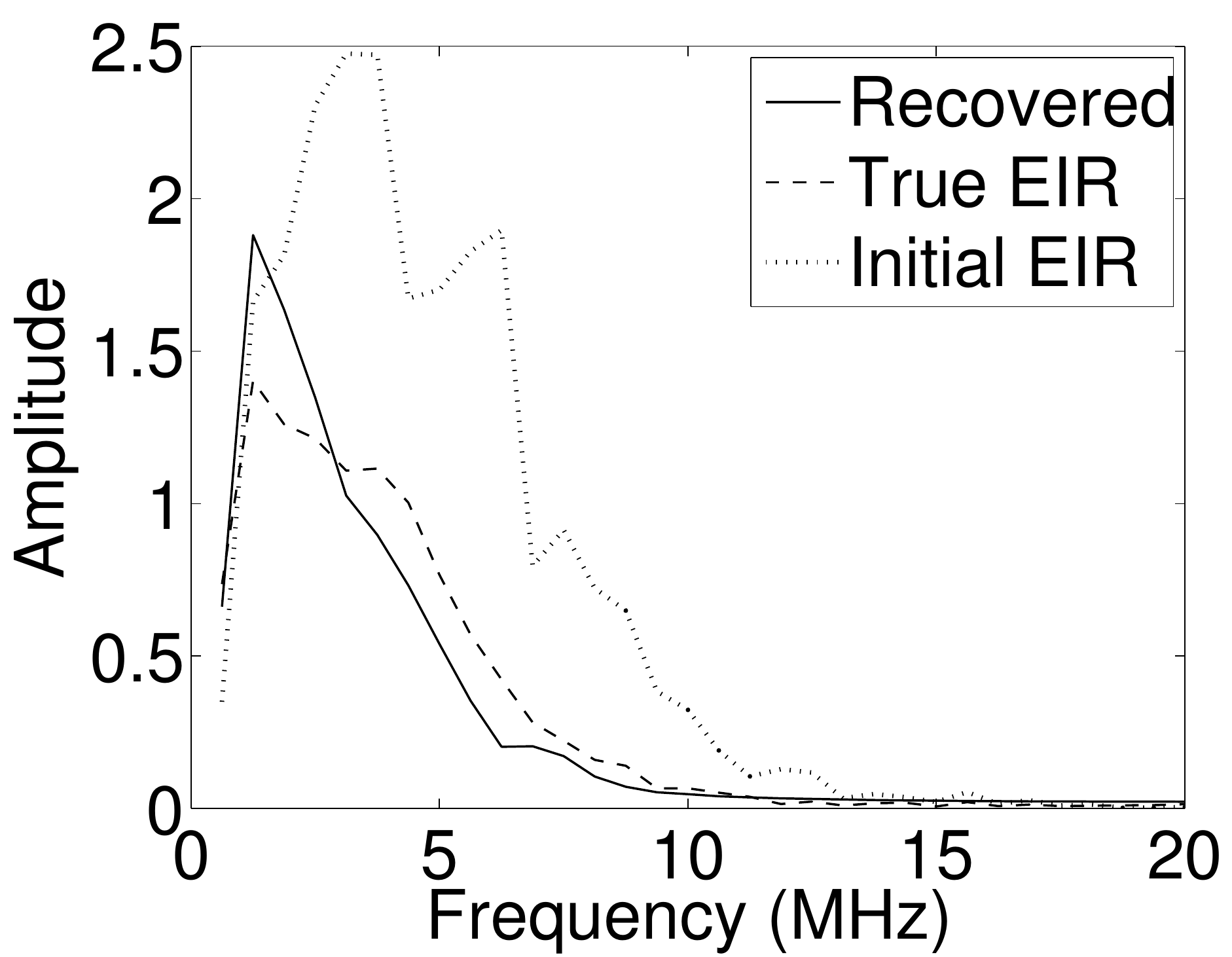}
  \caption{$\lambda=10^{-1}$, $\alpha=20000$}\label{3ns_B2_EIR_freq_lam1_1_alp2e4}
 \end{subfigure}
 \begin{subfigure}[!htb]{0.32\textwidth}
 \includegraphics[width=\textwidth]{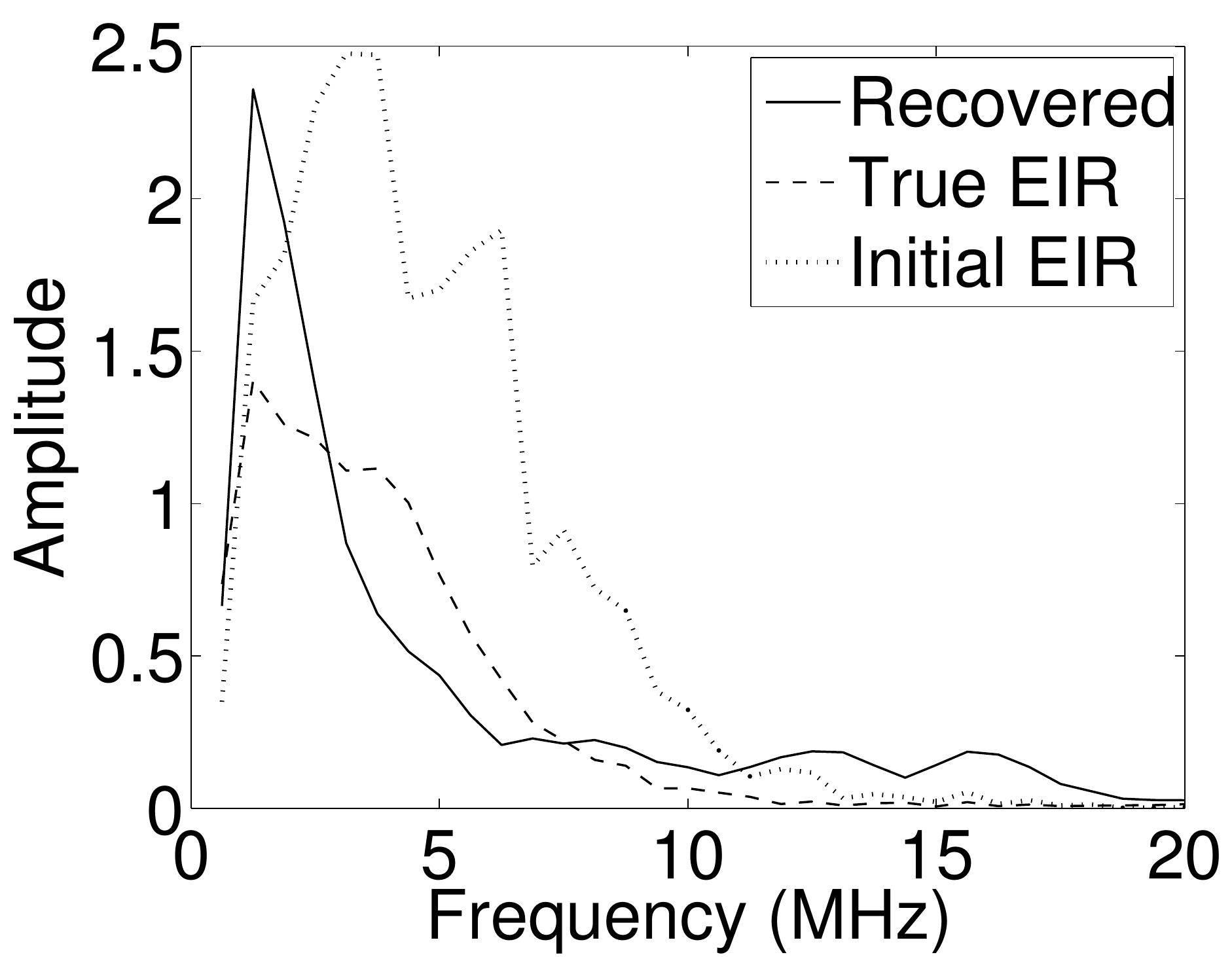}
  \caption{$\lambda=10^{-5}$, $\alpha=2.0\times 10^5$}\label{3ns_B2_EIR_freq_lam1_5_alp2e5}
 \end{subfigure}
 \begin{subfigure}[!htb]{0.32\textwidth}
 \includegraphics[width=\textwidth]{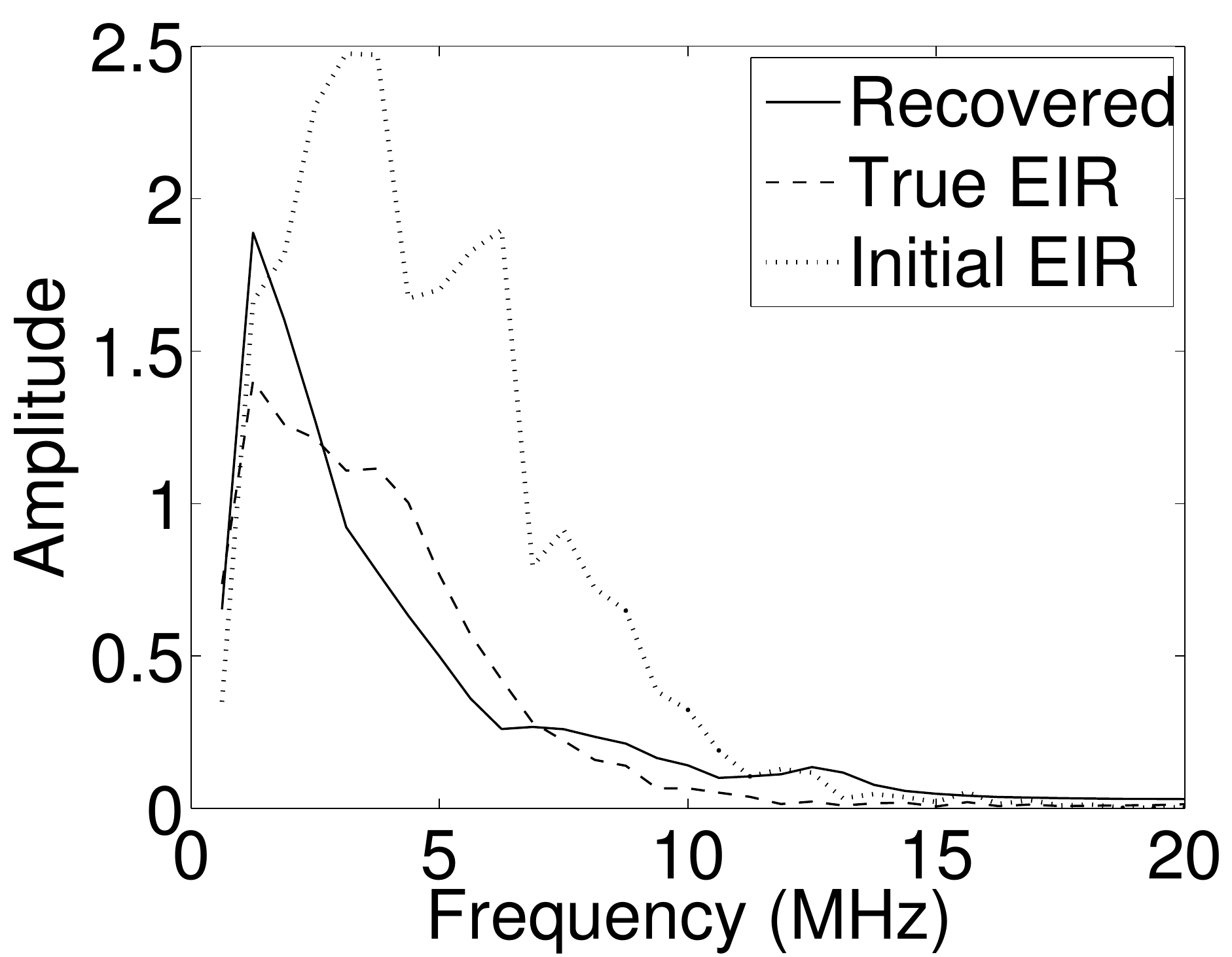}
  \caption{$\lambda=10^{-3}$, $\alpha=2.0\times 10^5$}\label{3ns_B2_EIR_freq_lam1_3_alp2e5}
 \end{subfigure}
 \begin{subfigure}[!htb]{0.32\textwidth}
 \includegraphics[width=\textwidth]{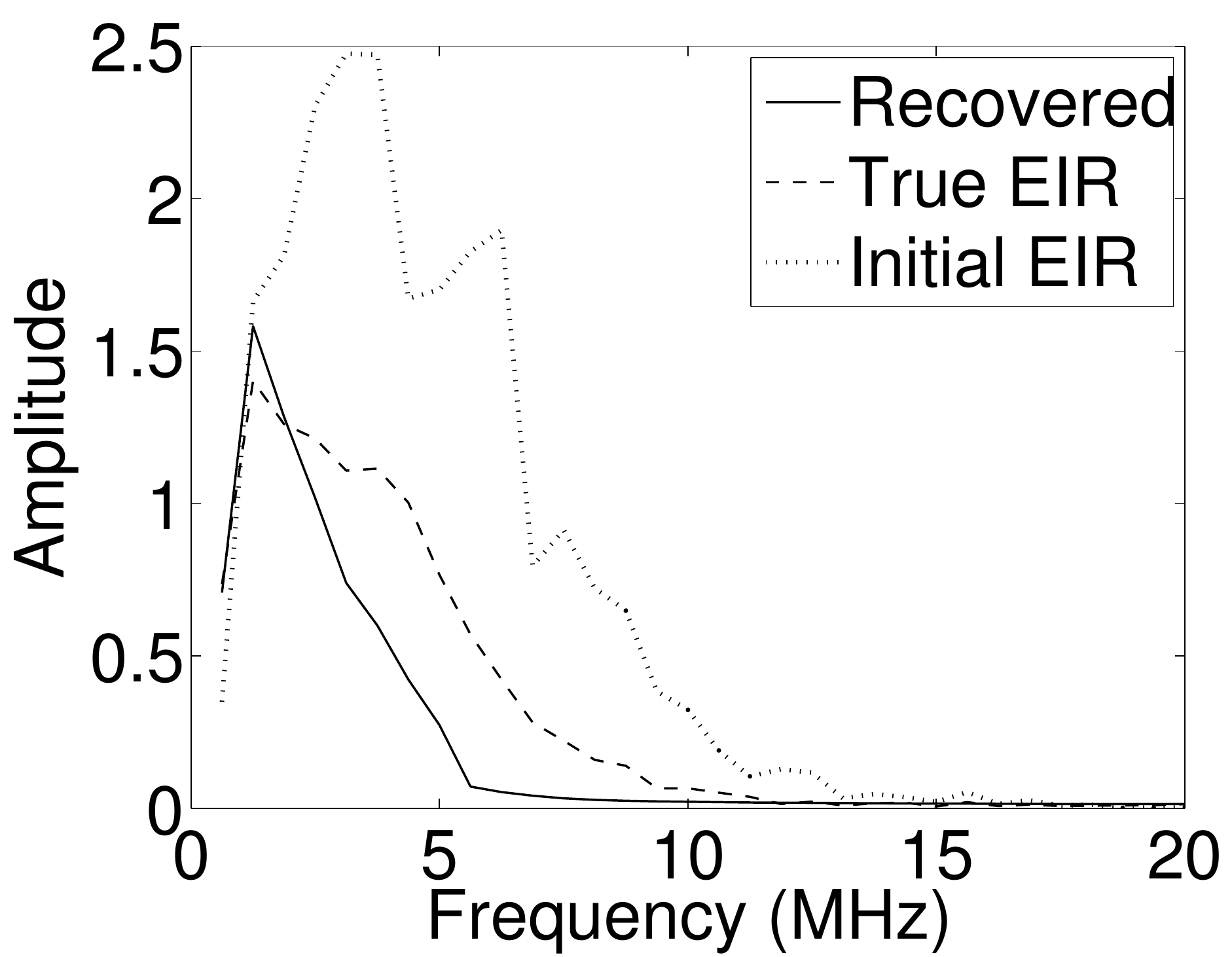}
  \caption{$\lambda=10^{-1}$, $\alpha=2.0\times 10^5$}\label{3ns_B2_EIR_freq_lam1_1_alp2e5}
 \end{subfigure}
\caption{The recovered EIRs in the frequency domain corresponding to the reconstructed images in Fig.~\ref{3ns_B2_reg_para}. $\lambda=\{10^{-5},10^{-3},10^{-1}\}$ and $\alpha=\{2.0\times 10^3, 2.0\times 10^4, 2.0\times 10^5\}$.}\label{3ns_B2_EIR_freq_reg_para}
\end{figure}

 Images reconstructed by use of the VP algorithm with different values of the
 regularization parameter values are shown in Figure~\ref{3ns_B2_reg_para}. The recovered EIRs and their corresponding Fourier spectra are shown in Figures~\ref{3ns_B2_EIR_reg_para} and \ref{3ns_B2_EIR_freq_reg_para}, respectively. The RMSE values are computed and displayed together with the corresponding images.
As expected, the images reconstructed with smaller values of $\lambda$ contain higher noise levels, while images using larger $\lambda$ possess a reduced noise level. However, larger values of $\lambda$ also caused artifacts in the reconstructed images. The same observation can be made for the effect of the regularization parameter $\alpha$ on the recovered EIR. One also observes that the reconstructed images and EIRs depend continuously on the regularization parameters $\lambda$ and $\alpha$, i.e.~small changes in the regularization parameters cause minor changes in the reconstructed images and EIRs.

\section{Experimental validation}\label{Sect:Exp}
The proposed algorithm was further investigated by use of experimental data acquired from a 512-element full-ring-array photoacoustic computed tomography system \cite{XCMG2012}.

\subsection{Phantom objects}
The first phantom was comprised of a single black needle of diameter $0.25$ mm and length $20$ mm  embedded in an agar gel. The second phantom
was comprised of a mouse kidney embedded in an agar gel. In both experiments, the phantom and the transducer array were aligned so that the object of interest laid in the focal plane of the transducer array.


\subsection{Data acquisition}
The illumination light source was a tunable optical parametric oscillator (OPO) laser (basiScan, Spectral Physics) pumped by a Nd:YAG laser (Brilliant B, Quantel) with 12 ns pulse duration and 10 Hz pulse repetition rate. 
{Before reaching the sample surface, the laser beam was homogenized using an optical diffuser (EDC-5 Photonics) to form a 25-mm-diameter circular light beam}. For both experiments, we tuned the OPO laser output to a wavelength of $610$ nm . The maximum light intensity at the surface of the sample was approximately $5$ mJ/cm$^2$, which is well below the American National Standard Institute (ANSI) limit at 
$610$ nm ($20\;\text{mJ}/\text{cm}^2$) \cite{ANSI2000}.

The PA signals were detected by a $512$-element full-ring transducer array with $5$ MHz central frequency ($80\%$ bandwidth) and $50$ mm ring diameter. Each element in the array was mechanically shaped into an arc to produce an axial focal {length} of $19$ mm. At each transducer location, 1300 temporal samples were acquired at a sampling rate of $40$ MHz. Accordingly, the dimension of the measured data set was $1300\times 512$. Additional details of the imaging system can be found in \cite{XCMG2012,XHMA2013}.

\subsection{Image reconstruction}

{Two numerical imaging models were employed in the studies involving experimental data. 
Both models are described in Appendix A.
  In the first, the SIR effect was not considered and a 2D interpolation-based D-D imaging model 
 was employed. Most of the presented results were reconstructed by use of this imaging model. For the needle phantom, the size of the reconstructed region was 22.0 x 22.0 mm$^2$, which was represented by 440 x 440 pixels. For the mouse kidney, the size of the reconstructed region was 14.0 x 14.0 mm$^2$, which was represented by 280 x 280 pixels. For both objects, the pixel size was 0.05 x 0.05 mm$^2$. The needle and kidney data were processed using a single core of an Intel Core i7-3770 CPU (4 cores, 3.4 GHz). It required approximately 10.5 s and 5 s to complete one iteration for the needle and kidney data sets, respectively.
	The second imaging model included SIR effects and was
 based on a 3D spherical voxel imaging model \cite{WESB2011,MKA2014}. This model was only applied to the kidney data. In this case, only a single slice through the object at the focal plane was reconstructed. The object was represented by 280 x 280 x 1 voxels of dimension 0.05 x 0.05 x 0.05 mm$^3$. In order to model SIR effects, each transducer surface was divided into four equal parts in the elevation direction, and a far-field approximation was employed to calculated the measured pressure for each sub-element, also referred to as a 'patch' \cite{MKA2014}. We selected 4 patches because the maximum phase error in this case was less than one-eighth of the wavelength corresponding to the central frequency of the transducer. The algorithm was implemented by use of the CUDA parallel programming framework and executed on a single GPU (Tesla K20c). Processing the kidney data required
 15 minutes per iteration. Details on how the 3D model was constructed can be found in Appendix A.}

{	In both cases, images were reconstructed by use of both the VP algorithm and the conventional algorithm described previously. The VP algorithm was terminated after 120 iterations, while the conventional method was terminated after 50 iterations. The initial guess for the EIR was an experimentally-measured EIR from an element in the PACT system, and the initial guess for $\boldsymbol{\theta}$ was all zeros. The regularization functions
 employed corresponded to those in Eqns.\ (19) and (20). The values of the regularization parameters
 $\lambda$ and $\alpha$ were determined empirically. We swept the values
of these parameters over wide ranges with a small step size. Instead of attempting
to identify optimal regularization parameter values, which are application dependent,
 we investigated how the regularization parameter values affect the reconstructed images.}

\subsection{Results: Needle phantom}
\begin{figure}[!htb]
 \centering
\includegraphics[width=0.6\textwidth]{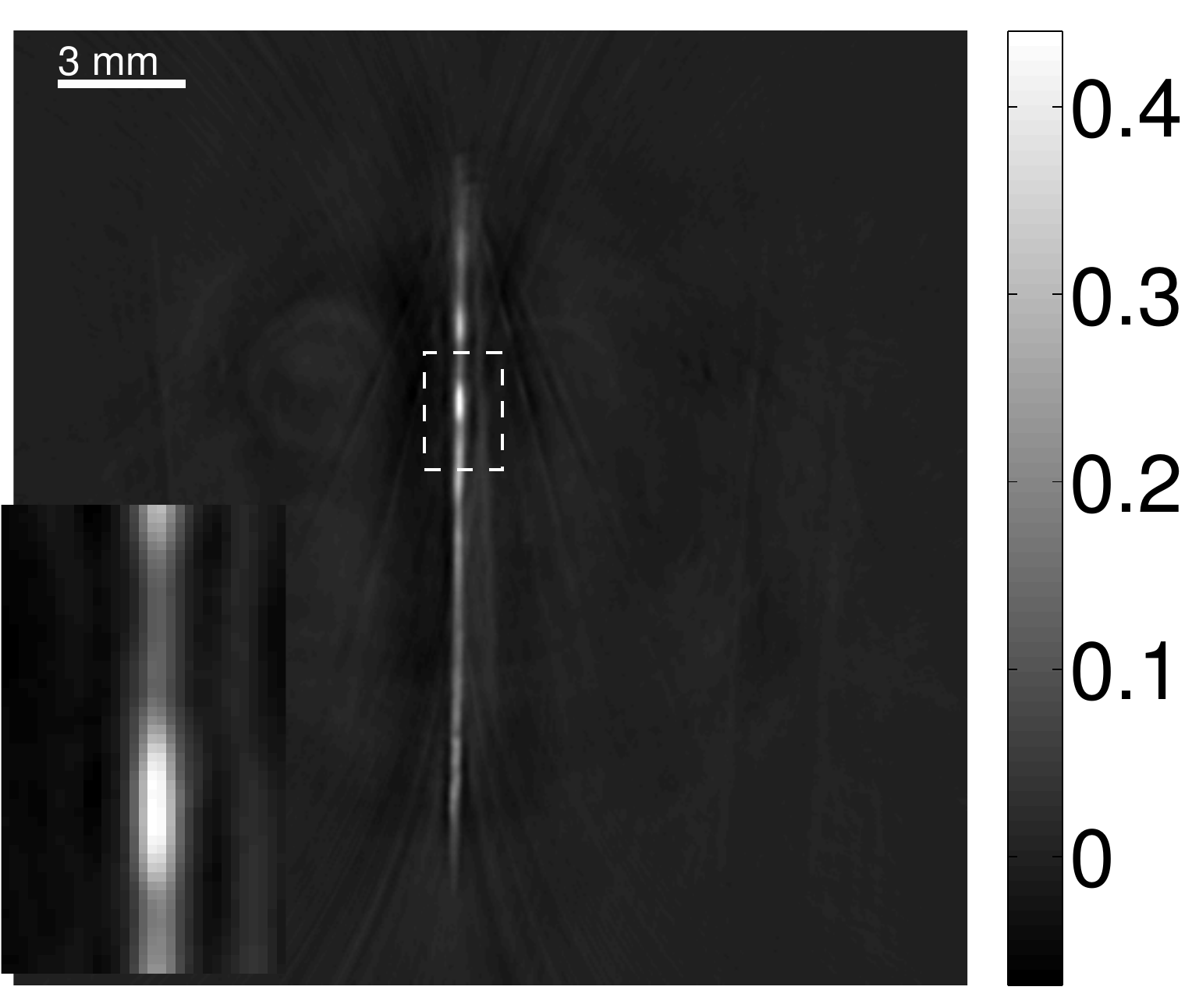}
\caption{Image of a needle phantom reconstructed using the backprojection.
The zoomed-in image corresponds to the ROI of the dashed rectangle.
}\label{2D_needle_BP}
\end{figure}

\begin{figure}[!htb]
 \centering
 \begin{subfigure}[!h]{0.47\textwidth}
 \includegraphics[width=\textwidth]{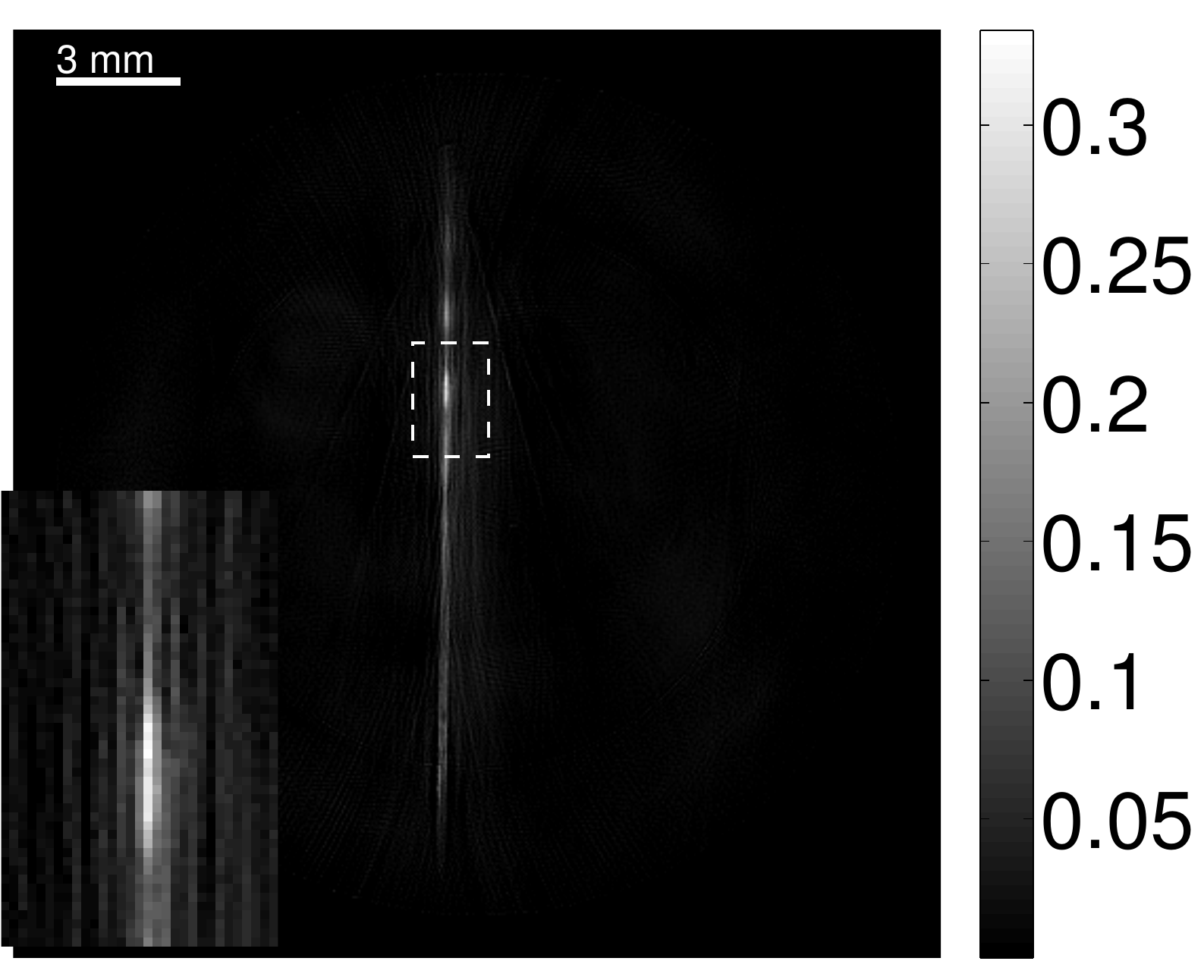}
 \caption{$\lambda=1.0\times 10^{-5}$}\label{2D_needle_EIR_noVPM_l1e_5}
 \end{subfigure}
 \begin{subfigure}[!h]{0.47\textwidth}
 \includegraphics[width=\textwidth]{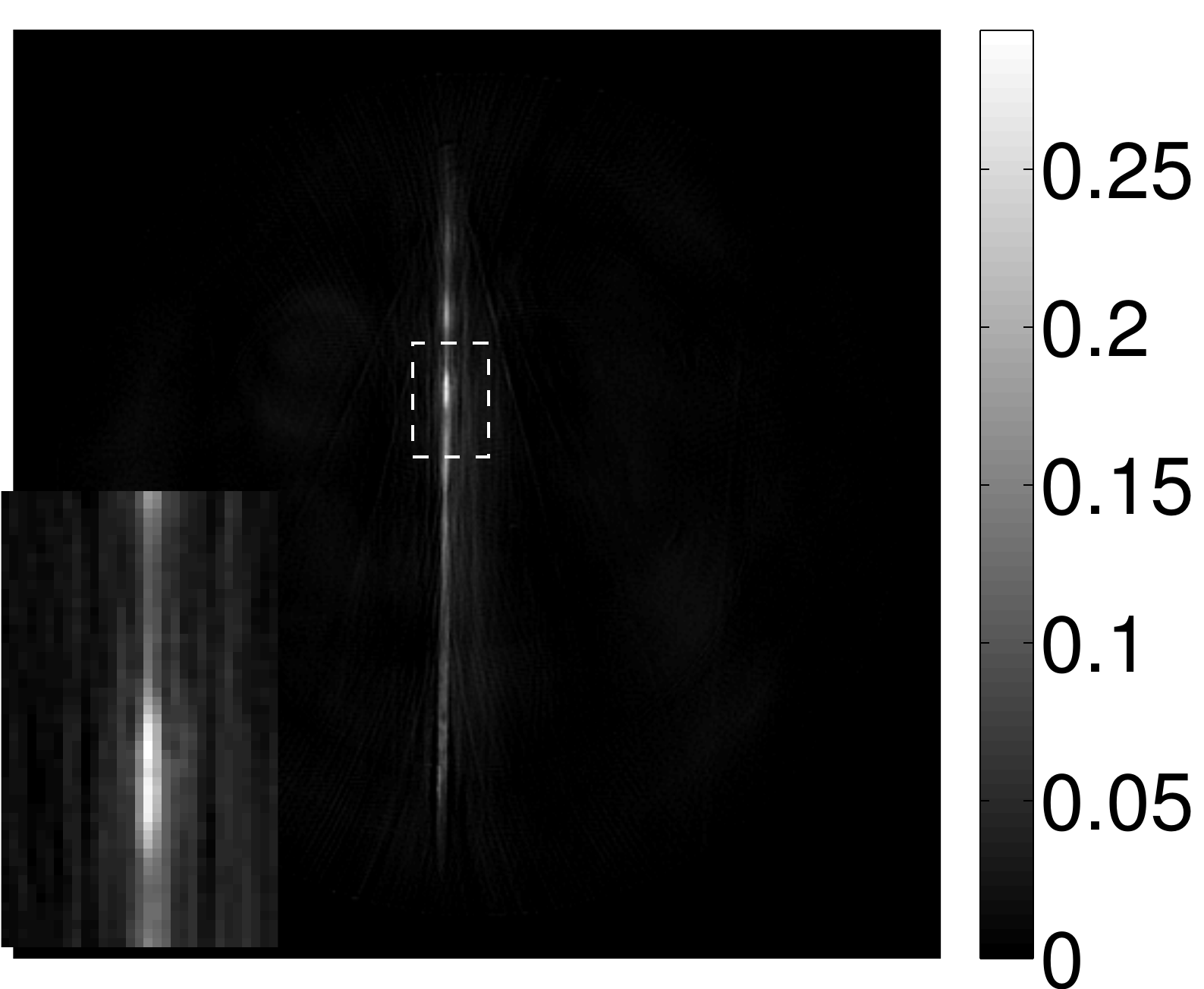}
  \caption{$\lambda=1.0\times 10^{-3}$}\label{2D_needle_EIR_noVPM_l1e_3}
 \end{subfigure}
\caption{ Reconstructed needle phantom image using the conventional iterative method with the non-negativity constraint. (a) $\lambda=1.0\times 10^{-5}$ and (b) $\lambda=1.0\times 10^{-3}$.
The zoomed-in image corresponds to the ROI of the dashed rectangle.
} \label{2D_needle_CIM}
\end{figure}

\begin{figure}[!htb]
 \centering
  \begin{subfigure}[!h]{0.5\textwidth}
 \includegraphics[width=\textwidth]{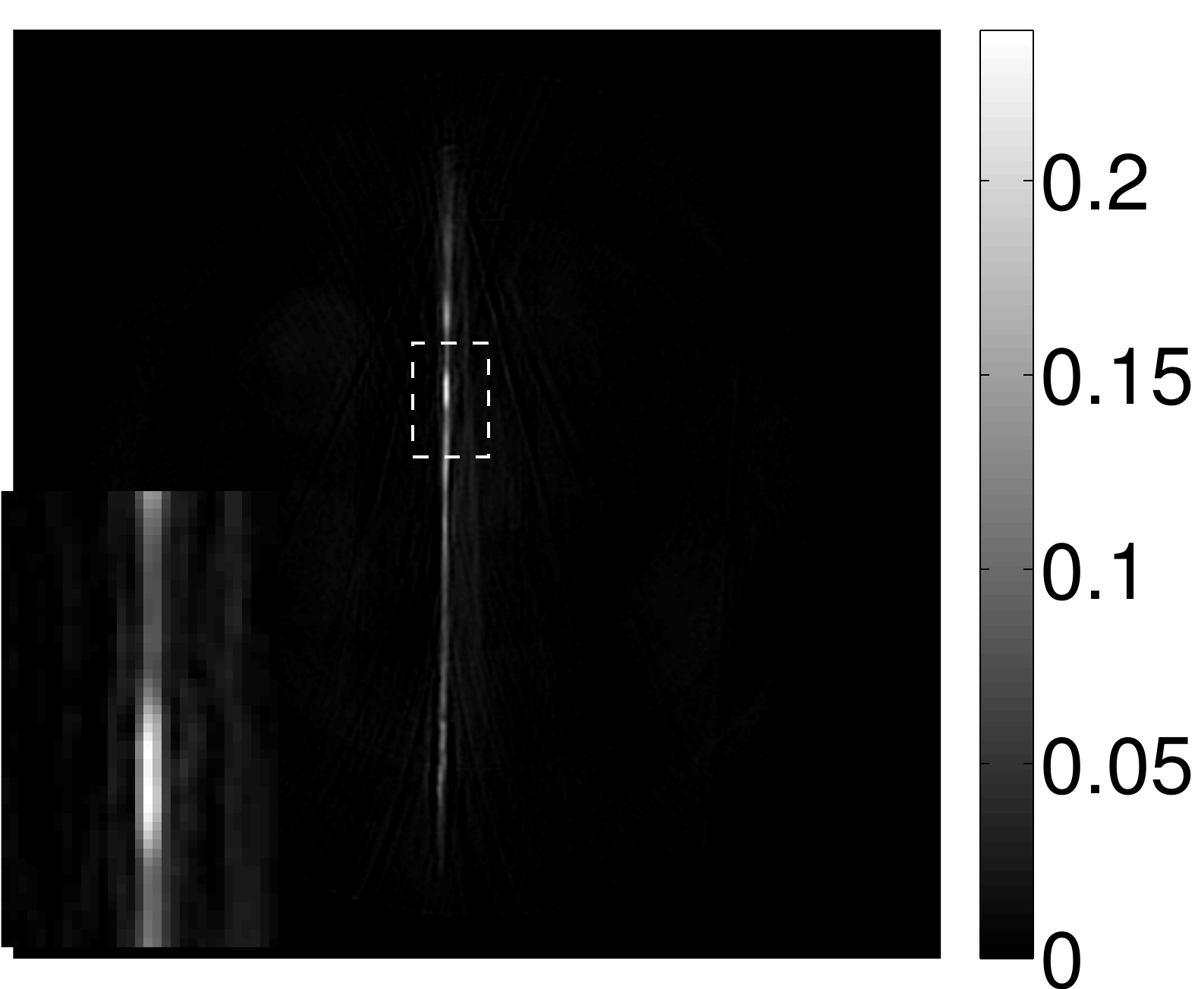}
 \caption{$\lambda=1.0\times 10^{-5}, \alpha=5000$}
 \end{subfigure}
 \begin{subfigure}[!h]{0.5\textwidth}
 \includegraphics[width=\textwidth]{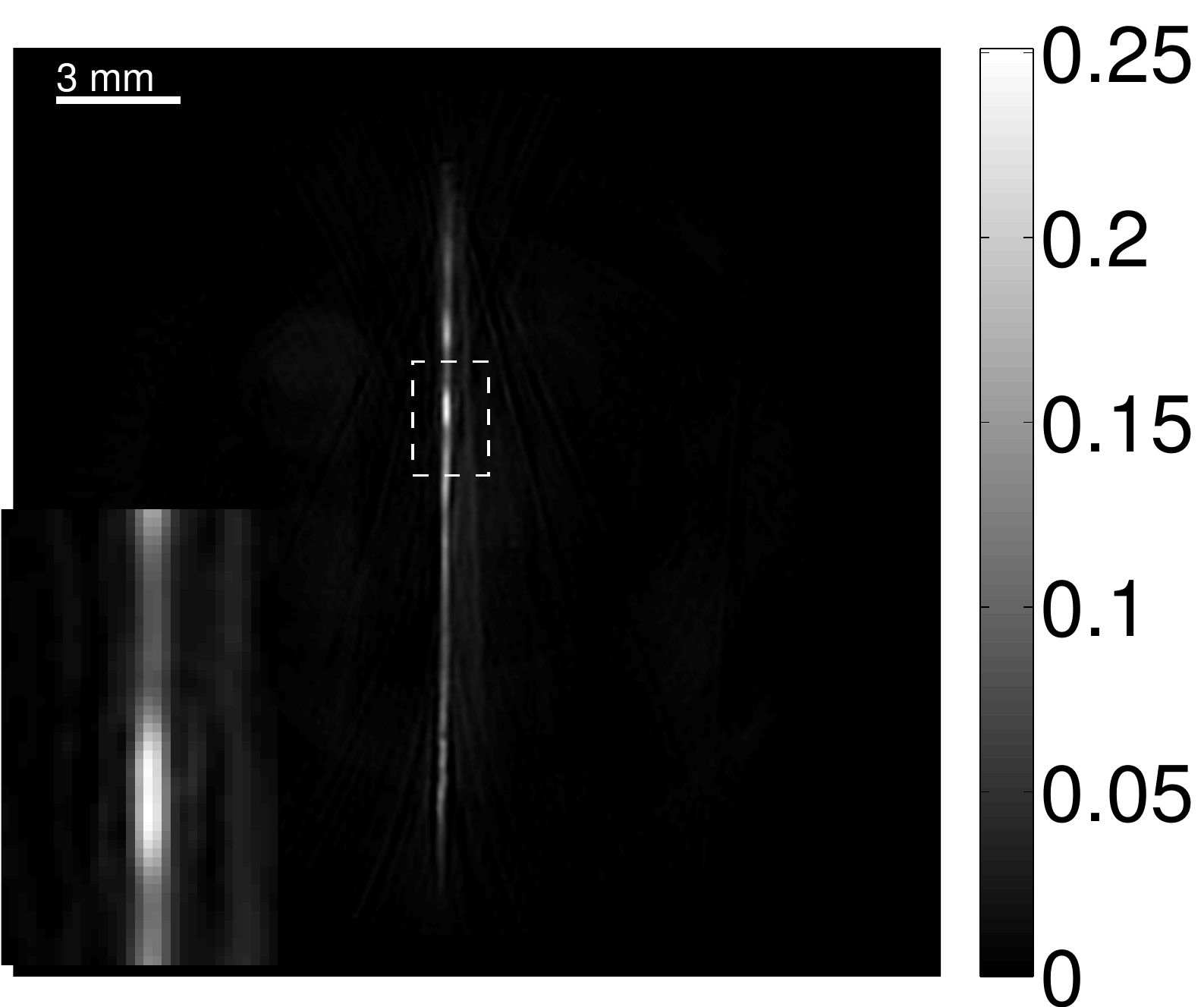}
 \caption{$\lambda=1.0\times 10^{-4}, \alpha=5000$}\label{2D_needle_vpm_init_hexp}
 \end{subfigure}
 \begin{subfigure}[!h]{0.45\textwidth}
 \includegraphics[width=\textwidth]{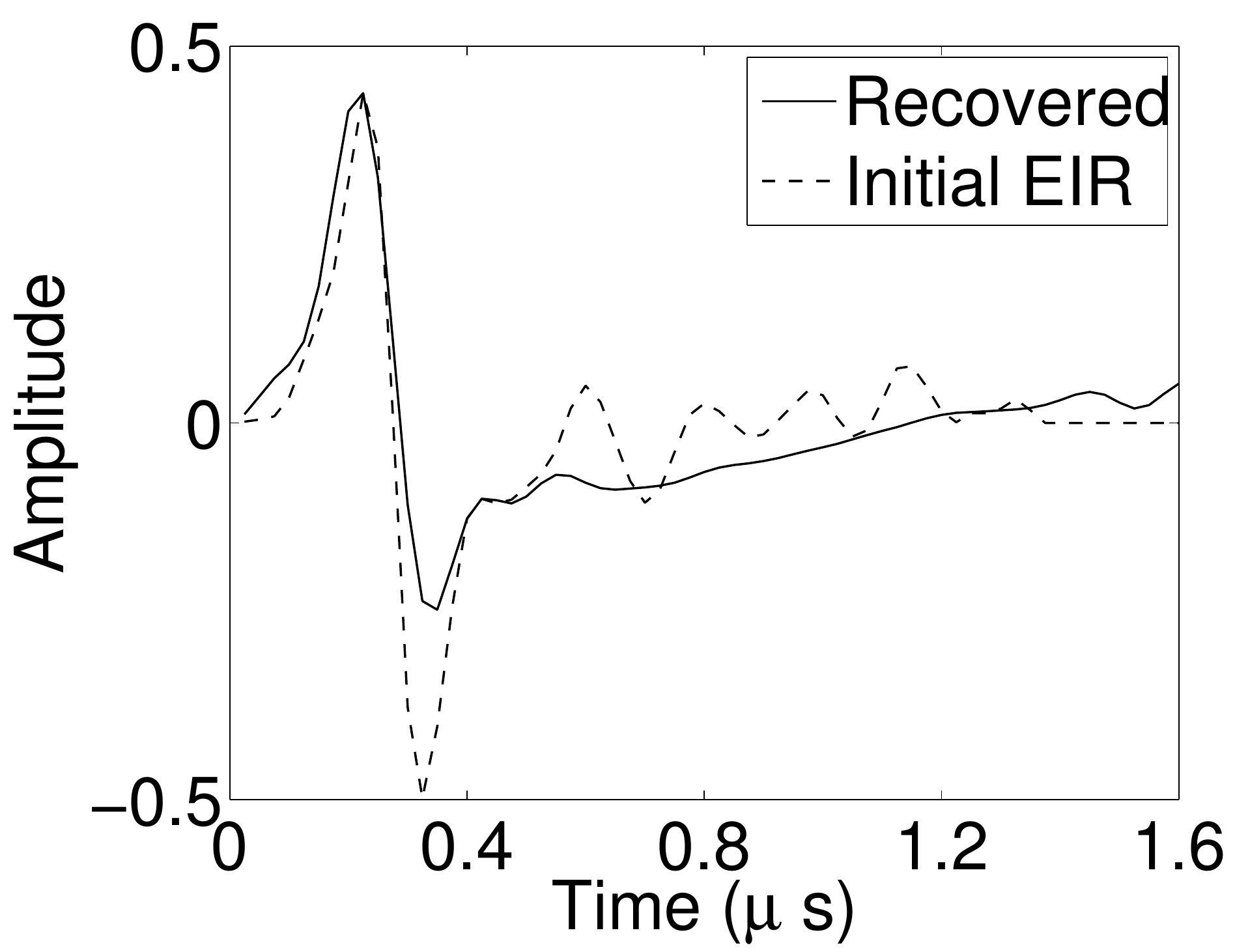}
  \caption{$\lambda=1.0\times 10^{-4}, \alpha=5000$}\label{2D_needle_vpm_init_hexp_eir}
 \end{subfigure}
\caption{(a) and (b) Images reconstructed by use of the VP algorithm corresponding
to the initial EIR guesses  shown in the plots of Fig.~\ref{2D_needle_vpm_init_hexp_eir}. (a) $\lambda=1.0\times 10^{-5}$ and (b) $\lambda=1.0\times 10^{-3}$. The zoomed-in image corresponds to the ROI of the dashed rectangle.
Images are displayed in their full dynamic ranges respectively.
Fig.~\ref{2D_needle_vpm_init_hexp_eir} shows the recovered and initial EIRs.} \label{2D_needle_VPM}
\end{figure}

Figure~\ref{2D_needle_BP} displays images of the needle phantom reconstructed by use of the simple backprojection method \cite{BM2003}. Figures~\ref{2D_needle_CIM} and \ref{2D_needle_VPM} display the images reconstructed by use of the conventional iterative method and VP algorithm, respectively.

\begin{figure}[!htb]
 \centering
 \includegraphics[width=0.6\textwidth]{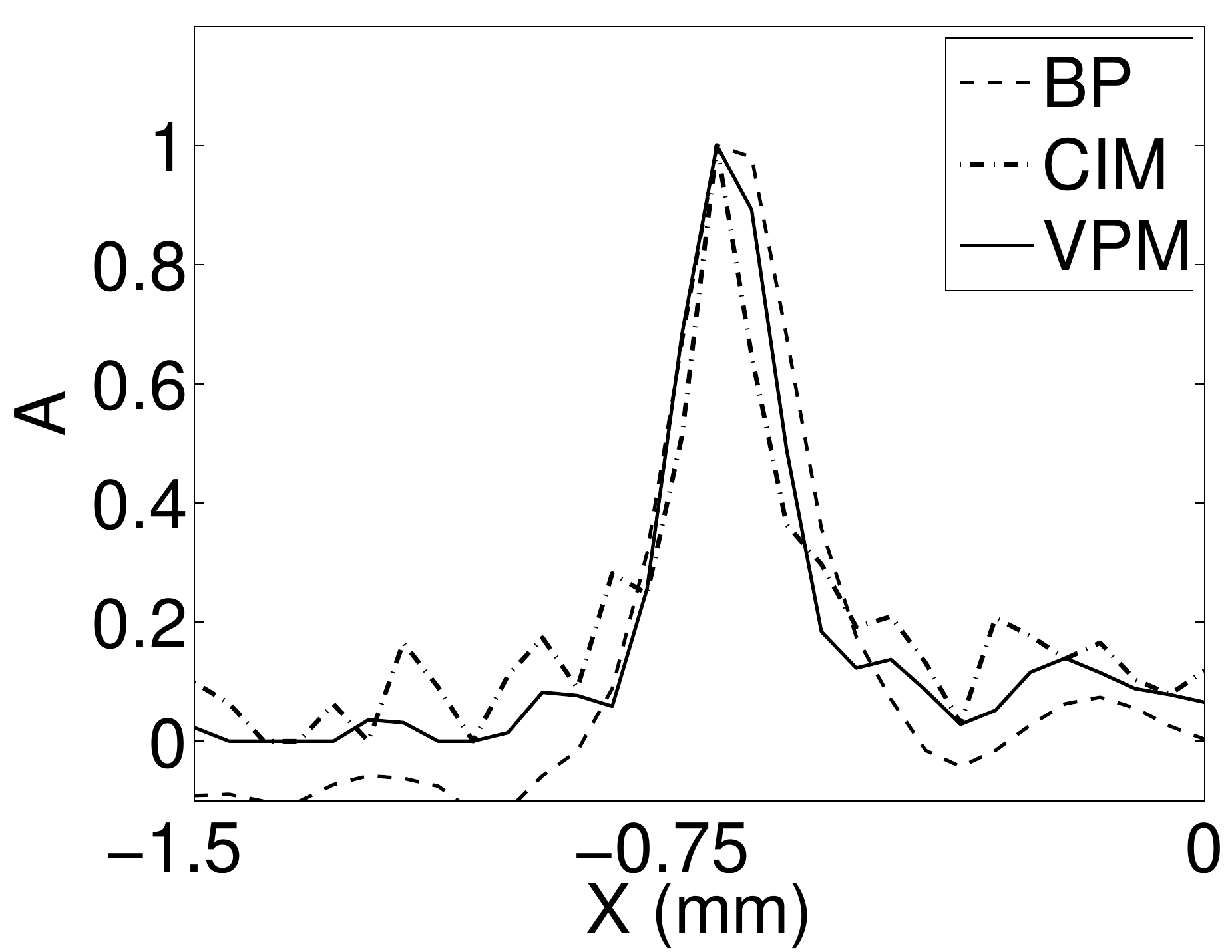}
 \caption{Image profiles at $y=2.4$ mm along the $x$-axis from $-1.5$ mm to $0$ mm extracted from images ($22.0\times 22.0$ mm$^2$ and centered at $(0.0,0.0)$) reconstructed by the backprojection algorithm (dashed line), the conventional iterative method (CIM) with $\lambda=1.0\times 10^{-4}$, and the VP algorithm with $\lambda=1.0\times 10^{-4}$ and $\alpha=5000$.} \label{2D_needle_profile}
\end{figure}
Figures~\ref{2D_needle_CIM} and \ref{2D_needle_VPM}  show that the width of the needle in the reconstructed image increases as the regularization parameter $\lambda$ increases for both the conventional iterative method and the VP algorithm. 
The images reconstructed by use of the VP algorithm appear to have a reduced noise level compared to the images reconstructed by the backprojection and conventional iterative methods, regardless of the choice of the regularization parameter values. The profile plots corresponding to these three methods are shown in Figure~\ref{2D_needle_profile}. Since the image of the coefficient vector $\boldsymbol\theta$ and the EIR $\mathbf{h}$ are recoverable only up to a multiplicative constant, every profile was normalized for comparison. 
These plots demonstrate that the image reconstructed by use of the VP algorithm possessed a more uniform background than those obtained by the backprojection and the conventional iterative methods. 

\subsection{Results: kidney phantom}
\begin{figure}[!htb]
 \centering
 \begin{subfigure}[!h]{0.49\textwidth}
 \includegraphics[width=\textwidth]{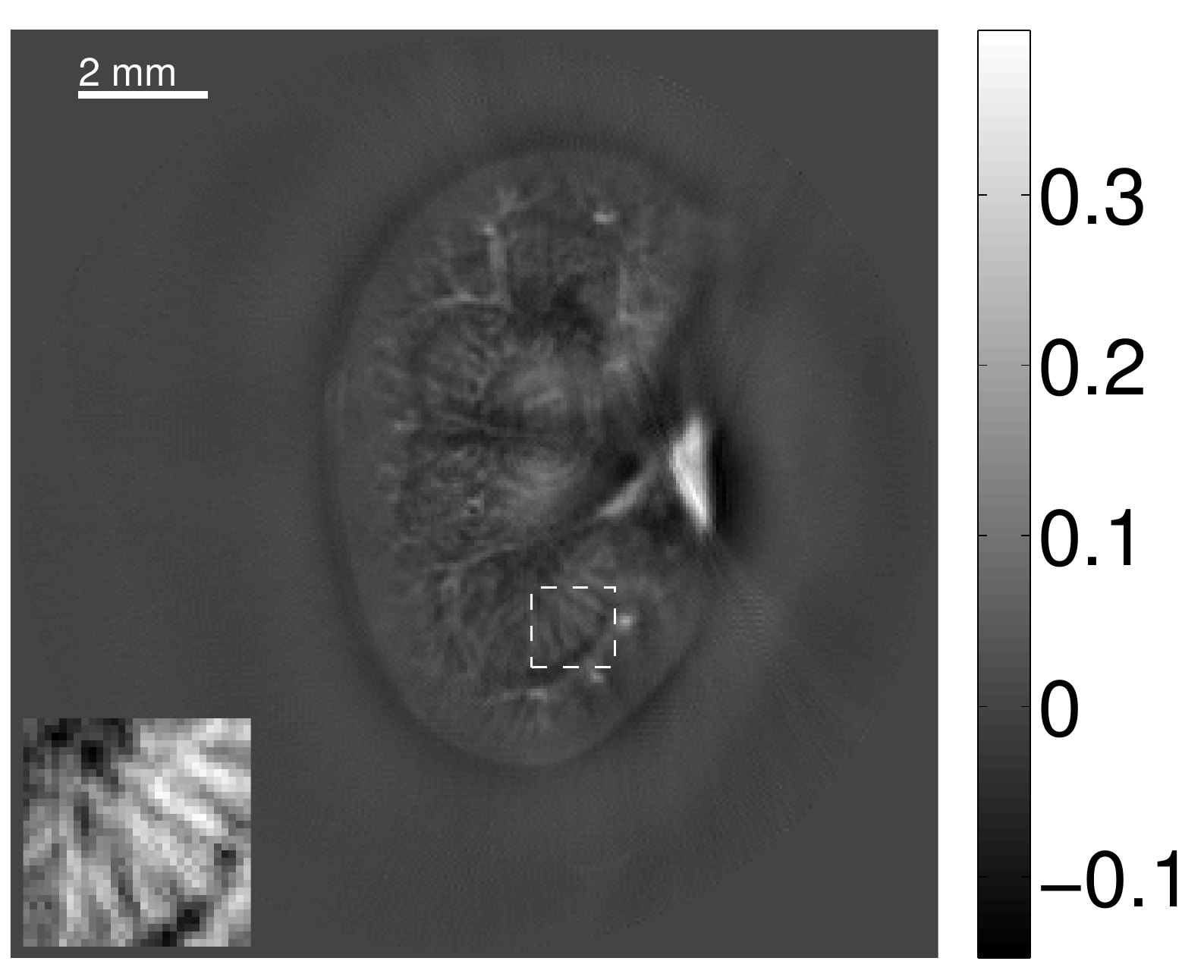}
  \caption{$\lambda=0.0$}\label{2D_mouse_kidney2_EIR_noVPM_NoPos}
 \end{subfigure}
  \begin{subfigure}[!h]{0.49\textwidth}
 \includegraphics[width=\textwidth]{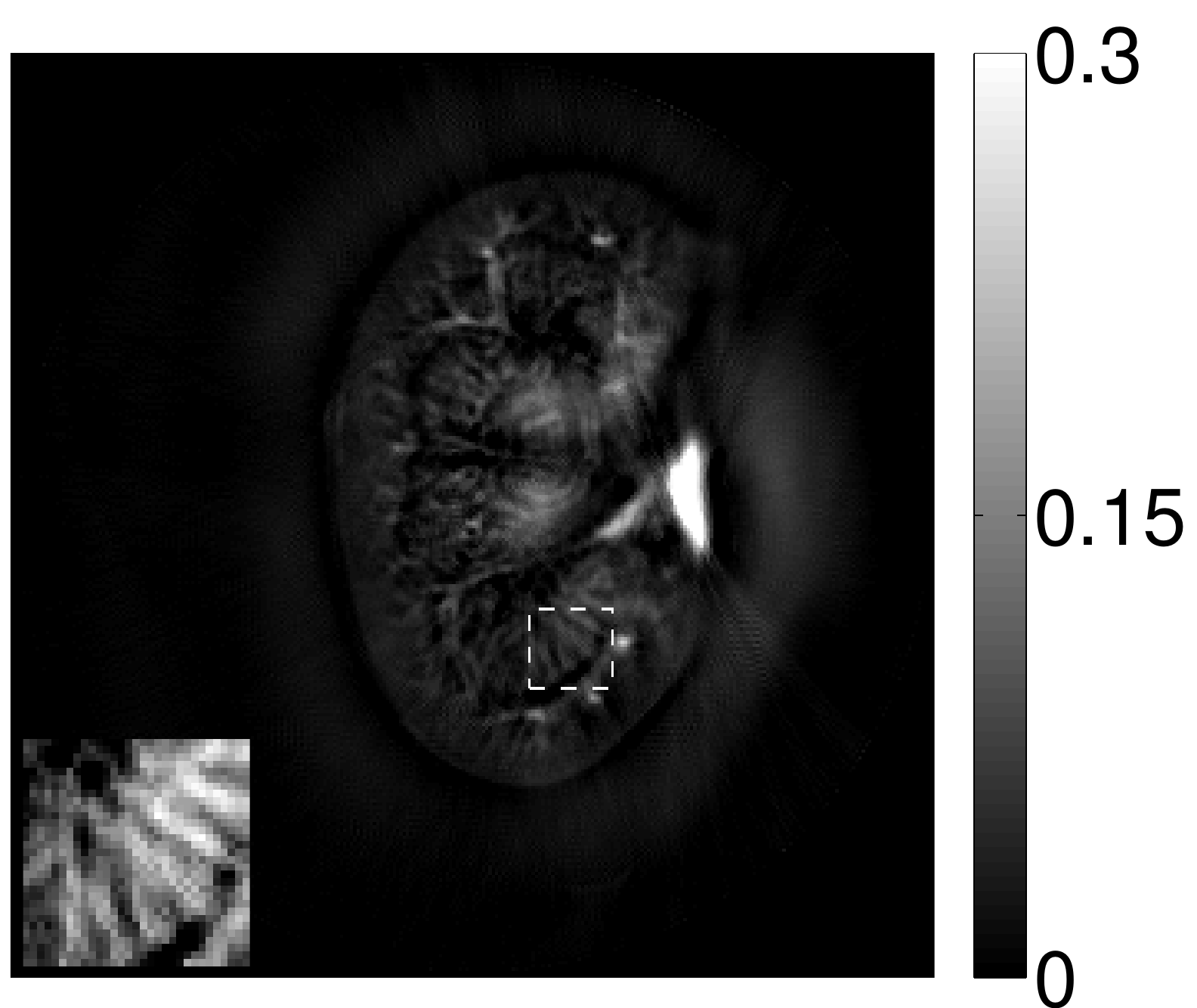}
  \caption{$\lambda=0.0$}\label{2D_mouse_kidney2_EIR_noVPM}
 \end{subfigure}
 \begin{subfigure}[!h]{\textwidth}
 \includegraphics[width=0.49\textwidth]{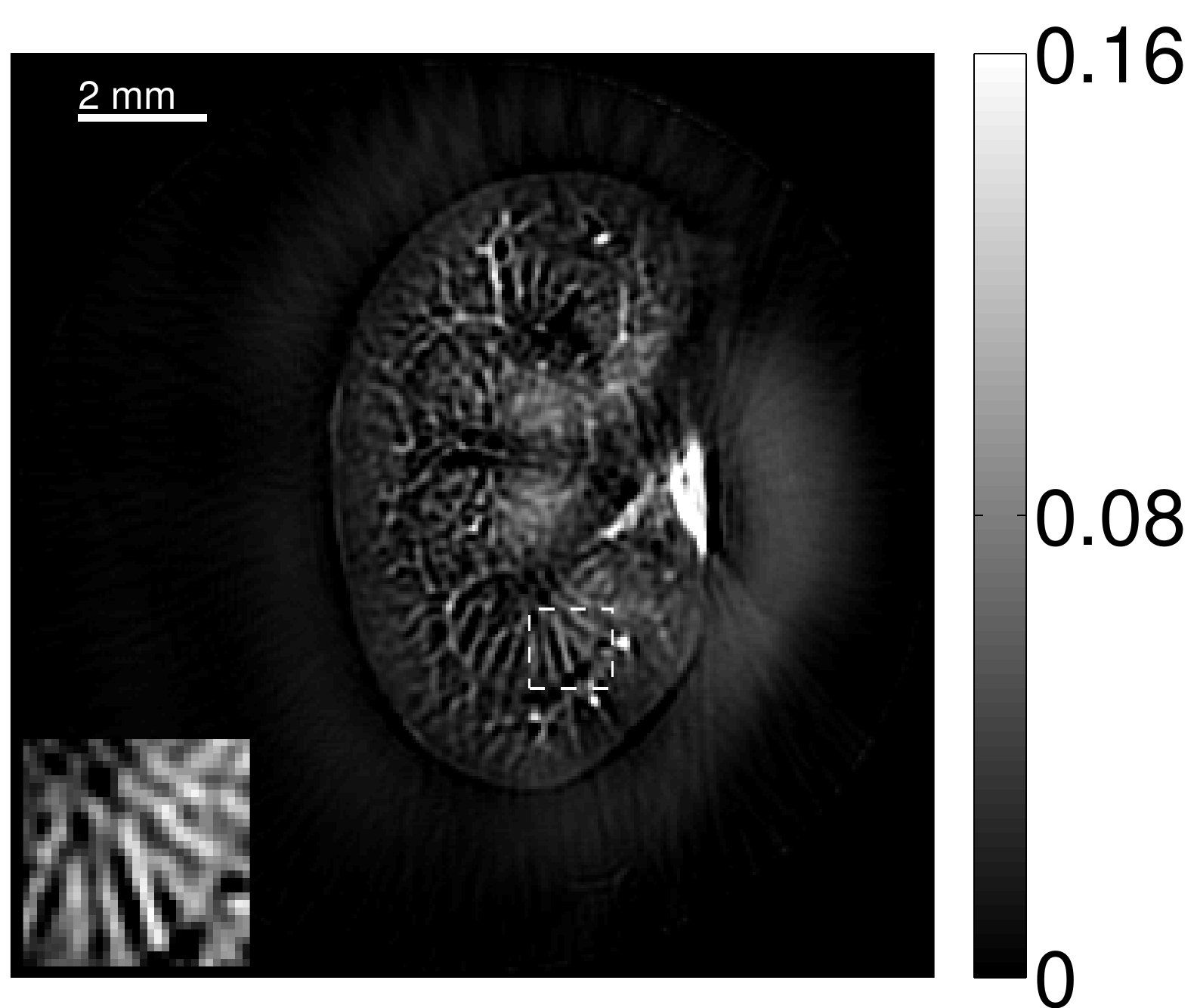}
 \includegraphics[width=0.49\textwidth]{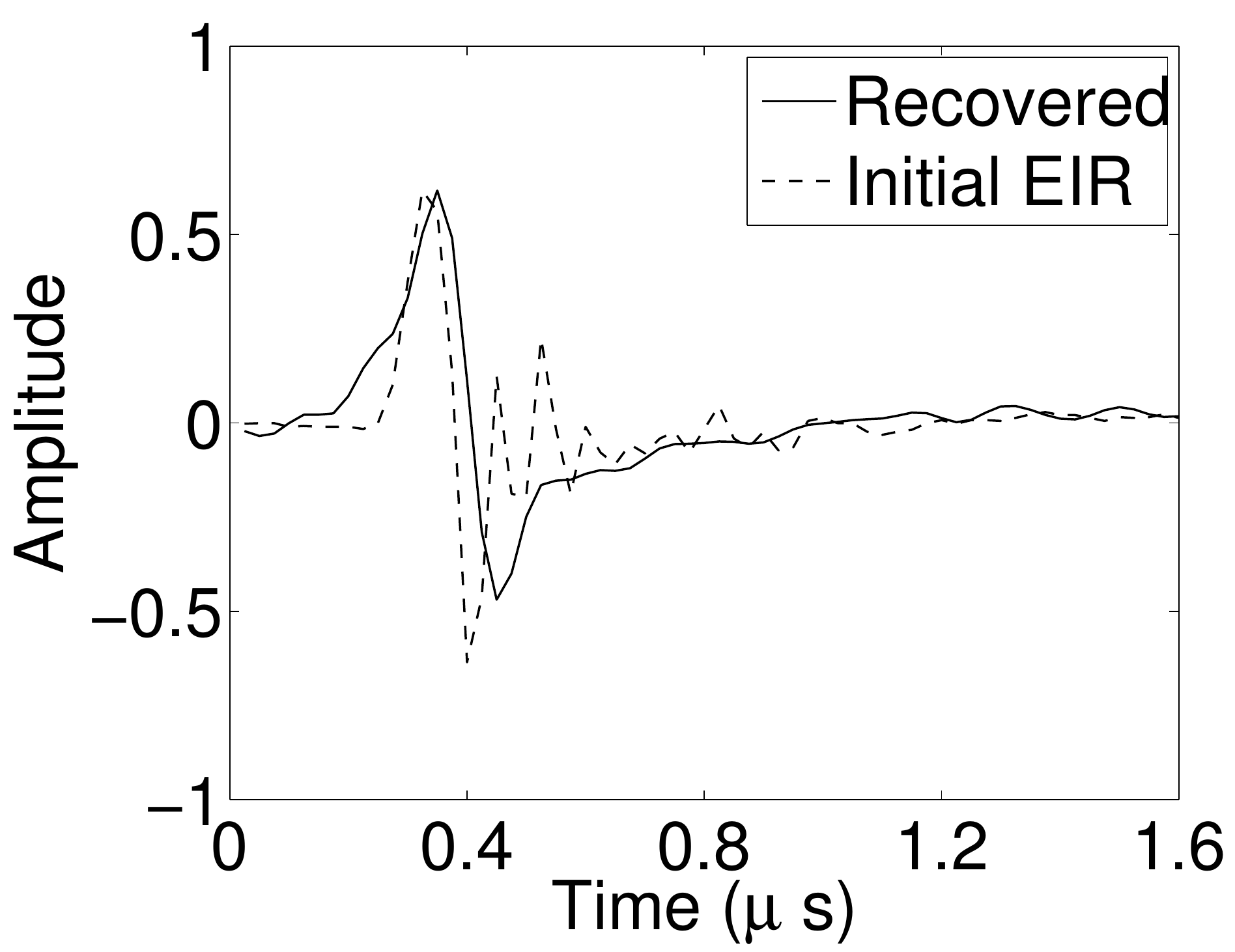}
  \caption{$\lambda=1.0\times 10^{-5}, \alpha=5000$
  }\label{2D_mouse_kidney2_vpm_init_hmouse}
 \end{subfigure}
\caption{(a)
 Image reconstructed by use of the  conventional iterative
 method without the non-negativity constraint.  
(b) Image reconstructed by use of the conventional iterative method with the non-negativity constraint.  
(c) Image reconstructed by use of the VP algorithm, with the initial guess of the EIR shown in the right plots of subfigure (c).  
The zoomed-in image corresponds to the ROI of the dashed rectangle. 
\hl{ Images are displayed in their full dynamic ranges respectively.}
The right plot shows the recovered and initial EIR.  The SIR was ignored in these studies.
}\label{2D_mouse_kidney2_EIR_mouse}
\end{figure}

\begin{figure}[!htb]
 \centering
  \begin{subfigure}[!h]{\textwidth}
 \includegraphics[width=0.43\textwidth]{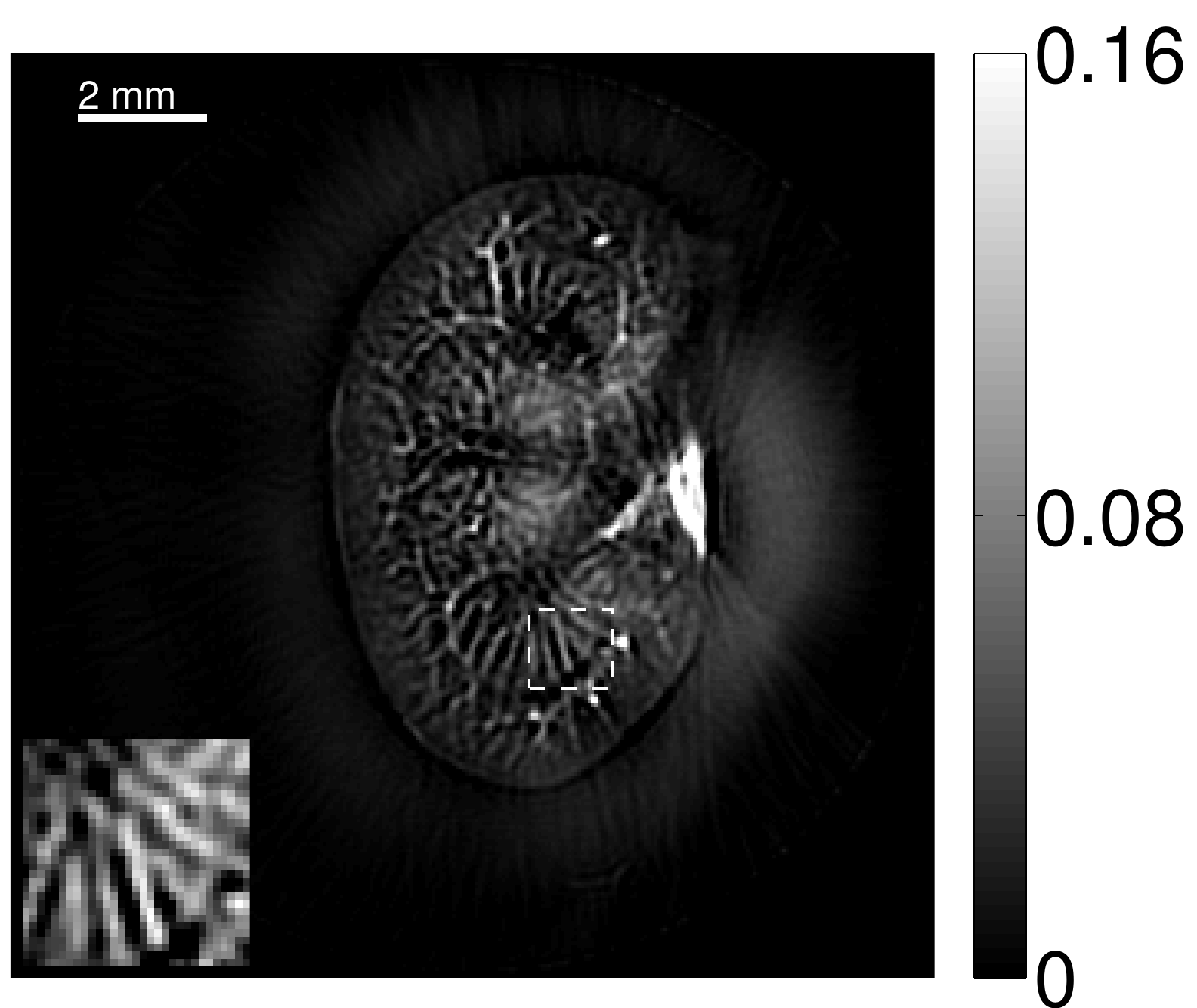}
 \includegraphics[width=0.43\textwidth]{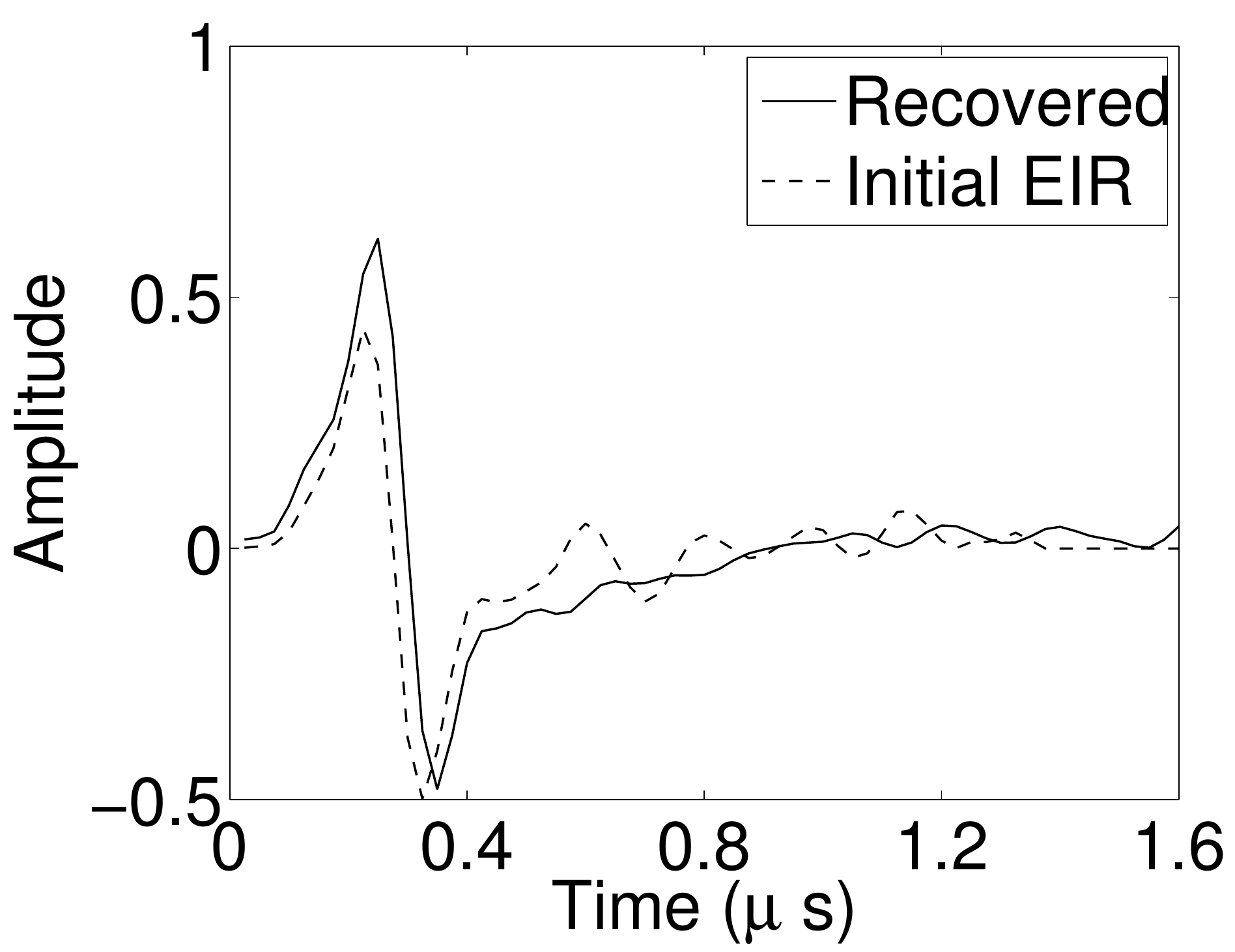}
  \caption{$\lambda=1.0\times 10^{-5},\alpha=5000$}\label{2D_mouse_kidney2_vpm_init_exp_l1e_5_a5000}
 \end{subfigure}
   \begin{subfigure}[!h]{\textwidth}
 \includegraphics[width=0.43\textwidth]{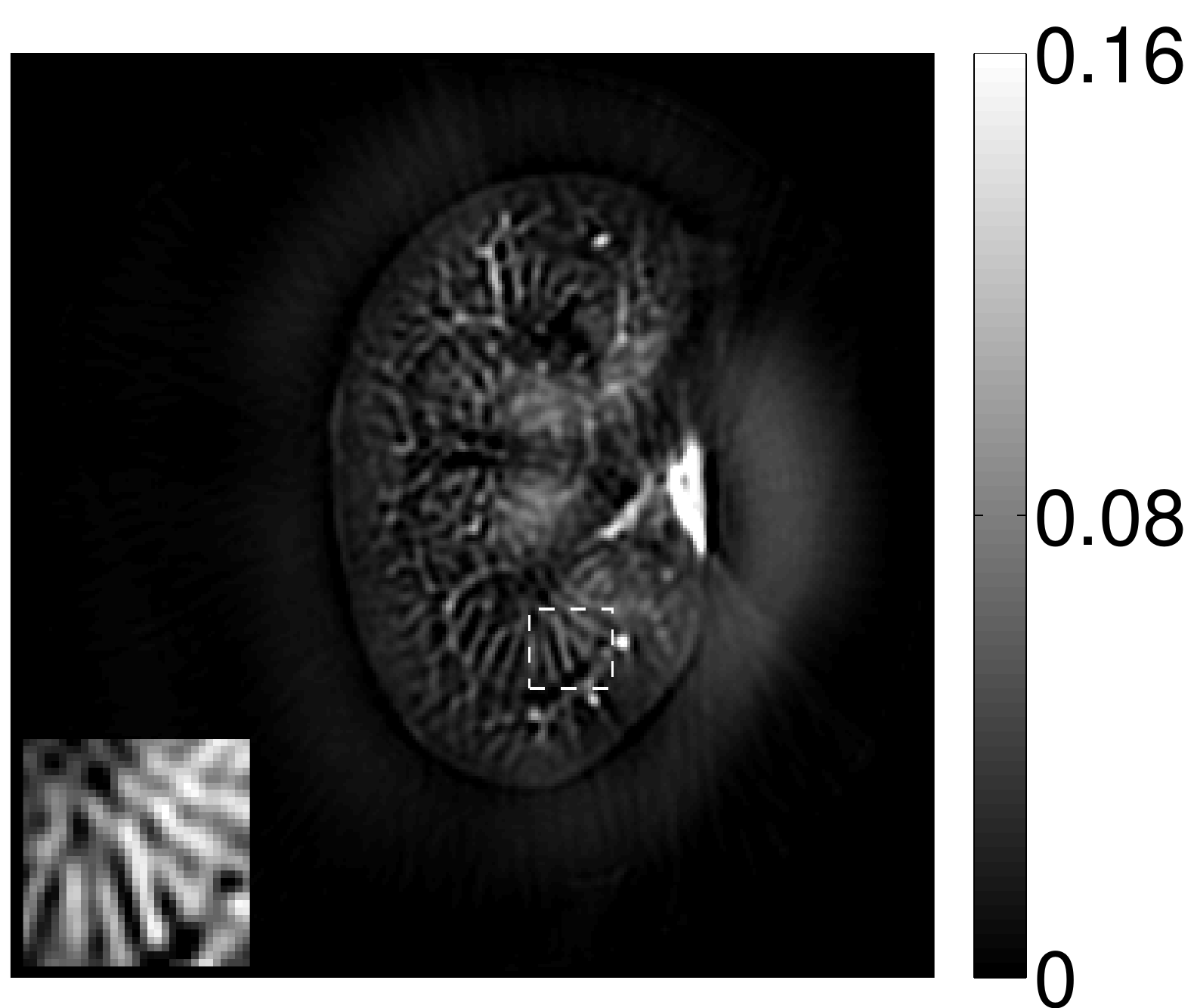}
 \includegraphics[width=0.43\textwidth]{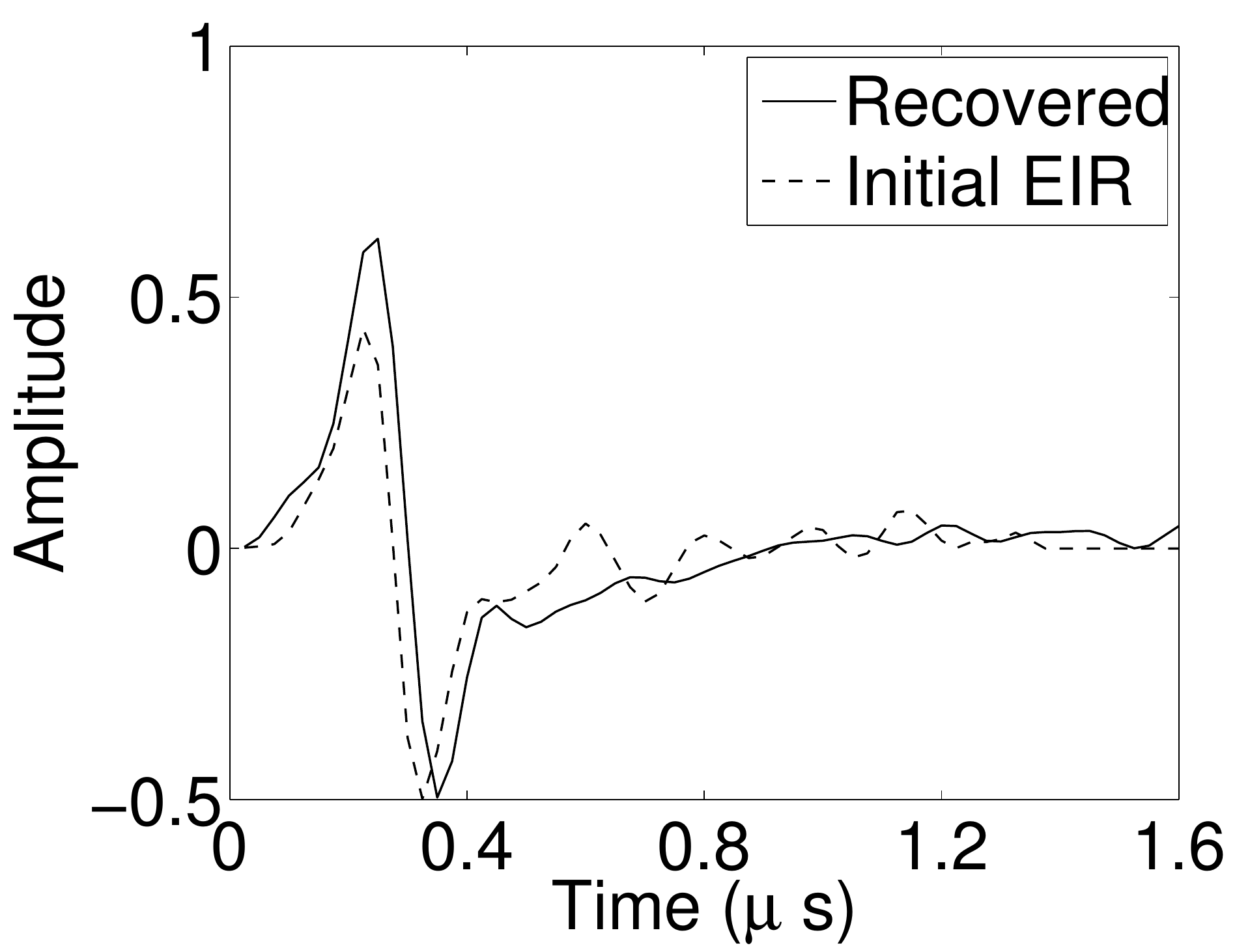}
  \caption{$\lambda=1.0\times 10^{-4},\alpha=5000$}\label{2D_mouse_kidney2_vpm_init_exp_l1e_4_a5000}
 \end{subfigure}
   \begin{subfigure}[!h]{\textwidth}
 \includegraphics[width=0.43\textwidth]{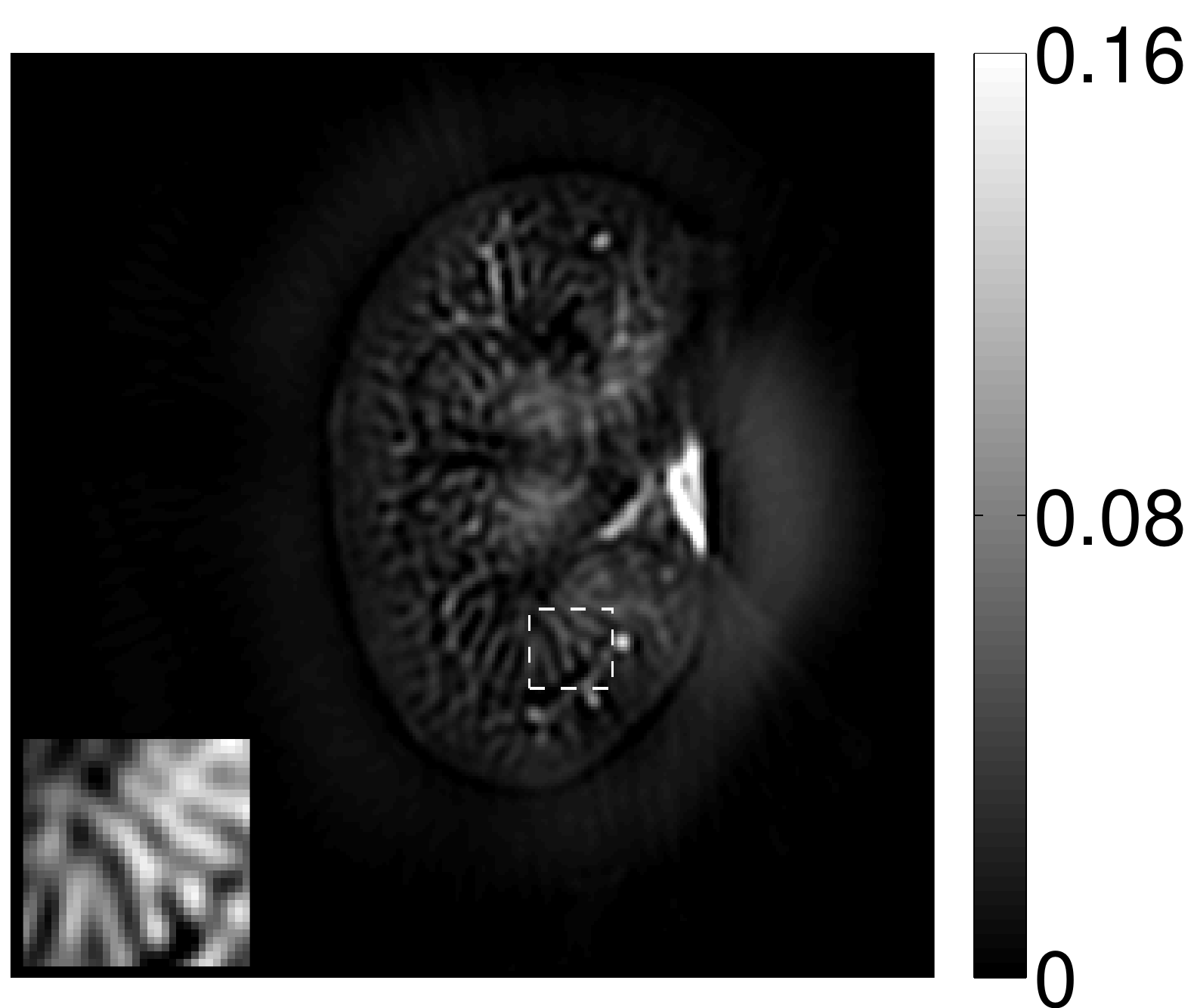}
 \includegraphics[width=0.43\textwidth]{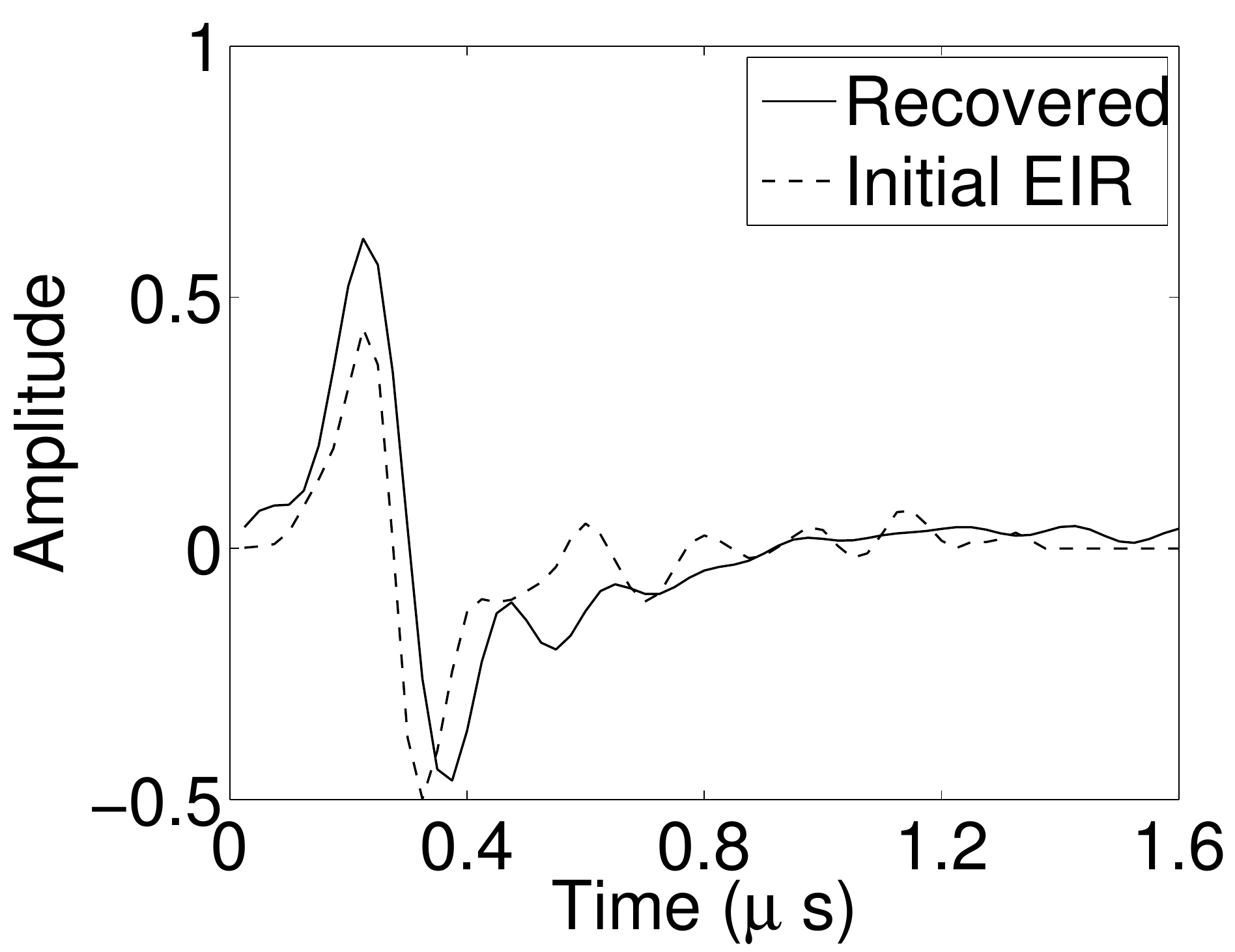}
  \caption{$\lambda=1.0\times 10^{-3},\alpha=5000$}\label{2D_mouse_kidney2_vpm_init_exp_l1e_3_a5000}
 \end{subfigure}
\caption{The left panels of subfigures (a), (b), and (c) display
 images reconstructed by use of the VP algorithm and different regularization parameters with the initial guess of the EIR shown in the corresponding right panels.  The SIR was ignored in all cases. The zoomed-in image corresponds to the ROI of the dashed rectangle. 
The grayscale windows were $[0,0.17]$.}\label{2D_mouse_kidney2}
\end{figure}

The images and EIRs reconstructed by use of the VP algorithm that was based on
the 2D imaging model that neglected the SIR are
 shown in Figures~\ref{2D_mouse_kidney2_EIR_mouse} and \ref{2D_mouse_kidney2}.
The latter figure contains results corresponding to different values for the regularization parameter $\lambda$.
From Fig.\ \ref{2D_mouse_kidney2_EIR_mouse}, it can be observed that
use of the conventional iterative method that utilized the measured EIR resulted in distortions and loss of details in the reconstructed images. Use of the VP algorithm improved
the contrast and the details in the reconstructed images  (Fig.~\ref{2D_mouse_kidney2_vpm_init_hmouse} and \ref{2D_mouse_kidney2_vpm_init_exp_l1e_5_a5000}). Furthermore, the images reconstructed by use of the VP algorithm had a more uniform background.

\begin{figure}[!htb]
 \centering
  \begin{subfigure}[!h]{0.49\textwidth}
 \includegraphics[width=\textwidth]{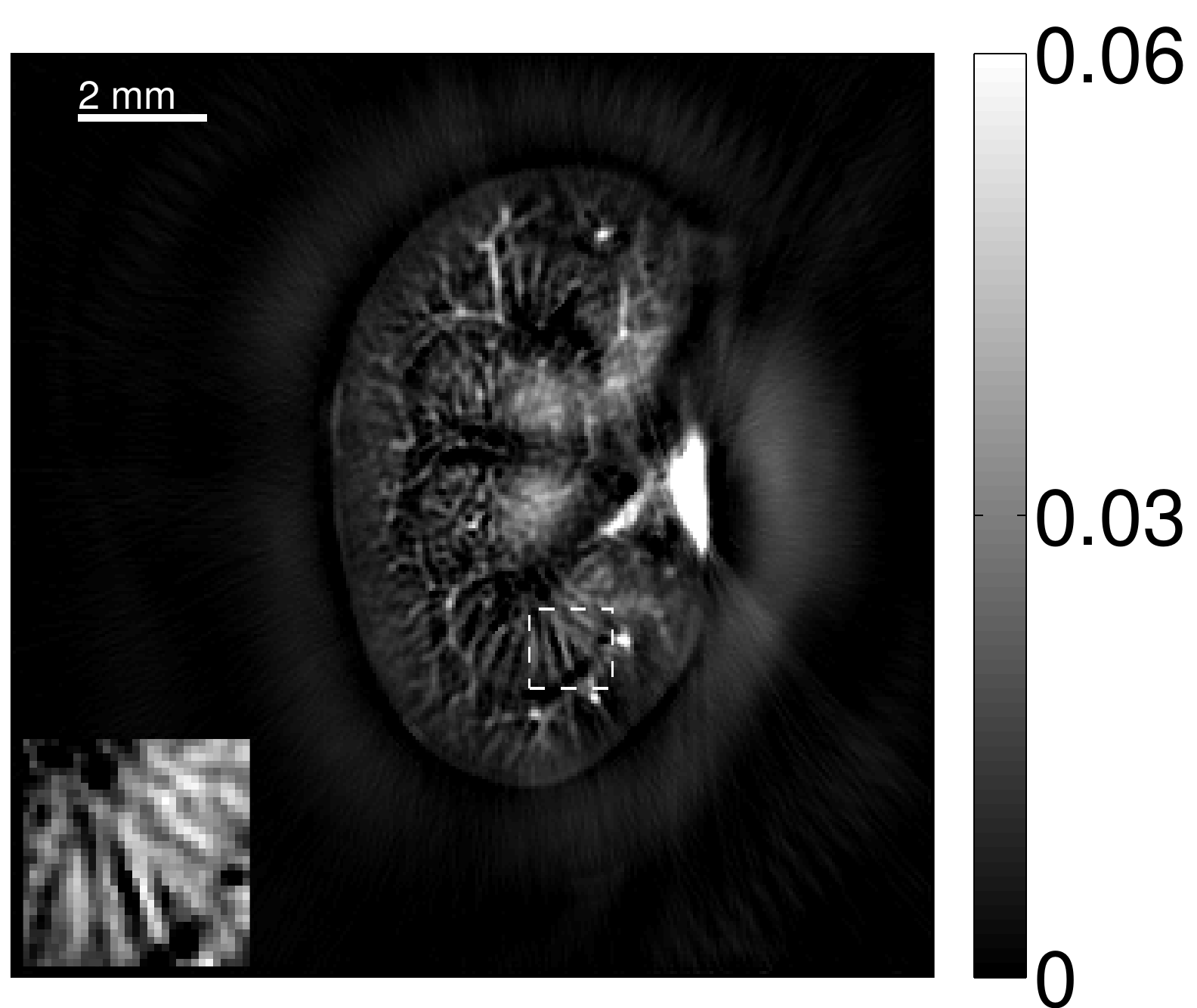}
  \caption{$\lambda=1.0\times 10^{-5}$}\label{2D_mouse_kidney2_SIR_noVPM}
 \end{subfigure}
 \begin{subfigure}[!h]{\textwidth}
 \includegraphics[width=0.49\textwidth]{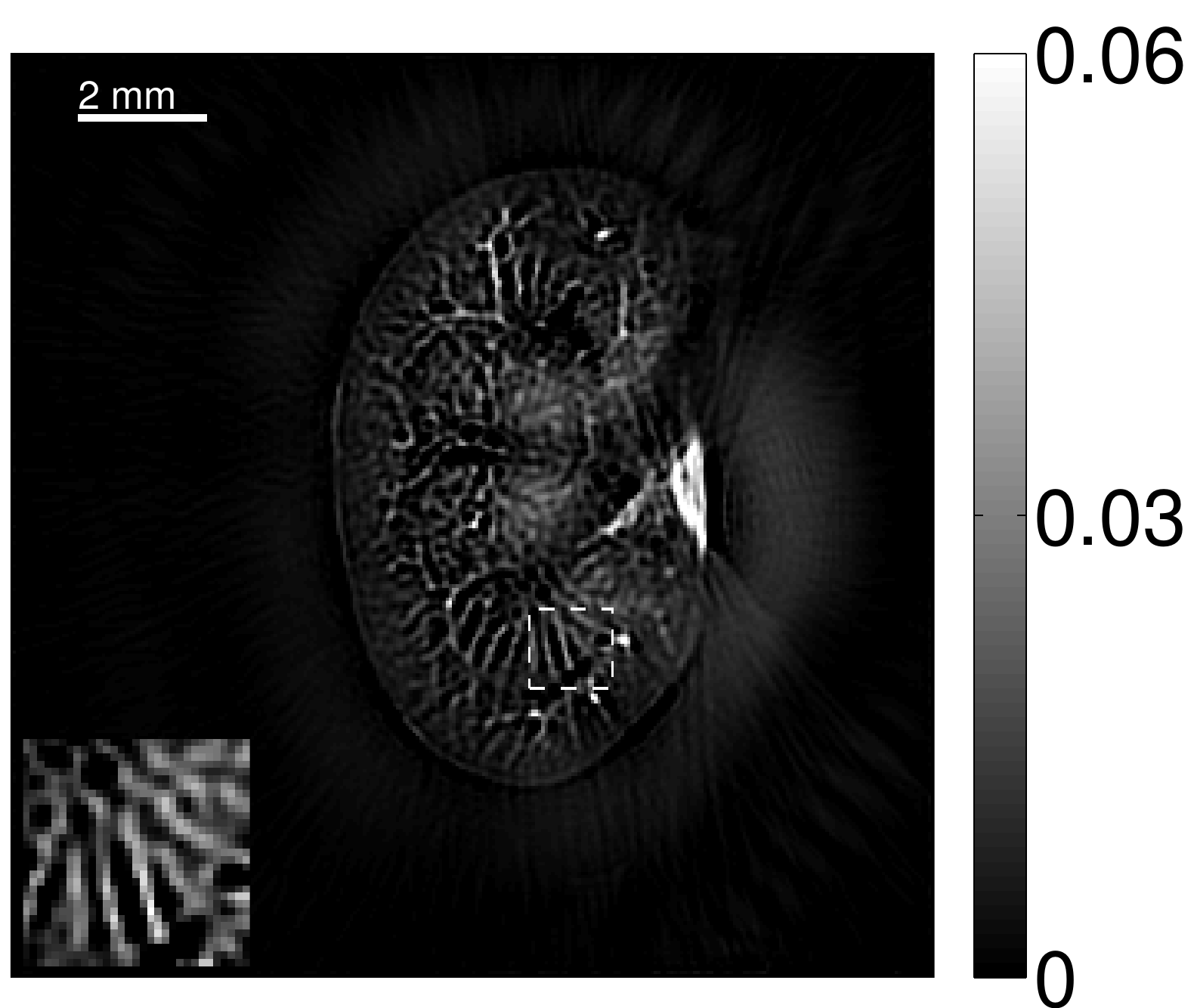}
 \includegraphics[width=0.49\textwidth]{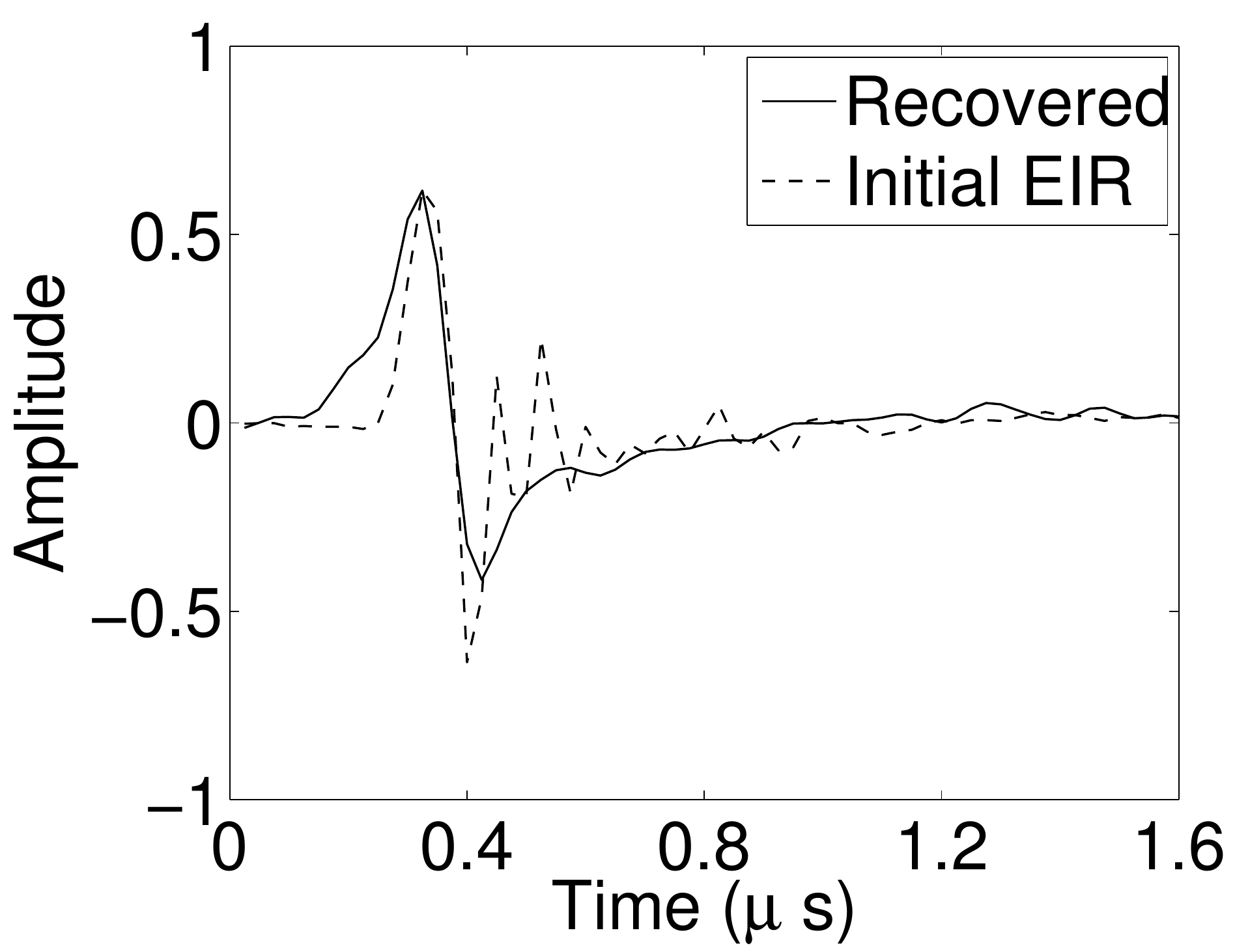}
  \caption{$\lambda=1.0\times 10^{-5}, \alpha=5000$
  }\label{2D_mouse_kidney2_SIR_VPM}
 \end{subfigure}
\caption{(a) Image reconstructed by use of the conventional iterative method with a
 non-negativity constraint. The grayscale window was $[0, 0.06]$. (b) Image reconstructed 
by use of the VP algorithm, with the initial EIR guess shown in the right panel
 of subfigure (b). The zoomed-in image corresponds to the ROI of the dashed rectangle. The grayscale window was $[0, 0.06]$. The right plot shows the recovered and initial EIR. The SIR was accounted for in both cases.}\label{2D_mouse_kidney2_SIR_EIR_mouse}
\end{figure}

\begin{figure}[!htb]
 \centering
  \begin{subfigure}[!h]{\textwidth}
 \includegraphics[width=0.32\textwidth]{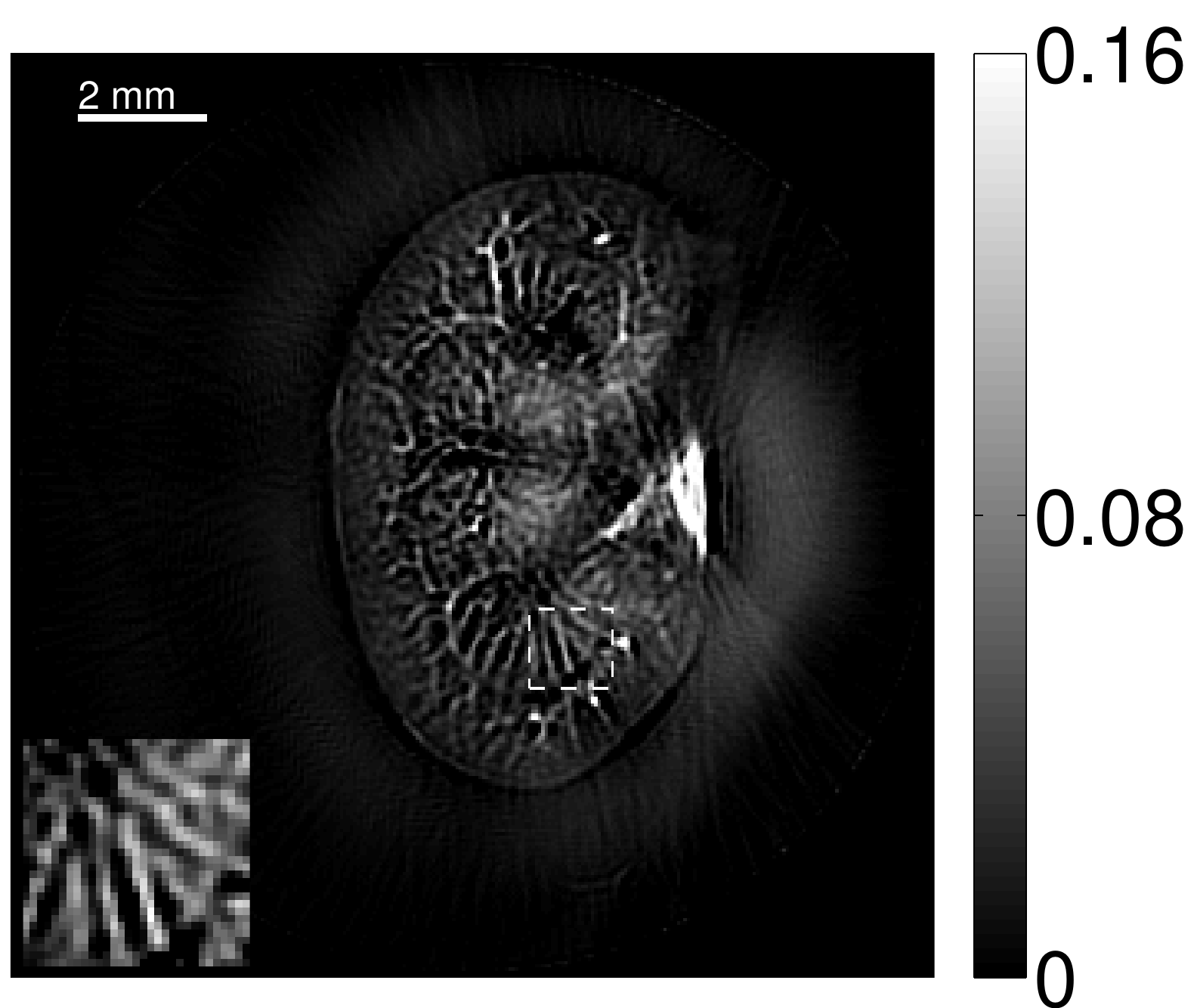}
 \includegraphics[width=0.32\textwidth]{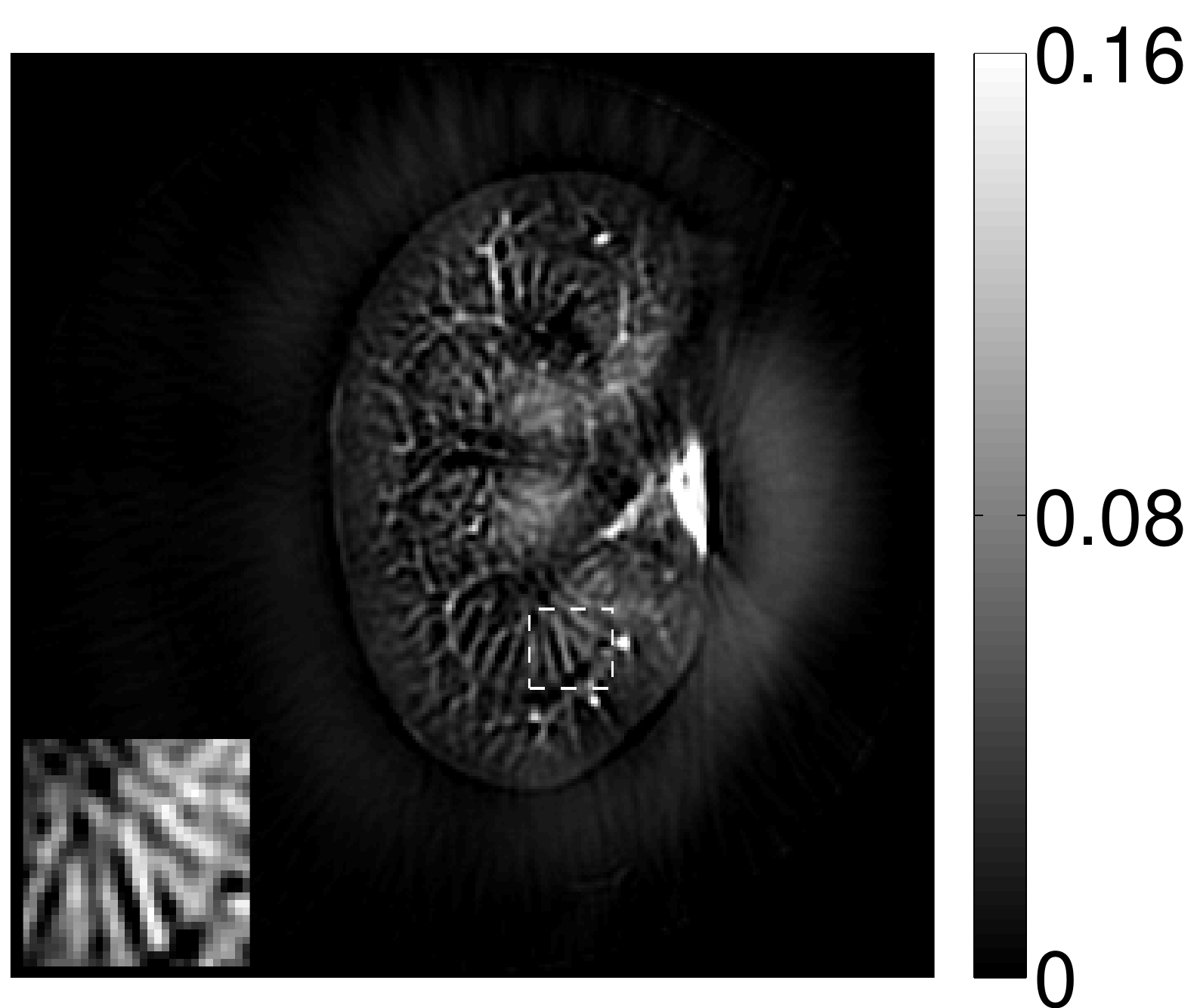}
 \includegraphics[width=0.32\textwidth]{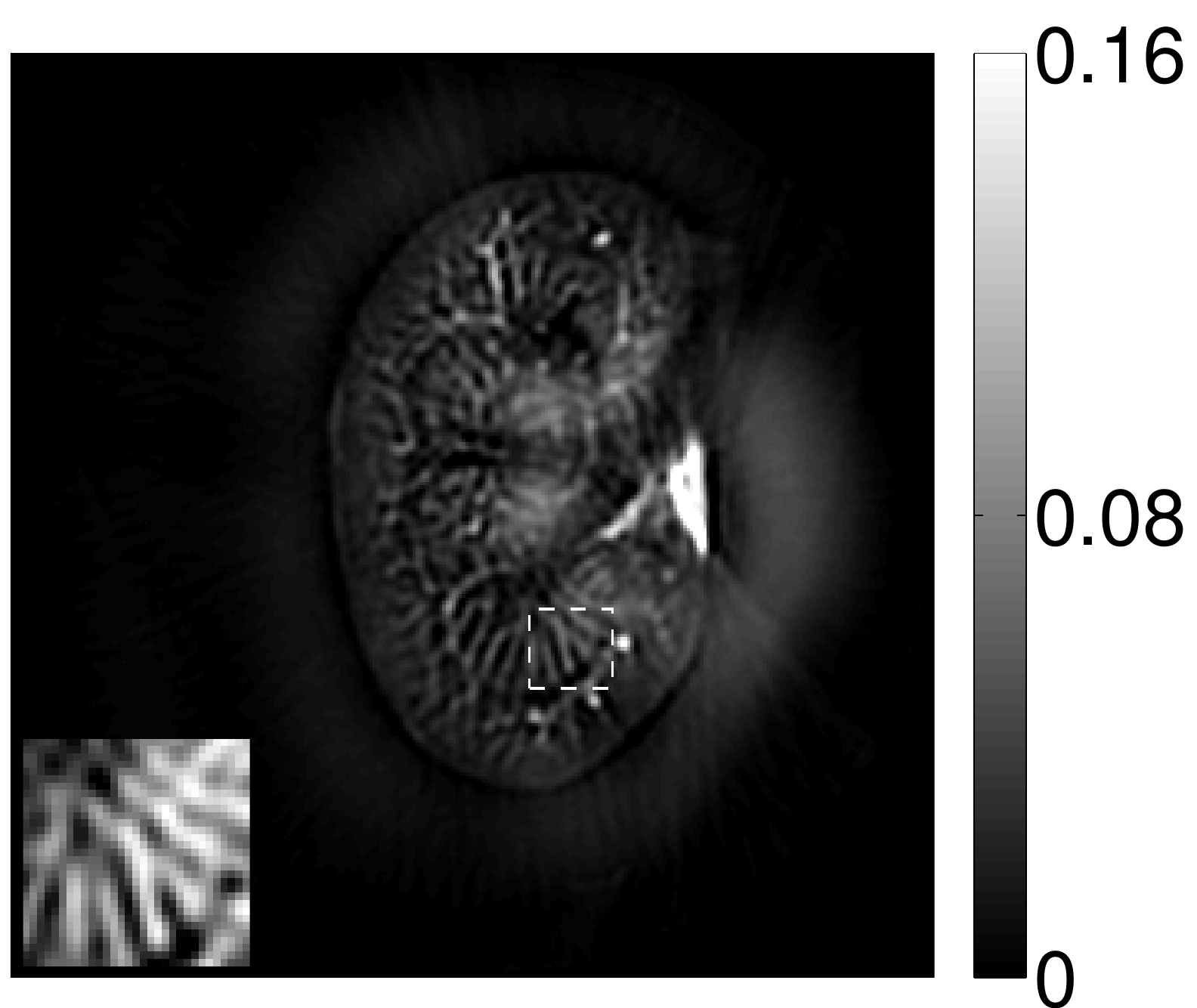}
  \caption{$\lambda=1.0\times 10^{-5}$, $\lambda=1.0\times 10^{-4}$,  $\lambda=1.0\times 10^{-3}$, $\alpha=1000$}\label{2D_mouse_kidney2_vpm_init_mouse_l1e_5_a1000}
 \end{subfigure}
   \begin{subfigure}[!h]{\textwidth}
 \includegraphics[width=0.31\textwidth]{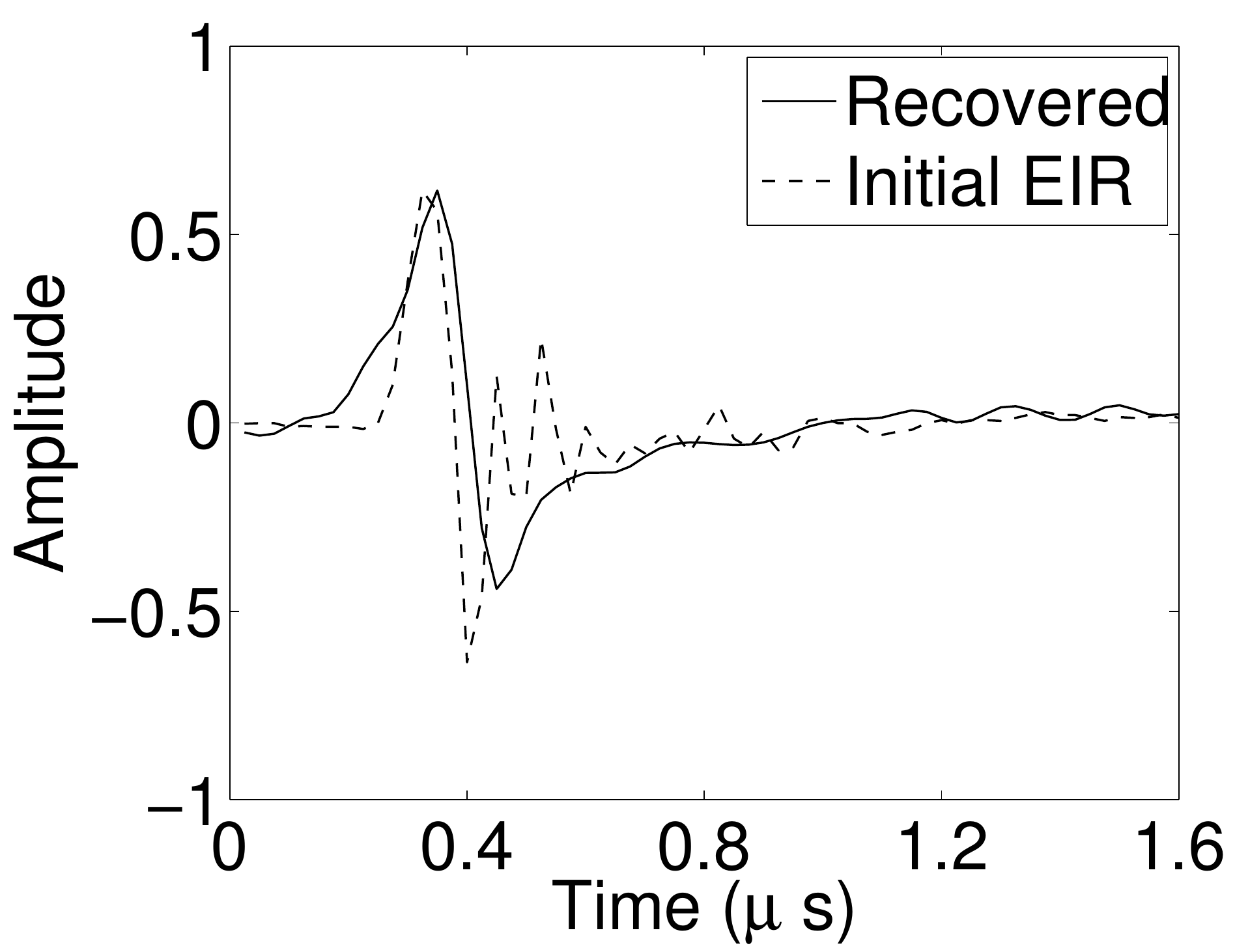}
 \includegraphics[width=0.31\textwidth]{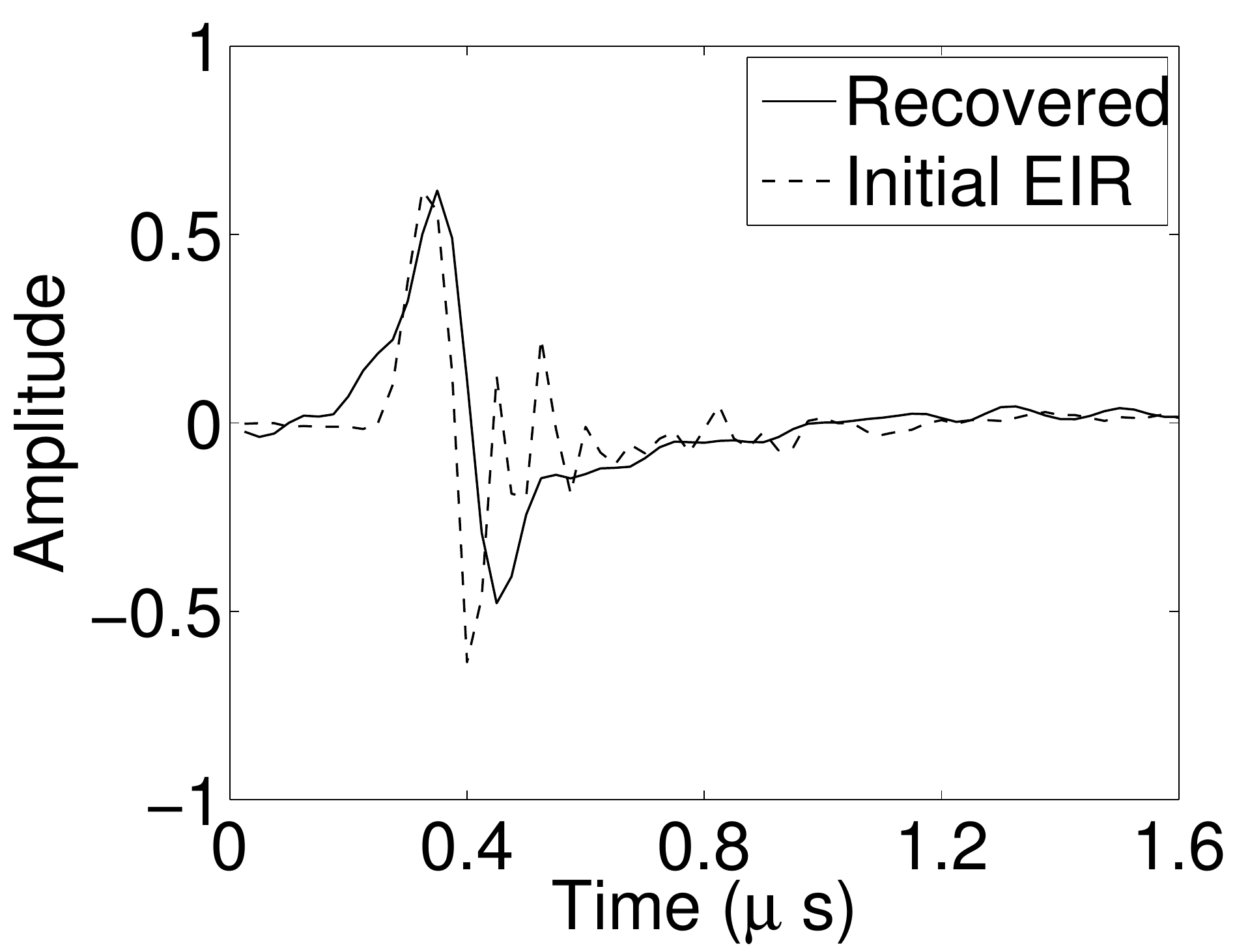}
 \includegraphics[width=0.31\textwidth]{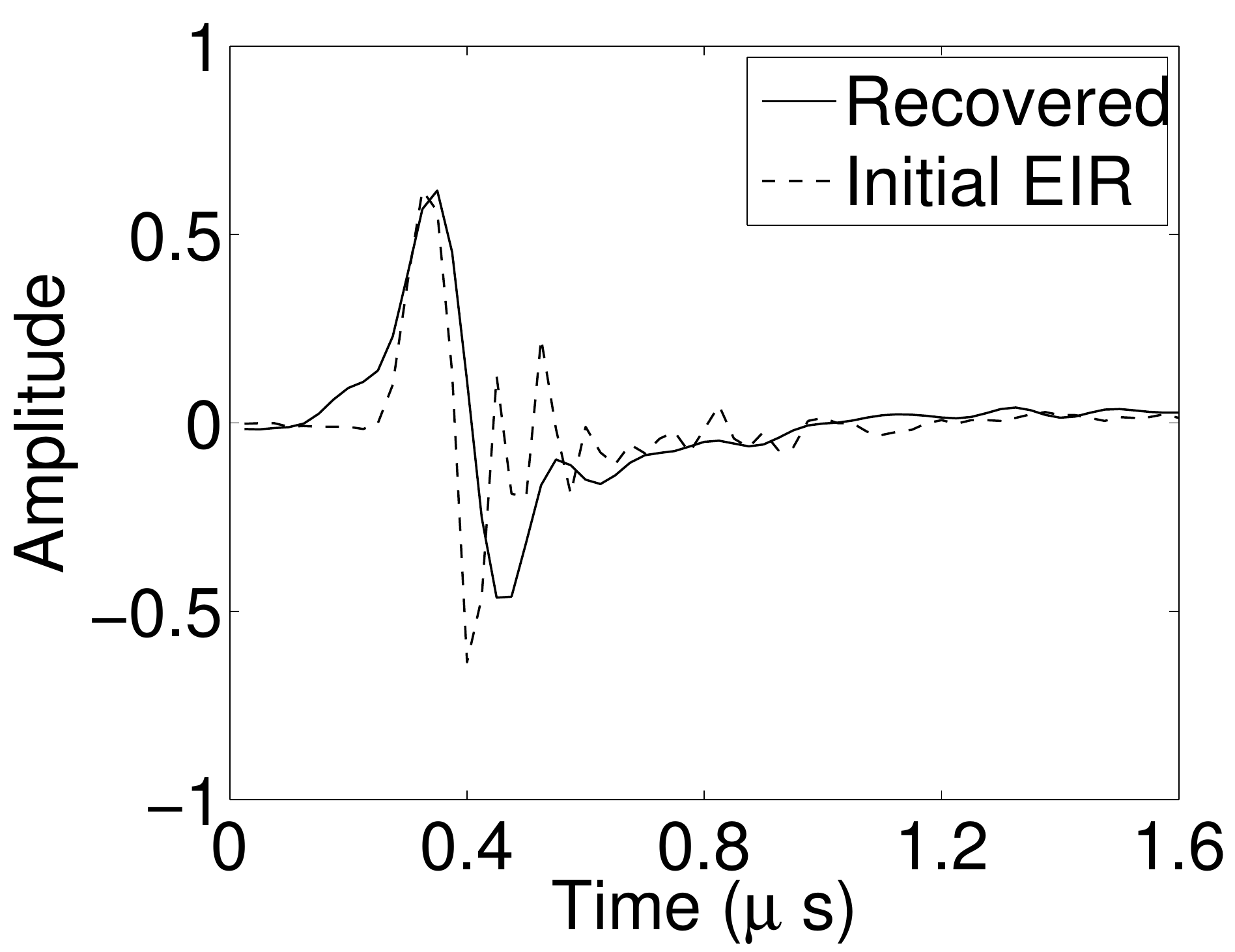}
  \caption{$\lambda=1.0\times 10^{-5}$, $\lambda=1.0\times 10^{-4}$,  $\lambda=1.0\times 10^{-3}$, $\alpha=1000$}\label{2D_mouse_kidney2_vpm_init_mouse_l1e_4_a1000}
 \end{subfigure}
 \begin{subfigure}[!h]{\textwidth}
 \includegraphics[width=0.367\textwidth]{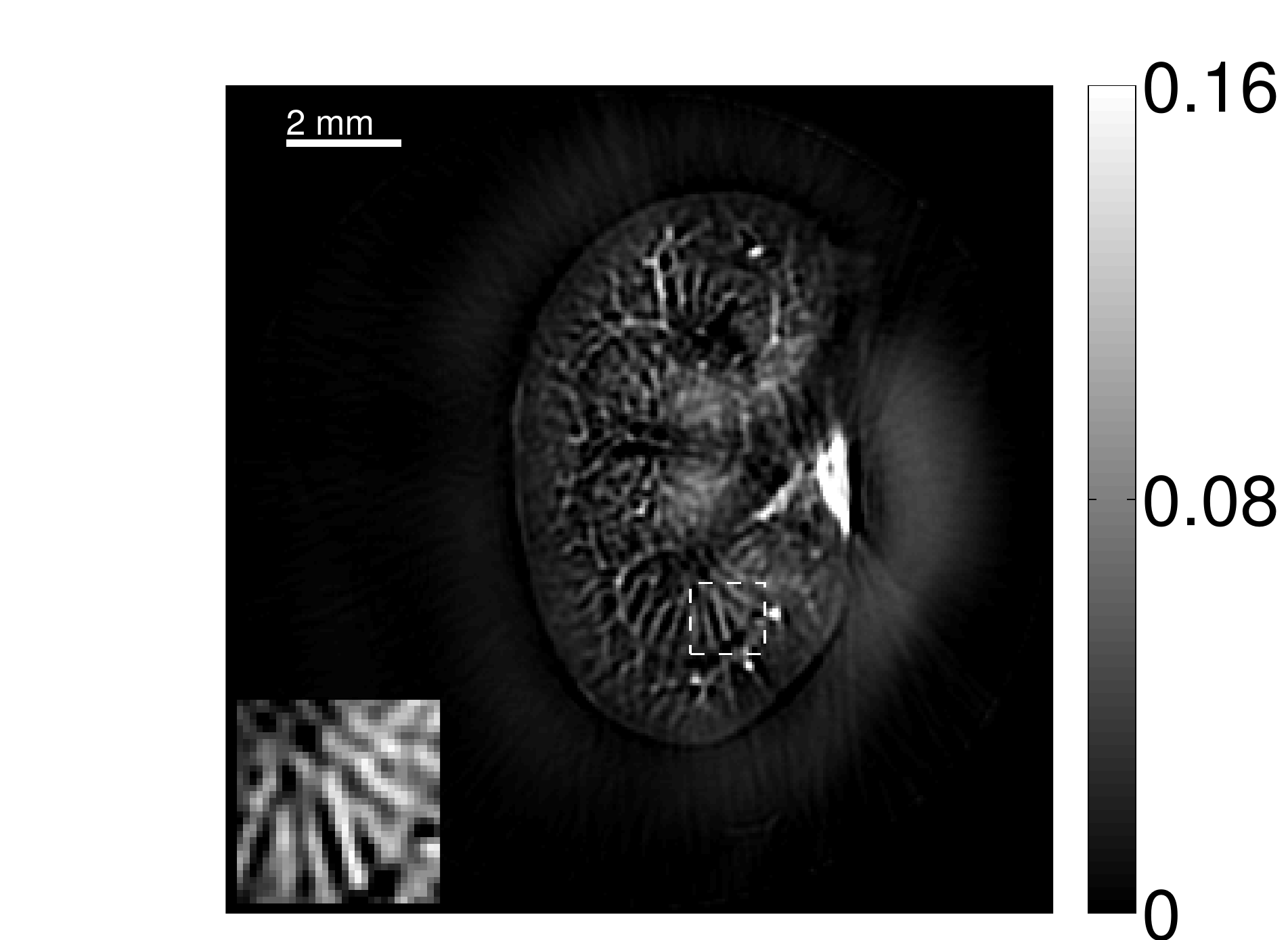}
 \includegraphics[width=0.31\textwidth]{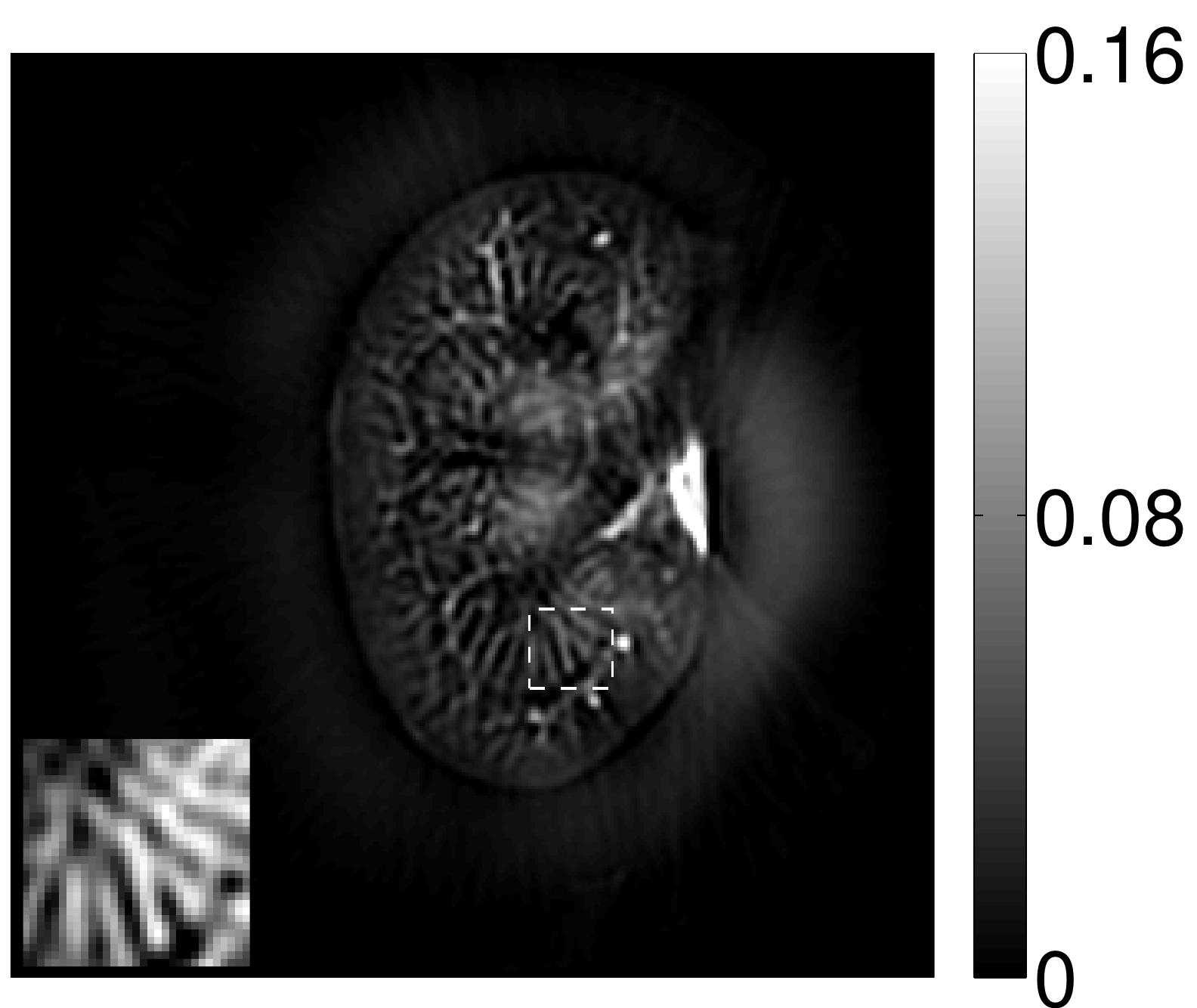}
 \includegraphics[width=0.31\textwidth]{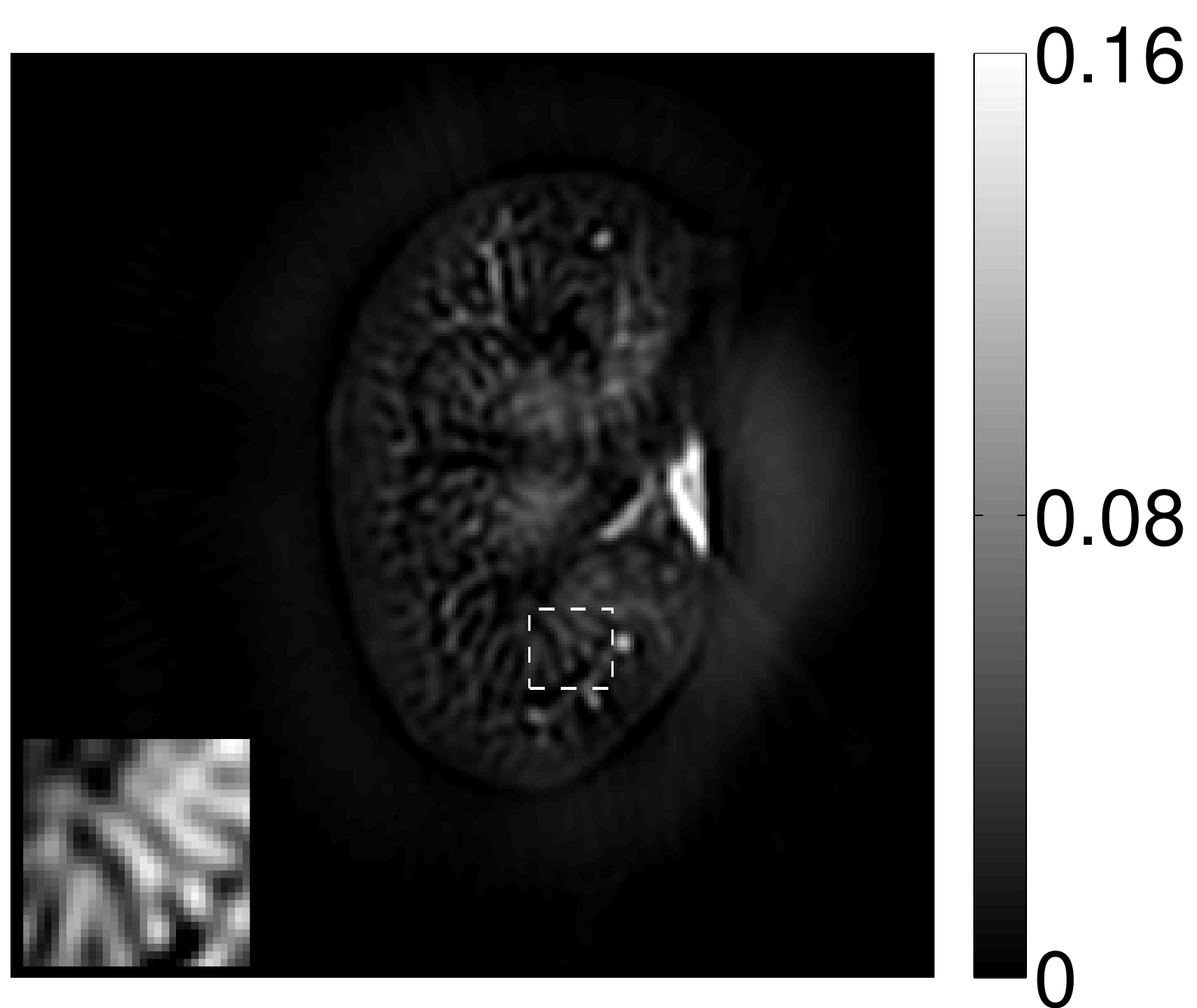}
  \caption{$\lambda=1.0\times 10^{-5}$, $\lambda=1.0\times 10^{-4}$,  $\lambda=1.0\times 10^{-3}$, $\alpha=10000$}\label{2D_mouse_kidney2_vpm_init_mouse_l1e_5_a10000}
 \end{subfigure}
 \begin{subfigure}[!h]{\textwidth}
 \includegraphics[width=0.31\textwidth]{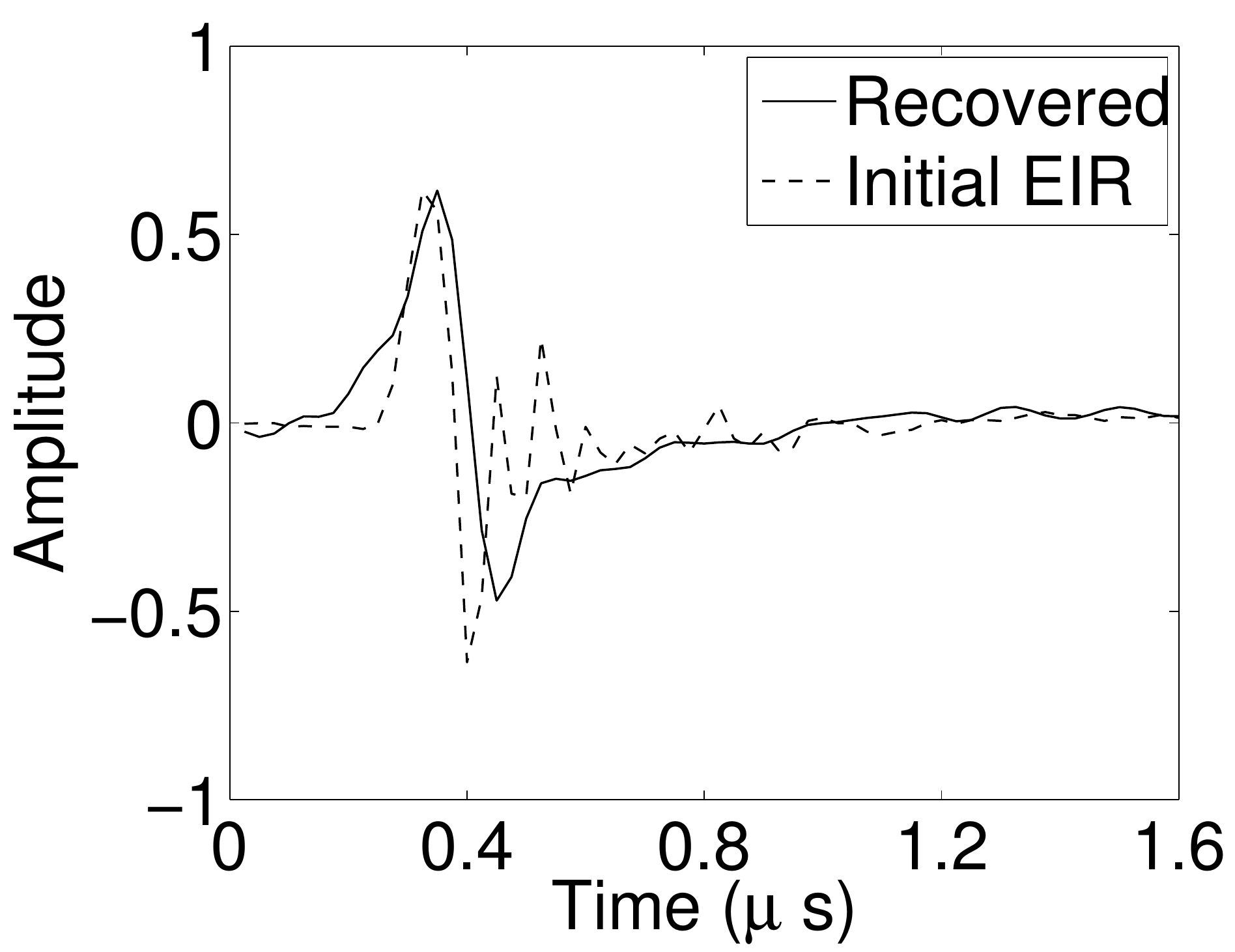}
 \includegraphics[width=0.31\textwidth]{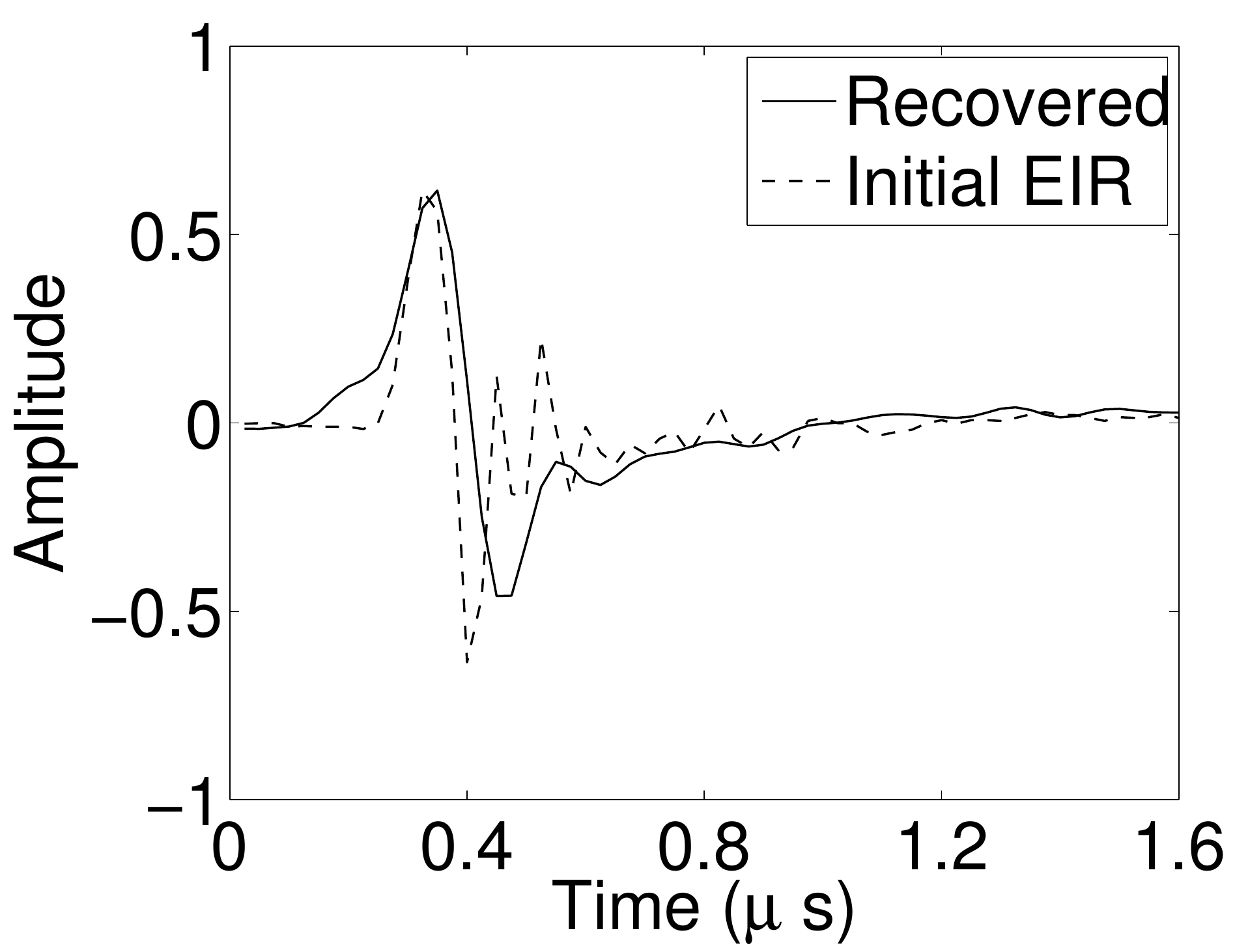}
 \includegraphics[width=0.31\textwidth]{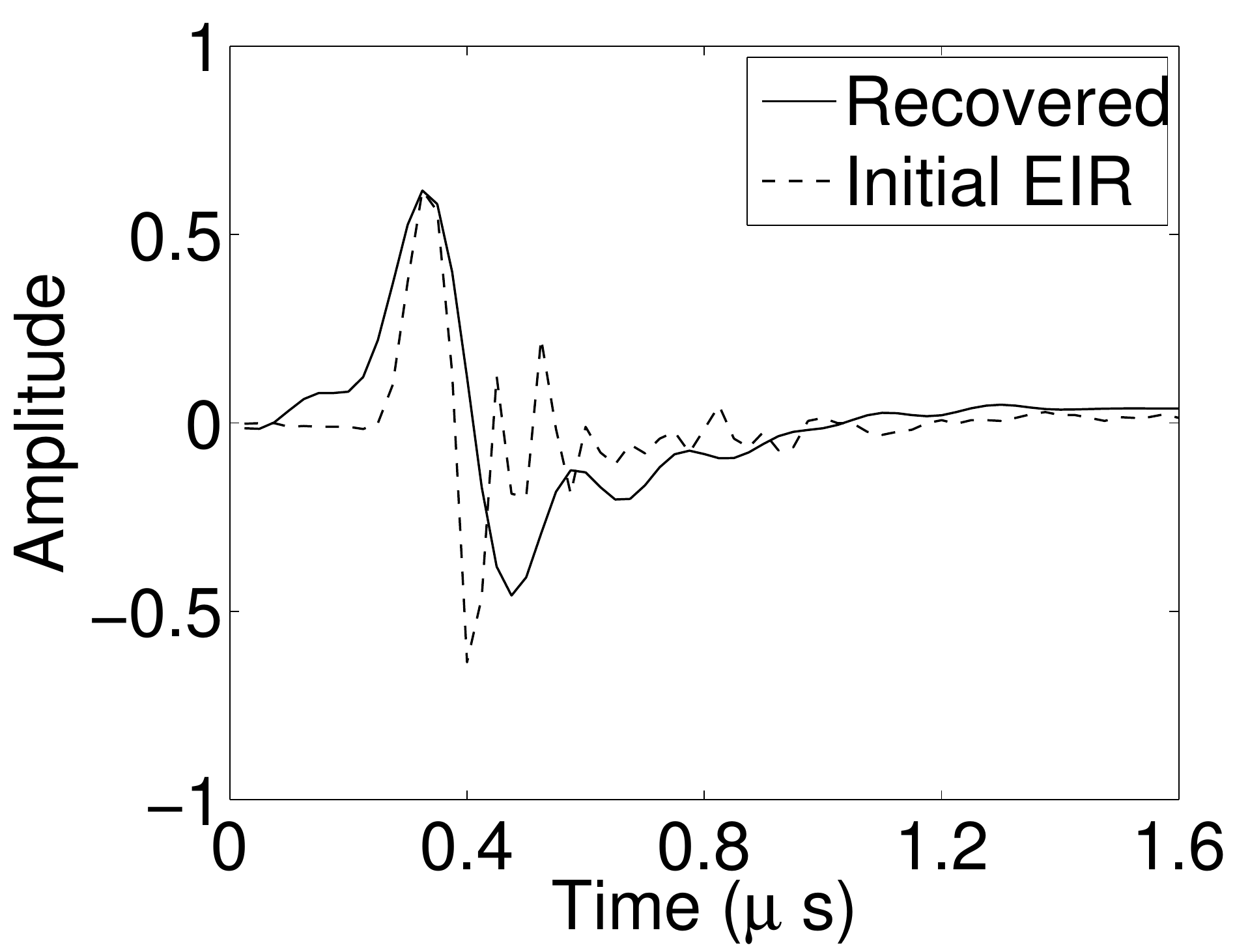}
  \caption{$\lambda=1.0\times 10^{-5}$, $\lambda=1.0\times 10^{-4}$,  $\lambda=1.0\times 10^{-3}$, $\alpha=10000$}\label{2D_mouse_kidney2_vpm_init_mouse_l1e_4_a10000}
 \end{subfigure}
\caption{(a) and (c) Images reconstructed using the VP algorithm and different regularization parameters $\lambda$ with the same initial guess of EIR. $\alpha=1000$ for all reconstructions in (a), and $\alpha=10000$ for (c). The zoomed-in image corresponds to the ROI of the dashed rectangle. 
The grayscale windows were $[0,0.16]$. (b) and (d) show the corresponding recovered EIR.
The SIR was accounted for in all cases.}\label{2D_mouse_kidney2_EIR_mouse_Reg}
\end{figure}

In Figure~\ref{2D_mouse_kidney2_SIR_EIR_mouse}, the results corresponding to use of
the 3D imaging model that incorporated SIR effects are shown.
 The EIR estimated by the VP algorithm is also shown.
 In Figure \ref{2D_mouse_kidney2_EIR_mouse_Reg}, images and EIRs reconstructed by use of the VP algorithm with different regularization parameters values are shown.
Similar to the case described above where the transducer SIR was neglected, 
these results reveal that use of the VP algorithm can produce images with a cleaner background
and enhanced spatial resolution than yielded by use of a conventional iterative algorithm that
employed the measured EIR.
For example, detailed information regarding the  vessels near the organ's periphery
 was better preserved by the VP algorithm than by the conventional iterative algorithm. These images corroborate our assertion that the VP algorithm can significantly reduce the artifacts and distortions in the reconstructed image.
 It is also worth pointing out that, unlike the numerical phantom studies, the artifacts and distortions in the images may be caused not only by the inaccurate EIR but also by
 other factors,
 such as neglecting 
acoustic heterogeneities and the variation of the EIRs among the elements of the transducer array.
In such cases, the EIR estimated by the VP algorithm represents an effective system impulse
response that  minimizes the inconsistency
between the measured data and the imaging model.

\subsection{Auto-focus capabilities}  
\begin{figure}[!htb]
 \centering
 \begin{subfigure}[!h]{\textwidth}\centering
 \includegraphics[width=0.32\textwidth]{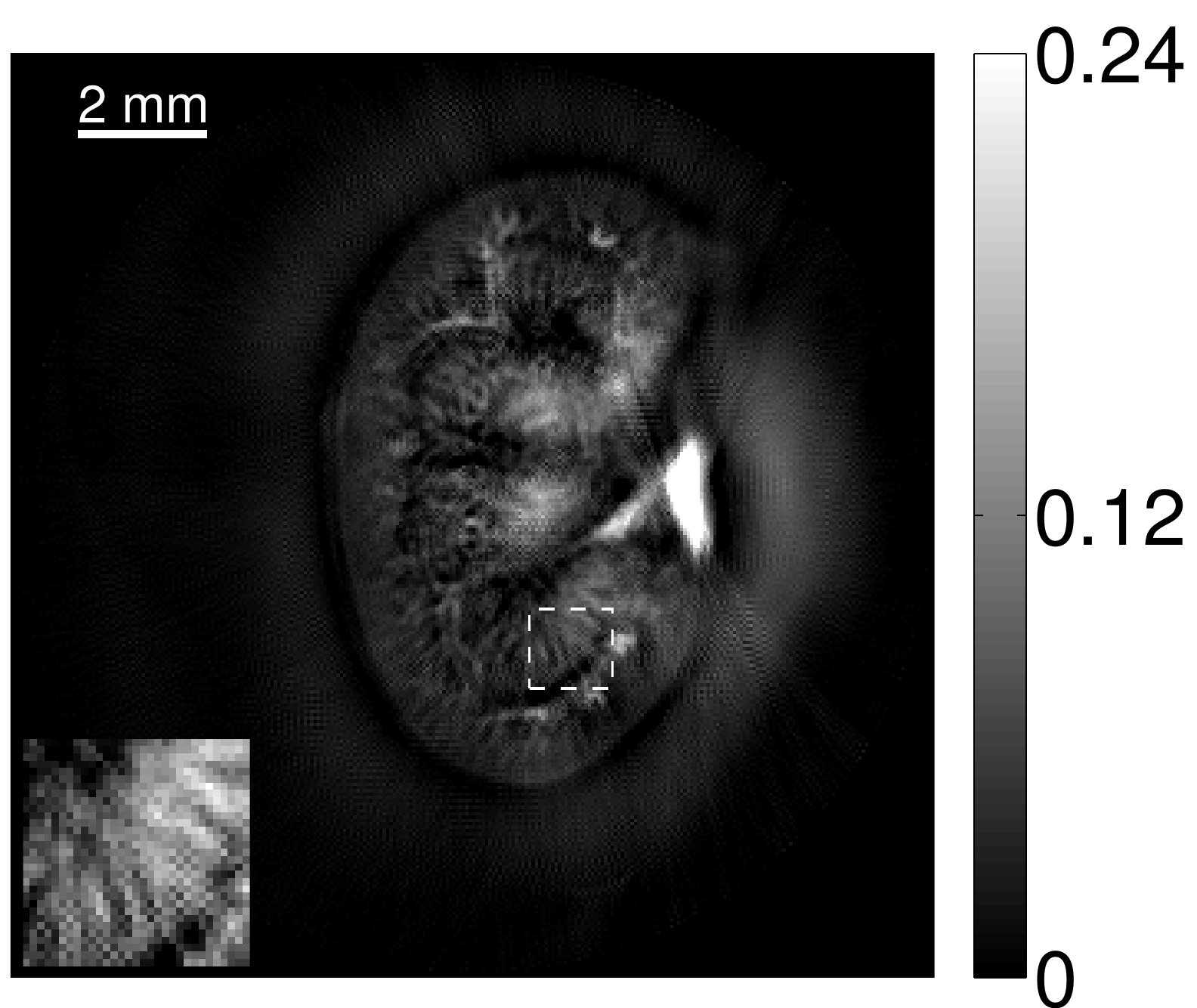}
 \includegraphics[width=0.32\textwidth]{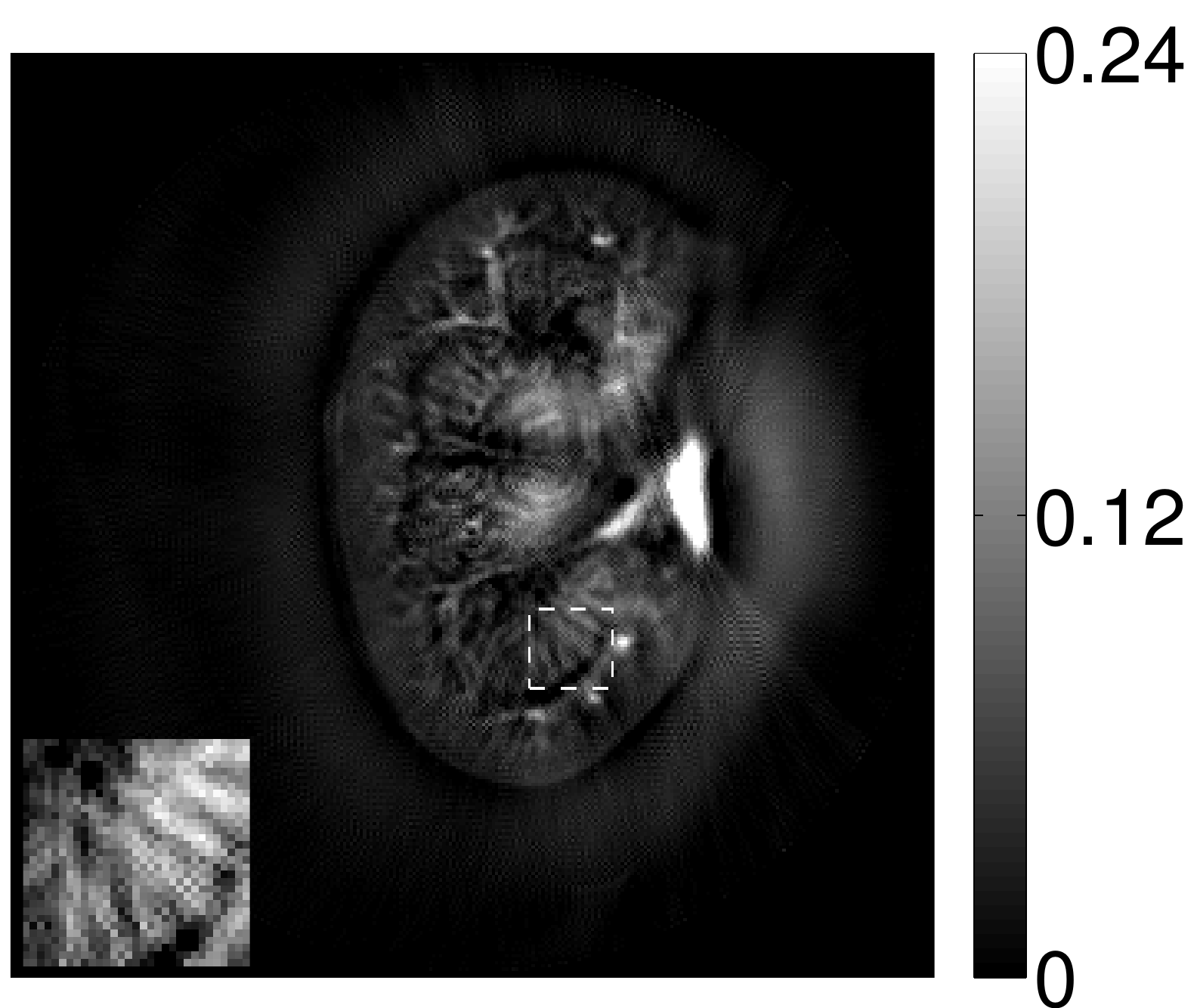}
 \includegraphics[width=0.32\textwidth]{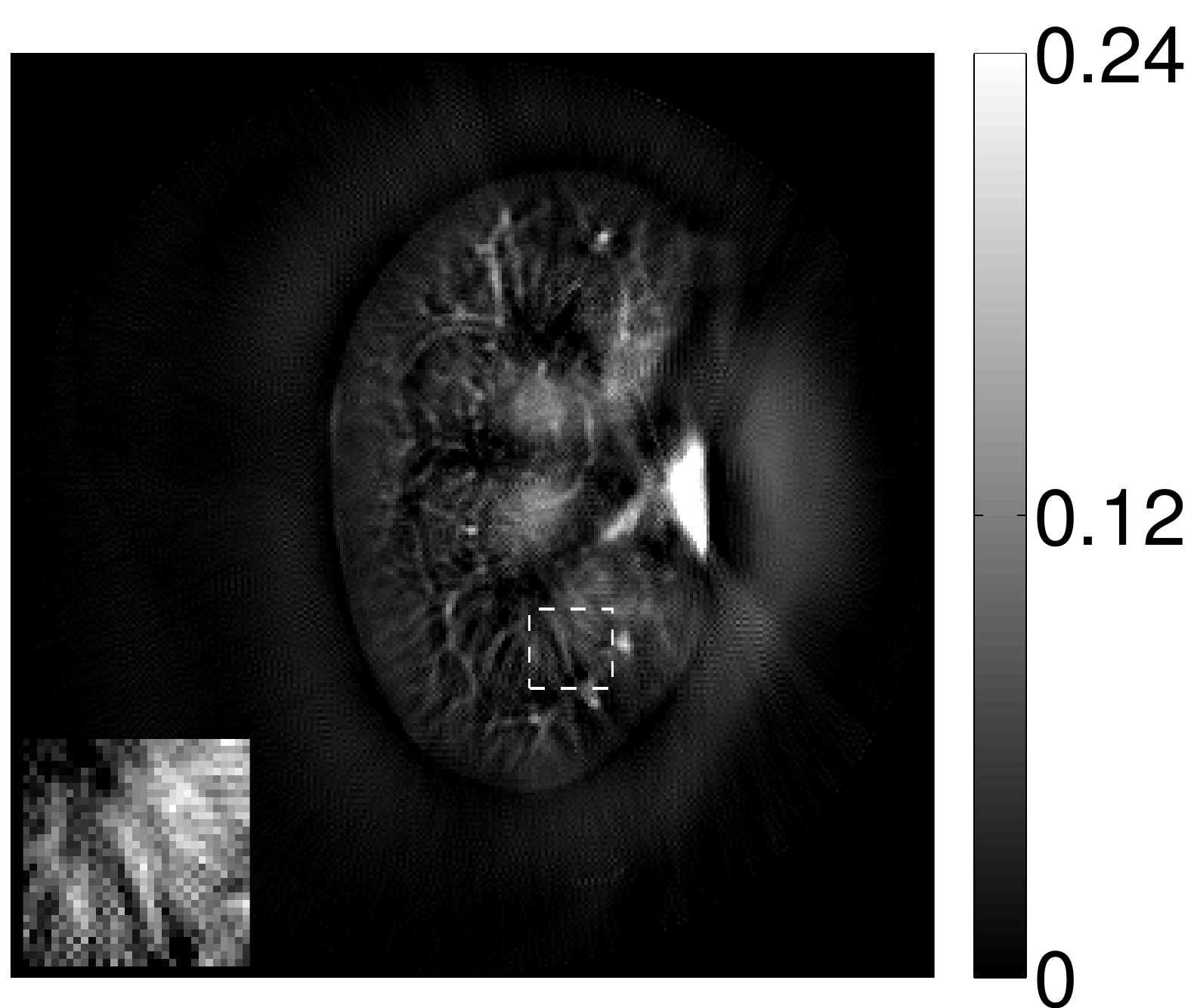}
  \caption{Images reconstructed by use of the conventional iterative method with different speeds of sound. The speeds of sound employed in the first, second, and third images are $c=1.542$, $c=1.54$, and $c=1.535$, respectively. $\lambda=1.0\times 10^{-5}$ for all reconstructions. The grayscale windows were $[0,0.24]$.}\label{2D_mouse_kidney2_sos_noVPM}
 \end{subfigure}
  \begin{subfigure}[!h]{\textwidth}\centering
 \includegraphics[width=0.32\textwidth]{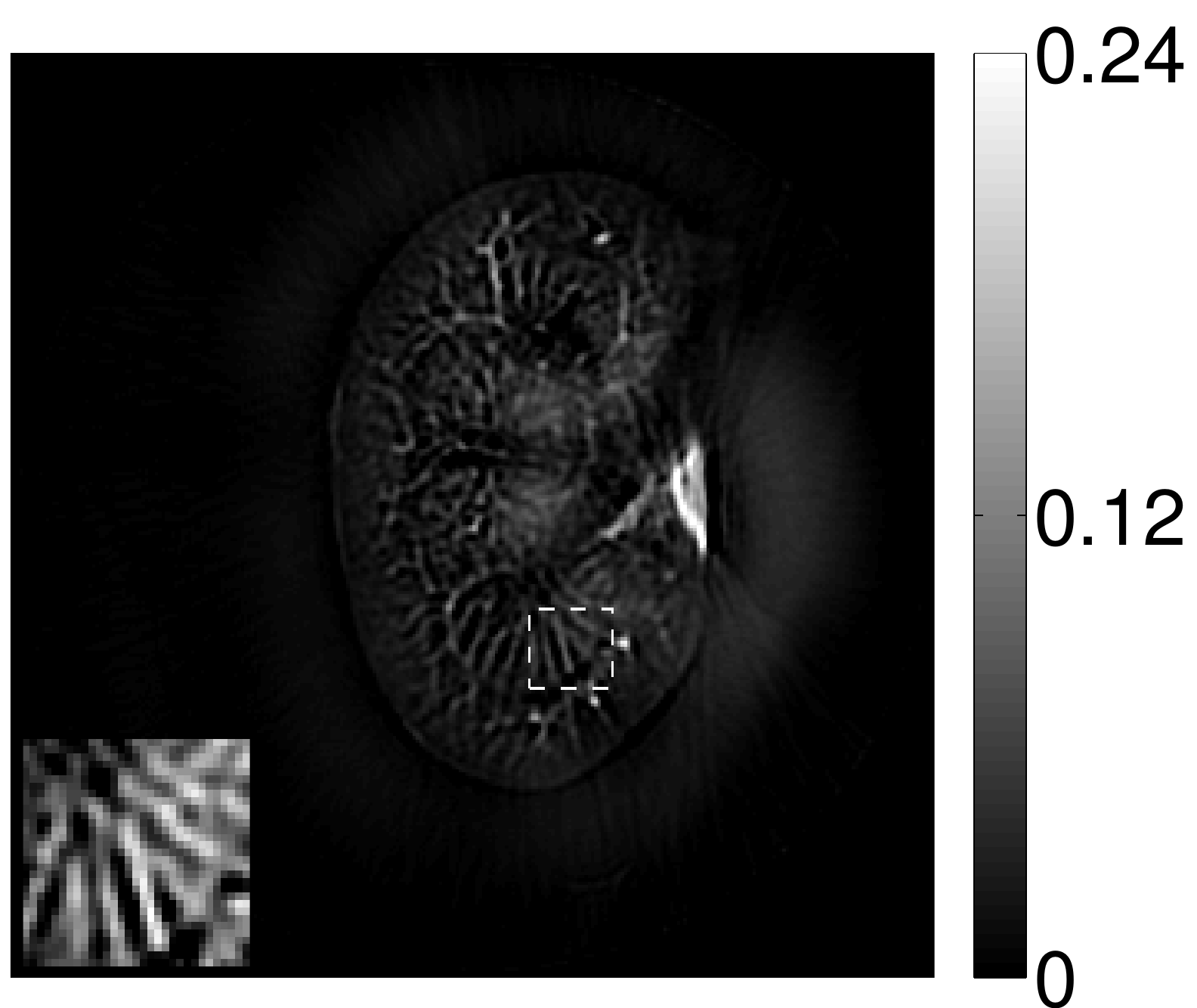}
 \includegraphics[width=0.32\textwidth]{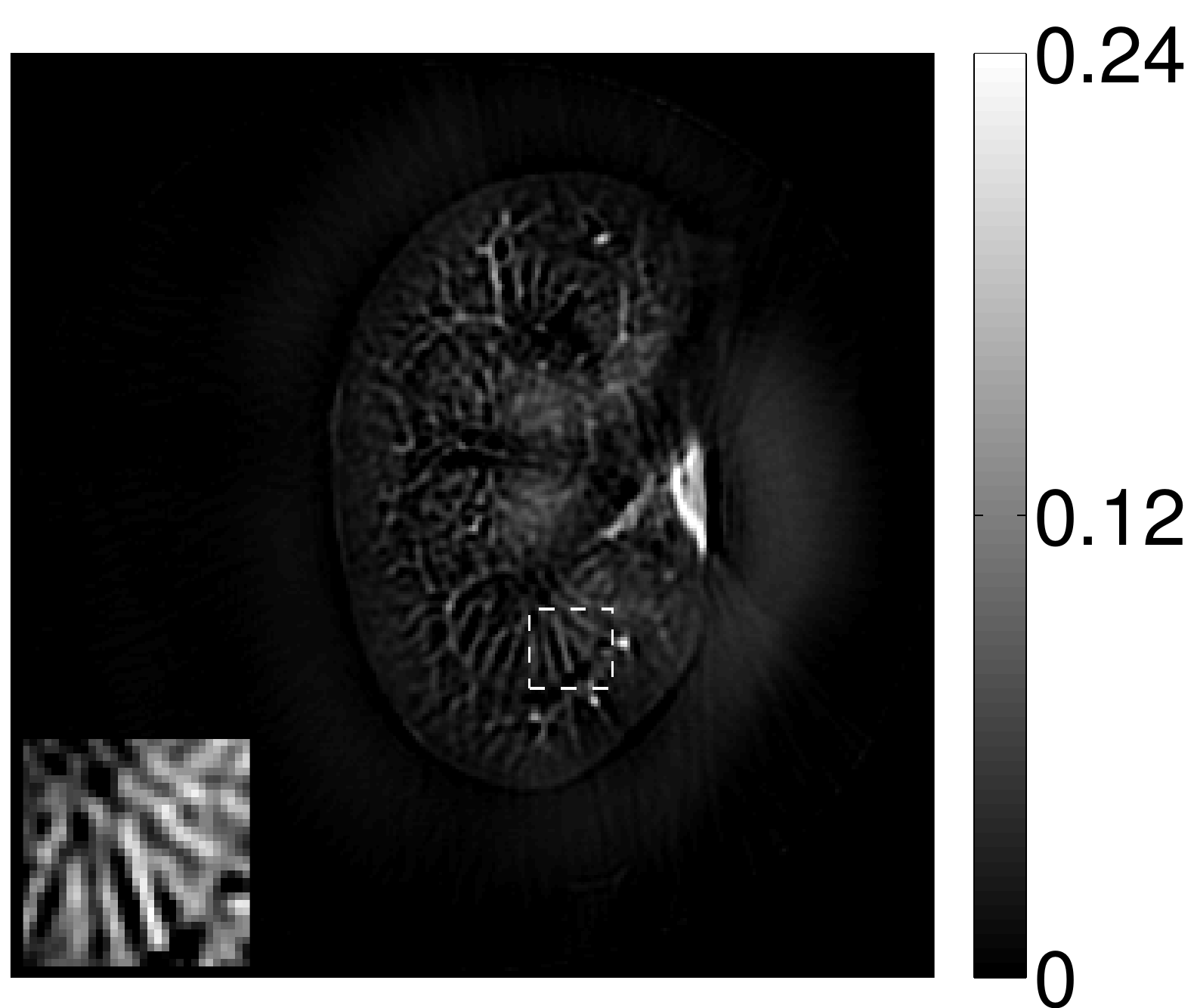}
 \includegraphics[width=0.32\textwidth]{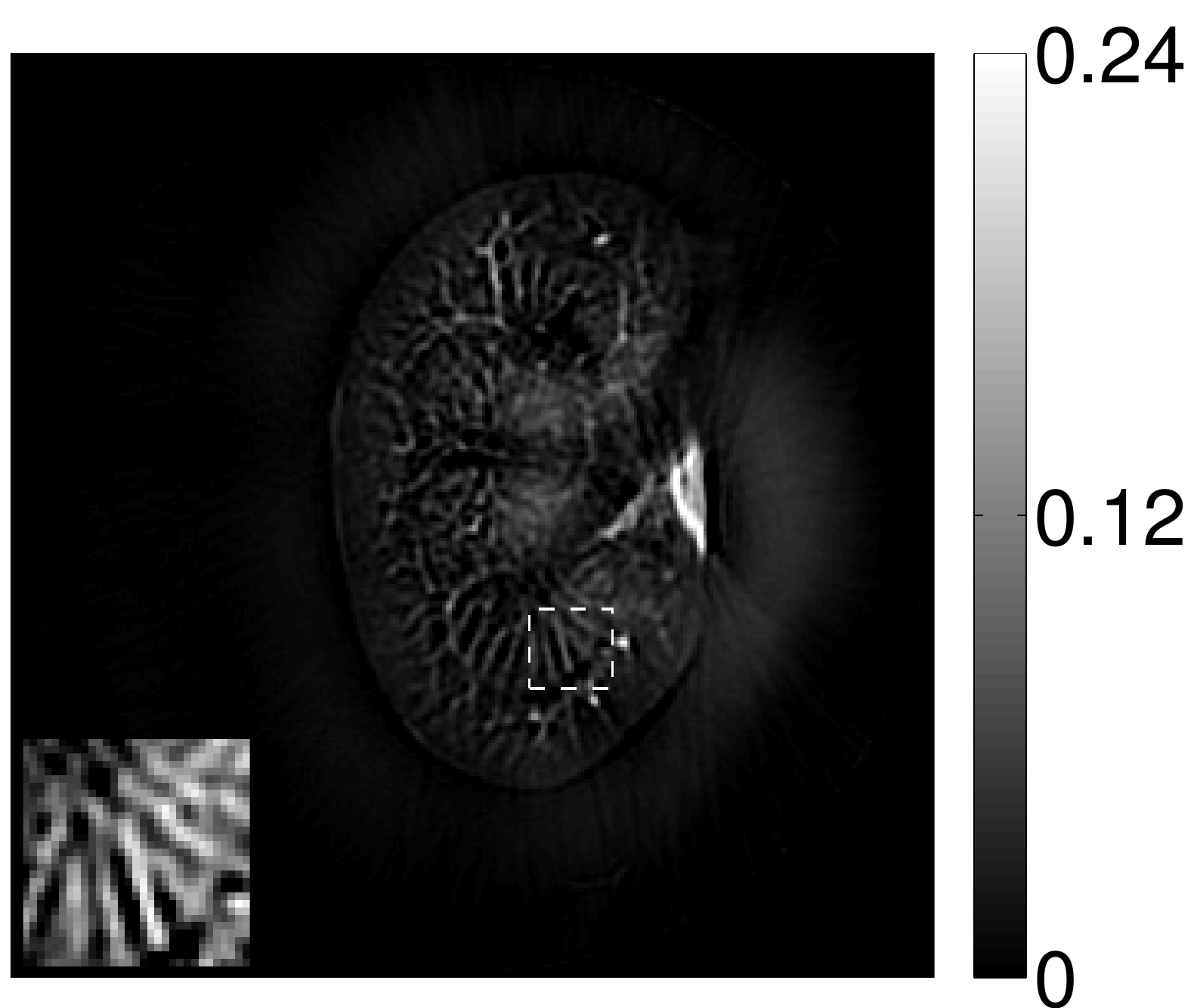}
  \caption{Images reconstructed by use of the VP algorithm with different speeds of sound. The speeds of sound employed in the first, second, and third images are $c=1.542$, $c=1.54$, and $c=1.535$, respectively. $\lambda=1.0\times 10^{-5}$, and $\alpha=5000$ for all reconstructions. The grayscale windows were $[0,0.24]$.}\label{2D_mouse_kidney2_sos_a}
 \end{subfigure}

 \begin{subfigure}[!h]{\textwidth} \centering
 \includegraphics[width=0.6\textwidth]{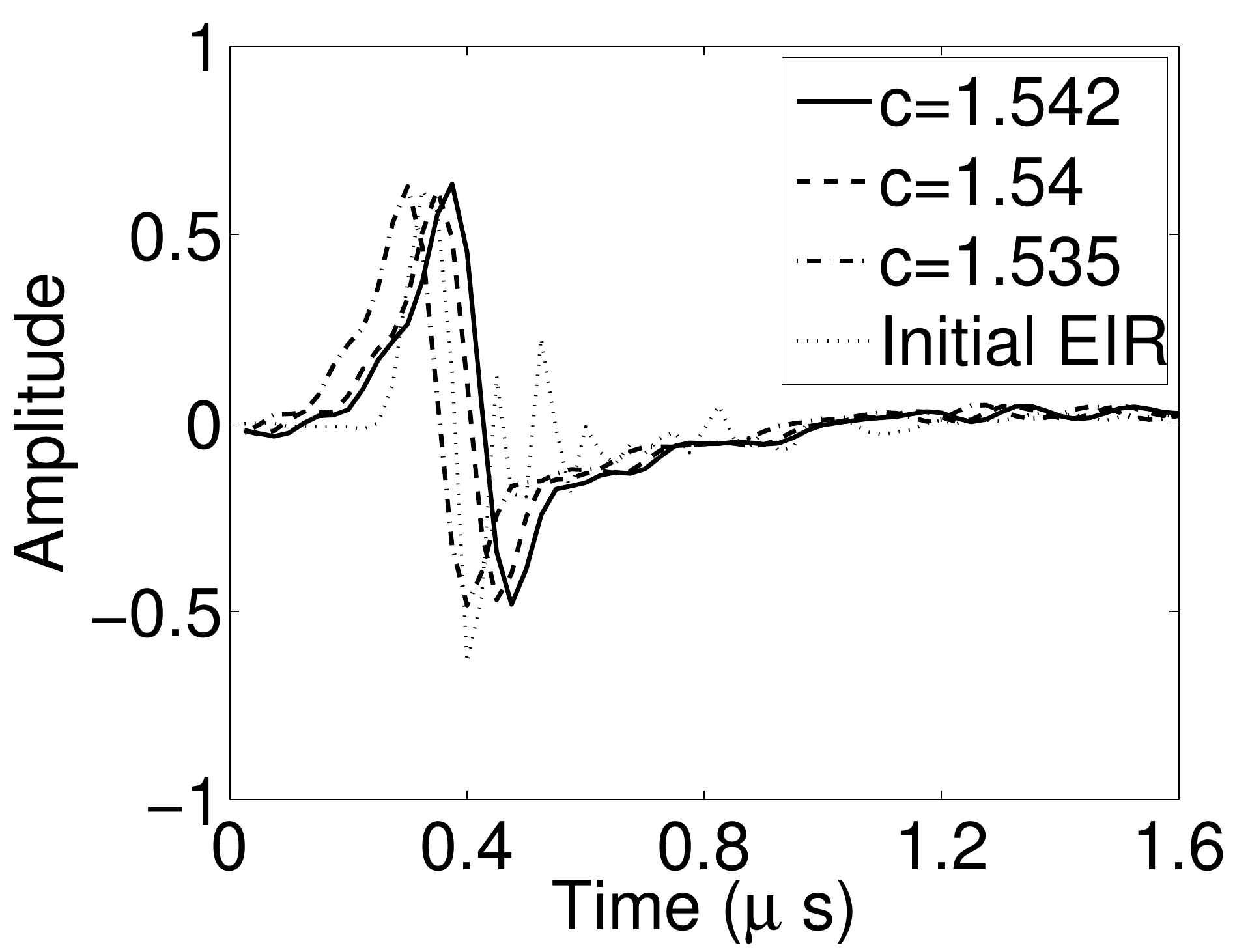}
 \caption{Recovered EIR}\label{2D_mouse_kidney2_sos_b}
 \end{subfigure}
\caption{Reconstructed images and EIRs using different speeds of sound. The zoomed-in image corresponds to the ROI of the dashed rectangle. 
}\label{2D_mouse_kidney2_c}
\end{figure}

Conventional PACT reconstruction algorithms assume that the medium is described by a constant speed-of-sound (SOS) value. In practice, this value may not be known precisely and can be tuned \cite{TVZL2011} to maximize the spatial resolution of the reconstructed images. The effect of an incorrect SOS value can  sometimes be compensated for by use of the VP algorithm due to modification of the EIR during the joint estimation. Figures~\ref{2D_mouse_kidney2_sos_noVPM} and \ref{2D_mouse_kidney2_sos_a} show images reconstructed by use of the conventional iterative method and the VP algorithm, respectively,
when different constant SOS values are assumed.
The 2D imaging model that ignored the SIR was employed.
  Nearly identical images were reconstructed by use of the VP algorithm, even though the assumed SOS values were different in each case.  The images contained reduced artifact levels as compared to those reconstructed
by use of the conventional method.  The recovered EIRs differed by a time shift (as displayed in Fig.~\ref{2D_mouse_kidney2_sos_b}). Since the object was located near the center of the transducer array and was small compared to the radius of the array, the scaling effect due to the inaccurate SOS can be approximated by the shift of the EIR, which explains how the recovered EIR compensates for the error in SOS value.

\section{Conclusions and discussion}\label{Sect:Summary}
In this study, we proposed a joint reconstruction approach for PACT that mitigates artifacts in the reconstructed images caused by use of an inaccurate EIR. A nonlinear least squares minimization problem was formulated, which exploited the bi-linear structure of the imaging model, and a VP algorithm was employed to solve the minimization problem.
The numerical properties of the  VP algorithm were also {investigated}. The results demonstrate that the joint reconstruction approach for estimating both the system response and the absorbed optical energy density can increase the fidelity of the reconstructed image.
Although not presented,  we also conducted computer-simulation {studies} based upon an existing three-dimensional small animal imaging system \cite{KRAM2012,BSFE2009}, and the results were consistent with those presented.

{It should be emphasized that the recovered EIR, in general, is not equivalent to the actual EIR of a system. Instead, the VP algorithm finds the linear temporal filter that best matches the measured pressure to the modeled pressure in a penalized least squares sense. If the EIR is the only source of model error, the filter will correspond to the EIR. However, if other system inconsistencies, such as sound speed variations, acoustic absorption, or the spatial impulse responses of the transducers, are present, the VP algorithm will produce an estimated filter that attempts to mitigate these sources of model error. In practice, it can be difficult or overly time-consuming to explicitly account for all these potential sources of inconsistency in a PACT reconstruction algorithm. Further, including them can result in a tremendous increase in the computational cost of the algorithm. Since the VP algorithm can provide a rough correction for these effects, it can serve as a cheap and effective way to compensate for model mismatch.}
 

The minimization problem defined in Eqn.~\eqref{min_prob} is non-convex. Hence, the optimization algorithm may converge to a local minimum. However, the literature \cite{DB2013} suggests that the VP algorithm is more likely to converge to the global minimum than other algorithms such as  block-coordinate descent algorithms. 
Our computer-simulation studies revealed that the VP algorithm consistently converged to accurate solutions, suggesting that utilizing proper regularization methods and good initial guesses will improve the ability of the algorithm to avoid local minima. 
 The experimental results confirmed the effectiveness of the proposed algorithm for mitigating image artifacts and distortions.

There remain several topics for future investigation.
Our current implementation involves two regularization parameters.  Although numerical methods---such as the $L$-curve method \cite{hansen1993use} --- have been proposed for determining reasonable values
for these parameters, these methods do not work perfectly in all applications. 
In this study, to reveal the impact of parameter settings on reconstruction algorithm
performance, we reconstructed a collection of images using different regularization parameter values.
The optimal regularization parameter values should depend on a specified diagnostic
task and observer \cite{BM2003} and their determination represents a topic for future investigation. 

There also remains a need to investigate methods for incorporating additional \emph{a priori}
information regarding the EIR into the reconstruction problem.
If we assume the EIR is sufficiently smooth, spline functions are a natural choice to parameterize the EIR and reduce the number of unknowns in the minimization problem. We conducted numerical studies to evaluate this. Although not shown here, the results suggest that, depending on the interpolation points, the number of unknowns employed to represent the EIR can be reduced (from 64 to 32 in this study). However, the reconstructed images and EIRs were similar to the results obtained without using the spline functions. Besides, no computational advantages (such as time and memory usage) were observed. It is also possible to employ an analytic parameter-based EIR model \cite{LKM1971}. To accurately model a realistic transducer, tens of parameters are needed. How to effectively solve the associated minimization problem remains a topic for future work. Non-smooth sparsity-promoting penalties, such as TV penalties \cite{ROF1992}, can be applied to the absorbed energy density \cite{KRAM2012}. In the VP algorithm, the updating scheme for $\boldsymbol{\theta}$ is based on a gradient-descent method that exploits the differentiability of the smoothness penalty (i.e.~Eqn.~\eqref{penalty1}). When non-smooth penalties are adopted, this gradient-descent method can potentially be replaced by a proximal gradient algorithm \cite{BT2009}.

\section*{ACKNOWLEDGMENTS}
This work was supported in part by NIH awards CA1744601 and EB01696301.

\appendices

\section{Explicit forms of system matrices}
\label{Sect:AppendixA}

\subsection{System matrix based on interpolation expansion functions}
In the 2D computer-simulation studies, an interpolation-based image model was employed. In  interpolation-based D-D imaging model the coefficient vector is defined as samples
of the object function on the nodes of a uniform Cartesian grid:
\begin{equation}\label{eqn:coeffint}
  \big[\boldsymbol\theta\big]_n = \int_V\!\!d
           \mathbf r\, \delta(\mathbf r-\mathbf r_n)
           A(\mathbf r),\quad n=0,1,\cdots,N-1,
\end{equation}
where $\mathbf r_n=(x_n,y_n)^T$ specifies the location of the $n$-th node of the uniform Cartesian grid. The definition of the expansion function depends on the choice of interpolation method.
If a trilinear interpolation method is employed, the expansion function can be
expressed as 
\begin{equation}\label{eqn:expfunI}
  \phi_n(\mathbf r)=\left\{\begin{array}{ll}
                    (1-\frac{|x-x_n|}{\Delta_s})
                    (1-\frac{|y-y_n|}{\Delta_s}), & \text{if}\, |x-x_n|, |y-y_n| \leq  \Delta_s\\
                0, & \text{otherwise}
                  \end{array}\right.,
\end{equation}
where $\Delta_s$ is the distance between two neighboring grid points.

In principle, the interpolation-based D-D imaging model can be constructed by substituting Eqns.~\eqref{eqn:coeffint} and \eqref{eqn:expfunI} into Eqn.~\eqref{eq_CD}. In practice, however, implementation of the surface integral over $S_q$ is difficult for the choice of expansion functions in Eqn.~\eqref{eqn:expfunI}. Therefore, utilization of the interpolation-based D-D model commonly assumes the transducers to be point-like.

Since Eqn.~\eqref{wave_eq_sol} can be reformulated as the well-known spherical Radon transform (SRT) 
\begin{equation}\label{eqn:srt}
  g(\mathbf r_0,t)=\int_V\!\! d \mathbf r\,
     A(\mathbf r) \delta(c_0t-|\mathbf r_0-\mathbf r |),
\end{equation}
where the function $g(\mathbf r_0, t)$ is related to $p(\mathbf r_0, t)$ as
\begin{equation}\label{eqn:g2p}
  p(\mathbf r_0, t)=\frac{\beta}{4\pi C_p}
     \frac{\partial}{\partial t}\Big(\frac{g(\mathbf r_0, t)}{t}\Big),
\end{equation}
the implementation of $\mathbf H$ is decomposed as a three-step operation:
\begin{equation}
  \mathbf u = \mathbf H\boldsymbol{\theta}
     \equiv \mathbf H^e \mathbf D \mathbf G \boldsymbol{\theta},
\end{equation}
where $\mathbf G$, $\mathbf D$, and  $\mathbf H^e$ are discrete approximations of the SRT (Eqn.\ \eqref{eqn:srt}), the differential operator (Eqn.\eqref{eqn:g2p}), and the operator that implements a temporal convolution with EIR, respectively.
$\mathbf G$ was implemented in a way
that is similar to the `ray-driven' implementation of Radon transform in X-ray CT, 
i.e., for each data sample, we accumulated the contributions from the voxels that resided on the spherical shell specified by the data sample.
By use of Eqns. \eqref{rep_A}, \eqref{eqn:coeffint}, \eqref{eqn:srt}, and
\eqref{eqn:expfunI}, one obtains
\begin{equation}\label{eqn:dis_srt}
\big[\mathbf G \boldsymbol\theta\big]_{qS+s}
      = \Delta_s^2
      \sum_{n=0}^{N-1}
      \big[\boldsymbol \theta\big]_n
      \sum_{i=0}^{N_i-1}
      \sum_{j=0}^{N_j-1}
      \phi_n(\mathbf r_{s,i,j})
      \equiv
  \big[\mathbf g\big]_{qS+s} ,
\end{equation}
where $[\mathbf g]_{qS+s} \approx g(\mathbf r_q, t)|_{t=k\Delta_t}$
with $\mathbf r_q$ specifying the location of the $q$-th point-like transducer,
and $N_i$ and $N_j$ denote the numbers of divisions over the two angular coordinates of a local spherical coordinate system.
The differential operator in Eqn.\ \eqref{eqn:g2p} is approximated as
\begin{equation}\label{eqn:dis_g2p}
\big[\mathbf D \mathbf g\big]_{qK+k}
= \frac{\beta}{8\pi C_p\Delta_t^2}
 \Big(\frac{[\mathbf g]_{qK+k+1}}{k+1}
     -\frac{[\mathbf g]_{qK+k-1}}{k-1}
 \Big)
    \equiv
\big[\mathbf p\big]_{qK+k},
\end{equation}
where $[\mathbf p]_{qS+s} \approx p(\mathbf r_q, t)|_{t=s\Delta_t}$.
Finally, the continuous temporal convolution is approximated by a discrete linear convolution as 
\begin{equation}\label{eqn:EIR}
  \big[\mathbf H^e \mathbf p\big]_{qS+s} =
  \sum_{\kappa = 0}^{S-1}  [ \mathbf h ]_{s-1-\kappa}  [\mathbf p]_{qS+\kappa}
  \equiv [\mathbf u]_{qS+s},
\end{equation}
where
$[\mathbf h]_{s} = \Delta_t h(t)|_{t=s\Delta_t}$.

\subsection{3D spherical voxel-based imaging model including SIR}

{
For the 3D spherical voxel-based model, the expansion functions were defined as
\begin{equation}\label{sphere}
 \phi_n(\mathbf{r})=\begin{cases}
                     1, &\text{if } |\mathbf{r}-\mathbf{r}_n|\leq \epsilon,\\
                     0, &\text{otherwise.}
                    \end{cases}
\end{equation}
where $\mathbf{r}=(x_n,y_n,z_n)^T$ specifies the coordinate of the $n$-th grid point of a uniform Cartesian lattice, $(\cdot)^T$ denotes the transpose of a vector, and $\epsilon$ is the half spacing between lattice points. The coefficient vector $\boldsymbol{\theta}$ es defined as
\begin{equation}
 [\boldsymbol{\theta}]_n=\frac{V_{cube}}{V_{voxel}}\int_V \phi_n(\mathbf{r})A(\mathbf{r})\,d\mathbf{r},
\end{equation}
where $V_{cube}$ and $V_{voxel}$ are the volumes of a cubic voxel of dimension $2\epsilon$ and $\phi_n(\mathbf{r})$, respectively.
}
{
Let $u^a_q(t)$ denote the pre-sampled voltage signal that would be produced by  $A_a(\mathbf{r})$,
where  $A_a(\mathbf{r})$ is the approximation of  $A(\mathbf{r})$ established by use of
the chosen expansion functions.
 By use of Eqns.~\eqref{eq_CD}, \eqref{rep_A}, and \eqref{sphere}, it can be verified that
\begin{equation}\label{pre_DD}
 u^a_q(t)=h^e(t)*_t p_0(t) *_t \frac{1}{\Omega_q}\sum^{N-1}_{n=0}[\boldsymbol{\theta}]_n h^s_q(\mathbf{r}_n,t).
\end{equation}
Here, $p_0(t)$ is the `N'-shaped profile produced by a uniform sphere of radius $\epsilon$:
\begin{equation}\label{p_0}
 p_0(t)=-\frac{\beta c^3\pi}{C_p}\bigg[H\left(t+\frac{\epsilon}{c_0}\right)-H\left(t-\frac{\epsilon}{c_0}\right)\bigg],
\end{equation}
where $H(t)$ is the Heaviside step function and
\begin{equation}\label{h_s}
 h^s_q(\mathbf{r}_n,t)=\int_{\Omega_q} \frac{\delta(t-\frac{|\mathbf{r}'-\mathbf{r}_{n}|}{c_0})}{2\pi|\mathbf{r}'-\mathbf{r}_{n}|} \,d\mathbf{r}'
\end{equation}
is the SIR of the $q$-th transducer. By temporally sampling \eqref{pre_DD} and employing the approximation $[\mathbf{u}]_{qS+s}\approx u^a_q(t)\big|_{t=s\Delta T}$,
the D-D imaging model in Eqn.\ (\ref{DDD}) is established \cite{KRAM2012}, where
\begin{equation}
  [\mathbf{H}]_{qS+s,n}=h^e(t)*_t p_0(t) *_t \frac{1}{\Omega_q}h^s_q(\mathbf{r}_n,t)\bigg|_{t=s\Delta T}.
\end{equation}
Here, $[\cdot]_{m,n}$ denote the entry in the $m$-th row and $n$-th column of the matrix.
When the transducer has a flat and rectangular detecting surface of area $a\times b$, under the far-field assumption, the temporal Fourier transform of the SIR is given by \cite{KRAM2012}
\begin{equation}
 \tilde{h}^s_q(\mathbf{r}_n,t)=\frac{ab\exp(-j2\pi f\frac{|\mathbf{r}'_q-\mathbf{r}_n|}{c_0})}{2\pi|\mathbf{r}'_q-\mathbf{r}_n|}\sinc\Big(\pi f\frac{ax^{tr}_{nq}}{c_0|\mathbf{r}'_q-\mathbf{r}_n|}\Big)\sinc\Big(\pi f\frac{bx^{tr}_{nq}}{c_0|\mathbf{r}'_q-\mathbf{r}_n|}\Big),
\end{equation}
where $x^{tr}_{nq}$ and $x^{tr}_{nq}$ specify the transverse coordinates in a local coordinated system that is
 centered about the $q$th transducer.
}

{
Since the surfaces of the focused transducers employed in the reported
experimental studies  are curved, direct use of the far-field approximation assuming a flat transducer
can result in patterned image artifacts. To alleviate this limitation, we adopt a simple divide-and-integrate algorithm \cite{MKA2014}, where each transducer element face
 is divided into $m\times 1$ identical patches. Each patch is considered to be flat and described by the far-field approximation. Let $\tilde{h}^s_{q,i}(\mathbf{r}_n,t)$ be the resulting SIRs that are specified by the patch index $i=1, \cdots, m$. The SIR for the original transducer face $\tilde{h}^s_q(\mathbf{r}_n,t)$ is then approximated by averaging the patch SIRs over all patches:
\begin{equation}
 \tilde{h}^s_q(\mathbf{r}_n,t)=\frac{1}{m}\sum^m_{i=1}\tilde{h}^s_{q,i}(\mathbf{r}_n,t).
\end{equation}
}

\section{An equivalent reformulation of the imaging model}\label{appd:B}
First observe that a D-D model without considering EIR 
can be derived as
\begin{equation}\label{ddm}
 \mathbf{p}=\left[\begin{array}{c}
                                \mathbf{p}_0\\
                                \mathbf{p}_1\\
                                \vdots\\
                                \mathbf{p}_{Q-1}
                               \end{array}\right]
                               =\mathbf{H}_p\boldsymbol{\theta}.
\end{equation}
Here, the vector $\mathbf{p}\in \mathbb{R}^P$ represents a lexicographically ordered representation of the sampled pressure data, the dimension $P$ is defined by the product of the number of pressure temporal samples ($T$) acquired at each transducer location and the number of transducer locations ($Q$), and $\mathbf{p}_q\in \mathbb{R}^T$ ($q=0,1,\cdots,Q-1$) is the sampled pressure data corresponding to location index $q$. The system matrix $\mathbf{H}_p$, without considering EIR, is of dimension $P\times N$, whose elements are
defined by Eqn.~\eqref{eqn:dis_srt} and \eqref{eqn:dis_g2p}.
To update $\mathbf{h}$ using \eqref{eqn:ParaModel}, another equivalent formulation of the D-D image model \eqref{DD} can be established as
\begin{align}
 \mathbf{u}&= \mathbf{H_p}\boldsymbol{\theta}*_t \mathbf{h}= \left[\begin{array}{c}
                                \mathbf{p}_0\\
                                 \mathbf{p}_1\\
                                \vdots\\
                                \mathbf{p}_{Q-1}
                               \end{array}\right]*_t\mathbf{h} := \left[\begin{array}{c}
                                 \mathbf{p}_0*_t\mathbf{h}\\
                                 \mathbf{p}_1*_t\mathbf{h}\\
                                \vdots\\
                                 \mathbf{p}_{Q-1}*_t\mathbf{h}
                               \end{array}
\right]=\left[\begin{array}{c}
                                 \mathbf{P}_0\mathbf{h}\\
                                 \mathbf{P}_1\mathbf{h}\\
                                \vdots\\
                                 \mathbf{P}_{Q-1}\mathbf{h}
                               \end{array}
\right]\\
&\hskip 5cm =\begin{bmatrix}
        \mathbf{P}_0\\
        \mathbf{P}_1 \\
        \vdots \\
        \mathbf{P}_{Q-1} 
       \end{bmatrix}\mathbf{h}=\mathbf{P}\mathbf{h}\label{eqn:matrix_P}
.
\end{align}
Here, $*_t$ denotes the discrete temporal convolution and $\mathbf{P_i}$ is the convolution matrix corresponding to $\mathbf{p_i}$. Matrix $\mathbf{P}$ is defined by Eqn.~\eqref{eqn:matrix_P}. The number of temporal samples $S$ (of each transducer), $I$ (of EIR), and $T$ (of the pressure) satisfy the relation $S=I+T-1$. With this reformulation, one has $f(\boldsymbol{\theta},\mathbf{h})=\|\mathbf{u}-\mathbf{P}\mathbf{h}\|^2$.

\bibliographystyle{IEEEtran}
\bibliography{VPM}

\end{document}